\documentstyle[11pt]{article}
\catcode`@11
\def\seceqaa{\@addtoreset{equation}{section}
\def\theequation{A\arabic{equation}}}
\def\seceqbb{\@addtoreset{equation}{section}
\def\theequation{B\arabic{equation}}}
\def\seceqcc{\@addtoreset{equation}{section}
\def\theequation{C\arabic{equation}}}
\def\seceqdd{\@addtoreset{equation}{section}
\def\theequation{D\arabic{equation}}}
\def\seceqee{\@addtoreset{equation}{section}
\def\theequation{E\arabic{equation}}}
\catcode`@=11
\begin{document}
\begin{titlepage}
\begin{center}
{\Large \bf Swiss Cheese D3-D7 Soft SUSY Breaking}
\vskip 0.1in { Aalok Misra\footnote{e-mail: aalokfph@iitr.ernet.in} and
Pramod Shukla\footnote{email: pmathdph@iitr.ernet.in}\\
Department of Physics, Indian Institute of Technology,
Roorkee - 247 667, Uttarakhand, India}
\end{center}
\thispagestyle{empty}
\begin{abstract}
We address issues related to (i) a proposal for resolving a long-standing tension between large volume cosmology and phenomenology as regards reconciliation of requirements of different gravitino masses within the same string-theoretic framework, as well as (ii) evaluation of soft supersymmetry breaking terms and open-string moduli masses in the context of type IIB large volume compactifications involving orientifolds of the Swiss-Cheese Calabi-Yau ${\bf WCP}^4[1,1,1,6,9]$ with a single mobile space-time filling $D3$-brane  and stacks of $D7$-branes wrapping the ``big" divisor $\Sigma_B$ as well as supporting $D7$-brane fluxes. In addition, we also include perturbative $\alpha^\prime$-corrections  and non-perturbative world-sheet instanton corrections to the K\"{a}hler potential as well as Euclidean D3-instanton superpotential. First, using the toric data for the aforementioned Swiss-Cheese Calabi-Yau and GLSM techniques, we obtain in the large volume limit, the geometric K\"{a}hler potential for the big (and small) divisor(s) in terms of derivatives of genus-two Siegel theta functions. Next, we show that as the mobile space-time filling $D3$-brane moves from a particular non-singular elliptic curve embedded in the Swiss-Cheese Calabi-Yau to another non-singular elliptic curve, it is possible to obtain $10^{12}GeV$ gravitino during the primordial inflationary era as well as, e.g., a $TeV$ gravitino in the present era, within the same set up for the same volume of the Calabi-Yau stabilized at around $10^6l_s^6$. Then by constructing local (i.e. localized around the location of the mobile $D3$-brane in the Calabi-Yau) appropriate involutively-odd harmonic one-form on the big divisor that lies in $coker\left(H^{(0,1)}_{{\bar\partial},-}(CY_3)\stackrel{i^*}{\rightarrow}H^{(0,1)}_{{\bar\partial},-}(\Sigma_B)\right)$ and  extremizing the potential, we show that it is possible to obtain an ${\cal O}(1)$ $g_{YM}$ from the wrapping of $D7$-branes on the big divisor due to competing contributions from the Wilson line moduli relative to the divisor volume modulus. To permit gaugino condensation, we take the rigid limit of the big divisor by considering zero sections of the normal bundle of the same - the same being justified by the extremization of the potential. For the purposes of calculation of the gaugino masses, matter moduli masses and soft supersymmetry breaking parameters, we restrict the mobile $D3$-brane to the big divisor - this has the additional advantage of nullification of the superpotential generated from gaugino condensation. With the inclusion of the matter moduli corresponding to the position moduli of the mobile $D3$-brane and the Wilson line moduli corresponding to the $D7$-branes, we obtain gaugino masses of the order of gravitino mass and the matter fields' masses to be enhanced relative to the gravitino mass. The anomaly-mediated gaugino masses are found to be suppressed relative to the gravity-mediated gaugino masses by the standard loop factor. New non-zero contributions to the $\mu$-terms, though sub-dominant in the large volume limit, are obtained from section of the (small) divisor bundle encoding information about the $ED3$-instanton fluctuation determinant.
filling $D3$-brane and the Wilson line moduli.  There is  a (near) universality in the masses, $\hat{\mu}$-parameters, Yukawa couplings  and the $\hat{\mu}B$-terms for the $D3$-brane position moduli - the Higgs doublet in our construction - and a hierarchy in the same set and a universality in the $A$ terms on inclusion of the $D7$-brane Wilson line moduli.
\end{abstract}
\end{titlepage}

\section{Introduction}

In the context of string compactifications, obtaining $dS$ vacua and realizing the Standard Model have been the two major issues for a long time. In the context of realizing $dS$ vacua,  the complex structure moduli and the axion-dilaton modulus were stabilized with the inclusion of fluxes \cite{fluxesGiddindsetal,Granafluxreview} and the K\"{a}hler moduli could be stabilized only with inclusion of non-perturbative effects. A supersymmetric $AdS$ minimum was obtained in Type IIB orientifold compactification which was uplifted to a non-supersymmetric metastable $dS$ by adding $\overline D3$-brane, in \cite{KKLT}. Subsequently, several other uplifting mechanisms were proposed \cite{otherupliftings}. In a different approach with more than one K\"{a}hler modulus  in the context of the Type IIB orientifold compactification in the large volume scenarios, a non-supersymmetric $AdS$ was realized with the inclusion of perturbative ${\alpha^{\prime}}^3$ correction to the K\"{a}hler potential which was then uplifted to $dS$ vacuum \cite{Balaetal2}. Followed by this, again in the context of Type IIB orientifold compactification in large volume scenarios, it was shown in \cite{dSetal} that with the inclusion of (non-)perturbative $\alpha^{\prime}$ corrections to the K\"{a}hler potential and instanton corrections to the superpotential, one can realize {\it non}-supersymmetric metastable $dS$ solution in a more natural way without having to add an uplifting term (via inclusion of $\overline D3$-brane).

On the way of embedding (MS)SM  and realizing its matter content from string phenomenology, the questions of supersymmetry breaking and its transmission to the visible sector are among the most challenging issues - the first being mainly controlled by the moduli potentials while the second one by the coupling of supersymmetry-breaking fields to the visible sector matter fields. The breaking of supersymmetry which  is encoded in soft terms, is supposed to occur in a hidden sector and then  communicated to the visible sector (MS)SM via different mediation processes (e.g. gravity mediation, anomaly mediation, gauge mediation) among which although none is clearly preferred, gravity mediation is the most studied one due to its efficient computability. However there was a problem of non-universality in gravity mediation of supersymmetry breaking to the visible sector that has been addressed (see \cite{mirrormediation,conloncal}) with the arguments that the K\"{a}hler moduli sector (which controls the supersymmetry-breaking) and the complex structure moduli sector (which sources the flavor) are decoupled at least at the tree level resulting the flavour universal soft-terms, though it has been argued that the non-universality can appear at higher order. Further it has been discussed that the small experimental FCNC constraints at low energy can be translated to the non-universal soft scalar masses at energy $\sim$ GUT scale \cite{FCNC}.

The study of supersymmetry-breaking in string theory context has been initiated long back \cite{susyinitials} and a lot of work has been done in this direction (see \cite{conloncal,Quevedosusy2,beckerhack,towardsrealvacua,susybreakingKKLT,susykklt2} and references therein). A more controlled study of supersymmetry-breaking has been possible only after all moduli could be stabilized with inclusion of fluxes along with non-perturbative effects.  Since it is possible to embed the chiral gauge sectors (like that of the (MS)SM) in D-brane Models with fluxes, the study of $D$-brane Models have been fascinating  since the discovery of D-branes \cite{SMreview,SM2,QuevedoMSSM,SM3}. In a generic sense, the presence of fluxes generate the soft supersymmetry-breaking terms, the soft terms in various models in the context of gauge sectors realized on fluxed D-branes have been calculated \cite{susybreakingKKLT,susykklt2,granagrimm,ibanezuranga,Ibanez,Lustetal}. In the context of $dS$ realized in the KKLT setup, the uplifting term from the $\overline D3$- brane causes the soft supersymmetry-breaking; (also see \cite{susybreakingKKLT,susykklt2} for KKLT type models).

Similar to the context of $dS$ realization and its cosmological implications, the models in L(arge) V(olume) S(cenarios) have been realized to be exciting steps towards realistic supersymmetry-breaking \cite{Balaetal2,conloncal,towardsrealvacua,balaetal2,conlonLVSsusy,quevedojan09} with some natural advantages such as the large volume not only suppresses the string scale but also the gravitino mass and results in the hierarchically small scale of supersymmetry-breaking. Also unlike the KKLT models in which the anomaly mediated soft terms are equally important to that of the gravity mediated one \cite{susybreakingKKLT}, in some of the Large Volume models, it has been found that the gaugino mass contribution coming from gravity mediation dominates to the anomaly mediation one (the same being suppressed by the standard loop factor) \cite{conloncal,towardsrealvacua} and the same can be expected for the other soft masses as well. Further the study of LVS models in the context of $N=1$ type IIB orientifold compactification in the presence of D7-branes, has been quite attractive and promising for the phenomenological purposes as in such models, D7-brane wrapping the smaller cycle produces the qualitatively similar gauge coupling as  that of the Standard Model and also with the magnetized D7-branes, the Standard Model chiral matter can be realized from strings stretching between stacks of D7-branes \cite{conloncal,Lustetal,conlonLVSsusy,ibanezfont,jockersetal}. In one of such models, RG evolutions  of soft-terms to the weak scale have been studied  to have a low energy spectra by using the RG equations of MSSM (assuming that only charged matter content below the string scale is the MSSM) and it was found that with D7 chiral matter fields, low energy supersymmetry-breaking could be realized at a small hierarchy between the gravitino mass and soft supersymmetry-breaking terms \cite{conloncal}. A much detailed study with fluxed $D3/D7$ branes has been done in the context of $N=1$ type IIB orientifold compactification \cite{Lustetal,conlonLVSsusy,jockersetal} and it has been found that the $N=1$ coordinates get modified with the inclusion of $D3$ and $D7$-branes. The gauge coupling of $D7$-brane wrapping a 4-cycle  depends mainly on the size modulus of the wrapped 4-cycle and also on the complex structure as well as axion-dilaton modulus  after including the loop-corrections, which in the diluted flux limit (without loop-corrections) was found to be dominated by the size modulus of the wrapping 4-cycle \cite{conlonLVSsusy,berghack}. In the models having branes at  singularities, it has been argued that at the leading order, the soft terms vanish for the no-scale structure  which gets broken at higher orders with the inclusion of (non-)perturbative $\alpha^{\prime}$ and loop-corrections to the K\"{a}hler potential resulting in the non-zero soft-terms at higher orders. In the context of LVS phenomenology in such models with D-branes at singularities, it has been argued that all the leading order contributions to the soft supersymmetry-breaking (with gravity as well as anomaly mediation processes) still vanish and the non-zero soft terms have been calculated in the context of gravity mediation with inclusion of loop-corrections \cite{towardsrealvacua}. In the context of type IIB LVS Swiss-Cheese orientifold compactifications within $D3/D7$-branes setup, soft terms have been calculated in \cite{conlonLVSsusy}. Recently  supersymmetry breaking with both D-term and F-term and some cosmological issues have been discussed in \cite{quevedojan09}.

Further, there has been a tension between  LVS cosmology and LVS phenomenology studied so far. The scale required by cosmological/astrophysical experiments is nearly the same order as the GUT scale ($\sim 10^{16}$ GeV) while in LVS phenomenology, the supersymmetry-breaking at $TeV$ scale requires the string scale to be some intermediate scale of the order of $10^{11}$ GeV. In this way there is a hierarchy in scales involved on both sides making it impossible to fulfill both requirements in the same string theory setup. Although LVS limits of Type IIB Swiss-Cheese orientifold compactifications have been exciting steps in the search for  realistic models on both cosmology as well as phenomenology sides, this hierarchy is reflected in LVS setups, as a hierarchy of compactification volume requirement from $ \cal{V}$ $\sim 10^6$ (for cosmology requirement, e.g. see \cite{kaehlerinflation}) to $\cal{V}$ $\sim 10^{14}$ (for phenomenology requirement \footnote{In a recent paper \cite{quevedoftheorysusy}, the authors have realized soft terms $\sim$ TeV with $ {\cal V}\sim O(10^6-10^7)$ in the context of String/F-theory models with SM supported on a del Pezzo surface, but with very heavy gravitino.}, e.g. see \cite{conloncal}) and  the tension has remained unresolved in a single string theoretic setup with the Calabi-Yau volume stabilized at a particular value\footnote {There has been a proposal \cite{tension1}, which involves a small CY volume for incorporating high-scale inflation and then evolves the volume modulus over a long range and finally stabilizes it in the large volume minimum with TeV gravitino mass after inflation.}. Now in the present LHC era equipped with PAMELA and PLANCK, string theoretic models with numbers, which could match with experimental-data are yet to come; and several phenomenologically motivated steps have also been initiated in this direction \cite{quevedojan09,abdussalam2,LHCpheno,KKLMMT,LargeVcons,largefNL_r}.

The paper is organized as follows. In section {\bf 2}, we start off with a summary of our previous work on obtaining a metastable dS vacuum in type IIB compactifications on orientifolds of a particular type of Swiss-Cheese Calabi-Yau with the inclusion of perturbative $\alpha^\prime$ corrections and their modular completion along with $D1$- and $D3$-instanton contributions to the non-perturbative superpotential, but without having to add anti-$D3$-branes. We then summarize our work pertaining to the applications of this setup to study cosmological aspects namely axionic slow-roll inflation and ${\cal O}(1)$ $f_{NL}$ pertaining to non-Gaussianities in curvature perturbations. We then discuss the appropriate ${\cal N}=1$ coordinates required on inclusion of a mobile $D3$-brane and a $D7$-brane, the latter wrapping a divisor inside the aforementioned Calabi-Yau. Finally, we discuss the construction of local involutively-odd harmonic one-forms on the aforementioned divisor to enable getting an ${\cal O}(1)$
$g_{YM}$ on the world-volume of a stack of $D7$-branes wrapping the divisor. Section {\bf 3} has a detailed discussion on obtaining the geometric K\"{a}hler potential for the Calabi-Yau and in particular, the abovementioned (``big") divisor using toric geometry, GLSM techniques and results by Umemura and Zhivkov. We also write out the complete moduli-space K\"{a}hler potential in terms of the closed-string moduli as well as the open-string moduli or matter fields, the latter being the position moduli of the mobile $D3$-brane and the Wilson-line moduli on the $D7$-brane(s). Section 4 is about resolution of a long-standing problem in large volume string phenomenology and cosmology - giving a mechanism that would generate a $10^{12}GeV$ gravitino in the early inflationary epoch of the universe and then a $TeV$ gravitino at the present times to possibly be detected at the LHC, for the same value of the volume modulus of the Calabi-Yau at around $10^6l_s^6$. Section 5 has the details of calculations of the gaugino and matter fields' masses, soft SUSY breaking parameters: the physical mu terms ($\hat{\mu}$), the $\hat{\mu}B$-terms, the Yukawa couplings and the $A$-terms. Section 6 has a summary of the results and a discussion. There are five appendices - appendix A provides a justification for stabilizing the Wilson-line moduli at values which combined with the involutively-odd harmonic one-forms on the big divisor enable an ${\cal O}(1)$ $g_{YM}$; appendix B has expressions for the first and second geometric K\"{a}hler potentials on the ``big" and ``small" divisors; appendix C lists intermediate steps relevant to obtaining the quadratic terms in the expansion of the complete K\"{a}hler potential as a power series in (fluctuations of) the $D3$-brane position moduli  and the $D7$-brane Wilson-line moduli (about their extremum values); appendix D has the the first and second derivatives of the quadratic components as well as the determinant generated by the latter, in the abovementioned expansion of the complete K\"{a}hler potential in a power series with one holomorphic and one anti-holomorphic matter fields relevant to the evaluation of the soft SUSY breaking parameters; appendix E has the first and second derivatives of the quadratic components in the aforementioned expansion of the complete K\"{a}hler potential involving only holomorphic (or anti-holomorphic) matter fields - these again are relevant to the evaluation of the soft SUSY breaking parameters.

\section{The Setup}

Let us first summarize the results of our previous works on type IIB compactifications on a Swiss-Cheese Calabi-Yau orientifold (section {\bf 4} of \cite{dSetal}) and its cosmological implications (\cite{largefNL_r,axionicswisscheese}). The term ``Swiss cheese" (See \cite{conlonLVSsusy}) is used to denote those Calabi-Yau's whose volume can be written as: ${\cal V}=(\tau^B + \sum_{i\neq B} a_i\tau^S_i)^{\frac{3}{2}} - (\sum_{j\neq B}b_j\tau^S_j)^{\frac{3}{2}} - ...$, where $\tau^B$ is the volume of the big divisor and $\tau^S_i$ are the volumes of the $h^{1,1}-1$ (corresponding to the (1,$h^{1,1}-1$)-signature of the Hessian) small divisors. The big divisor governs the size of the Swiss cheese and the small divisors control the size of the holes of the same Swiss cheese. Calabi-Yau three-fold obtained as a resolution of the degree-18 hypersurface in ${\bf WCP}^4[1,1,1,6,9]$:
\begin{equation}
\label{eq:hypersurface}
x_1^{18} + x_2^{18} + x_3^{18} + x_4^3 + x_5^2 - 18\psi \prod_{i=1}^5x_i - 3\phi x_1^6x_2^6x_3^6 = 0.
\end{equation}
Similar to the explanation given in \cite{Kachruetal}, it is understood that only two complex structure
moduli $\psi$ and $\phi$ are retained in (\ref{eq:hypersurface}) which are invariant under the group $G={\bf Z}_6\times{\bf Z}_{18}$ (${\bf Z}_6:(0,1,3,2,0,0); {\bf Z}_{18}:(1,-1,0,0,0)$ (See \cite{Candelasetal}), setting the other invariant complex structure moduli appearing at a higher order (due to invariance under $G$)
 at their values at the origin.

 With the inclusion of perturbative (using \cite{BBHL}) and non-perturbative (using \cite{Grimm}) $\alpha^\prime$-corrections as well as the loop corrections (using \cite{berghack,loops}), the K\"{a}hler potential for the two-parameter ``Swiss-Cheese" Calabi-Yau expressed as a projective variety in ${\bf WCP}^4[1,1,1,6,9]$, can be shown to be given by:
\begin{eqnarray}
\label{eq:nonpert81}
& & K = - ln\left(-i(\tau-{\bar\tau})\right) -ln\left(-i\int_{CY_3}\Omega\wedge{\bar\Omega}\right)\nonumber\\
 & & - 2\ ln\Biggl[{\cal V} + \frac{\chi(CY_3)}{2}\sum_{m,n\in{\bf Z}^2/(0,0)}
\frac{({\bar\tau}-\tau)^{\frac{3}{2}}}{(2i)^{\frac{3}{2}}|m+n\tau|^3}\nonumber\\
& & - 4\sum_{\beta\in H_2^-(CY_3,{\bf Z})} n^0_\beta\sum_{m,n\in{\bf Z}^2/(0,0)}
\frac{({\bar\tau}-\tau)^{\frac{3}{2}}}{(2i)^{\frac{3}{2}}|m+n\tau|^3}cos\left((n+m\tau)k_a\frac{(G^a-{\bar G}^a)}{\tau - {\bar\tau}}
 - mk_aG^a\right)\Biggr]\nonumber\\
 & & +\frac{C^{KK\ (1)}_s(U_\alpha,{\bar U}_{\bar\alpha})\sqrt{\tau_s}}{{\cal V}\left(\sum_{(m,n)\in{\bf Z}^2/(0,0)}\frac{\frac{(\tau-{\bar\tau})}{2i}}{|m+n\tau|^2}\right)} + \frac{C^{KK\ (1)}_b(U_\alpha,{\bar U}_{\bar\alpha})\sqrt{\tau_b}}{{\cal V}\left(\sum_{(m,n)\in{\bf Z}^2/(0,0)}\frac{\frac{(\tau-{\bar\tau})}{2i}}{|m+n\tau|^2}\right)}.
\end{eqnarray}
In (\ref{eq:nonpert81}), the first line and $-2\ ln({\cal V})$ are the tree-level contributions. The second (excluding the volume factor in the argument of the logarithm) and third lines are the perturbative and non-perturbative $\alpha^\prime$ corrections. $\{n^0_\beta\}$ are the genus-zero Gopakumar-Vafa invariants that count the number of genus-zero rational curves. The fourth line is the 1-loop contribution; $\tau_s$ is the volume of the ``small" divisor and $\tau_b$ is the volume of the ``big" divisor. The loop-contributions arise from KK modes corresponding to closed string or 1-loop open-string exchange between $D3$- and $D7$-(or $O7$-planes)branes wrapped around the ``s" and ``b" divisors. Note that the two divisors for
${\bf WCP}^4[1,1,1,6,9]$, do not intersect (See \cite{Curio+Spillner}) implying that there is no contribution from winding modes corresponding to strings winding non-contractible 1-cycles in the intersection locus corresponding to stacks of intersecting $D7$-branes wrapped around the ``s" and ``b" divisors. One sees from (\ref{eq:nonpert81}) that in the LVS limit, loop corrections are sub-dominant as compared to the perturbative and non-perturbative $\alpha^\prime$ corrections.

To summarize the result of section 4 of \cite{dSetal}, one gets the following potential:
\begin{eqnarray}
\label{eq:nonpert21}
& & V\sim\frac{{\cal V}\sqrt{ln {\cal V}}}{{\cal V}^{2}}e^{-2\phi}\frac{\left(\sum_{n^s}n^s\sum_{m^a}\frac{e^{-\frac{m^2}{2g_s} + \frac{m_ab^a n^s}{g_s} + \frac{n^s\kappa_{1ab}b^ab^b}{2g_s}}}{{\cal V}^{n_s}}\right)^2}{\left|f(\tau)\right|^2}
\nonumber\\
& & + \sum_{n^s}\frac{W ln { \cal V} }{{\cal V}^{n^s+2}}\left(\frac{\theta_{n^s}({\bar\tau},{\bar G})}{f({\bar\tau})}
\right)e^{-in^s(-\tilde{\rho_1}+\frac{1}{2}\kappa_{1ab}
\frac{{\bar\tau}G^a-\tau{\bar G}^a}{({\bar\tau}-\tau)}\frac{(G^b-{\bar G}^b)}{({\bar\tau}-\tau)} -
\frac{1}{2}\kappa_{1ab}\frac{G^a(G^b-{\bar G}^b)}{(\tau-{\bar\tau})})}+c.c.\nonumber\\
& & +
\sum_{k^1,k^2}\frac{|W|^2}{{\cal V}^3}\left(\frac{3k_2^2+k_1^2}{k_1^2-k_2^2}\right)
\frac{\left|\sum_c\sum_{n,m\in{\bf Z}^2/(0,0)}e^{-\frac{3\phi}{2}}A_{n,m,n_{k^c}}(\tau) sin(nk.b+mk.c)\right|^2}
{\sum_{c^\prime}\sum_{m^\prime,n^\prime\in{\bf Z}^2/(0,0)} e^{-\frac{3\phi}{2}}|n+m\tau|^3
|A_{n^\prime,m^\prime,n_{k^{c^{\prime}}}}(\tau)|^2 cos(n^\prime k.b+m^\prime k.c)}+\frac{\xi|W|^2}{{\cal V}^3},
\nonumber\\
& &
\end{eqnarray}
where ${\cal V}$ is the overall volume of the Swiss-Cheese Calabi-Yau, $f(\tau)$ is an approoriate apriori unknown modular function, $n^s$ is the $D3$-brane instanton quantum number and $m^a$'s are the $D1$ instanton numbers. The expressions for ${\cal Y}$, the holomorphic Jacobi theta function $\theta_{n^\alpha}(\tau,G)$ and $A_{n,m,n_{k^c}}(\tau)$ are defined as:

\begin{eqnarray}
\label{eq:nonpert14}
&& {\cal Y}\equiv {\cal V}_E + \frac{\chi}{2}\sum_{m,n\in{\bf Z}^2/(0,0)}
\frac{(\tau - {\bar\tau})^{\frac{3}{2}}}{(2i)^{\frac{3}{2}}|m+n\tau|^3} \nonumber\\
& &
- 4\sum_{\beta\in H_2^-(CY_3,{\bf Z})}n^0_\beta\sum_{m,n\in{\bf Z}^2/(0,0)}
\frac{(\tau - {\bar\tau})^{\frac{3}{2}}}{(2i)^{\frac{3}{2}}|m+n\tau|^3}cos\left((n+m\tau)k_a\frac{(G^a-{\bar G}^a)}{\tau - {\bar\tau}}
 - mk_aG^a\right),\nonumber\\
& & \theta_{n^\alpha}(\tau,G)=\sum_{m_a}e^{\frac{i\tau m^2}{2}}e^{in^\alpha G^am_a},\ A_{n,m,n_{k^c}}(\tau)\equiv \frac{(n+m\tau)n_{k^c}}{|n+m\tau|^3}.
\end{eqnarray}
Also, $G^a$ are defined by $G^a\equiv c^a-\tau b^a$ (where $c^a$'s and $b^a$'s are defined through the real RR two-form potential $C_2=c_a\omega^a$ and the real NS-NS two-form potential $B_2=b_a\omega^a$).
Note that summing over all values of $m^a$ and for any given value or set of values of $n^s$, the potential of (\ref{eq:nonpert21}) is symmetric under an NS-NS axionic shift symmetry as well as a discrete subgroup of $SL(2,{\bf Z})$ that survives the process of orientifolding - see \cite{Grimm}.
We are hence justified in choosing a particular value or set of values  for $n^s$.  This will also be used in the remainder of the paper.

On comparing (\ref{eq:nonpert21}) with the analysis of \cite{Balaetal2}, one sees that for generic values of
the moduli $\rho_\alpha, G^a, k^{1,2}$ and ${\cal O}(1)\ W_{c.s.}$, and $n^s$(the $D3$-brane instanton quantum number)=1, analogous to \cite{Balaetal2}, the second term
dominates; the third term is a new term. However, as in KKLT scenarios (See \cite{KKLT}), $W_{c.s.}<<1$; we would henceforth assume that the fluxes and complex structure moduli have been so fine tuned/fixed that $W\sim W_{n.p.}$. We assume that the fundamental-domain-valued $b^a$'s satisfy: $\frac{|b^a|}{\pi}<1$\footnote{If one puts in appropriate powers of the Planck mass $M_p$, $\frac{|b^a|}{\pi}<1$ is equivalent to $|b^a|<\pi M_p$, i.e., NS-NS axions are sub-Planckian in units of $\pi M_p$.}. This implies that for $n^s>1$, the first term in (\ref{eq:nonpert21}) - $|\partial_{\rho^s}W_{np}|^2$ - a positive definite term, is the most dominant. In the same, $\rho^s$ is the volume of the small divisor complexified by RR 4-form axions. Hence, if a minimum exists, it will be positive. As shown in \cite{dSetal}, the potential can be extremized along the locus:
\begin{equation}
\label{eq:ext_locus}
mk.c + nk.b = N_{(m,n;,k^a)}\pi
\end{equation}
with $n^s>1$ and for all values of the $D1$-instanton quantum numbers $m^a$.\footnote{Considering the effect of axionic shift symmetry (of the $b^a$ axions) on the $D1$-instanton superpotential ($W_{D1-instanton}$), one can see that $m^a$ is valued in a lattice with coefficients being integral multiple of $2\pi$.} As shown in section {\bf 3} of \cite{axionicswisscheese}, it turns out that the locus $nk.b + mk.c = N\pi$ for $|b^a|<\pi$ and $|c^a|<\pi$ corresponds to a flat saddle point with the NS-NS axions providing a flat direction.
For all directions in the moduli space with  $W_{c.s.}\sim {\cal O}(1)$ and away from $D_iW_{cs}=D_\tau W=0=\partial_{c^a}V=\partial_{b^a}V=0$, the ${\cal O}(\frac{1}{{\cal V}^2})$ contribution
of $\sum_{\alpha,{\bar\beta}\in{c.s.}}(G^{-1})^{\alpha{\bar\beta}}D_\alpha W_{cs}{\bar D}_{\bar\beta}{\bar W}_{cs}$  dominates over (\ref{eq:nonpert21}),
ensuring that that there must exist a minimum, and given the positive definiteness of the potential, this will be a dS minimum. There has been no need to add any $\overline{D3}$-branes as in KKLT to generate a dS vacuum.

In \cite{axionicswisscheese}, we discussed the possibility of getting slow roll inflation along a flat direction provided by the NS-NS axions starting from a saddle point and proceeding to the nearest dS minimum. In what follows, we will assume that the volume moduli for the small and big divisors and the axion-dilaton modulus have been stabilized. All calculations henceforth will be in the axionic sector - $\partial_a$ will imply $\partial_{G^a}$ in the remainder of this paragraph.
On evaluation of the slow-roll inflation parameters (in $M_p=1$ units)
$\epsilon\equiv\frac{{\cal G}^{ab}\partial_aV\partial_bV}{2V^2},\ \eta\equiv$ the most negative eigenvalue of the matrix $N^a_{\ b}\equiv\frac{{\cal G}^{ac}\left(\partial_c\partial_bV - \Gamma^d_{bc}\partial_dV\right)}{V}$ with $\Gamma^a_{bc}$ being the affine connection components, we found that $\epsilon\sim\frac{(n^s)^2e^{-\frac{2\alpha}{g_s}}{\cal V}}{\sum_{\beta\in H_2} n^0_\beta}$ and $\eta\sim\frac{{\cal V}n^sg_s\kappa}{\sum_{\beta\in H_2} n^0_\beta}$, \footnote{The $g_s$-dependence of $\epsilon$ and $\eta$ was missed in \cite{axionicswisscheese}. The point is that the extremization of the potential w.r.t.$b^a$'s and $c^a$'s in the large volume limit yields a saddle point at $sin(nk.b+mk.c)=0$ and for those degree-$k^a$ holomorphic curves $\beta$ for which $b^a\sim -m^a/{\kappa}$ (assuming that $\frac{nk.m}{\pi\kappa}\in{\bf Z}$). The latter corresponds to the small values of $m^a$ (as $b^a$'s are sub-Planckian). Large values of $m^a$'s (which are also permitted by induced shift symmetry of $m^a$'s due to that of axions in $W_{D1-instanton}$) although don't satisfy $b^a\sim -m^a/{\kappa}$, are damped because of $\exp({-m^2/2g_s})$, especially in the $g_s<<1$ limit, the weak coupling limit in which the LVS scenarios are applicable.}, where $\alpha\sim\xi$ - See \cite{LargeVcons}. We will choose ${\cal V}$ to be such that ${\cal V}\sim e^{\frac{2\alpha}{g_s}}$ (See \cite{LargeVcons}), implying that $\epsilon\sim\frac{(n^s)^2}{\sum_{\beta\in H_2}n^0_\beta}$ and $\eta\sim\frac{{\cal V}n^s\kappa}{ln {\cal V}\sum_{\beta\in H_2}n^0_\beta}$. Using Castelnuovo's theory of study of moduli spaces that are fibrations of Jacobian of curves over the moduli space of their deformations, for compact Calabi-Yau's expressed as projective varieties in weighted complex projective spaces (See \cite{Klemm_GV}) one sees that for appropriate degrees of the holomorphic curve, the genus-0 Gopakumar-Vafa invariants can be very large. Hence the slow-roll conditions can be satisfed, and in particular, there is no ``$\eta$"-problem. By investigating the eigenvalues of the Hessian, we showed (in \cite{axionicswisscheese}) that one could identify a linear combination of the NS-NS axions (``$k_2b^2+k_1b^1$" with $k_{1,2}$ corresponding to the degrees of rational curves with the largest value of $n^0_\beta$ for a given involution) with the inflaton and the slow-roll inflation starts from the aforementioned saddle-point and ends when the slow-roll conditions were violated, which most probably corresponded to the nearest dS minimum\footnote{The $g_s$-dependence does not influence the calculations largely because $e^{-\frac{\alpha}{g_s}}$ enters only odd number of derivatives of $V$; the Hessian involves the second derivative of $V$.}. To evaluate the number of e-foldings $N_e$, defining the inflaton ${\cal I}\sim k_2b^2+k_1b^1$, one can show that (in $M_p=1$ units)
\begin{equation}
\label{eq:Ne def}
N_e=-\int_{{\rm in:\ Saddle\ Point}}^{{\rm fin:\ dS\ Minimum}}\frac{1}{\sqrt{\epsilon}}d{\cal I}\sim
  \frac{\sqrt{\sum_{\beta\in H_2}n^0_\beta}}{n^s}.
\end{equation}
One gets the required 60 e-foldings for a choice of the involution such that degree-$\{k^a\}$ genus-zero curves correspond to maximum values of the genus-zero Gopakumar-Vafa invariants $n^0_\beta\sim(60n^s)^2$. Hence, $\eta\sim\frac{\kappa{\cal V}}{3600 n^s ln {\cal V}}$.

 Using the ``$\delta N$" formalism as developed in \cite{Yokoyamafnl,Yokoyama}, it was shown in \cite{largefNL_r} that to get the non-linear parameter ``$f_{NL}$" relevant to studies of non-Gaussianities,
to be ${\cal O}(1)$, one would need a very large value for $n^s$, about the order of $10^4$. Now, as we had assumed that the overall volume ${\cal V}$ of the Swiss Cheese Calabi-Yau ${\bf WCP}^4[1,1,1,6,9]$ had to be chosen in such a way that the Gopakumar-Vafa invariants $n^0_\beta\sim (60 n^s)^2$ - this ensured that we got 60 e-foldings \cite{axionicswisscheese}, the contribution to the non-linear parameter $f_{NL}$  would therefore be $\sim\frac{g_se^{\frac{\alpha}{g_s}}}{3600}\sim\frac{\sqrt{\cal V}}{N^2 ln {\cal V}}$.  One gets the non linear parameter $f_{NL}\sim {\cal O}(1)$ for ${\cal V}\sim10^7$ (which is also consistent with COBE measurements - see \cite{kaehlerinflation}). To get $\eta<1$, one would need $n^s\sim 10^4$, as was mentioned earlier. In this way we had shown the possibility of getting finite values of the non-linear parameter $f_{NL}$ (hence for getting large detectable non-Gaussianities) in the LVS limit of type IIB Swiss Cheese orientifold compactifications. For slow-roll inflation, $H^2\sim\frac{V}{3}$; the Friedmann equation implies that
$H^2>\frac{V}{3}$ when slow-roll conditions are violated. The number of e-foldings away from slow-roll
is given by: $N\sim\int\frac{db^a}{||\frac{db^b}{dN}||}$, which using the Friedmann equation implies
$N\sim\int\frac{db^a}{\sqrt{1-\frac{V}{3H^2}}}$.
Unlike the slow-roll inflation
scenario discussed earlier, we will assume that the involution is such that the degrees of the holomorphic curve being summed over in the non-perturbative $\alpha^\prime$-corrections involving the genus-zero Gopakumar-Vafa invariants correspond to the largest  $n^0_\beta\stackrel{<}{\sim}{\cal V}$ [instead of $n^0_\beta\sim(60n^s)^2$ as was assumed for the slow-roll inflationary scenarios]. Further, {\it we will assume $n^s$ to be ${\cal O}(1)$ but more than 1;} there will be no $e^{-\frac{\alpha}{g_s}}$ away from slow-roll trajectories. This implies that ${\cal G}_{ab}$ which is
$\sim\frac{\sum_{\beta\in H_2}n^0_\beta}{{\cal V}}$ (and ${\cal G}^{ab}$ which is
$\frac{{\cal V}}{\sum_{\beta\in H_2}n^0_\beta}$)$\sim{\cal O}(1)$,
$\frac{db^a}{dN}\sim\sqrt{1-\frac{V}{3H^2}}\sim{\cal O}(1)$. Note that when the slow-roll conditions are violated, $sin(nk.b + mk.c)$ is away from zero, and consequently $\Gamma^a_{bc}\sim{\cal O}(1)$, and so are the curvature components $R^a_{\ bcd}$. This was then used to obtain $f_{NL}\sim {\cal O}(1)$ in \cite{largefNL_r}.

In this paper, we address phenomenological aspects of Swiss-Cheese Calabi-Yau orientifolds in type IIB compactifications, pertaining to soft supersymmetry breaking. The appropriate ${\cal N}=1$ coordinates in the presence of a single $D3$-brane and a single $D7$-brane
wrapping the ``big" divisor $\Sigma^B$  along with $D7$-brane fluxes are
given as under (See \cite{jockersetal,Jockers_thesis}):
\begin{eqnarray}
\label{eq:N=1_coords}
& & S = \tau + \kappa_4^2\mu_7{\cal L}_{A{\bar B}}\zeta^A{\bar\zeta}^{\bar B}\nonumber\\
& & {\cal G}^a = c^a - \tau {\cal B}^a\nonumber\\
& & T_\alpha=\frac{3i}{2}(\rho_\alpha - \frac{1}{2}\kappa_{\alpha bc}c^b{\cal B}^c) + \frac{3}{4}\kappa_\alpha + \frac{3i}{4(\tau - {\bar\tau})}\kappa_{\alpha bc}{\cal G}^b({\cal G}^c
- {\bar G}^c) \nonumber\\
& & + 3i\kappa_4^2\mu_7l^2C_\alpha^{I{\bar J}}a_I{\bar a_{\bar J}} + \frac{3i}{4}\delta^B_\alpha\tau Q_{\tilde{f}} + \frac{3i}{2}\mu_3l^2(\omega_\alpha)_{i{\bar j}} \Phi^i\left({\bar\Phi}^{\bar j}-\frac{i}{2}{\bar z}^{\tilde{a}}({\bar{\cal P}}_{\tilde{a}})^{\bar j}_l\Phi^l\right)\nonumber\\
& & \tau=l+ie^{-\phi},
\end{eqnarray}
where
\begin{itemize}

\item
for future reference in the remainder of the paper, one defines: ${\cal T}_\alpha\equiv\frac{3i}{2}(\rho_\alpha - \frac{1}{2}\kappa_{\alpha bc}c^b{\cal B}^c) + \frac{3}{4}\kappa_\alpha + \frac{3i}{4(\tau - {\bar\tau})}\kappa_{\alpha bc}{\cal G}^b({\cal G}^c
- {\bar G}^c)$,

\item
\begin{equation}
  {\cal L}_{A{\bar B}}=\frac{\int_{\Sigma^B}\tilde{s}_A\wedge\tilde{s}_{\bar B}}{\int_{CY_3}\Omega\wedge{\bar\Omega}},
  \end{equation}
$\tilde{s}_A$ forming a basis for $H^{(2,0)}_{{\bar\partial},-}(\Sigma^B)$,

\item
the fluctuations of $D7$-brane in the $CY_3$ normal to $\Sigma^B$ are denoted by $\zeta\in H^0(\Sigma^B,N\Sigma^B)$, i.e., they are the space of global sections of the normal bundle $N\Sigma^B$,

\item
${\cal B}\equiv b^a - lf^a$, where $f^a$ are the components of elements of two-form fluxes valued in $i^*\left(H^2_-(CY_3)\right)$, the immersion map being defined as:
$i:\Sigma^B\hookrightarrow CY_3$,

\item
$C^{I{\bar J}}_\alpha=\int_{\Sigma^B}i^*\omega_\alpha\wedge A^I\wedge A^{\bar J}$, $\omega_\alpha\in H^{(1,1)}_{{\bar\partial},+}(CY_3)$ and $A^I$ forming a basis for $H^{(0,1)}_{{\bar\partial},-}(\Sigma^B)$,

\item
$a_I$ is defined via a Kaluza-Klein reduction of the $U(1)$ gauge field (one-form) $A(x,y)=A_\mu(x)dx^\mu P_-(y)+a_I(x)A^I(y)+{\bar a}_{\bar J}(x){\bar A}^{\bar J}(y)$, where $P_-(y)=1$ if $y\in\Sigma^B$ and -1 if $y\in\sigma(\Sigma^B)$,

\item
$z^{\tilde{a}}$ are $D=4$ complex scalar fields arising due to complex structure deformations of the Calabi-Yau orientifold defined via: $\delta g_{{\bar i}{\bar j}}(z^{\tilde{a}})=-\frac{i}{||\Omega||^2}z^{\tilde{a}}\left(\chi_{\tilde{a}}\right)_{{\bar i}jk}\left({\bar\Omega}\right)^{jkl}g_{l{\bar j}}$, where $\left(\chi_{\tilde{a}}\right)_{{\bar i}jk}$ are components of elements of $H^{(2,1)}_{{\bar\partial},+}(CY_3)$,

\item
$\left({\cal P}_{\tilde{a}}\right)^i_{\bar j}\equiv\frac{1}{||\Omega||^2}{\bar\Omega}^{ikl}\left(\chi_{\tilde{a}}\right)_{kl{\bar j}}$, i.e.,
${\cal P}:TCY_3^{(1,0)}\longrightarrow TCY_3^{(0,1)}$ via the transformation:
$\Phi\stackrel{\rm c.s.\ deform}{\longrightarrow}\Phi^i+\frac{i}{2}z^{\tilde{a}}\left({\cal P}_{\tilde{a}}\right)^i_{\bar j}{\bar\Phi}^{\bar j}$,

\item
$\Phi^i$ are scalar fields corresponding to geometric fluctuations of $D3$-brane inside the Calabi-Yau and defined via: $\Phi(x)=\Phi^i(x)\partial_i + {\bar\Phi}^{\bar i}({\bar x}){\bar\partial}_{\bar i}$,

and

\item
$Q_{\tilde{f}}\equiv l^2\int_{\Sigma^B}\tilde{f}\wedge\tilde{f}$, where $\tilde{f}\in\tilde{H}^2_-(\Sigma^B)\equiv{\rm coker}\left(H^2_-(CY_3)\stackrel{i^*}{\rightarrow}H^2_-(\Sigma^B)\right)$.

\end{itemize}

The constant $\kappa_{10}$ enters the ten-dimensional Newton's constant via $G^{(10}=\kappa_{10}^2=g_s^2\kappa^2$; $\kappa\sim(\alpha^\prime)^2$.
As the four-dimensional Newton's constant $G^{(4)}=\frac{G^{(10)}}{{\cal V}}\sim\frac{g_s^2\alpha^\prime}{\frac{\cal V}{(\alpha^\prime)^3}}\sim g_s^2\kappa_4^2$, where ${\cal V}$ is the volume of the Calabi-Yau in the string frame, thus, $\kappa_4^2\sim\alpha^\prime\frac{1}{\frac{\cal V}{(\alpha^\prime)^3}}$. Now, $\mu_7\sim\frac{1}{\kappa^2(\alpha^\prime)^4}$\cite{Jockers_thesis}. This hence implies $\kappa_4^2\mu_7\sim\frac{1}{\frac{{\cal V}}{(\alpha^\prime)^3}}\frac{1}{(\alpha^\prime)^3}$, i.e., $\kappa_4^2\mu_7\sim\frac{1}{{\cal V}}$ in $\alpha^\prime=1$ units.

Working in the $x_2=1$-coordinate patch throughout this paper, for definiteness, and defining $z_1=
\frac{x_1}{x_2},\ z_2=\frac{x_3}{x_2},\ z_3=
\frac{x_4}{x_2^6}$ and $z_4=\frac{x_5}{x_2^9}$ ($z_4$ for later use) therein,  let $\omega=\omega_1(z_1,z_2)dz_1\in H^{(1,0)}_{\partial,-}(\Sigma_B:x_5=0)$ - this implies that $\omega_1(z_1\rightarrow-z_1,z_2\rightarrow-z_2)=\omega_1(z_1,z_2)$. Then
$\partial(=dz_i\partial_i)\omega=0$ and $\omega$ must not be exact. Let
$\partial\omega=(1+z_1^{18}+z_2^{18}+z_3^3-\phi_0 z_1^6z_2^6)^2dz_1\wedge dz_2$ - it is exact on $\Sigma_B$ but not at any other point in the Calabi-Yau. This implies that restricted to $\Sigma_B$, $\frac{\partial\omega_1}{\partial z_2}|_{\Sigma_B}\sim(\phi_0 z_1^6z_2^6 - z_1^{18} - z_2^{18}-z_3^3)^2$. Taking $z_3$ to be around ${\cal V}^{\frac{1}{6}}$ - this would actually correspond to the location of the mobile $D3$-brane in the Calabi-Yau which for concrete calculations and its facilitation will eventually be taken to lie at $({\cal V}^{\frac{1}{36}}e^{i\theta_1},{\cal V}^{\frac{1}{36}}e^{i\theta_2},{\cal V}^{\frac{1}{6}}e^{i\theta_3})$ - one sees that {\bf in the LVS limit}
\begin{equation}
\label{eq:harm1formD5}
\omega_1(z_1,z_2;z_3\sim{\cal V}^{\frac{1}{6}})|_{\Sigma_B}
=-2\frac{\phi_0}{25}z_1^6z_2^{25}-(z_1^{18}+z_3^3)^2z_2 - \frac{z_2^{37}}{37} + \frac{\phi_0^2}{13}z_1^{12}z_2^{13} + 2(z_1^{18}+z_3^3)(\frac{z_2^{19}}{19}-\frac{\phi_0}{7}z_1^6z_2^7);
\end{equation}
 this indeed does satisfy the required involutive property of being even.

Now, the Wilson-line moduli term is: $i\kappa_4^2\mu_7\int_{\Sigma_B}i^*\omega\wedge A^I\wedge{\bar A}^{\bar J}a_I{\bar a}_{\bar J}$, where $\omega\in H^{(1,1)}_+(\Sigma_B)$ could be taken to be $i(dz_1\wedge d{\bar z_1} \pm dz_2\wedge d{\bar z_2})$. Hence,
\begin{eqnarray}
\label{eq:C11bar}
& &
 C^{1{\bar 1}}\sim\int_{\{3\phi_0z_1^6z_2^6-z_1^{18}-z_2^{18}\sim\sqrt{\cal V}\}\subset\Sigma_B}|\omega_1|^2dz_1\wedge d{\bar z_1}\wedge dz_2\wedge d{\bar z_2}|_{|z_3|\sim{\cal V}^{\frac{1}{6}}}\sim\nonumber\\
& & \hskip-0.86in\int_{\{3\phi_0z_1^6z_2^6-z_1^{18}-z_2^{18}\sim\sqrt{\cal V}\}\subset\Sigma_B}\!\!\!\!\!\!\!\arrowvert-2\frac{\phi_0}{25}z_1^6z_2^{25}-(z_1^{18}+\sqrt{\cal V})^2z_2 - \frac{z_2^{37}}{37} + \frac{\phi_0^2}{13}z_1^{12}z_2^{13} + 2(z_1^{18}+\sqrt{\cal V})(\frac{z_2^{19}}{19}-\frac{\phi_0}{7}z_1^6z_2^7)\arrowvert^2 dz_1\wedge d{\bar z_1}\wedge dz_2\wedge d{\bar z_2}\nonumber\\
& & \sim\arrowvert\int_{z_1\sim{\cal V}^{\frac{1}{36}}e^{i\theta_1}}z_1^{18}dz_1\arrowvert^2
\arrowvert\int_{z_2\sim{\cal V}^{\frac{1}{36}}e^{i\theta_2}}z_2^{19}dz_2\arrowvert^2\nonumber\\
& & \sim\left({\cal V}^{\frac{37}{36}}\right)^2\times \left({\cal V}^{\frac{1}{36}}\right)^4
\sim{\cal V}^{\frac{3}{2}}{\rm vol}(\Sigma_B).
\end{eqnarray}
 Hence, if the Wilson line modulus $a_1$ is stabilized at around ${\cal V}^{-\frac{1}{4}}$, then
$i\kappa_4^2\mu_7\int_{\Sigma_B}i^*\omega\wedge A^I\wedge{\bar A}^{\bar J}a_I{\bar a}_{\bar J}
\sim{\rm vol}(\Sigma^B)$. The fact that this is indeed possible, will be justified in appendix A. Even with a more refined evaluation of the integral in (\ref{eq:C11bar}) to obtain $C_{1{\bar 1}}$, the results on soft masses and soft SUSY parameters in the rest of the paper, would qualitatively remain the same. This implies that the (tree-level\footnote{See section {\bf 6} for a discussion on inclusion of renormalization effects and loop corrections - the result remains unaltered.}) gauge couplings corresponding to the gauge theory living on a stack of $D7$ branes wrapping $\Sigma^B\equiv D_5$ will be given by:
\begin{equation}
\label{eq:g_YM}
g_a^{-2}=Re(T_B)\sim \mu_3{\cal V}^{\frac{1}{18}}\sim ln{\cal V},
\end{equation}
implying a finite $g_a$. In the absence of $\alpha^\prime$-corrections, strictly speaking $g_a^{-2}=Re(T_B) - {\cal F}Re(i\tau)$,
${\cal F}\equiv {\cal F}^\alpha {\cal F}^\beta \kappa_{\alpha\beta} + \tilde{\cal F}^a\tilde{\cal F}^b\kappa_{ab}$ (refer to \cite{V_D7_fl}) where ${\cal F}^\alpha$ and
${\cal F}^a$ are the components of the $U(1)$ two-form flux on the world-volume of $D7$-branes wrapping $\Sigma_B$ expanded in the basis
$i^*\omega_\alpha, \omega_\alpha\in H^{(1,1)}(CY_3)$ and $\tilde{\omega}_a\in coker\left(H^{(1,1)}_-(CY_3)\stackrel{i^*}{\rightarrow}H^{(1,1)}_-(\Sigma_B)\right)$, and $\kappa_{\alpha\beta}=\int_{\Sigma_B}i^*\omega_\alpha\wedge i^*\omega_\beta$ and $\kappa_{ab}=\int_{\Sigma_B}\tilde{\omega}_a\wedge\tilde{\omega}_b$. In the ``dilute flux approximation", we disregard the contribution coming from ${\cal F}$ as compared to the ${\cal V}^{\frac{1}{18}}$ coming from $Re(T_B)$.

\section{The Complete Moduli Space K\"{a}hler Potential}

In this section we will derive the geometric K\"{a}ler potential for the Swiss-Cheese Calabi-Yau ${\bf WCP}^4_{[1,1,1,6,9]}[18]$ because of a $D3$-brane present in our setup. This will enable us to determine the complete K\"{a}hler potential corresponding to the closed string moduli $\sigma^\alpha, {\cal G}^a$ as well as the open string moduli or matter fields: $z_i, a^I$.

\subsection{Geometric K\"{a}hler Potential and Metric}

The one-dimensional cones in the toric fan of the desingularized ${\bf WCP}^{1,1,1,6,9}[18]$ are given by the following vectors (See \cite{Candelasetal}):
\begin{eqnarray}
\label{eq:toric_fan}
& & v_1=(-1,-1,-6,-9)\nonumber\\
& & v_2=(1,0,0,0)\nonumber\\
& & v_3=(0,1,0,0)\nonumber\\
& & v_4=(0,0,1,0)\nonumber\\
& & v_5=(0,0,0,1)\nonumber\\
& & v_6({\rm Exceptional\ divisor})=(0,0,-2,-3).
\end{eqnarray}

The allowed charges under two $C^*$ actions are given by solutions to:
\begin{equation}
\label{eq:chargesEqn}
\sum_{i=1}^6Q^a_iv_i=0, a=1,2.
\end{equation}
The solution to (\ref{eq:chargesEqn}) are of the type:
\begin{equation}
\label{eq:chargesSoln}
(q_1,q_1,q_1,2q_6+6q_1,3q_6+9q_1,q_6).
\end{equation}

The $(C^*)^2$ charges will be taken as under:
\begin{equation}
\label{eq:charges}
\begin{array}{c|cccccc}
&\Phi_1&\Phi_2&\Phi_3&\Phi_4&\Phi_5&\Phi_6 \\ \hline
Q^1&0&0&0&2&3&1\\
Q^2&1&1&1&0&0&-3
\end{array}
\end{equation}
Hence, one can construct the following inhomogeneous coordinates: $z_1=\frac{\Phi_1}{\Phi_2},
z_2=\frac{\Phi_3}{\Phi_2}, z_3=\frac{\Phi_4}{\Phi_\epsilon^2\Phi_2^6}, z_4=\frac{\Phi_5}{\Phi_\epsilon^3\Phi_2^9}$.

The NLSM for a two-dimensional ${\cal N}=2$ supersymmetric gauge theory whose target space
is the toric variety corresponding to (\ref{eq:toric_fan}) with (anti-)chiral superfields $({\bar\Phi_i})\Phi_i$,
Fayet-Iliopoulos parameters $r_a$  and vector superfields $V_a$,  is specified by the K\"{a}hler potential:
\begin{equation}
\label{eq:NLSM_K}
\int d^4\theta K=\int d^4\theta\left(\sum_{i=1}^6{\bar\Phi_i}e^{\sum_{a=1}^22Q^a_iV_a}\Phi_i-2r_aV_a\right).
\end{equation}
Substituting (\ref{eq:charges}) in (\ref{eq:NLSM_K}), one sees that:
\begin{equation}
\label{eq:K1}
K=\left(|\Phi_1|^2+|\Phi_2|^2+|\Phi_3|^2\right)e^{2V_2}+|\Phi_4|^2e^{4V_1}
+|\Phi_5|^2e^{6V_1}+|\Phi_\epsilon|^2e^{2V_1-6V_2}-2r_1V_1-2r_2V_2.
\end{equation}
Now, $\int d^4\theta K$ can be regarded as the IR limit of the GLSM Lagrangian - the gauge kinetic terms hence
decouple in this limit. One hence gets a supersymmetric NLSM Lagrangian ${\cal L}_{\rm NLSM}=\int d^4\theta K$ wherein the gauge superfields act as auxiliary fields and can be eliminated by their equations of motion - see \cite{Kimura}.
One can show that the variation of the NLSM Lagrangian w.r.t. the vector superfields $V_a$ yield:
\begin{eqnarray}
\label{eq:eoms1}
& & \frac{\partial{\cal L}_{\rm NLSM}}{\partial V_1}=0\Leftrightarrow 2|\Phi_1|^2e^{2V_1}+3|\Phi_5|^2e^{6V_1}+|\Phi_\epsilon|^2
e^{2V_1-6V_2}=r_1\nonumber\\
& & \frac{\partial{\cal L}_{\rm NLSM}}{\partial V_2}=0\Leftrightarrow |\Phi_1|^2e^{2V_2}+|\Phi_2|^2e^{2V_2}
+|\Phi_3|^2e^{2V_2}-3|\Phi_\epsilon|^2e^{2V_1-6V_2}=r_2.\nonumber\\
& &
\end{eqnarray}
Defining $x\equiv e^{2V_1}, y\equiv e^{2V_2}$, (\ref{eq:eoms1}) can be rewritten as:
\begin{eqnarray}
\label{eq:eoms2}
& & a_1x^2+b_1x^3+c_1xy^{-3}=r_1,\nonumber\\
& & a_2y+c_2xy^{-3}=r_2,
\end{eqnarray}
where
\begin{equation}
\label{eq:defs}
a_1\equiv2|\Phi_4|^2, c_1\equiv|\Phi_\epsilon|^2, a_2\equiv|\Phi_1|^2+|\Phi_2|^2+|\Phi_3|^2,
c_2\equiv-3|\Phi_\epsilon|^2.
\end{equation}
We would now be evaluating the K\"{a}hler potential for the divisor $D_5:\Phi_5=0$ or equivalently $z_4=0$, in
the large volume limit of the Swiss-Cheese Calabi-Yau. In the $\Phi_2=\Phi_\epsilon=1$-coordinate patch,
the defining hypersurface for $D_5$ is $1+z_1^{18}+z_2^{18}+z_3^3-3\phi z_1^6z_2^6=0$. In the LVS limit,
we would assume a scaling: $z_3\sim{\cal V}^{\frac{1}{6}}, z_{1,2}\sim{\cal V}^{\frac{1}{36}}$. Further,
the FI parameters, $r_{1,2}$ taken to scale like the big and small two-cycle areas $t_5$ and $t_4$ respectively,
i.e. like ${\cal V}^{\frac{1}{3}}$ and $\sqrt{ln {\cal V}}$.

The system of equations (\ref{eq:eoms2}) is equivalent to the following octic - we will not be careful with numerical
factors in the following:
\begin{equation}
\label{eq:octic}
\hskip-1.3in P(z)\equiv|z_3|^2(1+|z_1|^2+|z_2|^2)^2y^8+|z_3|^2(1+|z_1|^2+|z_2|^2)r_2y^7
+|z_3|^2r_2^2y^6
+(1+|z_1|^2+|z_2|^2)y-4r_1=0.
\end{equation}
Using Umemura's result \cite{Umemura} on expressing the roots of an algebraic polynomial in terms of Siegel theta functions
of genus $g>1$ -  $\theta\left[\begin{array}{c} \mu\\
\nu
\end{array}\right](z,\Omega)$ for $\mu,\nu\in{\bf R}^g, z\in {\bf C}^g$ and $\Omega$ being a complex symmetric
$g\times g$ period matrix with $Im(\Omega)>0$ defined as follows:
\begin{equation}
\label{eq:Siegel_theta}
\theta\left[\begin{array}{c} \mu\\
\nu
\end{array}\right](z,\Omega)=\sum_{n\in{\bf Z}^g}e^{i\pi(n+\mu)^T\Omega(n+\mu)+2i\pi(n+\mu)^T(z+\nu)}.
\end{equation}
The degree $n$ of the polynomial is related to the genus $g$ of the Riemann surface via
$g=\left[\frac{n+2}{2}\right]$. Hence for an octic, one needs to use Siegel theta functions of genus five.
The period matrix $\Omega$ will be defined as follows:
\begin{equation}
\label{eq:Omega}
\left(\begin{array}{ccccc}
\Omega_{11} & \Omega_{12} & \Omega_{13} & \Omega_{14} & \Omega_{15}\\
\Omega_{12} & \Omega_{22} & \Omega_{23} & \Omega_{24} & \Omega_{25}\\
\Omega_{13} & \Omega_{23} & \Omega_{33} & \Omega_{34} & \Omega_{35}\\
\Omega_{14} & \Omega_{24} & \Omega_{34} & \Omega_{44} & \Omega_{45}\\
\Omega_{15} & \Omega_{25} & \Omega_{35} & \Omega_{45} & \Omega_{55}\\
\end{array}
\right)=\left(\begin{array}{ccccc}
\sigma_{11} & \sigma_{12} & \sigma_{13} & \sigma_{14} & \sigma_{15} \\
\sigma_{21} & \sigma_{22} & \sigma_{23} & \sigma_{24} & \sigma_{25} \\
\sigma_{31} & \sigma_{32} & \sigma_{33} & \sigma_{34} & \sigma_{35} \\
\sigma_{41} & \sigma_{42} & \sigma_{43} & \sigma_{44} & \sigma_{45} \\
\sigma_{51} & \sigma_{52} & \sigma_{53} & \sigma_{54} & \sigma_{55} \\
\end{array}\right)^{-1}
\left(\begin{array}{ccccc}
\rho_{11} & \rho_{12} & \rho_{13} & \rho_{14} & \rho_{15} \\
\rho_{21} & \rho_{22} & \rho_{23} & \rho_{24} & \rho_{25} \\
\rho_{31} & \rho_{32} & \rho_{33} & \rho_{34} & \rho_{35} \\
\rho_{41} & \rho_{42} & \rho_{43} & \rho_{44} & \rho_{45} \\
\rho_{51} & \rho_{52} & \rho_{53} & \rho_{54} & \rho_{55} \\
\end{array}
\right),
\end{equation}
where $\sigma_{ij}\equiv\oint_{A_j}dz \frac{z^{i-1}}{\sqrt{z(z-1)(z-2)P(z)}}$ and
$\rho_{ij}\equiv\oint_{B_j}\frac{z^{i-1}}{\sqrt{z(z-1)(z-2)P(z)}}$,
$\{A_i\}$ and $\{B_i\}$ being a canonical basis of cycles satisfying: $A_i\cdot A_j=B_i\cdot B_j=0$ and
$A_i\cdot B_j=\delta_{ij}$. Umemura's result then is that a root of (\ref{eq:octic}) can be written as:
\begin{eqnarray}
\label{eq:soln_octic}
& & \frac{1}{2\left(\theta\left[\begin{array}{ccccc}
\frac{1}{2} & 0 & 0 & 0 & 0 \\
0 & 0 & 0 & 0 & 0  \end{array}\right](0,\Omega)\right)^4
\left(\theta\left[\begin{array}{ccccc}
\frac{1}{2} & \frac{1}{2} & 0 & 0 & 0 \\
0 & 0 & 0 & 0 & 0  \end{array}\right](0,\Omega)\right)^4}\nonumber\\
& & \times\Biggl[\left(\theta\left[\begin{array}{ccccc}
\frac{1}{2} & 0 & 0 & 0 & 0 \\
0 & 0 & 0 & 0 & 0  \end{array}\right](0,\Omega)\right)^4
\left(\theta\left[\begin{array}{ccccc}
\frac{1}{2} & \frac{1}{2} & 0 & 0 & 0 \\
0 & 0 & 0 & 0 & 0  \end{array}\right](0,\Omega)\right)^4\nonumber\\
& & + \left(\theta\left[\begin{array}{ccccc}
0 & 0 & 0 & 0 & 0 \\
0 & 0 & 0 & 0 & 0  \end{array}\right](0,\Omega)\right)^4
\left(\theta\left[\begin{array}{ccccc}
0 & \frac{1}{2} &  0 & 0 & 0 \\
0 & 0 & 0 & 0 & 0  \end{array}\right](0,\Omega)\right)^4\nonumber\\
& & - \left(\theta\left[\begin{array}{ccccc}
0 & 0 & 0 & 0 & 0 \\
\frac{1}{2} & 0 & 0 & 0 & 0  \end{array}\right](0,\Omega)\right)^4
\left(\theta\left[\begin{array}{ccccc}
0 & \frac{1}{2} & 0 & 0 & 0 \\
\frac{1}{2} & 0 & 0 & 0 & 0 \end{array} \right](0,\Omega)\right)^4\Biggr].\nonumber\\
& &
\end{eqnarray}

Now, if $|z_3|^2r_1^2y^6\sim r_1\sim\sqrt{ln {\cal V}}$, then this suggests that $y\sim \left(ln {\cal V}\right)^{\frac{1}{12}}{\cal V}^{-\frac{1}{6}}$.
Substituting this estimate for $y$ into the octic and septic terms, one sees that the same are of
${\cal O}\left(\left(ln {\cal V}\right)^{\frac{2}{3}}{\cal V}^{-\frac{8}{9}}\right)$ and
${\cal O}\left(\left(ln {\cal V}\right)^{\frac{7}{12}}{\cal V}^{-\frac{4}{9}}\right)$ respectively which
are both suppressed w.r.t. to the sextic term. Hence, in the LVS limit (\ref{eq:octic}) reduces to the
following sextic:
\begin{equation}
\label{eq:sextic}
y^6+\alpha y+\beta=0.
\end{equation}
Umemura's result would require the use of genus-four Siegel theta functions. However, using the results of
\cite{Zhivkov}, one can express the roots of a sextic in terms of genus-two Siegel theta functions as follows:
\begin{eqnarray}
\label{eq:roots}
& & \left[\frac{\sigma_{22}\frac{d}{dz_1}\theta\left[\begin{array}{cc}
\frac{1}{2}&\frac{1}{2} \\
0&\frac{1}{2}
\end{array}\right]\left((z_1,z_2),\Omega\right)
- \sigma_{21}\frac{d}{dz_2}\theta\left[\begin{array}{cc}
\frac{1}{2}&\frac{1}{2} \\
0&\frac{1}{2}
\end{array}\right]\left((z_1,z_2),\Omega\right) }
{\sigma_{12}\frac{d}{dz_1}\theta\left[\begin{array}{cc}
\frac{1}{2}&\frac{1}{2} \\
0&\frac{1}{2}
\end{array}\right]\left((z_1,z_2),\Omega\right)
- \sigma_{12}\frac{d}{dz_2}\theta\left[\begin{array}{cc}
\frac{1}{2}&\frac{1}{2} \\
0&\frac{1}{2}
\end{array}\right]\left((z_1,z_2),\Omega\right)}\right]_{z_1=z_2=0}, \nonumber\\
& & \left[\frac{\sigma_{22}\frac{d}{dz_1}\theta\left[\begin{array}{cc}
0&\frac{1}{2} \\
0&\frac{1}{2}
\end{array}\right]\left((z_1,z_2),\Omega\right)
- \sigma_{21}\frac{d}{dz_2}\theta\left[\begin{array}{cc}
0&\frac{1}{2} \\
0&\frac{1}{2}
\end{array}\right]\left((z_1,z_2),\Omega\right) }
{\sigma_{12}\frac{d}{dz_1}\theta\left[\begin{array}{cc}
0&\frac{1}{2} \\
0&\frac{1}{2}
\end{array}\right]\left((z_1,z_2),\Omega\right)
- \sigma_{12}\frac{d}{dz_2}\theta\left[\begin{array}{cc}
0&\frac{1}{2} \\
0&\frac{1}{2}
\end{array}\right]\left((z_1,z_2),\Omega\right)}\right]_{z_1=z_2=0}, \nonumber\\
& & \left[\frac{\sigma_{22}\frac{d}{dz_1}\theta\left[\begin{array}{cc}
0&\frac{1}{2} \\
\frac{1}{2}&\frac{1}{2}
\end{array}\right]\left((z_1,z_2),\Omega\right)
- \sigma_{21}\frac{d}{dz_2}\theta\left[\begin{array}{cc}
0&\frac{1}{2} \\
\frac{1}{2}&\frac{1}{2}
\end{array}\right]\left((z_1,z_2),\Omega\right) }
{\sigma_{12}\frac{d}{dz_1}\theta\left[\begin{array}{cc}
0&\frac{1}{2} \\
\frac{1}{2}&\frac{1}{2}
\end{array}\right]\left((z_1,z_2),\Omega\right)
- \sigma_{12}\frac{d}{dz_2}\theta\left[\begin{array}{cc}
0&\frac{1}{2} \\
\frac{1}{2}&\frac{1}{2}
\end{array}\right]\left((z_1,z_2),\Omega\right)}\right]_{z_1=z_2=0}, \nonumber\\
& & \left[\frac{\sigma_{22}\frac{d}{dz_1}\theta\left[\begin{array}{cc}
\frac{1}{2}&0 \\
\frac{1}{2}&\frac{1}{2}
\end{array}\right]\left((z_1,z_2),\Omega\right)
- \sigma_{21}\frac{d}{dz_2}\theta\left[\begin{array}{cc}
\frac{1}{2}&0 \\
\frac{1}{2}&\frac{1}{2}
\end{array}\right]\left((z_1,z_2),\Omega\right) }
{\sigma_{12}\frac{d}{dz_1}\theta\left[\begin{array}{cc}
\frac{1}{2}&0 \\
\frac{1}{2}&\frac{1}{2}
\end{array}\right]\left((z_1,z_2),\Omega\right)
- \sigma_{12}\frac{d}{dz_2}\theta\left[\begin{array}{cc}
\frac{1}{2}&0 \\
\frac{1}{2}&\frac{1}{2}
\end{array}\right]\left((z_1,z_2),\Omega\right)}\right]_{z_1=z_2=0}, \nonumber\\
& & \left[\frac{\sigma_{22}\frac{d}{dz_1}\theta\left[\begin{array}{cc}
\frac{1}{2}&0 \\
\frac{1}{2}&0
\end{array}\right]\left((z_1,z_2),\Omega\right)
- \sigma_{21}\frac{d}{dz_2}\theta\left[\begin{array}{cc}
\frac{1}{2}&0 \\
\frac{1}{2}&0
\end{array}\right]\left((z_1,z_2),\Omega\right) }
{\sigma_{12}\frac{d}{dz_1}\theta\left[\begin{array}{cc}
\frac{1}{2}&0 \\
\frac{1}{2}&0
\end{array}\right]\left((z_1,z_2),\Omega\right)
- \sigma_{12}\frac{d}{dz_2}\theta\left[\begin{array}{cc}
\frac{1}{2}&0 \\
\frac{1}{2}&0
\end{array}\right]\left((z_1,z_2),\Omega\right)}\right]_{z_1=z_2=0}, \nonumber\\
& & \left[\frac{\sigma_{22}\frac{d}{dz_1}\theta\left[\begin{array}{cc}
\frac{1}{2}&\frac{1}{2} \\
\frac{1}{2}&0
\end{array}\right]\left((z_1,z_2),\Omega\right)
- \sigma_{21}\frac{d}{dz_2}\theta\left[\begin{array}{cc}
\frac{1}{2}&\frac{1}{2} \\
\frac{1}{2}&0
\end{array}\right]\left((z_1,z_2),\Omega\right) }
{\sigma_{12}\frac{d}{dz_1}\theta\left[\begin{array}{cc}
\frac{1}{2}&\frac{1}{2} \\
\frac{1}{2}&0
\end{array}\right]\left((z_1,z_2),\Omega\right)
- \sigma_{12}\frac{d}{dz_2}\theta\left[\begin{array}{cc}
\frac{1}{2}&\frac{1}{2} \\
\frac{1}{2}&0
\end{array}\right]\left((z_1,z_2),\Omega\right)}\right]_{z_1=z_2=0}.\nonumber\\
& &
\end{eqnarray}
One can show that:
\begin{equation}
\label{eq:dertheta}
\hskip-0.5in\frac{d}{dz_i}\theta\left[\begin{array}{cc}
\mu_1 & \mu_2 \\
\nu_1 & \nu_2
\end{array}\right]\left((z_1,z_2),\Omega\right)_{z_1=z_2=0}=-2\pi\sum_{n_1,n_2\in{\bf Z}}
(-)^{2\nu_1n_1+2\nu_2n_2}(n_i+\mu_i)e^{i\pi\Omega_{11}(n_1+\mu_1)^2
+2i\pi\Omega_{12}(n_1+\mu_1)(n_2+\mu_2)+i\pi\Omega_{22}(n_2+\mu_2)^2},
\end{equation}
where $\mu_i$ and $\nu_i$ are either 0 or $\frac{1}{2}$. The symmetric period matrix corresponding to the hyperelliptic
curve $w^2=P(z)$ is given by:
\begin{equation}
\label{eq:Omega_g=2}
\left(\begin{array}{cc}
\Omega_{11} & \Omega_{12} \\
\Omega_{12} & \Omega_{22}
\end{array}\right)=\frac{1}{\sigma_{11}\sigma_{22}-\sigma_{12}\sigma_{21}}\left(\begin{array}{cc}
\sigma_{22} & -\sigma_{12} \\
-\sigma_{21} & \sigma_{11}
\end{array}\right)\left(\begin{array}{cc}
\rho_{11} & \rho_{12} \\
\rho_{21} & \rho_{22}
\end{array}\right),
\end{equation}
where $\sigma_{ij}=\int_{z_*{A_j}}\frac{z^{i-1}dz}{\sqrt{P(z)}}$ and
$\rho_{ij}=\int_{z_*{B_j}}\frac{z^{i-1}dz}{\sqrt{P(z)}}$ where $z$ maps the $A_i$ and $B_j$ cycles to the
$z-$plane (See \cite{Zhivkov}). Now, for $y^6\sim\beta$ in (\ref{eq:sextic}), one can show that the term
$\alpha y\sim\frac{\beta}{{\cal V}^{\frac{1}{9}}}$ and hence can be dropped in the LVS limit. Further,
along $z_*(A_i)$ and $z_*(B_j)$, $y^6\sim\beta$ and thus:
\begin{eqnarray}
\label{eq:ints}
& & \int_{A_i\ {\rm or}\ B_j}\frac{dy}{\sqrt{y^6+\beta}}\sim\beta^{-\frac{1}{3}}
\ _2F_1\left(\frac{1}{3},\frac{1}{2};\frac{4}{3};1\right),\nonumber\\
& & \int_{A_i\ {\rm or}\ B_j}\frac{y dy}{\sqrt{y^6+\beta}}\sim\beta^{-\frac{1}{6}}
\ _2F_1\left(\frac{1}{6},\frac{1}{2};\frac{7}{6};1\right).
\end{eqnarray}
Hence,
\begin{equation}
\label{eq:period_appr}
\Omega\sim\beta^{\frac{1}{2}}\left(\begin{array}{cc}
\beta^{-\frac{1}{6}} & \beta^{-\frac{1}{3}} \\
\beta^{-\frac{1}{6}} & \beta^{-\frac{1}{3}}
\end{array}\right)\left(\begin{array}{cc}
\beta^{-\frac{1}{3}} & \beta^{-\frac{1}{3}} \\
\beta^{-\frac{1}{6}} & \beta^{-\frac{1}{6}}
\end{array}\right)
\sim\left(\begin{array}{cc}
{\cal O}(1) & {\cal O}(1) \\
{\cal O}(1) & {\cal O}(1)
\end{array}\right).
\end{equation}
Hence, one can {\it ignore the $D3-$brane moduli dependence of the period matrix $\Omega$ in the LVS
limit}. Substituting (\ref{eq:period_appr}) into (\ref{eq:dertheta}), one sees that
\begin{equation}
\label{eq:sol_y}
e^{2V_2}\sim\left(\zeta\frac{1}{r_1|z_3|^2}\right)^{\frac{1}{6}},
\end{equation}
in the LVS limit where $\zeta$ encodes the information about the {\it exact} evaluation of
the period matrix. Substituting (\ref{eq:sol_y}) into the second equation of (\ref{eq:eoms2}),
one obtains:
\begin{equation}
\label{eq:sol_x}
e^{2V_1}=\frac{\left(r_2 - \left(1+|z_1|^2+|z_2|^2\right)\left(\frac{\zeta}{
r_1|z_3|^2}\right)^{\frac{1}{6}}\right)\sqrt{\zeta}}{3\sqrt{r_1|z_3|^2}}.
\end{equation}

The geometric K\"{a}hler potential for the divisor $D_5$ in the LVS limit is hence given by:
\begin{eqnarray}
\label{eq:Kaehler_D_5}
& & \hskip -1.3in K|_{D_5} =  r_2 - 4\frac{\left(r_2 - \left(1+|z_1|^2+|z_2|^2\right)\left(\frac{\zeta}{
r_1|z_3|^2}\right)^{\frac{1}{6}}\right)\sqrt{\zeta}}{3}+|z_3|^2\left(\frac{\left(r_2 - \left(1+|z_1|^2+|z_2|^2\right)\left(\frac{\zeta}{
r_1|z_3|^2}\right)^{\frac{1}{6}}\right)\sqrt{\zeta}}{3\sqrt{r_1|z_3|^2}}\right)^2\nonumber\\
& & - r_1 ln\left[\frac{\left(r_2 - \left(1+|z_1|^2+|z_2|^2\right)\left(\frac{\zeta}{
r_1|z_3|^2}\right)^{\frac{1}{6}}\right)\sqrt{\zeta}}{3\sqrt{r_1|z_3|^2}}\right]
-r_2 ln\left[\left(\zeta\frac{1}{r_1|z_3|^2}\right)^{\frac{1}{6}}\right]\nonumber\\
& & \sim\frac{{\cal V}^{\frac{2}{3}}}{\sqrt{ln {\cal V}}}.
\end{eqnarray}

The first order and (mixed) second order derivatives of $K|_{D_5}$ are given in appendix B.

The extremization of the NLSM Lagrangian w.r.t. the vector superfields corresponding to the divisor $D_4$ -  $z_4=0$ or equivalently $\Phi_4=0$ - yield the following pair of equations:

\begin{eqnarray}
\label{eq:D4_I}
& & b_1x^3 + c_1xy^{-3} = r_1\nonumber\\
& & a_2y + c_2xy^{-3} = r_2.
\end{eqnarray}
where $a_2\equiv|\Phi_1|^2+|\Phi_2|^2+|\Phi_3|^2$, $b_1\equiv3|\Phi_5|^2, c_1\equiv|\Phi_\epsilon|^2, c_2\equiv-3|\Phi_\epsilon|^2, x\equiv e^{2V_1}, y\equiv e^{2V_2}$. In the $\Phi_\epsilon=1$-patch, one gets the following degree-12 equation in $y$:
\begin{equation}
\label{eq:D4_II}
\frac{1}{9}|\Phi_5|^2\left(r_2^3y^9-3r_2^2a_2y^{10}+3r_2a_2^2y^{11}+a_2^3y^{12}\right)-\frac{1}{3}\left[r_2-\left(|\Phi_1|^2+|\Phi_2|^2+|\Phi_3|^2\right)y\right]
=r_1.
\end{equation}
Choosing a scaling (in $z_\epsilon=z_2=1$-patch): $z_4\sim z_{1,2}^9\sim\left(ln {\cal V}\right)^{\frac{1}{4}}$,
$r_1\sim\sqrt{ln {\cal V}}, r_2\sim{\cal V}^{\frac{1}{3}}$, (\ref{eq:D4_II}) would imply:
\begin{equation}
\label{eq:D4_III}
\frac{1}{9}\left({\cal V}y^9-3{\cal V}^{\frac{2}{3}}\left(ln {\cal V}\right)^{\frac{1}{18}}y^{10}+3{\cal V}^{\frac{1}{3}}\left(ln {\cal V}\right)^{\frac{1}{9}}y^{11}+\left(ln{\cal V}\right)^{\frac{1}{6}}y^{12}\right)
-\frac{1}{3}\left({\cal V}^{\frac{1}{3}}-(ln {\cal V})^{\frac{1}{18}}y\right)\sim\sqrt{ln {\cal V}}.
\end{equation}

Hence, if $y\sim\left[\left(ln {\cal V}\right)^{-\frac{1}{2}}{\cal V}^{-\frac{2}{3}}\right]^{\frac{1}{9}}$, i.e., if the $y^9$-term is the dominant term on the LHS of (\ref{eq:D4_III}), then the same is justified as the $y,y^{10}, y^{11}, y^{12}$-terms are respectively ${\cal V}^{-\frac{2}{27}},{\cal V}^{-\frac{2}{27}},{\cal V}^{-\frac{13}{27}}, {\cal V}^{-\frac{8}{9}}$, and hence are sub-dominant w.r.t. the $y^9$ terms and will hence be dropped. One thus obtains:
\begin{equation}
\label{eq:D4_V}
y=e^{2V_2}\sim\left[\frac{3}{r_2^2|z_1^{18}+z_2^{18}-3\phi z_1^6z_2^6|^2}\right]^{\frac{1}{9}},
\end{equation}
which hence yields:

$K|_{D_4}=
\frac{3^{\frac{1}{9}}\,\left( 1 + |{z_1}|^2 +
       |{z_2}|^2 \right) }{{\left( {{r_2}}^2\,
        |z_1^{18}+z_2^{18}-3\phi z_1^6z_2^6|^2 \right)  }^{\frac{1}{9}}} +
  \frac{{r_2} - \frac{3^{\frac{1}{9}}\,\left( 1 + |{z_1}|^2 +
          |{z_2}|^2 \right) }{{\left( {{r_2}}^2\,
           |z_1^{18}+z_2^{18}-3\phi z_1^6z_2^6|^2 \right) }^{\frac{1}{9}}}}{3} +
               \frac{{\left( {r_2} - \frac{3^{\frac{1}{9}}\,
           \left( 1 + |{z_1}|^2 +
             |{z_2}|^2 \right) }{{\left( {{r_2}}^2\,
              |z_1^{18}+z_2^{18}-3\phi z_1^6z_2^6|^2 \right) }^{\frac{1}{9}}} \right) }^3}{9\,{{r_2}}^2} -$

              $
   \frac{{r_2}\,\log (\frac{3}
       {{{r_2}}^2\,|z_1^{18}+z_2^{18}-3\phi z_1^6z_2^6|^2 })}{9} -
  {r_1}\,\log \Biggl(\frac{3^{\frac{1}{3}}\,\left( {r_2} -
         \frac{3^{\frac{1}{9}}\,\left( 1 + |{z_1}|^2 +
              |{z_2}|^2 \right) }{{\left( {{r_2}}^2\,
               |z_1^{18}+z_2^{18}-3\phi z_1^6z_2^6|^2 \right) }^{\frac{1}{9}}} \right) }{{\left(
          {{r_2}}^2\,|z_1^{18}+z_2^{18}-3\phi z_1^6z_2^6|^2 \right) }^{\frac{1}{3}}}\Biggl)$

The mixed second order derivatives of $K|_{D_4}$ are given in appendix B.

\subsection{The $\sigma^\alpha,{\cal G}^a,z_i,{\cal A}^I$-Moduli Space Metric}

The K\"{a}hler potential is given as under:
\begin{eqnarray}
\label{eq:K}
& & K = - ln\left(-i(\tau-{\bar\tau})\right) - ln\left(i\int_{CY_3}\Omega\wedge{\bar\Omega}\right)\nonumber\\
 & & - 2 ln\Biggl[a\left(T_B + {\bar T}_B - \gamma K_{\rm geom}\right)^{\frac{3}{2}}-a\left(T_S + {\bar T}_S - \gamma K_{\rm geom}\right)^{\frac{3}{2}} + \frac{\chi}{2}\sum_{m,n\in{\bf Z}^2/(0,0)}
\frac{({\bar\tau}-\tau)^{\frac{3}{2}}}{(2i)^{\frac{3}{2}}|m+n\tau|^3}\nonumber\\
& &  - 4\sum_{\beta\in H_2^-(CY_3,{\bf Z})} n^0_\beta\sum_{m,n\in{\bf Z}^2/(0,0)}
\frac{({\bar\tau}-\tau)^{\frac{3}{2}}}{(2i)^{\frac{3}{2}}|m+n\tau|^3}cos\left(mk.{\cal B} + nk.c\right)\Biggr],
\end{eqnarray}
where $n^0_\beta$ are the genus-0 Gopakumar-Vafa invariants for the curve $\beta$ and
$k_a=\int_\beta\omega_a$ are the degrees of the rational curve. Further, to work out the moduli-space metric, one needs to complexify the Wilson line moduli via sections of $N\Sigma_B$ (See \cite{Kachruetal}). Allowing for the possibility of gaugino condensation requires $\Sigma_B$ to be rigid - we hence consider only zero sections of $N\Sigma_B$, i.e., we set $\zeta^A=0$. The complexified Wilson line moduli would then be ${\cal A}_I=ia_I$. For a stack of $N D7$-branes wrapping $D_5$, stricly speaking $\zeta^A$ and $a_I$ are $U(N)$ Lie algebra valued, which implies that they can be written as:
$\zeta^A=(\zeta^A_1)_a U^a + (\zeta_2^A)_{ab}e^{ab}$ and similarly for $a_I$ (See \cite{Quevedo_etal_explicit_Y}) where $U^a$ and $e^{ab}$ are the generators of the $U(N)$ algebra. For reasons given in appendix A, restricting the mobile $D3$ brane to $D_5$ guarantees nullification of the non-perturbative superpotential from gaugino condensation for all values of $N>1$. Hence, we are justified in setting $N=1$ - ruling out gaugino condensation in our setup - $\zeta^A$ and $a_I$ are hence not matrix-valued. We then assume that $\zeta^A$  and all components save one of $a_I$ can be stabilized to a zero value; the non-zero component $a_1$ can be stabilized at around ${\cal V}^{-\frac{1}{4}}$. This is justified in a self-consistent manner, in appendix A.

There is the issue of using the modular completion of \cite{Grimm} for our setup which includes a $D7$ brane - or a stack of $D7$ branes \footnote{One of us (PS) thanks M.Bianchi for raising this issue with him.}. First, in our analysis, it is the large contribution from the world-sheet instantons - proportional to the genus-zero Gopakumar-Vafa invariants - that are relevant and not its appropriate form invariant (if  at all) under (a discrete subgroup of) $SL(2,{\bf Z})$ as in \cite{dSetal,Grimm}. Second, we could think of the $D7$ brane as a $(p,q,r)$ seven-brane satisfying the constraint: $pq=\left(\frac{r}{2}\right)^2$, which as per \cite{Bergshoeff_etal} would ensure $SL(2,{\bf Z})$ invariance.

Though the contribution from the matter fields ``$C_{37}$" coming from open strings stretched between the $D3$ and $D7$ branes wrapping $\Sigma_B(\equiv D_5$) for Calabi-Yau orientifolds is not known, but based on results for orientifolds of $\left(T^2\right)^3$ - see \cite{Lustetal} - we guess the following expression:
\begin{equation}
\label{eq:C_37}
\frac{|C_{37}|^2}{\sqrt{T_B}}\sim{\cal V}^{-\frac{1}{36}}|C_{37}|^2,
\end{equation}
which for sub-Planckian $C_{37}$ (implying that they get stabilized at ${\cal V}^{-c_{37}},\ c_{37}>0$) would be sub-dominant relative to contributions from world-sheet instantons, for instance, in (\ref{eq:K}). We will henceforth ignore (\ref{eq:C_37}).

It is justified in appendix A that the potential is extremized at values of the complexified Wilson line moduli at around ${\cal V}^{-\frac{1}{4}}$. We will use the same in (\ref{eq:G_wilson}) when evaluating the soft supersymmetry parameters in the next section.

\section{Resolution of the Tension between LVS Cosmology and LVS Phenomenology}

We need to figure a way of obtaining a TeV gravitino when dealing with LVS phenemenology and a $10^{12}$ GeV gravitino when dealing with LVS cosmology within the same setup for the same value of the volume modulus:
${\cal V}\sim10^6$ (in $l_s=1$ units). In this section we give a proposal to do the same.

The gravitino mass is given by:
$m_{\frac{3}{2}}=e^{\frac{K}{2}}W M_p\sim\frac{W}{{\cal V}}M_p$ in the LVS limit. Choose the complex-structure moduli-dependent superpotential to be negligible as compared to the non-perturbative superpotential. Consider now a single $ED3-$instanton obtained by an $n^s$-fold wrapping of $D_4$ by a single $ED3-$brane. The holomorphic prefactor appearing in the non-perturbative superpotential that depends on the mobile $D3$ brane's position moduli, has to be a section class of the divisor bundle $[D_4]$ - and should have a zero of degree $n_s$ at the location of the $ED3$ instanton\footnote{One of us(AM) is grateful to O.Ganor for very useful clarifications on this point. } - see \cite{Ganor1_2}. This will contribute a superpotential of the type:
$W\sim \left(1 + z_1^{18} + z_2^{18} + z_3^2 - 3\phi_0z_1^6z_2^6\right)^{n_s}e^{in^sT_s}\Theta_{n^s}({\cal G}^a,\tau)\sim
\frac{\left(1 + z_1^{18} + z_2^{18} + z_3^2 - 3\phi_0z_1^6z_2^6\right)^{n^s}}{{\cal V}^{n^s}}$. The main idea will be that for a volume modulus fixed at ${\cal V}\sim10^6l_s^6$, during early stages of cosmological evolution, the geometric location of the mobile $D3$-brane on a non-singular elliptic curve embedded within the Swiss-Cheese $CY_3$ that we are considering was sufficient to guarantee that the gravitino was super-massive with a mass of $10^{12}$ GeV as required by cosmological data, e.g., density perturbations. Later, as the $D3$-brane moved  to another non-singluar elliptic curve within the $CY_3$ with the same value of the volume, in the present epoch, one obtains a TeV gravitino as required. Let $z_{i,(0)}$ denote the position moduli of the mobile $D3$-brane. Consider fluctuations about the same given by $\delta z_{i,(0)}$. Defining $P(\{z_{i,(0)}\})\equiv 1 + z_{1,(0)}^{18} + z_{2,(0)}^{18} + z_{3,(0)}^2 - 3\phi_0 z_{1,(0)}^6z_{2,(0)}^6$, one obtains:
\begin{eqnarray}
\label{eq:Wfluc}
& & W\sim\frac{\left(P(\{z_{i,(0)}\}) + \sum_{i=1,2}a_i\delta z_{i,(0)}\right)^{n^s}}{{\cal V}^{n^s}}e^{in^sT_s({\cal G}^a,{\bar{\cal G}}^a;\tau,{\bar\tau}) + i\mu_3l^2\left(z_{i,(0)}{\bar z}_{{\bar j},(0)}a_{i{\bar j}} + z_{i,(0)}z_{j,(0)}\tilde{a}_{ij}\right) + i\mu_3l^2\left(\sum_i\alpha_i\delta z_{i,(0)} + \sum_{\bar i}\beta_{{\bar i}}\delta {\bar z}_{{\bar i}}\right)}\nonumber\\
& & \sim {\cal V}^{\alpha n^s - n^s}\left(1 + \frac{\sum_i a_i\delta z_{i,(0)}}{P(\{z_{i,(0)}\})}\right)^{n^s},
\end{eqnarray}
where one assumes $P(\{z_{i,(0)}\})\sim {\cal V}^\alpha$. This yields
$m_{\frac{3}{2}}\equiv e^{\frac{\hat{K}}{2}}|\hat{W}|\sim{\cal V}^{n^s(\alpha - 1) - 1}$.

\begin{enumerate}
\item
\underline{LVS Cosmology} Assume that one is a point in the Swiss-Cheese $CY_3:P(\{z_{i,(0)}\})\sim{\cal V}^{\alpha_{\rm cosmo}}$. Hence,  what we need is:
$10^{18 +6(n^s\alpha_{\rm cosmo} - n^s - 1)}\sim10^{12}$, or $\alpha_{\rm cosmo} = 1$ ($n^s\geq2$ to ensure a metastable dS minimum in the LVS limit - see \cite{dSetal}).
Now, either $z_{1,2}^{18}\sim{\cal V}$, i.e., $z_{1,2}\sim{\cal V}^{\frac{1}{18}}<{\cal V}^{\frac{1}{6}}$(as $z_{1,2,3}\leq{\cal V}^{\frac{1}{6}}$) and is hence alright, or
$z_3^2\sim{\cal V}$, i.e., $z_3\sim\sqrt{\cal V}>{\cal V}^{\frac{1}{6}}$ and hence is impossible. Therefore, geometrically if one is at a point $(z_1,z_2,z_3)\sim({\cal V}^{\frac{1}{18}},{\cal V}^{\frac{1}{18}},z_3)$ where $z_3$ (in an appropriate coordinate patch) using (\ref{eq:hypersurface}) satisfies:
\begin{equation}
\label{eq:elliptic1}
\psi_0{\cal V}^{\frac{1}{9}}z_3z_4 - z_3^2 - z_4^3 \sim {\cal V},
\end{equation}
one can generate a $10^{12}$GeV gravitino at ${\cal V}\sim 10^6$!!! Note that (\ref{eq:elliptic1}) is a non-singular elliptic curve embedded in the Calabi-Yau. On redefining $iz_3\equiv y$ and $z_4\equiv x$,  one can compare (\ref{eq:elliptic1}) with the following elliptic
curve over {\bf C}:
\begin{equation}
\label{eq:elliptic2}
y^2 + a_1xy+a_3y=x^3+a_2x^2+a_4x+a_6,
\end{equation}
for which the $j$-invariant is defined as: $j=\frac{(a_1^2+4a_2)^2 -
24(a_1a_3 + a_4)}{\Delta}$ where the discriminant $\Delta$ is defined as follows - see \cite{Husemoeller} -
\begin{equation}
\label{eq:discdef}
\hskip-0.5in\Delta\equiv
-(a_1^2+4a_2)^2(a_1^2a_6 - a_1a_3a_4 + a_2a_3^2+4a_2a_6 - a_4^2) +
9(a_1^2+4a_2)(a_1a_3+2a_4)(a_3^2+4a_6) -
8(a_1a_3+2a_4)^3-27(a_3^2+4a_6)^2.
\end{equation}
 The discriminant works out to $-\psi_0^4{\cal V}^{\frac{37}{9}}  - 432{\cal V}^2\neq0$, implying that (\ref{eq:elliptic2}) is non-singular.

\item
\underline{LVS Phenomenology} A similar analysis would require: $6(n^s\alpha_{\rm pheno} - n^s - 1) + 18\sim 3$, or $\alpha_{\rm pheno}=1-\frac{3}{2n^s}$, which for $n^s=2$ ($n^s\geq2$ to ensure a metastable dS minimum in the LVS limit) yields $\alpha_{\rm pheno}=\frac{1}{4}$. So, either $z_{1,2}\sim{\cal V}^{\frac{1}{72}}<{\cal V}^{\frac{1}{6}}$ which is fine or $z_3\sim{\cal V}^{\frac{1}{8}}<{\cal V}^{\frac{1}{6}}$ and hence also alright. However, for ${\cal V}\sim10^7l_s^6$, one can show that one ends up with a non-singular elliptic curve
 embedded inside the Calabi-Yau given by: $\psi_0{\cal V}^{\frac{1}{21}}z_3z_4 - (z_3^3 + z_4^2)\sim{\cal V}^{\frac{3}{7}}$. It is hence more natural to thus choose $z_{1,2}\sim{\cal V}^{\frac{1}{72}}$ over $z_3\sim{\cal V}^{\frac{1}{8}}$. Hence, the mobile $D3$ brane moves to the elliptic curve embedded inside the Swiss-Cheese Calabi-Yau:
\begin{equation}
\label{eq:elliptic3}
\psi_0{\cal V}^{\frac{1}{36}}z_3z_4 - (z_3^2 + z_4^3)\sim{\cal V}^{\frac{1}{4}},
\end{equation}
one obtains a TeV gravitino. One can again see that the discriminant
corresponding to (\ref{eq:elliptic3}) is $-\psi_0^4{\cal V}^{\frac{13}{36}} - 432\sqrt{\cal V}\neq0$ implying that the elliptic curve (\ref{eq:elliptic3}) is non-singular.
\end{enumerate}

The volume of the Calabi-Yau can be extremized at one value - $10^6$ - for varying positions of the mobile $D3$-brane as discussed above for the following three reasons. Taking the small divisor's volume modulus and the Calabi-Yau volume modulus as independent variables, (a) the $D3$-brane position moduli enter the holomorphic prefactor - the section of the divisor bundle - and hence the overall potential will be proportional to the modulus square of the same and the latter does not influence the extremization condition of the volume modulus,  (b) in consistently taking the large volume limit as done in this paper, the superpotential is independent of the Calabi-Yau volume modulus, and (c) $vol(\Sigma_S)\geq\mu_3{\cal V}^\beta$ for values of $\beta$ taken in this paper corresponding to different positions of the $D3$-brane. Combining these three reasons, one can show that the extremization condition for the volume modulus is independent of the position of the $D3$-brane position moduli.

\section{Soft Supersymmetry Breaking Parameters}

In this section we evaluate the gravitino mass, the masses of the metter fields, the $\mu$ and the physical $\hat{\mu}$ parameters, the Yukawa couplings $Y_{ijk}$ and the physical Yukawa couplings $A_{ijk}$, and
the $\hat{\mu}B$-parameters, in the large volume limit in our setup.

The soft supersymmetry parameters are related to the expansion of the K\"{a}hler potential and superpotential for the open- and closed-string moduli (henceforth referred to as the complete K\"{a}hler potential and superpotential) as a power series in the open-string(the ``matter fields") moduli.  The power series is conventionally about zero values of the matter fields. In our setup, the matter fields - the mobile space-time filling $D3$-brane position moduli in the Calabi-Yau (restricted for convenience to the big divisor $D_5$) and the complexified Wilson line moduli arising due to the wrapping of $D7$-brane(s) around four-cycles - take values (at the extremeum of the potential) respectively of order
${\cal V}^{\frac{1}{36}}$ and ${\cal V}^{-\frac{1}{4}}$, which are finite. We will consider the soft supersymmetry parameters corresponding to expansions of the complete K\"{a}hler potential and the superpotential as a power series in fluctuations about the aforementioned extremum values of the open-string moduli.

The fluctuations around the extremum values of $z_{1,2}$ and ${\cal A}_1$ are:
\begin{eqnarray}
\label{eq:fluctuations}
& & z_{1,2}={\cal V}^{\frac{1}{36}} + \delta z_{1,2},\nonumber\\
& & {\cal A}_1={\cal V}^{-\frac{1}{4}}+\delta {\cal A}_1.
\end{eqnarray}

Using (\ref{eq:K_fluc_1}) - similar to \cite{mirage} -  one arrives at the following expression for the K\"{a}hler potential (not being careful about ${\cal O}(1)$ numerical factors):
\begin{eqnarray}
\label{eq:K2}
& & K \left(\left\{\sigma^b,{\bar\sigma}^B;\sigma^S,{\bar\sigma}^S;{\cal G}^a,{\bar{\cal G}}^a;\tau,{\bar\tau}\right\};\left\{z_{1,2},{\bar z}_{1,2};{\cal A}_1,{\bar{\cal A}_1}\right\}\right) \sim - ln\left(-i(\tau-{\bar\tau})\right) - ln\left(i\int_{CY_3}\Omega\wedge{\bar\Omega}\right) - 2\ ln\ \Xi\nonumber\\
& & + \left(|\delta z_1|^2 + |\delta z_2|^2 + \delta z_1{\bar\delta z_2} + \delta z_2{\bar\delta z_1}\right)K_{z_i{\bar z}_j} + \left((\delta z_1)^2 + (\delta z_2)^2\right)Z_{z_iz_j} + c.c\nonumber\\
   & & + |\delta{\cal A}_1|^2K_{{\cal A}_1\bar{\cal A}_1} + (\delta {\cal A}_1)^2 Z_{{\cal A}_1{\cal A}_1} + c.c + \left(\delta z_1\delta{\bar{\cal A}_1} + \delta z_2\delta{\bar{\cal A}_1} \right)
K_{z_i\bar{\cal A}_1} + c.c + (\delta z_1\delta{\cal A}_1 + \delta z_2\delta{\cal A}_1 ) Z_{z_i{\cal A}_1} +  c.c. + ...\nonumber\\
& &
\end{eqnarray}
where $K_{z_i{\bar z}_j}, Z_{z_iz_j},K_{{\cal A}_1\bar{\cal A}_1},Z_{{\cal A}_1{\cal A}_1},K_{z_i\bar{\cal A}_1}$ and $Z_{z_i{\cal A}_1}$ are defined in appendix C and appendix E.

With $\gamma\sim\kappa_4^2T_3$(See \cite{Maldaetal_Wnp_pref}) $\sim\frac{1}{\cal V}$, the matrix $\hat{K}_{i{\bar j}}\equiv\frac{\partial^2 K \left(\left\{\sigma^b,{\bar\sigma}^B;\sigma^S,{\bar\sigma}^S;{\cal G}^a,{\bar{\cal G}}^a;\tau,{\bar\tau}\right\};\left\{\delta z_{1,2},{\bar\delta}{\bar z}_{1,2};\delta{\cal A}_1,{\bar\delta}{\bar{\cal A}_1}\right\}\right)}{\partial C^i{\bar\partial} {\bar C}^{\bar j}}|_{C^i=0}$ - the matter field fluctuations denoted by $C^i\equiv \delta z_{1,2},\delta{\cal A}_1$ -  is therefore given by:
\begin{equation}
\label{eq:Khat}
\hat{K}_{i{\bar j}}\sim\left(\begin{array}{ccc}
A_{z_1z_1}\frac{{\cal V}^{\frac{1}{36}}}{\sum_\beta n^0_\beta} & A_{z_1z_2}\frac{{\cal V}^{\frac{1}{36}}}{\sum_\beta n^0_\beta} & A_{z_1a_1}\frac{{\cal V}^{\frac{11}{12}}}{\sum_\beta n^0_\beta}\\
{\bar A}_{z_1z_2}\frac{{\cal V}^{\frac{1}{36}}}{\sum_\beta n^0_\beta} & A_{z_1z_2}\frac{{\cal V}^{\frac{1}{36}}}{\sum_\beta n^0_\beta} & A_{z_2a_1}\frac{{\cal V}^{\frac{11}{12}}}{\sum_\beta n^0_\beta} \\
{\bar A}_{z_1a_1}\frac{{\cal V}^{\frac{11}{12}}}{\sum_\beta n^0_\beta} & {\bar A}_{z_2a_1}\frac{{\cal V}^{\frac{11}{12}}}{\sum_\beta n^0_\beta} & A_{a_1a_1}\frac{{\cal V}^{\frac{65}{36}}}{\sum_\beta n^0_\beta}
\end{array}\right).
\end{equation}
In (\ref{eq:Khat}) and the remainder of the paper, given the cancelation between the volume of $D_5$ and the quadratic term in the Wilson line moduli, appearing in $T_B$, one has used the following:
\begin{eqnarray}
\label{eq:Xi_T}
& & \Xi\sim\sum_\beta n^0_\beta,\nonumber\\
& & {\cal T}_B(\sigma^B,{\bar\sigma ^B};{\cal G}^a,{\bar{\cal G}^a};\tau,{\bar\tau}) + \mu_3{\cal V}^{\frac{1}{18}} + i\kappa_4^2\mu_7C_{1{\bar 1}}{\cal V}^{-\frac{1}{2}} - \gamma\left(r_2 + \frac{r_2^2\zeta}{r_1}\right)\sim \mu_3{\cal V}^{\frac{1}{18}},\nonumber\\
& & {\cal T}_S(\sigma^S,{\bar\sigma ^S};{\cal G}^a,{\bar{\cal G}^a};\tau,{\bar\tau}) + \mu_3{\cal V}^{\frac{1}{18}}  - \gamma\left(r_2 + \frac{r_2^2\zeta}{r_1}\right)\sim\mu_3{\cal V}^{\frac{1}{18}}\sim ln{\cal V}.
\end{eqnarray}

To work out the physical $\mu$ terms, Yukawa couplings, etc., one needs to diagonalize (\ref{eq:Khat}) and then work with corresponding diagonalized matter fields. To simplify, we have assumed $A_{z_iz_j}, A_{z_ia_1}$ to be real. One can show that (\ref{eq:Khat}) has the following two sets of eigenvalues, the second being two-fold degenerate:

$\underline{Eigenvalue\ 1}$:

\begin{equation}
\label{eq:ev1_1}
\hskip-0.3in\frac{1}{6} V^{-3 k-1} \Biggl(2 {A_{a_1a_1}} V^{2
   k+\frac{89}{36}}+\frac{2 \sqrt[3]{2} \left(V^{16/9}
   {A_{a_1a_1}}^2-({A_{z_1z_1}}+{A_{z_2z_2}}) {A_{a_1a_1}}+3
   {A_{z_1a_1}}^2+3 {A_{z_2a_1}}^2\right) V^{4
   k+\frac{19}{6}}}{\sqrt[3]{\Sigma}}+2^{2/3} \sqrt[3]{\Sigma}\Biggr),
   \end{equation}

   where
   $$\hskip-0.3in\Sigma\equiv 9 {A_{a_1a_1}} {A_{z_1a_1}}^2 V^{6
   k+\frac{203}{36}}+9 {A_{a_1a_1}} {A_{z_2a_1}}^2 V^{6
   k+\frac{203}{36}}-3 {A_{a_1a_1}}^2 {A_{z_1z_1}} V^{6
   k+\frac{203}{36}}-3 {A_{a_1a_1}}^2 {A_{z_2z_2}} V^{6
   k+\frac{203}{36}}+2 {A_{a_1a_1}}^3 V^{6
   k+\frac{89}{12}}$$

   $$\hskip-0.6in+\sqrt{-\left(3 {A_{z_1a_1}}^2+3
   {A_{z_2a_1}}^2-{A_{a_1a_1}} ({A_{z_1z_1}}+{A_{z_2z_2}})\right)^2
   V^{12 k+\frac{19}{2}} \left(3 V^{16/9} {A_{a_1a_1}}^2-4
   ({A_{z_1z_1}}+{A_{z_2z_2}}) {A_{a_1a_1}}+12 {A_{z_1a_1}}^2+12
   {A_{z_2a_1}}^2\right)}$$

In the LVS limit, the above expands out to yield:
\begin{equation}
\label{eq:ev1_2}
\frac{1}{\sum_\beta n^0_\beta}\left(A_{a_1a_1}{\cal V}^{\frac{65}{36}} + \frac{\alpha_4}{3A_{a_1a_1}}\right).
\end{equation}

$\underline{Eigenvalue\ 2}$

\begin{eqnarray}
\label{eq:ev2_1}
& & \frac{1}{12} V^{-3 k-1} \Biggl(4 \left({A_{a_1a_1}}
   V^{16/9}+{A_{z_1z_1}}+{A_{z_2z_2}}\right) V^{2
   k+\frac{25}{36}}\nonumber\\
& & -\frac{2 i \sqrt[3]{2} \left(-i+\sqrt{3}\right)
   \left(V^{16/9} {A_{a_1a_1}}^2-({A_{z_1z_1}}+{A_{z_2z_2}})
   {A_{a_1a_1}}+3 {A_{z_1a_1}}^2+3 {A_{z_2a_1}}^2\right) V^{4
   k+\frac{19}{6}}}{\sqrt[3]{\Sigma}}+i 2^{2/3} \left(i+\sqrt{3}\right)
   \sqrt[3]{\Sigma}\Biggr),\nonumber\\
   & &
   \end{eqnarray}
which in the LVS limit expands out to yield:
\begin{equation}
\label{eq:ev2_2}
\frac{{\cal V}^{\frac{1}{36}}}{\sum_\beta n^0_\beta}\left(4(A_{z_1z_1} + A_{z_2z_2}) - \frac{2\left(-({A_{z_1z_1}}+{A_{z_2z_2}}) {A_{a_1a_1}}+3
   {A_{z_1a_1}}^2+3 {A_{z_2a_1}}^2\right)}{A_{a_1a_1}}\right).
\end{equation}

$\underline{Eigenvalue\ 3}=Eigenvalue\ 2$

The eigenvectors are as under:

$\underline{Eigenvector1}$

$$\Biggl(\frac{1}{{A_{z_1a_1}}}\frac{1}{6} V^{-2 k-\frac{19}{12}} \Biggl[2 \left({A_{a_1a_1}}
   V^{16/9}+{A_{z_1z_1}}+{A_{z_2z_2}}\right) V^{2 k+\frac{25}{36}}-6 {A_{a_1a_1}} V^{2
   k+\frac{89}{36}}$$

   $$+\frac{2 \sqrt[3]{2} \left(V^{16/9} {A_{a_1a_1}}^2-({A_{z_1z_1}}+{A_{z_2z_2}})
   {A_{a_1a_1}}+3 {A_{z_1a_1}}^2+3 {A_{z_2a_1}}^2\right) V^{4 k+\frac{19}{6}}}{\sqrt[3]{\Sigma}}+2^{2/3} \sqrt[3]{\Sigma}\Biggr]-\frac{{A_{z_2a_1}} \left(\Sigma_2\right)}{\Sigma_3},\frac{\Sigma_2}{\Sigma_3},1\Biggr)^T,$$

where

$$\Sigma_2\equiv \frac{1}{6} {A_{z_1z_2}} \Biggl[2
   \left({A_{a_1a_1}} V^{16/9}+{A_{z_1z_1}}+{A_{z_2z_2}}\right) V^{2 k+\frac{25}{36}}-6 {A_{a_1a_1}}
   V^{2 k+\frac{89}{36}}$$

   $$+\frac{2 \sqrt[3]{2} \left(V^{16/9}
   {A_{a_1a_1}}^2-({A_{z_1z_1}}+{A_{z_2z_2}}) {A_{a_1a_1}}+3 {A_{z_1a_1}}^2+3 {A_{z_2a_1}}^2\right)
   V^{4 k+\frac{19}{6}}}{\sqrt[3]{\Sigma}}+2^{2/3}
   \sqrt[3]{\Sigma}\Biggr] V^{-2 k-\frac{2}{3}}$$

   $$+{A_{z_1a_1}} {A_{z_2a_1}}
   V^{65/36}\sim {\cal V}^{\frac{65}{36}},$$

$$\Sigma_3\equiv V^{8/9} \Biggl[\frac{1}{6} {A_{z_1a_1}} \Biggl(-6 {A_{z_2z_2}} V^{2
   k+\frac{25}{36}}+2 {A_{a_1a_1}} V^{2 k+\frac{89}{36}}+$$

   $$\frac{2 \sqrt[3]{2} \left(V^{16/9}
   {A_{a_1a_1}}^2-({A_{z_1z_1}}+{A_{z_2z_2}}) {A_{a_1a_1}}+3 {A_{z_1a_1}}^2+3 {A_{z_2a_1}}^2\right)
   V^{4 k+\frac{19}{6}}}{\sqrt[3]{\Sigma}}+2^{2/3}
   \sqrt[3]{\Sigma}\Biggr) V^{-2 k-\frac{2}{3}}$$

   $$+{A_{z_1z_2}} {A_{z_2a_1}}
   \sqrt[36]{V}\Biggr]\sim {\cal V}^{-\frac{8}{9}}.$$

   Hence, the eigenvector corresponding to the first eigenvalue (\ref{eq:ev1_2}) is:
   \begin{equation}
   \label{eq:eigenvector1}
   \left(\begin{array}{c} \beta_1{\cal V}^{-\frac{8}{9}} \\ \beta_2{\cal V}^{-\frac{8}{9}} \\ 1
   \end{array}\right).
   \end{equation}
In the LVS limit, this is already normalized to unity.

$\underline{Eigenvectors\ 2\ and\ 3}$

Given the two-fold degeneracy of the second eigenvalue (\ref{eq:ev2_2}), of the three equations implied
by:
\begin{equation}
\label{eq:ev2_eqn1}
\hat{K}\left(\begin{array}{c}
\alpha_1 \\ \alpha_2 \\ \alpha_3 \end{array}\right)=\Lambda_2\frac{{\cal V}^{\frac{1}{36}}}{\sum_\beta n^0_\beta}\left(\begin{array}{c}
\alpha_1 \\ \alpha_2 \\ \alpha_3 \end{array}\right),
\end{equation}
only one equation is independent, say:
\begin{equation}
\label{eq:ev2_eqn2}
\alpha_1(A_{z_1z_1} - \Lambda_2) + \alpha_2 A_{z_1z_2} + \alpha_3 A_{z_1a_1} {\cal V}^{\frac{8}{9}} = 0.
\end{equation}
Two independent solutions to (\ref{eq:ev2_eqn2}) are:
\begin{eqnarray}
\label{eq:ev2_eqn3}
& & \alpha_1=0, \alpha_2=\frac{A_{z_1a_1}}{A_{z_1z_2}}{\cal V}^{\frac{8}{9}}\alpha_3;\nonumber\\
& & \alpha_2=0, \alpha_1=\frac{A_{z_1a_1}}{(A_{z_1z_1} - \Lambda_2)}{\cal V}^{\frac{8}{9}}\alpha_3.
\end{eqnarray}
Thus, the following are the remaining two linearly independent eigenvectors of (\ref{eq:Khat}):
\begin{equation}
\label{eq:eigenvectors2_3}
\left(\begin{array}{c}0 \\ 1\\ \lambda_1{\cal V}^{-\frac{8}{9}}\end{array}\right),\ \left(\begin{array}{c}
1\\ 0\\ \lambda_2{\cal V}^{-\frac{8}{9}}\end{array}\right).
\end{equation}
In the LVS limit, (\ref{eq:eigenvector1}) and (\ref{eq:eigenvectors2_3}) form an orthonormal set of eigenvectors corresponding to $\hat{K}$. Hence, for evaluating the physical $\mu$ terms, Yukawa couplings, etc., we will work with the following set of (fluctuation) fields:
\begin{eqnarray}
\label{eq:diagonal}
& & \delta\tilde{{\cal A}_1}\equiv (\beta_1\delta z_1 + \beta_2\delta z_2){\cal V}^{-\frac{8}{9}} +
\delta{\cal A}_1,\nonumber\\
& & \delta{\cal Z}_1\equiv \delta z_1 + \lambda_2\delta{\cal A}_1{\cal V}^{-\frac{8}{9}},\nonumber\\
& & \delta{\cal Z}_2\equiv \delta z_2 + \lambda_1\delta{\cal A}_1{\cal V}^{-\frac{8}{9}}.
\end{eqnarray}
For purposes of evaluation of the physical $\mu$ terms, Yukawa couplings, etc., we will need the following expressions for the square-root of the elements of the diagonalized $\hat{K}$ in the basis of (\ref{eq:diagonal}):
\begin{eqnarray}
& & \sqrt{\hat{K}_{\tilde{{\cal A}_1}{\bar{\tilde{\cal A}_1}}}}\sim\sqrt{\frac{1}{\sum_\beta n^0_\beta}\left(A_{a_1a_1}{\cal V}^{\frac{65}{36}} + \frac{\alpha_4}{3A_{a_1a_1}}\right)}
\sim\frac{{\cal V}^{\frac{65}{72}}}{\sqrt{\sum_\beta n^0_\beta}},\nonumber\\
& & \hskip-1in\sqrt{\hat{K}_{Z_1{\bar Z}_1}}=\sqrt{\hat{K}_{Z_2{\bar Z}_2}}\sim\sqrt{\frac{{\cal V}^{\frac{1}{36}}}{\sum_\beta n^0_\beta}\left(4(A_{z_1z_1} + A_{z_2z_2}) - \frac{2\left(-({A_{z_1z_1}}+{A_{z_2z_2}}) {A_{a_1a_1}}+3
   {A_{z_1a_1}}^2+3 {A_{z_2a_1}}^2\right)}{A_{a_1a_1}}\right)}\sim\frac{{\cal V}^{\frac{1}{72}}}{\sqrt{\sum_\beta n^0_\beta}}.\nonumber\\
   & &
\end{eqnarray}

From (\ref{eq:K2}), one sees that the coefficients of the ``pure" terms, $Z_{ij}$ are as given in (\ref{eq:Zcoeffs}) in appendix E. Quite interestingly, one can show that
\begin{equation}
\label{eq:Z2}
Z\sim\frac{{\cal V}^{\frac{1}{12}}}{\sum_\beta n^0_\beta}\left(\begin{array}{ccc}
\frac{{\cal V}^{\frac{1}{36}}}{\sum_\beta n^0_\beta} & \frac{{\cal V}^{\frac{1}{36}}}{\sum_\beta n^0_\beta} & \frac{{\cal V}^{\frac{11}{12}}}{\sum_\beta n^0_\beta} \\
\frac{{\cal V}^{\frac{1}{36}}}{\sum_\beta n^0_\beta} & \frac{{\cal V}^{\frac{1}{36}}}{\sum_\beta n^0_\beta} & \frac{{\cal V}^{\frac{11}{12}}}{\sum_\beta n^0_\beta} \\
\frac{{\cal V}^{\frac{11}{12}}}{\sum_\beta n^0_\beta} & \frac{{\cal V}^{\frac{11}{12}}}{\sum_\beta n^0_\beta} & \frac{{\cal V}^{\frac{65}{36}}}{\sum_\beta n^0_\beta} \\
\end{array}\right).
\end{equation}
Hence, the eigenvectors corresponding to the diagonalized $Z_{ij}$ are the same as that for $\hat{K}_{i{\bar j}}$ - hence (\ref{eq:eigenvector1}) and (\ref{eq:eigenvectors2_3}) simultaneously diagonalize $\hat{K}_{i{\bar j}}$ and $Z_{ij}$! The eigenvalues of (\ref{eq:Z2}) corresponding to the diagonalized $Z$ are:
\begin{eqnarray}
\label{eq:diagonal_Z}
& & Z_{{\cal Z}_1{\cal Z}_1}=6\left(Z_{z_1z_1} + Z_{z_2z_2}\right) - \frac{6}{Z_{{\cal A}_1{\cal A}_1}}\left(Z^2_{z_1{\cal A}_1}+Z^2_{z_2{\cal A}_1}\right)\sim{\cal V}^{-\frac{17}{9}};\nonumber\\
& & Z_{\tilde{{\cal A}_1}\tilde{{\cal A}_1}}=Z_{{\cal A}_1{\cal A}_1}\sim{\cal V}^{-\frac{1}{9}}.
\end{eqnarray}

Before we proceed to read-off the soft SUSY breaking terms, we would like to point out the following. The non-perturbative superpotential corresponding to an $ED3-$instanton obtained as an
$n^s$-fold wrapping of $D_4$ by a single $ED3$-brane as well as a single $D7$-brane wrapping $D_5$ taking the rigid limit of the wrapping, along with a space-time filling $D3$-brane restricted for purposes of definiteness and calculational convenience, to $D_5$ will be given by:
\begin{equation}
\label{eq:WD5}
W\sim\frac{\left[1 + z_1^{18} + z_2^{18} + \left(3\phi_0z_1^6z_2^6-z_1^{18}-z_2^{18}-1\right)^{\frac{2}{3}} - 3\phi_0z_1^6z_2^6\right]^{n^s}}{{\cal V}^{n^s}},
\end{equation}
which for $(z_1,z_2)\sim({\cal V}^{\frac{1}{36}},{\cal V}^{\frac{1}{36}})$, yields ${\cal V}^{-\frac{n^s}{2}}$. Hence, the gravitino mass
$m_{\frac{3}{2}}=e^{\frac{K}{2}}W M_p\sim{\cal V}^{-\frac{n^s}{2} - 1}M_p,$ which for $n^s=2, {\cal V}\sim10^7$ gives about $10TeV$.

Substituting
(\ref{eq:fluctuations}) into (\ref{eq:WD5}) (and again not being careful about ${\cal O}(1)$ numerical factors), one obtains:
\begin{eqnarray}
\label{eq:W_exp}
& & W\sim{\cal V}^{\frac{n^s}{2}}\Theta_{n^s}(\tau,{\cal G}^a)e^{in^sT(\sigma^S,{\bar\sigma^S};{\cal G}^a,{\bar{\cal G}^a};\tau,{\bar\tau})}\Biggl[1 + (\delta z_1 + \delta z_2)\Biggl\{n^s{\cal V}^{-\frac{1}{36}} + (in^s\mu_3)^3{\cal V}^{\frac{1}{36}}\Biggr\}\nonumber\\
& & +\delta\tilde{{\cal A}}_1\left\{-[\lambda_1+\lambda_2](in^s\mu_3){\cal V}^{-\frac{31}{36}} - n^s[\lambda_1+\lambda_2]{\cal V}^{-\frac{11}{12}}\right\}\Biggr]+ \left((\delta z_1)^2 + (\delta z_2)^2\right)\mu_{z_iz_i}  + \delta z_1\delta z_2\mu_{z_1z_2}\nonumber\\
& & +
\left(\delta\tilde{{\cal A}}_1\right)^2\mu_{\tilde{\cal A}_1\tilde{\cal A}_1} + \delta z_1\delta\tilde{{\cal A}}_1\mu_{z_1\tilde{\cal A}_1} + \delta z_2\delta\tilde{{\cal A}}_1\mu_{z_2\tilde{\cal A}_1} + \left((\delta z_1)^3 + (\delta z_2)^3\right)Y_{z_iz_iz_i} + \left((\delta z_1)^2\delta z_2 + (\delta z_2)^2\delta z_1\right)Y_{z_iz_iz_j}\nonumber\\
& & + (\delta z_1)^2\delta{\tilde{\cal A}}_1Y_{z_iz_i\tilde{\cal A}_1} + \delta z_1(\delta\tilde{\cal A}_1)^2Y_{z_i\tilde{\cal A}_1\tilde{\cal A}_1} + \delta z_1\delta z_2\delta\tilde{{\cal A}}_1Y_{z_1z_2\tilde{\cal A}_1} + \left(\delta\tilde{\cal A}_1\right)^3Y_{\tilde{\cal A}_1\tilde{\cal A}_1\tilde{\cal A}_1}+
 ....
\end{eqnarray}
The $\mu$ terms ($\mu_{ij}$) and the Yukawa couplings $Y_{ijk}$ are spelt out in {\bf  5.2} and {\bf 5.3} respectively.

\subsection{Gauginos' and Matter Fields' Masses}

We first need to evaluate the $F^m$ terms where $F^m=e^{\frac{\hat{K}}{2}}\hat{K}^{m{\bar n}}{\bar D}_{\bar n}{\bar W}=e^{\frac{\hat{K}}{2}}\hat{K}^{m{\bar n}}\left({\bar\partial}_{\bar n}{\bar W} + {\bar W}{\bar\partial}_{\bar n}K\right)$, for which we first need to evaluate $\hat{K}^{m{\bar n}}$. Using (\ref{eq:K2}), in the LVS limit, one obtains:
\begin{equation}
\label{eq:Khat_metric}
\hat{K}_{m{\bar n}}\sim\left(\begin{array}{cccc}
\frac{1}{{\cal V}^{\frac{37}{36}}} & \frac{1}{{\cal V}^{\frac{35}{18}}} & \frac{\kappa_{B1a}}{({\cal G}^a,{\bar{\cal G}^a})}{{\cal V}^{\frac{37}{36}}} & \frac{\kappa_{B2a}}{({\cal G}^a,{\bar{\cal G}^a})}{{\cal V}^{\frac{37}{36}}}\\
 \frac{1}{{\cal V}^{\frac{35}{18}}} & \frac{1}{{\cal V}^{\frac{37}{36}}} & \frac{\kappa_{S1a}}{({\cal G}^a,{\bar{\cal G}^a})}{{\cal V}^{\frac{37}{36}}} & \frac{\kappa_{S2a}}{({\cal G}^a,{\bar{\cal G}^a})}{{\cal V}^{\frac{37}{36}}}\\
\frac{\kappa_{B1a}}{({\cal G}^a,{\bar{\cal G}^a})}{{\cal V}^{\frac{37}{36}}} & \frac{\kappa_{S1a}}{({\cal G}^a,{\bar{\cal G}^a})}{{\cal V}^{\frac{37}{36}}} & k_1^2 & k_1k_2 \\
\frac{\kappa_{B2a}}{({\cal G}^a,{\bar{\cal G}^a})}{{\cal V}^{\frac{37}{36}}} & \frac{\kappa_{S2a}}{({\cal G}^a,{\bar{\cal G}^a})}{{\cal V}^{\frac{37}{36}}} & k_1k_2 & k_2^2 \\
\end{array}\right),
\end{equation}
which - in the LVS limmit - hence yields:
\begin{equation}
\label{eq:Khat_metric_inv}
\hat{K}^{m{\bar n}}\sim\left(\begin{array}{cccc}
{\cal V}^{\frac{37}{36}} & {\cal V}^{\frac{1}{9}} & ({\cal G}^a,{\bar{\cal G}^a}) & ({\cal G}^a,{\bar{\cal G}^a})\\
{\cal V}^{\frac{1}{9}} & {\cal V}^{\frac{37}{36}} & ({\cal G}^a,{\bar{\cal G}^a}) & ({\cal G}^a,{\bar{\cal G}^a})\\
({\cal G}^a,{\bar{\cal G}^a}) & ({\cal G}^a,{\bar{\cal G}^a}) & {\cal O}(1) & {\cal O}(1) \\
({\cal G}^a,{\bar{\cal G}^a}) & ({\cal G}^a,{\bar{\cal G}^a}) & {\cal O}(1) & {\cal O}(1) \\
\end{array}\right).
\end{equation}
Therefore,
\begin{eqnarray}
\label{eq:Fs}
& & F^{\sigma^B}\sim\frac{e^{-i\mu_3l^2{\cal V}^{\frac{1}{18}} - \frac{n^s}{2}}}{\cal V}\left({\cal V}^{\frac{1}{18}}\right)\sim{\cal V}^{-\frac{n^s}{2}-\frac{17}{18}};\nonumber\\
& & F^{\sigma^S}\sim\frac{e^{-i\mu_3l^2{\cal V}^{\frac{1}{18}} - \frac{n^s}{2}}}{\cal V}\left({\cal V}^{\frac{37}{36}}(n^s+{\cal V}^{-\frac{35}{36}}) + ({\cal G}^a,{\bar{\cal G}^a})\left[n^s(m^a + \frac{({\cal G}^a,{\bar{\cal G}^a})}{ln {\cal V}}) + {\cal V}^{-\frac{1}{6}}\right]\right)\nonumber\\
& & \sim n^s{\cal V}^{-\frac{n^s}{2}+\frac{1}{36}};\nonumber\\
& & F^{{\cal G}^a}\sim\frac{e^{-i\mu_3l^2{\cal V}^{\frac{1}{18}} - \frac{n^s}{2}}}{\cal V}\left(({\cal G}^a,{\bar{\cal G}^a})(n^s + {\cal V}^{-\frac{35}{36}}) + k^ak^b\left[n^s(m^b + \frac{({\cal G}^a,{\bar{\cal G}^a})}{ln {\cal V}}) + {\cal V}^{-\frac{1}{6}}\right]\right)\nonumber\\
& & \sim n^sk.mk^a{\cal V}^{-\frac{n^s}{2} - 1}
\end{eqnarray}

From (\ref{eq:Fs}), we conclude that the gaugino masses will of given by
\begin{eqnarray}
\label{eq:gaugino_mass}
& & \frac{F^m\partial_m T_B}{2Re(T_B)}\sim\frac{{\cal V}^{-\frac{n^s}{2}}\left({\cal V}^{-\frac{17}{18}} + \frac{n^sk.mk.{\cal G}}{\cal V}\right)}{{\cal V}^{\frac{1}{18}}}\nonumber\\
& & \sim{\cal V}^{-\frac{n^s}{2}-1}\sim m_{\frac{3}{2}},
\end{eqnarray}
where we have used the fact that the gravitino mass $m_{\frac{3}{2}}\sim{\cal V}^{-\frac{n^s}{2} - 1}$. Hence, what we see is that like the claims in the literature (See \cite{conloncal}, etc.), with the inclusion of $D3$- and $D7$-brane moduli, the gaugino masses are of the order of gravitino mass  - however, given that we are keeping the volume stabilized at around $10^6$ (in $l_s=1$-units) such that for $n^s=2$, $m_{\frac{3}{2}}\sim 10^3TeV$.

The open-string moduli or matter fields' masses are given by:
\begin{eqnarray}
\label{eq:matter_masses_1}
& & m_i^2 = m_{\frac{3}{2}}^2 + V_0 - F^m{\bar F}^{\bar n}\partial_m{\bar\partial}_{\bar n}ln \hat{K}_{i{\bar i}}\nonumber\\
& & = m_{\frac{3}{2}}^2 + V_0 + F^m{\bar F}^{\bar n}\left(\frac{1}{\hat{K}_{i{\bar i}}^2}\partial_m\hat{K}_{i{\bar i}}{\bar\partial}_{\bar n}\hat{K}_{i{\bar i}} - \frac{1}{\hat{K}_{i{\bar i}}}\partial_m{\bar\partial}_{\bar n}\hat{K}_{i{\bar i}}\right).
\end{eqnarray}
Using (\ref{eq:F_terms}), we estimate
\begin{equation}
\label{eq:V_0}
V_0\sim e^KG^{\sigma^\alpha{\bar\sigma}^\alpha}D_{\sigma^\alpha}W{\bar D}_{{\bar\sigma}^\alpha}{\bar W}\sim\sim\frac{(n^s)^2|W|^2{\cal V}^{\frac{19}{18}}}{{\cal V}^2}\sim{\cal V}^{-n^s - 2 + \frac{19}{18}}\sim {\cal V}^{\frac{19}{18}}m^2_{\frac{3}{2}}.
\end{equation}
Further, using (\ref{eq:Fs}) and results of appendix D, we arrive at the following results:
\begin{eqnarray}
\label{eq:matter_masses_2}
& & F^m{\bar F}^{{\bar n}}\partial_m{\bar\partial}_{\bar n}\hat{K}_{{\cal Z}_i{\bar{\cal Z}}_i}\sim{\cal V}^{-n^s - \frac{1}{18}}\sim m^2_{\frac{3}{2}}{\cal V}^{\frac{35}{36}};
\nonumber\\
& & F^m{\bar F}^{{\bar n}}\partial_m{\bar\partial}_{\bar n}\hat{K}_{\tilde{\cal A}_1{\bar{\tilde{\cal A}}_1}}\sim{\cal V}^{-n^s + \frac{1}{36}}\sim{\cal V}^{\frac{73}{36}}m^2_{\frac{3}{2}}.
\end{eqnarray}
Substituting (\ref{eq:V_0}) and (\ref{eq:matter_masses_2}) into (\ref{eq:matter_masses_1}), one obtains:
\begin{eqnarray}
\label{eq:matter_masses_3}
& & m_{{\cal Z}_i}^2\sim m^2_{\frac{3}{2}}\left(1 + {\cal V}^{\frac{19}{18}} + {\cal V}^{\frac{35}{36}}\right)\sim m^2_{\frac{3}{2}}{\cal V}^{\frac{19}{18}};\nonumber\\
& & m_{\tilde{\cal A}_1}^2\sim m^2_{\frac{3}{2}}\left(1 + {\cal V}^{\frac{19}{18}} +
{\cal V}^{\frac{73}{36}}\right)\sim {\cal V}^{\frac{73}{36}}m^2_{\frac{3}{2}},
\end{eqnarray}
implying
\begin{equation}
\label{eq:matter_masses_4}
m_{{\cal Z}_i}\sim {\cal V}^{\frac{19}{36}}m_{\frac{3}{2}},\ m_{\tilde{\cal A}_1}\sim {\cal V}^{\frac{73}{72}}m_{\frac{3}{2}}.
\end{equation}

\subsection{Physical $\mu$ Terms}

To evaluate the canonical ``physical" $\mu$ terms - denoted by $\hat{\mu}$ - one needs to evaluate $F^m\partial_mZ_{{\cal Z}_i{\cal Z}_i}$ and
$F^m\partial_mZ_{{\cal A}_1{\cal A}_1}$. Therefore, using (\ref{eq:Fs}) and (\ref{eq:dZ_z}), one obtains:
\begin{equation}
\label{eq:F.dZ_z}
F^m\partial_m Z_{{\cal Z}_i{\cal Z}_i}\sim{\cal V}^{-\frac{n^s}{2}-\frac{17}{9}}.
\end{equation}
Similarly, using (\ref{eq:dZ_a}) and (\ref{eq:Fs}), one obtains:
\begin{equation}
\label{eq:F.dZ_a}
F^m\partial_mZ_{{\cal A}_1{\cal A}_1}\sim{\cal V}^{-\frac{n^s}{2} - \frac{19}{18}}.
\end{equation}
Now,
\begin{equation}
\label{eq:muhatdef}
\hat{\mu}_{ij}=\frac{\frac{{\bar{\hat{W}}}e^{\frac{\hat{K}}{2}}}{|\hat{W}|}\mu_{ij} + m_{\frac{3}{2}}Z_{ij}\delta_{ij} - {\bar F}^{\bar m}{\bar\partial}_{\bar m}Z_{ij}\delta_{ij}}{\sqrt{\hat{K}_{i{\bar i}}\hat{K}_{j{\bar j}}}}.
\end{equation}

From (\ref{eq:W_exp}), one obtains the following non-zero $\mu$-terms:
\begin{eqnarray}
\label{eq:mu}
& & \mu_{{\cal Z}_i{\cal Z}_i}\sim e^{-\frac{n^s}{2}}\left[\left(\frac{n^s(n^s - 1)}{2} + n^s\right){\cal V}^{-\frac{1}{18}} + (in^s\mu_3)^3{\cal V}^{\frac{1}{18}} + n^s(i\mu_3n^s)\right];\nonumber\\
& & \mu_{{\cal Z}_1{\cal Z}_2}\sim e^{-\frac{n^s}{2}}\left[n^s\phi_0{\cal V}^{-\frac{2}{9}} + (in^s\mu_3)^3{\cal V}^{\frac{1}{18}}+ n^s(i\mu_3n^s)\right];\nonumber\\
& & \mu_{\tilde{{\cal A}_1}\tilde{{\cal A}_1}}\sim e^{-\frac{n^s}{2}}\left[{\cal V}^{-\frac{33}{18}}[\lambda_1^2+\lambda_2^2]\left[\frac{n^s(n^s-1)}{2} + n^s\right]
+ [\lambda_1^2+\lambda_2^2](in^s\mu_3)^2{\cal V}^{-\frac{31}{18}} [\lambda_1+\lambda_2]^2n^s(in^s\mu_3){\cal V}^{-\frac{16}{9}}\right];\nonumber\\
& & \mu_{{\cal Z}_i\tilde{{\cal A}_1}}\sim e^{-\frac{n^s}{2}}\left[-\lambda_j{\cal V}^{-\frac{17}{18}}\Biggl\{n^s+\frac{n^s(n^s-1)}{2} + n^s(in^s\mu_3)[\lambda_1+\lambda_2]{\cal V}^{-\frac{8}{9}}\Biggr\} + \lambda_j(in^s\mu_3)^3{\cal V}^{-\frac{5}{6}}\right],\ j\neq i(=1,2),\nonumber\\
& &
\end{eqnarray}
using which we get:
\begin{eqnarray}
\label{eq:muhat}
& & \hat{\mu}_{{\cal Z}_i{\cal Z}_j}\sim{\cal V}^{-\frac{n^s}{2} - 1}{\cal V}^{\frac{35}{36}}\sim{\cal V}^{\frac{37}{36}}m_{\frac{3}{2}};
\nonumber\\
& & \hat{\mu}_{{\cal A}_1{\cal Z}_i}\sim{\cal V}^{-\frac{3}{4}}m_{\frac{3}{2}};\nonumber\\
& & \hat{\mu}_{{\cal A}_1{\cal A}_1}\sim{\cal V}^{-\frac{33}{36}}m_{\frac{3}{2}}.
\end{eqnarray}

\subsection{Physical Yukawa couplings ($\hat{Y}_{ijk}$) and $A$-parameters ($A_{ijk}$) }

From (\ref{eq:W_exp}), one obtains the following non-zero Yukawa couplings:
\begin{eqnarray}
\label{eq:Ys}
& & Y_{{\cal Z}_i{\cal Z}_i{\cal Z}_i}\sim e^{-\frac{n^s}{2}}\left\{n^s{\cal V}^{\frac{1}{3}} + (in^s\mu_3)^3{\cal V}^{\frac{1}{12}} + n^s(in^s\mu_3)^2{\cal V}^{\frac{1}{36}} + \left[\frac{n^s(n^s-1)}{2}+n^s(in^s\mu_3){\cal V}^{-\frac{1}{36}}\right]\right\};\nonumber\\
& & Y_{{\cal Z}_i^2{\cal Z}_j}\sim e^{-\frac{n^s}{2}}\Biggl\{{\cal V}^{-\frac{1}{12}}\left(\frac{n^s(n^s-1)(n^s-2)}{6} + \frac{n^s(n^s-1)}{2}\right) + (in^s\mu_3)^3{\cal V}^{\frac{1}{12}}\nonumber\\
& & + {\cal V}^{\frac{1}{36}}(in^s\mu_3)^2n^s + {\cal V}^{-\frac{1}{36}}(in^s\mu_3)\Biggl[n^s+\frac{n^s(n^s-1)}{2}\Biggr]\Biggr\};\nonumber\\
& & Y_{\tilde{\cal A}_1\tilde{\cal A}_1\tilde{\cal A}_1}\sim e^{-\frac{n^s}{2}}\Biggl\{n^s{\cal V}^{-\frac{7}{3}}(\lambda_1^3+\lambda_2^3) + (in^s\mu_3)^3{\cal V}^{-\frac{31}{12}} + {\cal V}^{-\frac{95}{36}}n^s[\lambda-1+\lambda_2](in^s\mu_3)^2[\lambda_1^2+\lambda_2^2]\nonumber\\
 & & + {\cal V}^{-\frac{97}{36}}(in^s\mu_3)\Biggl[n^s+\frac{n^s(n^s-1)}{2}\Biggr][\lambda_1+\lambda_2][\lambda_1^2+\lambda_2^2]\Biggr\};\nonumber\\
& & \hskip-0.5inY_{{\cal Z}_i^2\tilde{{\cal A}}_1}\sim e^{-\frac{n^s}{2}}\left\{-\lambda_2{\cal V}^{-\frac{5}{9}} - \lambda_2{\cal V}^{-\frac{29}{36}}(in^s\mu_3)^3+ \Biggl[n^s+\frac{n^s(n^s-1)}{2}\Biggr](in^s\mu_3)[\lambda_1+\lambda_2]{\cal V}^{-\frac{11}{12}} + {\cal V}^{-\frac{31}{36}}[\lambda_1+\lambda_2]n^s(in^s\mu_3)^2\right\};\nonumber\\
& & Y_{(\tilde{{\cal A}}_1)^2{\cal Z}_i}\sim e^{-\frac{n^s}{2}}\Biggl\{-n^s\lambda_2^2{\cal V}^{-\frac{13}{9}} + (\lambda_{j(\neq i)}^2+2\lambda_1\lambda_2)(in^s\mu_3)^3{\cal V}^{-\frac{61}{36}} + {\cal V}^{-\frac{7}{4}}n^s(in^s\mu_3)^2[\lambda_1^2+\lambda_2^2]\nonumber\\
 & & + {\cal V}^{-\frac{65}{36}}[\lambda_1^2+\lambda_2^2]\Biggl[n^s+\frac{n^s(n^s-1)}{2}\Biggr](in^s\mu_3)\Biggr\};\nonumber\\
& & Y_{{\cal Z}_1{\cal Z}_2\tilde{{\cal A}}_1}\sim e^{-\frac{n^s}{2}}\Biggl\{-2(\lambda_1+\lambda_2){\cal V}^{-\frac{35}{36}}\left[\frac{n^s(n^s-1)(n^s-2)}{6} + \frac{n^s(n^s-1)}{2}\right] - 2(\lambda_1+\lambda_2){\cal V}^{-\frac{29}{36}}(in^s\mu_3)^3\nonumber\\
& & +{\cal V}^{-\frac{39}{36}}(n^s\phi_0)(in^s\mu_3)[\lambda_1+\lambda_2] + {\cal V}^{-\frac{31}{36}}n^s[\lambda_1+\lambda_2](in^s\mu_3)^2\nonumber\\
& & + {\cal V}^{-\frac{11}{12}}(in^s\mu_3)\Biggl[n^s+\frac{n^s(n^s-1)}{2}\Biggr][\lambda_1+\lambda_2]\Biggr\}.
\end{eqnarray}
Given the following definition of the physical Yukawa couplings:
\begin{equation}
\label{eq:physicalYdef}
\hat{Y}_{ijk}=\frac{e^{\frac{\hat{K}}{2}}Y_{ijk}}{\sqrt{\hat{K}_{i{\bar i}}\hat{K}_{j{\bar j}}\hat{K}_{k{\bar k}}}},
\end{equation}
one obtains the following non-zero physical Yukawa couplings $\hat{Y}_{ijk}$s:
\begin{eqnarray}
\label{eq:physical_Ys}
& & \hat{Y}_{{\cal Z}_i{\cal Z}_i{\cal Z}_i}\sim {\cal V}^{\frac{43}{24}}m_{\frac{3}{2}};\nonumber\\
& & \hat{Y}_{{\cal Z}_i^2{\cal Z}_j}\sim {\cal V}^{\frac{37}{24}}m_{\frac{3}{2}};\nonumber\\
& & \hat{Y}_{\tilde{\cal A}_1\tilde{\cal A}_1\tilde{\cal A}_1}\sim {\cal V}^{-\frac{85}{24}}m_{\frac{3}{2}};\nonumber\\
& & \hat{Y}_{{\cal Z}_i^2\tilde{\cal A}_1}\sim {\cal V}^{\frac{1}{72}}m_{\frac{3}{2}};\nonumber\\
& & \hat{Y}_{\tilde{\cal A}_1^2{\cal Z}_i}\sim {\cal V}^{-\frac{127}{72}}m_{\frac{3}{2}};\nonumber\\
& & \hat{Y}_{{\cal Z}_1{\cal Z}_2\tilde{\cal A}_1}\sim {\cal V}^{-\frac{17}{72}}m_{\frac{3}{2}}.
\end{eqnarray}

The $A$-terms are defined as:
\begin{equation}
\label{eq:Adef}
A_{ijk}=F^m\left[\hat{K}_m + \partial_m ln Y_{ijk} - \partial_m ln\left(\hat{K}_{i{\bar i}}\hat{K}_{j{\bar j}}\hat{K}_{k{\bar k}}\right)\right].
\end{equation}
Using:
\begin{eqnarray}
\label{eq:dervs_Ys}
& & \partial_{\sigma^B}Y_{ijk}\sim0;\nonumber\\
& & \partial_{\sigma^S}Y_{ijk}\sim n^sY_{ijk};\nonumber\\
& & \partial_{{\cal G}^a}Y_{ijk}\sim({\cal G}^a,{\bar{\cal G}}^a)Y_{ijk},
\end{eqnarray}
and (\ref{eq:Fs})one obtains:
\begin{equation}
\label{eq:F.dY}
F^m\partial_mY_{ijk}\sim n^s{\cal V}^{-\frac{n^s}{2}-\frac{35}{36}}Y_{ijk}\sim n^s{\cal V}^{\frac{1}{36}}m_{\frac{3}{2}}.
\end{equation}
Using:
\begin{eqnarray}
\label{eq:ders_Khat}
& & \partial_{\sigma^\alpha}\hat{K}\sim \frac{\sqrt{{\cal T}_S(\sigma^\alpha,{\bar\sigma^\alpha};{\cal G}^a,{\bar{\cal G}^a};\tau,{\bar\tau}) + \mu_3{\cal V}^{\frac{1}{18}}  - \gamma\left(r_2 + \frac{r_2^2\zeta}{r_1}\right)}}{\Xi}\sim{\cal V}^{-\frac{35}{36}};\nonumber\\
& & \partial_{{\cal G}^a}\hat{K}\sim\frac{1}{\Xi}\times\Biggl[\sum_\beta k^an^0_\beta sin(...) + \left({\cal G}^a,{\bar{\cal G}}^a\right)\nonumber\\
& & \hskip-1in\times\Biggl\{\sqrt{{\cal T}_B(\sigma^B,{\bar\sigma ^B};{\cal G}^c,{\bar{\cal G}^c};\tau,{\bar\tau}) + \mu_3{\cal V}^{\frac{1}{18}} + i\kappa_4^2\mu_7C_{1{\bar 1}}{\cal V}^{-\frac{1}{2}} - \gamma\left(r_2 + \frac{r_2^2\zeta}{r_1}\right)}\kappa_{Bac}\nonumber\\
& & -\sqrt{{\cal T}_S(\sigma^S,{\bar\sigma ^S};{\cal G}^a,{\bar{\cal G}^a};\tau,{\bar\tau}) + \mu_3{\cal V}^{\frac{1}{18}}  - \gamma\left(r_2 + \frac{r_2^2\zeta}{r_1}\right)}\kappa_{Sac}\Biggr\}\Biggr]\sim{\cal V}^{-\frac{1}{6}}\nonumber\\
& & ({\rm having\ taken}\ \sum_\beta k^an^0_\beta sin(...)\sim{\cal V}^{\frac{5}{6}}),
\end{eqnarray}
and (\ref{eq:Fs}), one obtains:
\begin{equation}
\label{eq:F.dKhat}
F^m\partial_m\hat{K}\sim n^s{\cal V}^{-\frac{n^s}{2}-\frac{17}{18}}\sim n^s{\cal V}^{\frac{1}{18}}m_{\frac{3}{2}}.
\end{equation}
Finally, using (\ref{eq:dKhat_z}) and (\ref{eq:Fs}):
\begin{equation}
\label{eq:F.dlnKhat_zz}
F^m\partial_m ln\hat{K}_{{\cal Z}_i{\cal Z}_i}\sim n^s{\cal V}^{-\frac{n^s}{2}-\frac{1}{36}}\sim n^s  {\cal V}^{\frac{35}{36}}m_{\frac{3}{2}},
\end{equation}
and using (\ref{eq:Fs}) and (\ref{eq:dKhat_a}):
\begin{equation}
\label{eq:F.dlnKhat_aa}
F^m\partial_m ln\hat{K}_{\tilde{{\cal A}_1}{\bar{\tilde{\cal A}_1}}}\sim n^s{\cal V}^{-\frac{n^s}{2}-\frac{17}{18}}\sim n^s {\cal V}^{\frac{1}{18}}m_{\frac{3}{2}}.
\end{equation}
Hence, substituting (\ref{eq:F.dY}), (\ref{eq:F.dKhat}), (\ref{eq:F.dlnKhat_zz}) and (\ref{eq:F.dlnKhat_aa}) into (\ref{eq:Adef}), one obtains:
\begin{eqnarray}
\label{eq:A}
& & A_{{\cal Z}_i{\cal Z}_j{\cal Z}_k}\sim n^s{\cal V}^{\frac{37}{36}}m_{\frac{3}{2}};\nonumber\\
& & A_{\tilde{{\cal A}_1}\tilde{{\cal A}_1}\tilde{{\cal A}_1}}\sim n^s{\cal V}^{\frac{37}{36}}m_{\frac{3}{2}};\nonumber\\
& & A_{\tilde{{\cal A}_1}^2{\cal Z}_i}\sim n^s{\cal V}^{\frac{37}{36}}m_{\frac{3}{2}};\nonumber\\
& & A_{\tilde{{\cal A}_1}{\cal Z}_i{\cal Z}_j}\sim n^s{\cal V}^{\frac{37}{36}}m_{\frac{3}{2}}.
\end{eqnarray}

\subsection{The $\hat{\mu}B$ Parameters}

The $\hat{\mu}B$-parameters are defined as under:
\begin{eqnarray}
\label{eq:muhatB_def}
& & (\hat{\mu}B)_{ij}=\frac{1}{\sqrt{\hat{K}_{i{\bar i}}\hat{K}_{j{\bar j}}}}\nonumber\\
& & \times\Biggl\{\frac{{\bar{\hat{W}}}}{|\hat{W}|}e^{\frac{\hat{K}}{2}}
\left[F^m\left(\hat{K}_m\mu_{ij} + \partial_m\mu_{ij} - \mu_{ij}\partial_m ln\left(\hat{K}_{i{\bar i}}\hat{K}_{j{\bar j}}\right)\right) - m_{\frac{3}{2}}\mu_{ij}\right]\nonumber\\
& & + \left(2m^2_{\frac{3}{2}} + V_0\right)Z_{ij}\delta_{ij} - m_{\frac{3}{2}}{\bar F}^{\bar m}{\bar\partial}_{\bar n}Z_{ij}\delta_{ij} + m_{\frac{3}{2}}\delta_{ij}F^m\left[\partial_m Z_{ij} - Z_{ij}\partial_m ln\left(\hat{K}_{i{\bar i}}\hat{K}_{j{\bar j}}\right)\right]
\nonumber\\
& & - \delta_{ij}{\bar F}^{\bar m}F^n\left[{\bar\partial}_{\bar m}\partial_nZ_{ij} - {\bar\partial}_{\bar m} Z_{ij}\partial_n ln\left(\hat{K}_{i{\bar i}}\hat{K}_{j{\bar j}}\right)\right]\Biggr\},
\end{eqnarray}
where $\delta_{ij}$ has been put in before the $Z_{ij}$-dependent terms to indicate that we are working with the diagonalized matter fields (\ref{eq:diagonal}).

Substituting (\ref{eq:F.dKhat}), (\ref{eq:F.dlnKhat_zz}), (derivatives w.r.t the closed string moduli $\sigma^\alpha, {\cal G}^a$ of) (\ref{eq:mu}),
(\ref{eq:Zcoeffs}), (\ref{eq:diagonal_Z}), (\ref{eq:Fs}), (\ref{eq:F.dZ_z}) and (\ref{eq:ddZ_z}):
\begin{eqnarray}
\label{eq:muhatB_zz_intermediate_1}
& & (a)\ \frac{{\bar{\hat{W}}}}{|\hat{W}|}e^{\frac{\hat{K}}{2}}F^m\partial_m\hat{K}\mu_{{\cal Z}_i{\cal Z}_i}\sim {\cal V}^{\frac{1}{9}}m_{\frac{3}{2}}^2,
\nonumber\\
& & (b)\  \frac{{\bar{\hat{W}}}}{|\hat{W}|}e^{\frac{\hat{K}}{2}}F^m\partial_m\mu_{{\cal Z}_i{\cal Z}_i}\sim{\cal V}^{\frac{13}{12}}m_{\frac{3}{2}}^2,\nonumber\\
& & (c)\ \frac{{\bar{\hat{W}}}}{|\hat{W}|}e^{\frac{\hat{K}}{2}}\mu_{{\cal Z}_i{\cal Z}_i}F^m\partial_m ln\left(\hat{K}_{{\cal Z}_i{\bar{\cal Z}}_i}\right)
\sim {\cal V}^{\frac{37}{36}}m_{\frac{3}{2}}^2,\nonumber\\
& & (d)\ \frac{{\bar{\hat{W}}}}{|\hat{W}|}e^{\frac{\hat{K}}{2}}\mu_{{\cal Z}_i{\cal Z}_i}\sim{\cal V}^{\frac{1}{18}}m_{\frac{3}{2}}^2,
\end{eqnarray}
which gives:
\begin{eqnarray}
\label{eq:muhat_zz_intermediate_2}
& & \frac{(a)+(b)-2(c)-(d)}{\hat{K}_{{\cal Z}_i{\cal Z}_i}}\sim{\cal V}^{\frac{37}{18}}m_{\frac{3}{2}}^2.
\end{eqnarray}
Further, using (\ref{eq:F_terms}),
\begin{eqnarray}
\label{eq:muhatB_zz_intermediate_3}
& & \frac{\left(m_{\frac{3}{2}}^2+V_0\right)Z_{{\cal Z}_i{\cal Z}_i}}{\hat{K}_{{\cal Z}_i{\cal Z}_i}}\sim{\cal V}^{\frac{5}{36}}m_{\frac{3}{2}}^2,
\nonumber\\
& & \frac{m_{\frac{3}{2}}{\bar F}^{\bar m}{\bar\partial}_{\bar m}Z_{{\cal Z}_i{\cal Z}_i}}{\hat{K}_{{\cal Z}_i{\cal Z}_i}}\sim{\cal V}^{\frac{1}{12}}
m_{\frac{3}{2}}^2,\nonumber\\
& & \frac{m_{\frac{3}{2}}\left(F^m\partial_mZ_{{\cal Z}_i{\cal Z}_i} - 2Z_{{\cal Z}_i{\cal Z}_i}F^m\partial_m ln\left(\hat{K}_{{\cal Z}_i{\bar{\cal Z}}_i}\right)\right)}{\hat{K}_{{\cal Z}_i{\cal Z}_i}}\sim{\cal V}^{\frac{1}{12}}m_{\frac{3}{2}}^2,\nonumber\\
& & \frac{\left[{\bar F}^{\bar m}F^n{\bar\partial}_{\bar m}\partial_nZ_{{\cal Z}_i{\cal Z}_i} - 2 {\bar F}^{\bar m}F^n\left({\bar\partial}_{\bar m}Z_{{\cal Z}_i{\cal Z}_i}\right)\left(\partial_n ln\left(\hat{K}_{{\cal Z}_i{\cal Z}_i}\right)\right)\right]}{\hat{K}_{{\cal Z}_i{\cal Z}_i}}
\sim {\cal V}^{\frac{223}{108}}m_{\frac{3}{2}}^2.
\end{eqnarray}
Note, when substituting in the first equation of (\ref{eq:F_terms}) as the extremum value of the potential $V_0$ in (\ref{eq:muhatB_zz_intermediate_3}), we have assumed the following. For a non-supersymmetric configuration, from \cite{V_D7_fl} we see that the tadpole cancelation guarantees that the contributions to the potential from all the $D3$-branes and $O3$-planes as well as the $D7$-branes and $O7$-planes cancel out. However, there is still a $D$-term contribution from the
$U(1)$-fluxes on the world-volume of the $D7$-branes wrapped around $D_5$ of the form
$\frac{\left({\cal F}^\beta\kappa_{\alpha\beta}\partial_{T_\alpha}K\right)^2}{\left(Re(T_B) - {\cal F}Re(i\tau)\right)}$ - we drop the same in the dilute flux approximation as  was done in section {\bf 2}.

From (\ref{eq:muhat_zz_intermediate_2}) and (\ref{eq:muhatB_zz_intermediate_3}), one obtains:
\begin{eqnarray}
\label{eq:muhatB_zz_final}
& & \left(\hat{\mu}B\right)_{{\cal Z}_i{\cal Z}_i}\sim{\cal V}^{\frac{223}{108}}m_{\frac{3}{2}}^2,\nonumber\\
& &  \left(\hat{\mu}B\right)_{{\cal Z}_1{\cal Z}_2}\sim{\cal V}^{\frac{37}{18}}m_{\frac{3}{2}}^2.
\end{eqnarray}

Similarly,
\begin{eqnarray}
\label{eq:muhatB_aa_intermediate_1}
& & (a)\ \frac{{\bar{\hat{W}}}}{|\hat{W}|}e^{\frac{\hat{K}}{2}}F^m\partial_m\hat{K}\mu_{\tilde{\cal A}_1\tilde{\cal A}_1}\sim {\cal V}^{\frac{-5}{3}}m_{\frac{3}{2}}^2,
\nonumber\\
& & (b)\  \frac{{\bar{\hat{W}}}}{|\hat{W}|}e^{\frac{\hat{K}}{2}}F^m\partial_m\mu_{\tilde{\cal A}_1\tilde{\cal A}_1}\sim{\cal V}^{-\frac{25}{36}}m_{\frac{3}{2}}^2,\nonumber\\
& & (c)\ \frac{{\bar{\hat{W}}}}{|\hat{W}|}e^{\frac{\hat{K}}{2}}\mu_{\tilde{\cal A}_1\tilde{\cal A}_1}F^m\partial_m ln\left(\hat{K}_{\tilde{{\cal A}_1}{\bar{\tilde{\cal A}_1}}}\right)
\sim {\cal V}^{-\frac{5}{3}}m_{\frac{3}{2}}^2,\nonumber\\
& & (d)\ \frac{{\bar{\hat{W}}}}{|\hat{W}|}e^{\frac{\hat{K}}{2}}\mu_{\tilde{\cal A}_1\tilde{\cal A}_1}\sim{\cal V}^{-\frac{13}{18}}m_{\frac{3}{2}}^2,
\end{eqnarray}
which gives:
\begin{eqnarray}
\label{eq:muhat_aa_intermediate_2}
& & \frac{(a)+(b)-2(c)-(d)}{\hat{K}s_{\tilde{{\cal A}_1}{\bar{\tilde{\cal A}_1}}}}\sim{\cal V}^{-\frac{3}{2}}m_{\frac{3}{2}}^2.
\end{eqnarray}
Further,
\begin{eqnarray}
\label{eq:muhatB_aa_intermediate_3}
& & \frac{\left(m_{\frac{3}{2}}^2+V_0\right)Z_{\tilde{\cal A}_1\tilde{\cal A}_1}}{\hat{K}_{\tilde{{\cal A}_1}{\bar{\tilde{\cal A}_1}}}}\sim{\cal V}^{\frac{5}{36}}m_{\frac{3}{2}}^2,
\nonumber\\
& & \frac{m_{\frac{3}{2}}{\bar F}^{\bar m}{\bar\partial}_{\bar m}Z_{\tilde{\cal A}_1\tilde{\cal A}_1}}{\hat{K}_{\tilde{{\cal A}_1}{\bar{\tilde{\cal A}_1}}}}\sim{\cal V}^{\frac{31}{36}}
m_{\frac{3}{2}}^2,\nonumber\\
& & \frac{m_{\frac{3}{2}}\left(F^m\partial_mZ_{\tilde{\cal A}_1\tilde{\cal A}_1} - 2Z_{\tilde{\cal A}_1\tilde{\cal A}_1}F^m\partial_m ln\left(\hat{K}_{\tilde{{\cal A}_1}{\bar{\tilde{\cal A}_1}}}\right)\right)}{\hat{K}_{\tilde{{\cal A}_1}{\bar{\tilde{\cal A}_1}}}}\sim{\cal V}^{\frac{31}{36}}m_{\frac{3}{2}}^2,\nonumber\\
& & \frac{\left[{\bar F}^{\bar m}F^n{\bar\partial}_{\bar m}\partial_nZ_{\tilde{\cal A}_1\tilde{\cal A}_1} - 2 {\bar F}^{\bar m}F^n\left({\bar\partial}_{\bar m}Z_{\tilde{\cal A}_1\tilde{\cal A}_1}\right)\left(\partial_n ln\left(\hat{K}_{\tilde{\cal A}_1\tilde{\cal A}_1}\right)\right)\right]}{\hat{K}_{\tilde{{\cal A}_1}{\bar{\tilde{\cal A}_1}}}}
\sim {\cal V}^{\frac{1}{9}}m_{\frac{3}{2}}^2.
\end{eqnarray}
From (\ref{eq:muhat_aa_intermediate_2}) and (\ref{eq:muhatB_aa_intermediate_3}), one obtains:
\begin{equation}
\label{eq:muhatB_aa_final}
\left(\hat{\mu}B\right)_{\tilde{\cal A}_1\tilde{\cal A}_1}\sim{\cal V}^{\frac{5}{36}}m_{\frac{3}{2}}^2.
\end{equation}
Finally,
\begin{eqnarray}
\label{eq:muhatB_za_intermediate_1}
& & (a)\ \frac{{\bar{\hat{W}}}}{|\hat{W}|}e^{\frac{\hat{K}}{2}}F^m\partial_m\hat{K}\mu_{{\cal Z}_i\tilde{\cal A}_1}\sim {\cal V}^{-\frac{16}{9}}m_{\frac{3}{2}}^2,
\nonumber\\
& & (b)\  \frac{{\bar{\hat{W}}}}{|\hat{W}|}e^{\frac{\hat{K}}{2}}F^m\partial_m\mu_{{\cal Z}_i\tilde{\cal A}_1}\sim{\cal V}^{-\frac{29}{36}}m_{\frac{3}{2}}^2,\nonumber\\
& & (c)\ \frac{{\bar{\hat{W}}}}{|\hat{W}|}e^{\frac{\hat{K}}{2}}\mu_{{\cal Z}_i\tilde{\cal A}_1}F^m\partial_m ln\left(\hat{K}_{\tilde{{\cal A}_1}{\bar{\tilde{\cal A}_1}}}\hat{K}_{{\cal Z}_i{\bar{\cal Z}}_i}\right)
\sim {\cal V}^{-\frac{31}{36}}m_{
\frac{3}{2}}^2,\nonumber\\
& & (d)\ \frac{{\bar{\hat{W}}}}{|\hat{W}|}e^{\frac{\hat{K}}{2}}\mu_{{\cal Z}_i\tilde{\cal A}_1}\sim{\cal V}^{-\frac{11}{6}}m_{\frac{3}{2}}^2,
\end{eqnarray}
which gives:
\begin{eqnarray}
\label{eq:muhat_za_intermediate_2}
& & \frac{(a)+(b)-(c)-(d)}{\sqrt{\hat{K}_{\tilde{{\cal A}_1}{\bar{\tilde{\cal A}_1}}}\hat{K}_{{\cal Z}_i
{\bar{\cal Z}}_i}}}=\left(\hat{\mu}B\right)_{{\cal Z}_i\tilde{{\cal A}_1}}\sim{\cal V}^{-\frac{13}{18}}m_{\frac{3}{2}}^2.
\end{eqnarray}

\section{Summary and Discussion}

In this article, we have discussed several phenomenological issues in the context of L(arge) V(olume) S(cenarios) Swiss-Cheese orientifold compactifications of type IIB with
the inclusion of a single mobile space-time filling $D3$-brane  and stack@ÏÏð8@vU€ wrapping the ``big" divisor along with  supporting $D7$-brane fluxes (on two-cycles homologously non-trivial within the big divisor, and not the Calabi-Yau). Interestingly we have found several phenomenological implications which have been different from the LVS studies done so far in the literature.

We have proposed a possible resolution for the long-standing tension between LVS cosmology and LVS phenomenology : to figure out a way of obtaining a TeV gravitino when dealing with LVS phenemenology and a $10^{12}$ GeV gravitino when dealing with LVS cosmology in the early inflationary epoch of the universe, within the same setup.  The holomorphic pre-factor coming from the space-time filling mobile D3-brane position moduli  - section of (the appropriate) divisor bundle - plays a crucial role and we have shown that as the mobile space-time filling $D3$-brane moves from a particular non-singular elliptic curve embedded in the Swiss-Cheese Calabi-Yau to another non-singular elliptic curve, it is possible to obtain $10^{12}GeV$ gravitino during the primordial inflationary era supporting the cosmological/astrophysical data as well as a $TeV$ gravitino in the present era supporting the required SUSY breaking at $TeV$ scale within the same set up, for the same volume of the Calabi-Yau stabilized at around $10^6$ (in $l_s=1$ units). \footnote{There has been a different proposal \cite{tension1}, which involves treating the volume modulus as an inflaton which starts off at a small value for incorporating slow-roll inflation and then evolves  over a long range and finally stabilizes to the large volume minimum with TeV gravitino mass after inflation.} This way the string scale involved for our case is $\sim O(10^{15})$ GeV which is nearly of the same order as GUT scale. In the context of soft SUSY breaking, we have obtained the gravitino mass $m_{3/2}\sim  O(1-10^3)$ TeV  with ${\cal V}\sim 10^{6} {l_s}^6$ in our setup. We have found the gravity mediated gaugino masses to be of the same order as the gravitino mass.

While realizing the Standard Model (SM) gauge coupling $g_{YM}$ $ \sim O(1)$ in the LVS models with D7-branes, usually models with the D7-branes wrapping the smaller divisor have been proposed so far, as D7-branes wrapping the big divisor would produce very small gauge couplings. In our setup, we have realized $\sim O(1)$ $g_{YM}$ (at the tree level) with D7-branes wrapping the big divisor in the rigid limit (i.e. considering zero sections of the normal bundle of the big divisor to prevent any obstruction to chiral matter resulting from adjoint matter - corresponding to fluctuations of the wrapped $D7$-branes within the Calabi-Yau - giving mass to open strings stretched between wrapped $D7$-branes)  implying the new possibility of supporting SM on D7-branes wrapping the big divisor. This has been possible because after constructing appropriate local  involutively-odd harmonic one-forms on the big divisor lying in the cokernel of the pullback of the immersion map applied to $H^{(1,0)}_-$ in the large volume limit, the Wilson line moduli provide a competing contribution to the gauge kinetic function as compared to the volume of the big divisor. This requires the complexified Wilson line moduli to be stabilized at around ${\cal V}^{-\frac{1}{4}}$ (which has been justified). Note, similar to the case of local models corresponding to wrapping of $D7$-branes around the small divisor, our model is also local in the sense that the involutively-odd one-forms are constructed locally around the location of the mobile $D3$-brane restricted to (the rigid limit of) $D_5$.

Also, from the first reference in \cite{susyinitials}, the effective gauge couplings $g_a^{-2}$ for an observable gauge group $G_a$ including renormalization and string-loop corrections, to all orders, at an energy scale $\nu>>m_{\frac{3}{2}}$ satisfies the following equation:
\begin{eqnarray}
\label{eq:g_a-EXACT}
& & g_a^{-2}\left(\Phi^m,{\bar\Phi}^{\bar m};\nu\right) = Re f_a + \frac{\sum_r n_rT_a(r) - 3 T_{\rm adj}(G_a)}{8\pi^2}ln\frac{M_p}{\nu} + \frac{\sum_r n_rT_a(r) -  T_{\rm adj}(G_a)}{16\pi^2}\hat{K}\left(\Phi^m,{\bar\Phi}^{\bar m}\right)\nonumber\\
 & & + \frac{T_{\rm adj}(G_a)}{8\pi^2} ln \left[g_a^{-2}\left(\Phi^m,{\bar\Phi}^{\bar m};\nu\right)\right]- \sum_r\frac{T_a(r)}{8\pi^2} ln\ {\rm det}\left[\hat{K}_{i{\bar j}}\left(\Phi^m,{\bar\Phi}^{\bar m}\right)\right]
\end{eqnarray}
$\Phi^m$ denoting the closed string moduli, $f_a$ being the gauge ($G_a$) kinetic function, $r$ denoting a representation for an observable gauge group $G_a$, $n_r$ denoting the number of matter fields transforming under the representation $r$ of $G_a$ and $T$ denoting the trace of the square of the generators in the appropriate representations. Given that we have been working in the approximation:
$\mu_3{\cal V}^{\frac{1}{18}}\sim ln{\cal V}$ (justified by ${\cal V}\sim10^6$), from (\ref{eq:K2}), (\ref{eq:Khat}) and (\ref{eq:detKhat}) one sees that the  third and fifth terms on the RHS of (\ref{eq:g_a-EXACT}) are proportional to $ln{\cal V}\sim\mu_3{\cal V}^{\frac{1}{18}}\sim f_a$, implying thereby that there are no major modifications in the tree-level results for the gauge couplings.

On the geometric side to enable us to work out the complete K\"{a}hler potential, we have calculated the geometric K\"{a}hler potential (of the two divisors $D_4$ and $D_5$) for Swiss-Cheese Calabi-Yau ${\bf WCP}^4[1,1,1,6,9]$ using its toric data and GLSM techniques in the large volume limit. The geometric K\"{a}hler potential is first expressed, using a general theorem due to Umemura, in terms of genus-five Siegel Theta functions or in the LVS limit genus-four Siegel Theta functions. Later using a result due to Zhivkov, for purposes of calculations for our paper, we express the same in terms of derivatives of genus-two Siegel Theta functions.

Further, let's look at the anomaly-mediated gaugino masses which are given by (See \cite{Bagger_et_al,Alwis}):
\begin{equation}
\label{eq:gaugino_mass_anomaly_med_1}
\frac{m_a}{g_a^2}=-\frac{\left[-\left(\sum_r n_rT_a(r) - 3 T_{\rm adj}(G_a)\right)m_{\frac{3}{2}} - \left(\sum_r n_rT_a(r) -  T_{\rm adj}(G_a)\right)F^m\hat{K}_m + 2\sum_r T_a(r)F^m\partial_m ln\ {\rm det}\left(\hat{K}_{i{\bar j}}\right)\right]}{8\pi^2}.
\end{equation}
Using $F^m\partial_m\hat{K}\sim m_{\frac{3}{2}}{\cal V}^{\frac{1}{18}}$ and (\ref{eq:F.dlndetKhat}), one sees that:
\begin{equation}
\label{eq:eq:gaugino_mass_anomaly_med_2}
\frac{m_a}{g_a^2}\sim\frac{{\cal V}^{\frac{1}{18}}m_{\frac{3}{2}}}{8\pi^2},
\end{equation}
which using $g_a^2\sim{\cal V}^{-\frac{1}{18}}$ (from (\ref{eq:g_a-EXACT})), implies
\begin{equation}
\label{eq:gaugino_mass_anomaly_med_3}
m_a\sim {\frac{1}{8\pi^2}}m_{\frac{3}{2}}.
\end{equation}
From (\ref{eq:gaugino_mass}) and (\ref{eq:gaugino_mass_anomaly_med_3}), one sees that similar to \cite{conloncal,towardsrealvacua}, the anomaly mediated gaugino masses are suppressed by the standard loop factor as compared to the gravity mediated gaugino masses.
It has been found that due to competing contributions from the Wilson line moduli, there is a non-universality in the F-terms
$F^{\sigma^B} \sim {\cal V}^{\frac{1}{18}}$ $ m_{\frac{3}{2}}$ which for ${\cal V}\sim10^6$ is approximately of the same order as
$F^{{\cal G}^{a}} \sim m_{\frac{3}{2}}$; $F^{\sigma^S} \sim {\cal V}^{\frac{37}{36}}$ $ m_{\frac{3}{2}}$ - a reverse non-universality as compared to, e.g., \cite{Quevedo+Conlon_supp_gaugino_mass}. This is attributable to the cancelation between the divisor volume corresponding to $D_5$ and the Wilson line moduli contribution in $``T_B"$.
Further, wherever there is a contribution from $F^{\sigma^S}$ to the soft parameters, there will be a hierarchy/non-universality.

The matter fields corresponding to the position moduli of the mobile $D3$-brane are heavier than the gravitino and show universality. However, Wilson line modulus mass is different. We obtain a hierarchy in the physical mu terms $\hat{\mu}$, the $\hat{\mu}B$-terms as well as the
physical Yukawa couplings $\hat{Y}$; however we obtain a universality for the $A$-terms - larger than $m_{\frac{3}{2}}$ - for the $D3$-brane
position moduli {\it and} the Wilson line moduli. However it can be easily seen from  table 1 that in the
physical $\hat{\mu}, \hat{Y}$ and $\hat{\mu} B $ terms, that main part of the non-universality appears from the Wilson moduli contributions while there is an approximate universality in the $D3$-brane position moduli  components for which the physical $\hat{\mu}, \hat{Y}$ and $\hat{\mu} B $ are heavier than gravitino.  Also, as the string scale in our setup is nearly of the same order as the GUT scale and the open string moduli  are more massive as compared to the $\sim$ TeV gravitino (and gauginos), one can expect (e.g. see \cite{FCNC}) that the presence of non-universality will be consistent with the  low energy FCNC constraints.
 Further we have found that ${\hat{\mu}}^2 \sim {\hat{\mu} B}$ for the $D3$-brane position moduli (which
 show universality of almost all the soft SUSY breaking parameters) consistent with the requirement of a stable vacuum spontaneously breaking supersymmetry - see \cite{Green_Weigand} -
 whereas ${\hat{\mu}}^2 \ll {\hat{\mu} B}$ for components with only Wilson line modulus as well as the
   same  mixed with the $D3$-brane position moduli. Also, the un-normalized physical mu-parameters
     for the $D3$-brane position moduli ($\hat{K}_{{\cal Z}_i{\bar{\cal Z}}_i}\hat{\mu}_{{\cal Z}_i{\cal Z}_i}$) are $\sim$ TeV, as required for having correct electroweak symmetry breaking \cite{Green_Weigand,mu1}.   Our results are summarized in Table 1.
\begin{table}[htbp]
\centering
\begin{tabular}{|l|l|}
\hline
Gravitino mass &  $ m_{\frac{3}{2}}\sim{\cal V}^{-\frac{n^s}{2} - 1}$ \\
Gaugino mass & $ M_{\tilde g}\sim m_{\frac{3}{2}}$\\ \hline
$D3$-brane position moduli mass & $ m_{{\cal Z}_i}\sim {\cal V}^{\frac{19}{36}}m_{\frac{3}{2}}$ \\
Wilson line moduli mass & $ m_{\tilde{\cal A}_1}\sim {\cal V}^{\frac{73}{72}}m_{\frac{3}{2}}$\\ \hline
& $A_{{\cal Z}_i{\cal Z}_j{\cal Z}_k}\sim n^s{\cal V}^{\frac{37}{36}}m_{\frac{3}{2}}$\\
A-terms & $A_{{\tilde{\cal A}_1}{\tilde{\cal A}_1}{\tilde{\cal A}_1}}\sim n^s{\cal V}^{\frac{37}{36}}m_{\frac{3}{2}}$\\
& $A_{{{\tilde{\cal A}_1}^2}{{\cal Z}_i}}\sim n^s{\cal V}^{\frac{37}{36}}m_{\frac{3}{2}}$\\
& $A_{{\tilde{\cal A}_1}{{\cal Z}_i}{{\cal Z}_j}}\sim n^s{\cal V}^{\frac{37}{36}}m_{\frac{3}{2}}$\\
\hline
Physical $\mu$-terms & $\hat{\mu}_{{\cal Z}_i{\cal Z}_j}\sim{\cal V}^{\frac{37}{36}}m_{\frac{3}{2}}$\\
& $\hat{\mu}_{{\cal A}_1{\cal Z}_i}\sim{\cal V}^{-\frac{3}{4}}m_{\frac{3}{2}}$\\
& $\hat{\mu}_{{\cal A}_1{\cal A}_1}\sim{\cal V}^{-\frac{33}{36}}m_{\frac{3}{2}}$\\
\hline
& $\hat{Y}_{{\cal Z}_i{\cal Z}_i{\cal Z}_i}\sim {\cal V}^{\frac{43}{24}}m_{\frac{3}{2}}$\\
& $\hat{Y}_{{\cal Z}_i^2{\cal Z}_j}\sim {\cal V}^{\frac{37}{24}}m_{\frac{3}{2}}$\\
Physical Yukawa couplings&$\hat{Y}_{{\cal Z}_i^2\tilde{\cal A}_1}\sim {\cal V}^{\frac{1}{72}}m_{\frac{3}{2}}$\\
&$\hat{Y}_{{\cal Z}_1{\cal Z}_2\tilde{\cal A}_1}\sim {\cal V}^{-\frac{17}{72}}m_{\frac{3}{2}}$\\
&$\hat{Y}_{\tilde{\cal A}_1^2{\cal Z}_i}\sim {\cal V}^{-\frac{127}{72}}m_{\frac{3}{2}}$\\
&$\hat{Y}_{\tilde{\cal A}_1\tilde{\cal A}_1\tilde{\cal A}_1}\sim {\cal V}^{-\frac{85}{24}}m_{\frac{3}{2}}$\\
\hline
&$\left(\hat{\mu}B\right)_{{\cal Z}_i{\cal Z}_i}\sim{\cal V}^{\frac{223}{108}}m_{\frac{3}{2}}^2$\\
$\hat{\mu}B$-terms & $\left(\hat{\mu}B\right)_{{\cal Z}_1{\cal Z}_2}\sim{\cal V}^{\frac{37}{18}}m_{\frac{3}{2}}^2$\\
 &$\left(\hat{\mu}B\right)_{\tilde{\cal A}_1\tilde{\cal A}_1}\sim{\cal V}^{\frac{5}{36}}m_{\frac{3}{2}}^2$\\
&$\left(\hat{\mu}B\right)_{{\cal Z}_i\tilde{{\cal A}_1}}\sim{\cal V}^{-\frac{13}{18}}m_{\frac{3}{2}}^2$\\
\hline
\end{tabular}
\caption{Results Summarized}
\end{table}

It will be interesting to see what happens to the couplings with the inclusion of higher derivative terms - one expects to include
$\frac{1}{48}\int_{\Sigma_B}\left(p_1\left(T\Sigma_B\right) - p_1\left(N\Sigma_B\right)\right)$ as an additive shift to ${\cal F}$ of section {\bf 2} (See \cite{V_D7_fl}).

\newpage

\section*{Acknowledgements}

One of us (PS), is supported by a Senior Research Fellowship from the CSIR, India. AM would like to thank the Harvard theory group (specially C.Vafa), the SNS at IAS, Princeton (specially J.Maldacena), the theory group at McGill university (specially K.Dasgupta) and the Abdus Salam ICTP (under the junior associateship program) for their kind hospitality and support where part of this work was done. He would also like to thank C.Vafa and specially J.Maldacena for useful discussions and R.Blumenhagen, T.Grimm  and specially O.Ganor and H.Jockers for useful clarifications.

\appendix
\section{Justification behind ${\cal A}_I\sim{\cal V}^{-\frac{1}{4}}$}
\setcounter{equation}{0} \seceqaa

In this section we justify that the Wilson line moduli can be stabilized, in a self-consistent manner, at values of the order of ${\cal V}^{-\frac{1}{4}}$. We evaluate the complete moduli space metric for arbitrary Wilson line moduli but close to ${\cal V}^{-\frac{1}{4}}$ - for simplicity we assume only one such modulus. This implies that we replace ${\cal T}_B(\sigma^B,{\bar\sigma ^B};{\cal G}^a,{\bar{\cal G}^a};\tau,{\bar\tau}) + \mu_3{\cal V}^{\frac{1}{18}} + i\kappa_4^2\mu_7C_{1{\bar 1}}{\cal V}^{-\frac{1}{2}} - \gamma\left(r_2 + \frac{r_2^2\zeta}{r_1}\right)$ with ${\cal V}^{\frac{1}{18}}$ (and the same for
${\cal T}_S(\sigma^S,{\bar\sigma ^B};{\cal G}^a,{\bar{\cal G}^a};\tau,{\bar\tau}) + \mu_3{\cal V}^{\frac{1}{18}} + i\kappa_4^2\mu_7C_{1{\bar 1}}{\cal V}^{-\frac{1}{2}} - \gamma\left(r_2 + \frac{r_2^2\zeta}{r_1}\right)$) with the understanding that there is a cancelation between the big divisor's volume and the quadratic term in the Wilson line moduli. This is only to simplify the calculation of the metric for arbitrary values of the Wilson line modulus - we would arrive at the same conclusion by starting out with a completely arbitrary value of the Wilson line modulus and stabilizing it by extremizing the potential. We assume that all the remaining moduli have been stabilized (the complex structure and axion-dilaton moduli via the covariant constancy of the superpotential, the closed string K\"{a}hler and the open string mobile $D3$ brane position moduli via extremization of the potential). We then show that the potential is identically an extremum for all values of the Wilson line modulus close to ${\cal V}^{-\frac{1}{4}}$.

As we are considering the rigid limit of wrapping of the $D7$-brane around $D_5$ (to ensure that there is no obstruction to a chiral matter spectrum), there will be no superpotential generated due to the fluxes on the world volume of the $D7$-brane \cite{jockersetal} - the same is given by $\kappa_4^2\mu_7 l\zeta^A\int_{\Sigma_B}\tilde{s}_A\wedge
\tilde{\cal F}$, $\tilde{s}_A\in H^2_{{\bar\partial},-}(\Sigma_B)$ and vanishes when $\zeta^A=0$. Further,  by restricting the mobile $D3$-brane to $D_5$, possible contribution to the non-perturbative superpotential due to gaugino condensation in the presence of a stack of $D7$-branes wrapping (a rigid) $D_5$, will be nullified. The reason is that the contribution to the non-perturbative superpotential due to gaugino condensation on a stack of $N$ $D7$-branes wrapping $D_5$ will be proportional to $\left(1+z_1^{18}+z_2^{18}+z_3^3-3\phi_0z_1^6z_2^6\right)^{\frac{1}{N}}$, which according to \cite{Ganor1_2}, vanishes whenever the mobile $D3$-brane touches the wrapped $D7$-brane. Hence, when the mobile $D3$-brane is restricted to $D_5$, the aforementioned contribution to the non-perturbative superpotential goes to zero. It is for this reason that we are justified in considering a single  wrapped $D7$-brane, which anyway can not effect gaugino condensation. Hence, again using the reasoning of \cite{dSetal,Grimm,Ganor1_2}, the superpotential, assuming as in KKLT scenarios a very small complex-structure moduli-dependent superpotential which (as in \cite{dSetal}) we disregard as compared to the non-perturbative
superpotential, will be given by:
\begin{equation}
\label{eq:W}
W\sim\left(1+z_1^{18}+z_2^{18}+z_3^2-3\phi_0z_1^6z_2^6\right)^{n^s}\sum_{m^a}\frac{e^{i\frac{\tau m^2}{2}+in^sm_aG^a+in^sT_s}}{f(\tau)},
\end{equation}
where $T_S$ is the $a_I=0$ limit of (\ref{eq:N=1_coords}) corresponding to an $ED3$-instanton wrapping
$D_4$ and $f(\tau)$ is some appropriate modular function, which we do not know. In the following, we assume that the complexified Wilson line moduli are given entirely in terms of the Wilson line moduli and verify this in a self-consistent manner by extremization of the potential.

To evaluate the potential, we would need to evaluate the inverse of the moduli space metric. As also stated in {\bf 3.2}, we then show then in a self-consistent manner that one can set all components of sections of $N\Sigma_B$ and all components save one of the Wilson line moduli ${\cal A}_1$ to zero - the non-zero Wilson line modulus can be consistently stabilized to ${\cal V}^{-\frac{1}{4}}$.
Now, the derivatives of $K$ relevant to the calculation of the moduli space metric $G_{A{\bar B}}$, assuming ${\cal A}_1$ to be in the neighborhood of ${\cal V}^{-\frac{1}{4}}$,  are given below:

\begin{enumerate}
\item
{\it Single Derivatives}

\begin{itemize}
\item
$\frac{\partial K}{\partial z_i}$

\begin{eqnarray}
\label{eq:singleder_z}
& & \hskip-1.5in\frac{\partial K}{\partial z_i}=
-\frac{2}{\cal Y}\Biggl[\frac{3a}{2}\left(2\tau_B + \mu_3l^2{\cal V}^{\frac{1}{18}} + ... -\gamma K_{\rm geom}\right)^{\frac{1}{2}}\Biggl\{3i\mu_3\l^2(\omega_B)_{i{\bar j}}{\bar z}^{\bar j}+\frac{3}{4}\mu_3l^2\left((\omega_B)_{i{\bar j}}{\bar z}^{\tilde{a}}({\cal P}_{\tilde{a}})^{\bar j}_lz^l + (\omega_B)_{l{\bar j}}z^l{\bar z}^{\tilde{a}}({\cal P})^{\bar j}_i\right)\nonumber\\
& & \hskip-1.5in-\gamma(ln {\cal V})^{-\frac{7}{12}}{\cal V}^{\frac{29}{36}}\Biggr\}
- \frac{3a}{2}\left(2\tau_S + \mu_3l^2{\cal V}^{\frac{1}{18}} + ... -\gamma K_{\rm geom}\right)^{\frac{1}{2}}\Biggl\{3i\mu_3\l^2(\omega_S)_{i{\bar j}}{\bar z}^{\bar j}+\frac{3}{4}\mu_3l^2\left((\omega_S)_{i{\bar j}}{\bar z}^{\tilde{a}}({\cal P}_{\tilde{a}})^{\bar j}_lz^l + (\omega_S)_{l{\bar j}}z^l{\bar z}^{\tilde{a}}({\cal P})^{\bar j}_i\right)\nonumber\\
& & \hskip-1.5in-\gamma(ln {\cal V})^{-\frac{7}{12}}{\cal V}^{\frac{29}{36}}\Biggr\} \Biggr]
\end{eqnarray}

\item
$\frac{\partial K}{\partial\sigma^\alpha}$

\begin{equation}
\label{eq:singleder_sigma}
\frac{\partial K}{\partial\sigma^\alpha}=-\frac{2}{\cal Y}\left[\frac{3a}{2}(2\tau_\alpha + \mu_3l^2{\cal V}^{\frac{1}{18}} + ... -\gamma K_{\rm geom})\right]^{\frac{1}{2}}.
\end{equation}

\item
$\frac{\partial K}{\partial {\cal G}^a}$

\begin{eqnarray}
\label{eq:singleder_G}
& & \hskip-1.3in\frac{\partial K}{\partial {\cal G}^a}=-\frac{2}{{\cal Y}}\Biggl[-\frac{3a}{2}\frac{\left(2\tau_B + \mu_3l^2{\cal V}^{\frac{1}{18}} + ... -\gamma K_{\rm geom}\right)^{\frac{1}{2}}}{(\tau-{\bar\tau})}\kappa_{Bac}({\cal G}^c-{\bar{\cal G}}^c)
+\frac{3a}{2}\frac{\left(2\tau_S + \mu_3l^2{\cal V}^{\frac{1}{18}} + ... -\gamma K_{\rm geom}\right)^{\frac{1}{2}}}{(\tau-{\bar\tau})}\kappa_{Sac}({\cal G}^c-{\bar{\cal G}}^c)\nonumber\\
& & + 4\sum_{\beta\in H_2^-(CY_3,{\bf Z})} n^0_\beta\sum_{m,n\in{\bf Z}^2/(0,0)}
\frac{({\bar\tau}-\tau)^{\frac{3}{2}}}{(2i)^{\frac{3}{2}}|m+n\tau|^3}sin\left(mk.{\cal B} + nk.c\right)\frac{\tau nk^a + mk^a}{(\tau-{\bar\tau}^a)}\Biggr]\nonumber\\
& &
\end{eqnarray}

\item
$\frac{\partial K}{\partial {\cal A}^I}$

\begin{equation}
\label{eq:singleder_a}
\hskip -1.3in\frac{\partial K}{\partial {\cal A}^I}=-\frac{2}{{\cal Y}}\left[\left(2\tau_B + \mu_3l^2{\cal V}^{\frac{1}{18}} + ... -\gamma K_{\rm geom}\right)^{\frac{1}{2}}.6i\kappa^4\mu_7(C_B)^{I{\bar K}}{\bar {\cal A}}_{\bar K} \right]
\end{equation}

\end{itemize}

\item
{\it Double Derivatives}

\begin{itemize}
\item
$\frac{\partial^2K}{{\bar\partial}{\bar\sigma}^B\partial\sigma^B}$

\begin{eqnarray}
\label{eq:double_sigmasigma}
& & \hskip -0.3in\frac{\partial^2K}{{\bar\partial}{\bar\sigma}^\alpha\partial\sigma^\alpha}
=\frac{2}{{\cal Y}^2}
\left[\frac{3a}{2}\sqrt{2\tau_\alpha + \mu_3l^2{\cal V}^{\frac{1}{18}} + ... -\gamma K_{\rm geom}}\right]^2 -\frac{3a}{2{\cal Y}}\frac{1}{\sqrt{2\tau_\alpha + \mu_3l^2{\cal V}^{\frac{1}{18}} + ... -\gamma K_{\rm geom}}}
\sim\frac{\mu_3l^2}{{\cal V}^{\frac{35}{36}}}\nonumber\\
& &
\end{eqnarray}

\item
$\frac{\partial^2K}{{\bar\partial}{\bar\sigma}^B\partial\sigma^S}$

\begin{eqnarray}
\label{eq:double_sigmaBsigmaS}
& & \hskip-0.3in\frac{\partial^2K}{{\bar\partial}{\bar\sigma}^S\partial\sigma^B}
=\frac{2}{{\cal Y}^2}
\left[\frac{3a}{2}\sqrt{2\tau_S + \mu_3l^2{\cal V}^{\frac{1}{18}} + ... -\gamma K_{\rm geom}}\right]
\left[\frac{3a}{2}\sqrt{2\tau_B + \mu_3l^2{\cal V}^{\frac{1}{18}} + ... -\gamma K_{\rm geom}}\right]
\sim\frac{\mu_3l^2}{{\cal V}^{\frac{35}{36}}}\nonumber\\
& &
\end{eqnarray}

\item
$\frac{\partial^2K}{\partial{{\cal G}^a}{\bar\partial}{{\cal G}^a}}$

\begin{eqnarray}
\label{eq:double_GG}
& & \hskip-1.2in\frac{\partial^2K}{\partial{{\cal G}^a}{\bar\partial}{{\cal G}^b}}
=\frac{2}{{\cal Y}^2}
\Biggl[-\frac{3a}{2}\frac{\sqrt{2\tau_B + \mu_3l^2{\cal V}^{\frac{1}{18}} + ... -\gamma K_{\rm geom}}}{(\tau-{\bar\tau})}
\kappa_{Bac}({\cal G}^c-{\bar{\cal G}}^c)
+\frac{3a}{2}\frac{\sqrt{2\tau_S + \mu_3l^2{\cal V}^{\frac{1}{18}} + ... -\gamma K_{\rm geom}}}{(\tau-{\bar\tau})}
\kappa_{Sac}({\cal G}^c-{\bar{\cal G}}^c)\nonumber\\
& & - 4\sum_{\beta\in H_2^-(CY_3,{\bf Z})} n^0_\beta\sum_{m,n\in{\bf Z}^2/(0,0)}
\frac{({\bar\tau}-\tau)^{\frac{3}{2}}}{(2i)^{\frac{3}{2}}|m+n\tau|^3}sin\left(mk.{\cal B} + nk.c\right)
\frac{{\bar\tau} nk^a + mk^a}{(\tau-{\bar\tau})}\Biggr]\nonumber\\
& & \hskip-0.5in\times\Biggl[-\frac{3a}{2}\frac{\sqrt{2\tau_B + \mu_3l^2{\cal V}^{\frac{1}{18}} + ... -\gamma K_{\rm geom}}}{(\tau-{\bar\tau})}
\kappa_{Bad}({\cal G}^d-{\bar{\cal G}}^d)
+\frac{3a}{2}\frac{\sqrt{2\tau_S + \mu_3l^2{\cal V}^{\frac{1}{18}} + ... -\gamma K_{\rm geom}}}{(\tau-{\bar\tau})}
\kappa_{Sad}({\cal G}^d-{\bar{\cal G}}^d)\nonumber\\
& & - 4\sum_{\beta\in H_2^-(CY_3,{\bf Z})} n^0_\beta\sum_{m,n\in{\bf Z}^2/(0,0)}
\frac{({\bar\tau}-\tau)^{\frac{3}{2}}}{(2i)^{\frac{3}{2}}|m+n\tau|^3}sin\left(mk.{\cal B} + nk.c\right)
\frac{\tau nk^a + mk^a}{(\tau-{\bar\tau})}\Biggr]\nonumber\\
& & -\frac{2}{{\cal Y}}\Biggl[\frac{3a}{2}\frac{\sqrt{2\tau_B + \mu_3l^2{\cal V}^{\frac{1}{18}} + ... -\gamma K_{\rm geom}}}{(\tau-{\bar\tau})}\kappa_{Bab}
+ \frac{3a}{2}\frac{\sqrt{2\tau_S + \mu_3l^2{\cal V}^{\frac{1}{18}} + ... -\gamma K_{\rm geom}}}{(\tau-{\bar\tau})}
\kappa_{Sac}\nonumber\\
& & - 4\sum_{\beta\in H_2^-(CY_3,{\bf Z})} n^0_\beta\sum_{m,n\in{\bf Z}^2/(0,0)}
\frac{({\bar\tau}-\tau)^{\frac{3}{2}}}{(2i)^{\frac{3}{2}}|m+n\tau|^3}cos\left(mk.{\cal B} + nk.c\right)
\frac{{\bar\tau} nk^a + mk^a}{(\tau-{\bar\tau})}
\frac{{\bar\tau} nk^b + mk^b}{(\tau-{\bar\tau})}\Biggr]\nonumber\\
& & \sim
\frac{\sum n^0_\beta cos(...)}{{\cal V}}
\end{eqnarray}

\item
$\frac{\partial^2K}{\partial{\sigma^\alpha}{\bar\partial}{{\cal G}^a}}$

\begin{eqnarray}
\label{eq:Double_GBS}
& & \hskip-1.6in(a)\frac{\partial^2K}{\partial{\sigma^B}{\bar\partial}{{\cal G}^a}}
=\frac{3a}{{\cal Y}^2}\sqrt{2\tau_B + \mu_3l^2{\cal V}^{\frac{1}{18}} + ... -\gamma K_{\rm geom}}
\Biggl[-\frac{3a}{2}\frac{\sqrt{2\tau_B + \mu_3l^2{\cal V}^{\frac{1}{18}} + ... -\gamma K_{\rm geom}}}{(\tau-{\bar\tau})}
\kappa_{Bac}({\cal G}^c-{\bar{\cal G}}^c)\nonumber\\
& &
+\frac{3a}{2}\frac{\sqrt{2\tau_S + \mu_3l^2{\cal V}^{\frac{1}{18}} + ... -\gamma K_{\rm geom}}}{(\tau-{\bar\tau})}
\kappa_{Sac}({\cal G}^c-{\bar{\cal G}}^c)\nonumber\\
& & \hskip-1.5in- 4\sum_{\beta\in H_2^-(CY_3,{\bf Z})} n^0_\beta\sum_{m,n\in{\bf Z}^2/(0,0)}
\frac{({\bar\tau}-\tau)^{\frac{3}{2}}}{(2i)^{\frac{3}{2}}|m+n\tau|^3}sin\left(mk.{\cal B} + nk.c\right)
\frac{{\bar\tau} nk^a + mk^a}{(\tau-{\bar\tau})}\Biggr]\nonumber\\
& & \hskip-1.5in+ \frac{9ai}{2\sqrt{2\tau_B + \mu_3l^2{\cal V}^{\frac{1}{18}} + ... -\gamma K_{\rm geom}}}\kappa_{Bac}\left({\cal G}^a-{\bar{\cal G}}^a\right)
\sim\frac{{\cal V}^{-\frac{37}{36}}\kappa_{Bac}{\cal G}^c}{\sqrt{\mu_3l^2}}\nonumber\\
& & (b)\ {\rm Similarly},\ \frac{\partial^2K}{\partial{\sigma^S}{\bar\partial}{{\cal G}^a}}\sim\sim\frac{{\cal V}^{-\frac{37}{36}}\kappa_{Sac}{\cal G}^c}{\sqrt{\mu_3l^2}}.
\end{eqnarray}

\item
$\frac{\partial^2K}{\partial{\bar z}^{\bar i}{\bar\partial}{{\cal G}^a}}$

\begin{eqnarray}
\label{eq:Double_Gz}
& & \frac{\partial^2K}{\partial{\bar z}^{\bar i}{\bar\partial}{{\cal G}^a}}=-\frac{2}{{\cal Y}}\Biggl[-\frac{3a}
{2(\tau-{\bar\tau})}\left(\frac{\kappa_{Bac}}{\sqrt{2\tau_B + \mu_3l^2{\cal V}^{\frac{1}{18}} + ... -\gamma K_{\rm geom}}}-\frac{\kappa_{Sac}}
{\sqrt{2\tau_S + \mu_3l^2{\cal V}^{\frac{1}{18}} + ... -\gamma K_{\rm geom}}}
\right)\Biggr]\nonumber\\
& & +\frac{2}{{\cal Y}^2}\left[\sqrt{2\tau_B + \mu_3l^2{\cal V}^{\frac{1}{18}} + ... -\gamma K_{\rm geom}}
\left\{\mu_3l^2{\cal V}^{\frac{1}{18}}\left\{\omega_B-\omega_S\right\}_{i{\bar j}}\xi^{\bar j}-\gamma\left(ln {\cal V}\right)^{-\frac{7}{12}}{\cal V}^{\frac{29}{36}}\right\}\right]\nonumber\\
& & \hskip-0.3in\Biggl[-\frac{3a}{2}\frac{\sqrt{2\tau_B + \mu_3l^2{\cal V}^{\frac{1}{18}} + ... -\gamma K_{\rm geom}}}{(\tau-{\bar\tau})}
\kappa_{Bac}({\cal G}^c-{\bar{\cal G}}^c)
+\frac{3a}{2}\frac{\sqrt{2\tau_S + \mu_3l^2{\cal V}^{\frac{1}{18}} + ... -\gamma K_{\rm geom}}}{(\tau-{\bar\tau})}
\kappa_{Sac}({\cal G}^c-{\bar{\cal G}}^c)\nonumber\\
& & \hskip-1.5in- 4\sum_{\beta\in H_2^-(CY_3,{\bf Z})} n^0_\beta\sum_{m,n\in{\bf Z}^2/(0,0)}
\frac{({\bar\tau}-\tau)^{\frac{3}{2}}}{(2i)^{\frac{3}{2}}|m+n\tau|^3}sin\left(mk.{\cal B} + nk.c\right)
\frac{{\bar\tau} nk^a + mk^a}{(\tau-{\bar\tau})}\Biggr]\nonumber\\
& & \sim\frac{\left\{\omega_B-\omega_S\right\}_{i{\bar j}}\xi^{\bar j}\sum_\beta k^a n^0_\beta sin(...)(\mu_3l^2)^{\frac{3}{2}}}{{\cal V}^{\frac{35}{18}}}\sim\frac{k^a\left\{\omega_B-\omega_S\right\}_{i{\bar j}}\xi^{\bar j}}{{\cal V}^{\frac{10}{9}}}
\end{eqnarray}

\item
$\frac{\partial^2K}{\partial {\cal A}^I{\bar\partial}{{\cal G}^a}}$

\begin{eqnarray}
\label{eq:Double_aG}
& & \frac{\partial^2K}{\partial {\cal A}^I{\bar\partial}{{\cal G}^a}}=-\frac{2}{\cal Y}
\Biggl[\frac{6i\kappa^2\mu_7(c_B)^{I{\bar K}}}{\sqrt{2\tau_B + \mu_3l^2{\cal V}^{\frac{1}{18}} + ... -\gamma K_{\rm geom}}}
\left(-\frac{3i}{(\tau-{\bar\tau})}
\kappa_{Bac}({\cal G}^c-{\bar G}^{\bar c})\right)\Biggr]\nonumber\\
& & \hskip-1.4in+\frac{2}{{\cal Y}}\Biggl[-\frac{3a}{2}\frac{\sqrt{2\tau_B + \mu_3l^2{\cal V}^{\frac{1}{18}} + ... -\gamma K_{\rm geom}}}{(\tau-{\bar\tau})}
\kappa_{Bac}({\cal G}^c-{\bar{\cal G}}^c)
+\frac{3a}{2}\frac{\sqrt{2\tau_S + \mu_3l^2{\cal V}^{\frac{1}{18}} + ... -\gamma K_{\rm geom}}}{(\tau-{\bar\tau})}
\kappa_{Sac}({\cal G}^c-{\bar{\cal G}}^c)\nonumber\\
& & \hskip-1.5in- 4\sum_{\beta\in H_2^-(CY_3,{\bf Z})} n^0_\beta\sum_{m,n\in{\bf Z}^2/(0,0)}
\frac{({\bar\tau}-\tau)^{\frac{3}{2}}}{(2i)^{\frac{3}{2}}|m+n\tau|^3}sin\left(mk.{\cal B} + nk.c\right)
\frac{\tau nk^a + mk^a}{(\tau-{\bar\tau})}\Biggr]\nonumber\\
& & \hskip-1.3in\times\Biggl[\frac{3a}{2}\left(2\tau_B + \mu_3l^2{\cal V}^{\frac{1}{18}} + ... -\gamma K_{\rm geom}\right)^{\frac{1}{2}}
.6i\kappa^2\mu_7(c_B)^{I{\bar K}}
{\bar {\cal A}}_{\bar K}\Biggr]\nonumber\\
& & \sim\frac{{\cal V}^{\frac{5}{36}}\kappa_{Bab}{\cal G}^b}{\sqrt{\mu_3l^2}}{\cal A}_1
\end{eqnarray}

\item
$\frac{\partial^2K}{\partial z_i{\bar\partial}{\bar z}_{\bar j}}$

\begin{eqnarray}
\label{eq:Double_zz}
& & \hskip-1in\frac{\partial^2K}{\partial z_i{\bar\partial} {\bar z}_{\bar j}}=
\frac{2}{{\cal Y}^2}\Biggl[\frac{3a}{2}\left(2\tau_B + \mu_3l^2{\cal V}^{\frac{1}{18}} + ... -\gamma K_{\rm geom}\right)^{\frac{1}{2}}\Biggl\{3i\mu_3\l^2
(\omega_B)_{i{\bar k}}{\bar z}^{\bar k}+\frac{3}{4}\mu_3l^2\left((\omega_B)_{i{\bar k}}{\bar z}^{\tilde{a}}
({\cal P}_{\tilde{a}})^{\bar k}_lz^l + (\omega_B)_{l{\bar k}}z^l{\bar z}^{\tilde{a}}({\cal P})^{\bar k}_i\right)\nonumber\\
& & \hskip-1.3in-\gamma(ln {\cal V})^{-\frac{7}{12}}{\cal V}^{\frac{29}{36}}\Biggr\}
- \frac{3a}{2}\left(2\tau_S + \mu_3l^2{\cal V}^{\frac{1}{18}} + ... -\gamma K_{\rm geom}\right)^{\frac{1}{2}}\Biggl\{3i\mu_3\l^2(\omega_S)_{i{\bar k}}
{\bar z}^{\bar k}+\frac{3}{4}\mu_3l^2\left((\omega_S)_{i{\bar k}}{\bar z}^{\tilde{a}}({\cal P}_{\tilde{a}})^{\bar j}_lz^l
+ (\omega_S)_{l{\bar k}}z^l{\bar z}^{\tilde{a}}({\cal P})^{\bar k}_i\right)\nonumber\\
& & \hskip-1in-\gamma(ln {\cal V})^{-\frac{7}{12}}{\cal V}^{\frac{29}{36}}\Biggr\} \Biggr]\nonumber\\
& & \hskip -1in\times\Biggl[\frac{3a}{2}\left(2\tau_B + \mu_3l^2{\cal V}^{\frac{1}{18}} + ... -\gamma K_{\rm geom}\right)^{\frac{1}{2}}\Biggl\{-3i\mu_3\l^2
(\omega_B)_{k{\bar j}}z^k-\frac{3}{4}\mu_3l^2\left((\omega_B)_{k{\bar j}}z^{\tilde{a}}
({\cal P}_{\tilde{a}})^k_{\bar l}{\bar z}^l + (\omega_B)_{{\bar l}k}{\bar z}^{\bar l}z^{\tilde{a}}
({\cal P})^k_{\bar i}\right)\nonumber\\
& & \hskip-1.3in-\gamma(ln {\cal V})^{-\frac{7}{12}}{\cal V}^{\frac{29}{36}}\Biggr\}
- \frac{3a}{2}\left(2\tau_S + \mu_3l^2{\cal V}^{\frac{1}{18}} + ... -\gamma K_{\rm geom}\right)^{\frac{1}{2}}\Biggl\{-3i\mu_3\l^2(\omega_S)_{k{\bar j}}
z^k-\frac{3}{4}\mu_3l^2\left((\omega_S)_{k{\bar j}}z^{\tilde{a}}({\cal P}_{\tilde{a}})^j_{\bar l}{\bar z}^{\bar l}
+ (\omega_S)_{k{\bar l}}{\bar z}^{\bar l}z^{\tilde{a}}({\cal P})^k_{\bar i}\right)\nonumber\\
& & \hskip-1in-\gamma(ln {\cal V})^{-\frac{7}{12}}{\cal V}^{\frac{29}{36}}\Biggr\} \Biggr]
-\frac{2}{{\cal Y}}\Biggl[\frac{3a}{2}\left(2\tau_B + \mu_3l^2{\cal V}^{\frac{1}{18}} + ... -\gamma K_{\rm geom}\right)^{\frac{1}{2}}\left\{
3i\mu_3l^2(\omega_B)_{i{\bar j}}-\gamma\left(ln {\cal V}\right)^{-\frac{7}{12}}{\cal V}^{\frac{5}{18}}\right\}\Biggr]\nonumber\\
& & \sim\frac{(\mu_3l^2)^3\left(\left\{\omega_B-\omega_S\right\}_{i{\bar j}}\xi^{\bar j}\right)^2}{{\cal V}^{\frac{17}{18}}}
\end{eqnarray}

\item
$\frac{\partial^2K}{\partial{a^I}{\bar\partial}{\bar z_i}}$

\begin{eqnarray}
\label{eq:Double_az}
& & \hskip-1in\frac{\partial^2K}{\partial{{\cal A}^I}{\bar\partial}{\bar z_i}}=\frac{2}{{\cal Y}^2}
\Biggl[\frac{3a}{2}\left(2\tau_B + \mu_3l^2{\cal V}^{\frac{1}{18}} + ... -\gamma K_{\rm geom}\right)^{\frac{1}{2}}\Biggl\{3i\mu_3\l^2
(\omega_B)_{i{\bar k}}{\bar z}^{\bar k}+\frac{3}{4}\mu_3l^2\left((\omega_B)_{i{\bar k}}{\bar z}^{\tilde{a}}
({\cal P}_{\tilde{a}})^{\bar k}_lz^l + (\omega_B)_{l{\bar k}}z^l{\bar z}^{\tilde{a}}({\cal P})^{\bar k}_i\right)\nonumber\\
& & \hskip-1.3in-\gamma(ln {\cal V})^{-\frac{7}{12}}{\cal V}^{\frac{29}{36}}\Biggr\}
- \frac{3a}{2}\left(2\tau_S + \mu_3l^2{\cal V}^{\frac{1}{18}} + ... -\gamma K_{\rm geom}\right)^{\frac{1}{2}}\Biggl\{3i\mu_3\l^2(\omega_S)_{i{\bar k}}
{\bar z}^{\bar k}+\frac{3}{4}\mu_3l^2\left((\omega_S)_{i{\bar k}}{\bar z}^{\tilde{a}}({\cal P}_{\tilde{a}})^{\bar j}_lz^l
+ (\omega_S)_{l{\bar k}}z^l{\bar z}^{\tilde{a}}({\cal P})^{\bar k}_i\right)\nonumber\\
& & -\gamma(ln {\cal V})^{-\frac{7}{12}}{\cal V}^{\frac{29}{36}}\Biggr\} \Biggr]\nonumber\\
& & \hskip-1in-\frac{2}{{\cal Y}}\Biggl[6i\kappa_4^2\mu_7(c_B)^{I{\bar J}}{\bar {\cal A}}_{\bar J}\times\frac{3a}{2}
\frac{\Biggl\{3i\mu_3\l^2
(\omega_B)_{i{\bar k}}{\bar z}^{\bar k}+\frac{3}{4}\mu_3l^2\left((\omega_B)_{i{\bar k}}{\bar z}^{\tilde{a}}
({\cal P}_{\tilde{a}})^{\bar k}_lz^l + (\omega_B)_{l{\bar k}}z^l{\bar z}^{\tilde{a}}({\cal P})^{\bar k}_i\right)
-\gamma(ln {\cal V})^{-\frac{7}{12}}{\cal V}^{\frac{29}{36}}\Biggr\}}{\left(2\tau_B + \mu_3l^2{\cal V}^{\frac{1}{18}} + ... -\gamma K_{\rm geom}\right)^{\frac{1}{2}}}\Biggr]\nonumber\\
& & \sim{\cal V}^{\frac{1}{6}}\sqrt{\mu_3 l^2}\left\{\omega_B-\omega_S\right\}_{i{\bar j}}\xi^{\bar j}{\cal A}_I
\end{eqnarray}

\item
$\frac{\partial^2K}{{\bar\partial}{\bar z_i}\partial\sigma^\alpha}$

\begin{eqnarray}
\label{eq:Double_BSz}
& &  \frac{\partial^2K}{{\bar\partial}{\bar z_i}\partial\sigma^\alpha}=
\frac{2}{{\cal Y}^2}\frac{3a}{2}\left(2\tau_B + \mu_3l^2{\cal V}^{\frac{1}{18}} + ... -\gamma K_{\rm geom}\right)^{\frac{1}{2}}\nonumber\\
& & \hskip-1.5in\times\Biggl[
\frac{3a}{2}\left(2\tau_B + \mu_3l^2{\cal V}^{\frac{1}{18}} + ... -\gamma K_{\rm geom}\right)^{\frac{1}{2}}
\Biggl\{3i\mu_3\l^2
(\omega_B)_{i{\bar k}}{\bar z}^{\bar k}+\frac{3}{4}\mu_3l^2\left((\omega_B)_{i{\bar k}}{\bar z}^{\tilde{a}}
({\cal P}_{\tilde{a}})^{\bar k}_lz^l + (\omega_B)_{l{\bar k}}z^l{\bar z}^{\tilde{a}}({\cal P})^{\bar k}_i\right)-\gamma\left(ln {\cal V}
\right)^{-\frac{7}{12}}{\cal V}^{\frac{29}{36}}\Biggr\}\nonumber\\
 & & \hskip-1.5in - \frac{3a}{2}\left(2\tau_S + \mu_3l^2{\cal V}^{\frac{1}{18}} + ... -\gamma K_{\rm geom}\right)^{\frac{1}{2}}
\Biggl\{3i\mu_3\l^2(\omega_S)_{i{\bar k}}
{\bar z}^{\bar k}+\frac{3}{4}\mu_3l^2\left((\omega_S)_{i{\bar k}}{\bar z}^{\tilde{a}}({\cal P}_{\tilde{a}})^{\bar j}_lz^l
+ (\omega_S)_{l{\bar k}}z^l{\bar z}^{\tilde{a}}({\cal P})^{\bar k}_i\right)-\gamma\left(ln {\cal V}
\right)^{-\frac{7}{12}}{\cal V}^{\frac{29}{36}}\Biggr\}\Biggr]\nonumber\\
& & \hskip -1.3in-\frac{3a}{2{\cal Y}}\frac{\Biggl\{3i\mu_3\l^2
(\omega_\alpha)_{i{\bar k}}{\bar z}^{\bar k}+\frac{3}{4}\mu_3l^2\left((\omega_\alpha)_{i{\bar k}}{\bar z}^{\tilde{a}}
({\cal P}_{\tilde{a}})^{\bar k}_lz^l + (\omega_\alpha)_{l{\bar k}}z^l{\bar z}^{\tilde{a}}({\cal P})^{\bar k}_i\right)-\gamma\left(ln {\cal V}
\right)^{-\frac{7}{12}}{\cal V}^{\frac{29}{36}}\Biggr\}}{\left(2\tau_\alpha + \mu_3l^2{\cal V}^{\frac{1}{18}} + ... -\gamma K_{\rm geom}\right)^{\frac{1}{2}}}
\sim\left\{\omega_B-\omega_S\right\}_{i{\bar j}}\xi^{\bar j}\frac{\mu_3l^2}{{\cal V}}.\nonumber\\
& &
\end{eqnarray}

\item
$\frac{\partial^2K}{{\bar\partial}{{\bar\sigma}^\alpha}{\partial}{{\cal A}_1}}$

\begin{eqnarray}
\label{eq:Double_BSa}
& & \hskip-1.65in(a)\frac{\partial^2K}{{\bar\partial}{{\bar\sigma}^B}{\partial}{{\cal A}^I}}=\frac{2}{{\cal Y}^2}\left(2\tau_B + \mu_3l^2{\cal V}^{\frac{1}{18}} + ... -\gamma K_{\rm geom}\right)^{\frac{1}{2}}
\Biggl[\frac{3a}{2}\left(2\tau_B + \mu_3l^2{\cal V}^{\frac{1}{18}} + ... -\gamma K_{\rm geom}\right)^{\frac{1}{2}}.6i\kappa^2\mu_7(c_B)^{I{\bar J}}{\cal A}_{\bar J}\Biggr]\nonumber\\
& & \hskip-1in-\frac{2}{{\cal Y}}\Biggl[\frac{3a}{4}\frac{6i\kappa^2\mu_7(c_B)^{I{\bar K}}{\bar {\cal A}}_{\bar K}}{\left(2\tau_B + \mu_3l^2{\cal V}^{\frac{1}{18}} + ... -\gamma K_{\rm geom}\right)^{\frac{1}{2}}}\Biggr]\sim\frac{{\cal V}^{\frac{5}{36}}}{\sqrt{\mu_3l^2}}{\cal A}_1;\nonumber\\
& & \hskip-1in(b)\ {\rm Similarly},\ \frac{\partial^2K}{{\bar\partial}{{\bar\sigma}^S}{\partial}{{\cal A}^I}}
\sim\frac{{\cal V}^{\frac{5}{36}}}{\sqrt{\mu_3l^2}}{\cal A}_1.
\end{eqnarray}

\item
$\frac{\partial^2K}{\partial{{\cal A}^I}{\bar\partial}{\bar{\cal A}_I}}$

\begin{eqnarray}
\label{eq:Double_aa}
& & \hskip -1.5in\frac{\partial^2K}{\partial{a^I}{\bar\partial}{\bar{\cal A}_I}}=-\frac{2}{{\cal Y}}\Biggl[\frac{3a}{4}\frac{(6i\kappa_4^2\mu_7)^2
(c_B)^{I{\bar K}}{\cal A}_{\bar K}(c_B)^{L{\bar J}}{\cal A}_L}{\left(2\tau_B + \mu_3l^2{\cal V}^{\frac{1}{18}} + ... -\gamma K_{\rm geom}\right)^{\frac{1}{2}}}
+ \frac{3a}{2}\left(2\tau_B + \mu_3l^2{\cal V}^{\frac{1}{18}} + ... -\gamma K_{\rm geom}\right)^{\frac{1}{2}}.6i\kappa^2\mu_7(c_B)^{I{\bar J}}
\nonumber\\
& & \hskip-1.5in +\frac{2}{{\cal Y}^2}\Biggl[\frac{3a}{2}\left(2\tau_B + \mu_3l^2{\cal V}^{\frac{1}{18}} + ... -\gamma K_{\rm geom}\right)^{\frac{1}{2}}.6i\kappa^2\mu_7(c_B)^{I{\bar K}}{\bar a}_{\bar K}
\Biggr] \times\Biggl[\frac{3a}{2}\left(2\tau_B + \mu_3l^2{\cal V}^{\frac{1}{18}} + ... -\gamma K_{\rm geom}\right)^{\frac{1}{2}}.6i\kappa^2\mu_7(c_B)^{L{\bar J}}
{\cal A}_L \Biggr]\nonumber\\
& & \sim\frac{{\cal V}^{\frac{47}{36}}}{\sqrt{\mu_3 l^2}}|{\cal A}_1|^2
\end{eqnarray}

\end{itemize}
\end{enumerate}

Hence, the combined closed- and open-string (matter field) moduli-space metric is given as under:
\begin{equation}
\label{eq:G_wilson}
\hskip-0.3in G_{A{\bar B}}\sim\left(
\begin{array}{lllllll}
 \frac{1}{z^{35/36}} & \frac{1}{z^{35/36}} & \frac{{A_{\sigma^B{\cal G}^1}}}{z^{37/36}} & \frac{{A_{\sigma^B{\cal G}^2}}}{z^{37/36}} &
   \frac{{A_{\sigma^Bz_1}}}{z} & \frac{{A_{\sigma^Bz_2}}}{z} & {A_{\sigma^\alpha{\cal A}_1}} {{\cal A}_1} z^{5/36} \\
 \frac{1}{z^{35/36}} & \frac{1}{z^{35/36}} & \frac{{A_{\sigma^S{\cal G}^1}}}{z^{37/36}} & \frac{{A_{\sigma^S{\cal G}^2}}}{z^{37/36}} &
   \frac{{A_{\sigma^Sz_1}}}{z} & \frac{{A_{\sigma^Sz_2}}}{z} & {A_{\sigma^\alpha{\cal A}_1}} {{\cal A}_1} z^{5/36} \\
 \frac{{A_{\sigma^B{\cal G}^1}}}{z^{37/36}} & \frac{{A_{\sigma^S{\cal G}^1}}}{z^{37/36}} & {A_{{\cal G}^1{\cal G}^1}} & {A_{{\cal G}^1{\cal G}^2}} & \frac{{A_{{\cal G}^1z_1}}}{z^{10/9}} &
   \frac{{A_{{\cal G}^1z_2}}}{z^{10/9}} & {A_{{\cal G}^1{\cal A}_1}} {{\cal A}_1} z^{5/36} \\
 \frac{{A_{\sigma^B{\cal G}^2}}}{z^{37/36}} & \frac{{A_{\sigma^S{\cal G}^2}}}{z^{37/36}} & {A_{{\cal G}^1{\cal G}^2}} & {A_{{\cal G}^2{\cal G}^2}} & \frac{{A_{{\cal G}^2z_1}}}{z^{10/9}} &
   \frac{{A_{{\cal G}^2z_2}}}{z^{10/9}} & {A_{{\cal G}^2{\cal A}_1}} {{\cal A}_1} z^{5/36} \\
 \frac{{A_{\sigma^Bz_1}}}{z} & \frac{{A_{\sigma^Sz_1}}}{z} & \frac{{A_{{\cal G}^1z_1}}}{z^{10/9}} & \frac{{A_{{\cal G}^2z_1}}}{z^{10/9}} &
   \frac{{A_{z_1z_1}}}{z^{17/18}} & \frac{{A_{z_1z_2}}}{z^{17/18}} & {A_{z_1{\cal A}_1}} {{\cal A}_1} \sqrt[6]{z} \\
 \frac{{A_{\sigma^Bz_2}}}{z} & \frac{{A_{\sigma^Sz_2}}}{z} & \frac{{A_{{\cal G}^1z_2}}}{z^{10/9}} & \frac{{A_{{\cal G}^2z_2}}}{z^{10/9}} &
   \frac{{A_{z_1z_2}}}{z^{17/18}} & \frac{{A_{z_2z_2}}}{z^{17/18}} & {A_{z_2{\cal A}_1}} {{\cal A}_1} \sqrt[6]{z} \\
 {A_{\sigma^\alpha{\cal A}_1}} {{\cal A}_1} z^{5/36} & {A_{\sigma^\alpha{\cal A}_1}} {{\cal A}_1} z^{5/36} & {A_{{\cal G}^1{\cal A}_1}} {{\cal A}_1} z^{5/36} & {A_{{\cal G}^2{\cal A}_1}}
   {{\cal A}_1} z^{5/36} & {A_{z_1{\cal A}_1}} {{\cal A}_1} \sqrt[6]{z} & {A_{z_2{\cal A}_1}} {{\cal A}_1} \sqrt[6]{z} & {{\cal A}_1}^2
   z^{47/36}
\end{array}
\right)
\end{equation}

We have assumed in this paper that the involution $\sigma$ is such that
$\frac{\sum_{\beta\in H_2^-(CY_3,{\bf Z})}n^0_\beta sin(...)}{{\cal V}^{\frac{1}{3}}}\sim{\cal V}^k$, where $k\left(\in\left(0,\frac{2}{3}\right)\right)=\frac{1}{2}$ and $\sum_\beta n^0_\beta cos(...)\sim{\cal V}$.
The components of $(G^{-1})^{A{\bar B}}$ are given as follows:
\begin{eqnarray}
\label{eq:inv_comps}
& & (G^{-1})^{\sigma^\alpha{\bar\sigma^\beta}}\sim{\cal V}^{\frac{19}{18}};\nonumber\\
& & (G^{-1})^{\sigma^\alpha{\bar z}^{\bar i}}\sim{\cal V};\nonumber\\
  & & (G^{-1})^{\sigma^\alpha{\bar G}^a}\sim{\cal V}^{\frac{1}{36}};\nonumber\\
  & & (G^{-1})^{\sigma^\alpha{\bar {\cal A}}_1}\sim\frac{{\cal V}^{\frac{5}{36}}}{{\cal A}_1};\nonumber\\
   & & (G^{-1})^{{\cal G}^a{\bar{\cal G}^b}}\sim{\cal V}^0;\nonumber\\
   & & (G^{-1})^{{\cal G}^a{\bar z}^{\bar i}}\sim{\cal V}^{-\frac{1}{36}};\nonumber\\
   & & (G^{-1})^{{\cal G}^a{\bar {\cal A}}_1}\sim \frac{{\cal V}^{-\frac{7}{6}}}{{\cal A}_1};\nonumber\\
   & & (G^{-1})^{z^i{\bar z^{\bar j}}}\sim{\cal V}^{\frac{17}{18}};\nonumber\\& & (G^{-1})^{z^i{\bar {\cal A}^1}}\sim \frac{{\cal V}^{-\frac{7}{36}}}{{\cal A}_1};\nonumber\\
   & & (G^{-1})^{{\cal A}_1{\bar {\cal A}_1}}\sim\frac{{\cal V}^{-\frac{47}{36}}}{{\cal A}_1^2}.
\end{eqnarray}

Now, restricted to $D_5$, using (\ref{eq:inv_comps}), (\ref{eq:W}), assuming that the complexified Wilson line moduli can be stabilized around ${\cal A}_1\sim{\cal V}^{-\frac{1}{4}}$ and:
\begin{eqnarray}
\label{eq:dersKW}
& & \partial_{\sigma^\alpha}K\sim\sqrt{\mu_3l^2}{\cal V}^{-\frac{35}{36}},\ \partial_{\sigma^B}W\sim0,\ \partial_{\sigma^S}W\sim n^sW;\nonumber\\
& & \partial_{{\cal G}^a}K\sim\frac{\sum_\beta n^0_\beta sin(...)}{\cal V}\sim{\cal V}^{-\frac{1}{6}},\
\partial_{{\cal G}^a}W\sim n^s(m_a+\frac{{\cal G}^a}{ln{\cal V}})W;\nonumber\\
& & \partial_{z^i}K|_{D_5}\sim \frac{(\mu_3l^2)^{\frac{3}{2}}\left\{\omega_B-\omega_S\right\}_{i{\bar j}}\xi^{\bar j}}{{\cal V}^{\frac{17}{18}}},\ \partial_{z^i}W|_{D_5}\sim \mu_3l^2\left\{\omega_S\right\}_{i{\bar j}}{\cal V}^{\frac{1}{36}}W;\nonumber\\
& & \partial_{{\cal A}_1}K\sim\frac{\sqrt{\mu_3l^2}(i\kappa_4^2\mu_7C_{1{\bar 1}})}{\cal V}{\cal A}_1\sim{\cal V}^{\frac{7}{36}}{\cal A}_1,\ \partial_{{\cal A}_1}W\sim0,
\end{eqnarray}
one obtains the following F-terms:
\begin{eqnarray}
\label{eq:F_terms}
& & e^KG^{\sigma^\alpha{\bar\sigma^{\bar\alpha}}}D_{\sigma^\alpha}W{\bar D}_{\bar\sigma^{\bar\alpha}}{\bar W}\sim\frac{(n^s)^2|W|^2{\cal V}^{\frac{19}{18}}}{{\cal V}^2}\equiv{\rm most\ dominant}\sim V_0(\equiv{\rm extremum\ value});\nonumber\\
& & e^KG^{{\cal G}^a{\bar{\cal G}^b}}D_{{\cal G}^a}WD_{\bar{\cal G}^b}{\bar W}\sim\frac{(n^s)^2m_am_b|W|^2}{{\cal V}^2};\nonumber\\
& & e^KG^{\sigma^\alpha{\bar{\cal G}^a}}D_{\sigma^\alpha}WD_{\bar{\cal G}^a}{\bar W}
\sim\frac{(n^s)^2m_a|W|^2{\cal V}^{\frac{1}{36}}}{{\cal V}^2};\nonumber\\
& & e^KG^{\sigma^\alpha{\bar z^{\bar i}}}D_{\sigma^\alpha}WD_{\bar z^{\bar i}}{\bar W}\sim
\frac{|W|^2n^s\mu_3l^2\left(\omega_\alpha\right)_{i{\bar j}}\xi^{\bar j}{\cal V}^{\frac{37}{36}}}{{\cal V}^2};\nonumber\\
& & e^KG^{\sigma^\alpha{\bar {\cal A}_1}}D_{\sigma^\alpha}WD_{\bar {\cal A}_1}{\bar W}\sim\frac{n^s|W|^2{\cal V}^{\frac{1}{18}}{\cal A}_1}{{\cal V}^2{\cal A}_1},\nonumber\\
& & e^KG^{{\cal G}^a{\bar z^{\bar i}}}D_{{\cal G}^a}WD_{\bar z^i}{\bar W}\sim
\frac{|W|^2n^s\left(m^a+\frac{{\cal G}^a}{ln {\cal V}}\right)\mu_3l^2\left(\omega_S\right)_{i{\bar j}}\xi^{\bar j}}{{\cal V}^2};\nonumber\\
& & e^KG^{{\cal G}^a{\bar {\cal A}_1}}D_{{\cal G}^a}WD_{\bar {\cal A}_1}{\bar W}\sim
\frac{n^sm^a|W|^2{\cal V}^{-\frac{35}{36}}{\cal A}_1}{{\cal V}^2{\cal A}_1},\nonumber\\
& & e^KG^{z^i{\bar z^j}}D_{z^i}WD_{\bar z^j}{\bar W}\sim\frac{|W|^2{\cal V}(\mu_3l^2)^2\left(\omega_S\right)_{i{\bar k}}\xi^{\bar k}\left(\omega_S\right)_{{\bar j}l}\xi^l}{{\cal V}^2},\nonumber\\
& & e^KG^{z^i{\bar {\cal A}_1}}D_{z^i}WD_{\bar {\cal A}_1}{\bar W}\sim
\frac{|W|^2\mu_3l^2{\cal V}^{\frac{1}{36}}\left(\omega_S\right)_{i{\bar j}}\xi^{\bar j}{\cal A}_1}{{\cal V}^2{\cal A}_1},\nonumber\\
& & e^KG^{{\cal A}_1{\bar {\cal A}_1}}D_{{\cal A}_1}WD_{\bar {\cal A}_1}{\bar W}\sim
\frac{{\cal V}^{-\frac{11}{12}}|{\cal A}_1|^2}{{\cal V}^2|{\cal A}_1|^2}.
\end{eqnarray}
We thus see the independence of the ${\cal N}=1$ potential in the LVS limit in a self-consistent way on ${\cal A}_1$ assuming it to be around ${\cal V}^{-\frac{1}{4}}$.
This justifies our assumption that one can taken the Wilson line moduli to be stabilized around ${\cal V}^{-\frac{1}{4}}$; we hence do get a competing contribution of the order of the volume of $D_5$ in $T_B$, which would hence guarantee  ${\cal O}(1)$
Yang-Mills coupling constant corresponding to the non-abelian gauge theory living on a stack of $D7$-branes wrapping $D_5$. Note that ${\cal A}_I=0$ is also an allowed extremum, which is in conformity with switching off of all but one Wilson line moduli for our analysis.

\section{Derivatives of $K|_{D_5}$ and $K|_{D_4}$}
\setcounter{equation}{0} \seceqbb

One needs the first and second derivatives of the geometric K\"{a}hler potential with respect to the position moduli of the mobile $D3$ brane, restricted in this paper, for convenience, to $D_5$. We also give, for completeness and for future work, the same for the geometric K\"{a}hler potential restricted to $D_4$.

The first and (mixed) second order derivatives of $K|_{D_5}$ are as follows:\begin{itemize}
\item
$\partial_{z_1}K|_{D_5}=$

$
\frac{-3\,{r2}\,}{3\,\left( z_3^3 \right) }\left( {{z_1}}^{17} - \phi\,{{z_1}}^5\,{{z_2}}^6 \right)  +
    4\,{\left( \frac{\zeta}
         {r_1\,|z_3|^2} \right) }^{\frac{1}{6}}\,
     \left( \left( -2\,\phi\,{{z_1}}^6\,{{z_2}}^6 + {{z_2}}^{18} \right) \,
        {\bar z_1} - {{z_1}}^5\,
        \left( {{z_1}}^{12} - \phi\,{{z_2}}^6 \right) \,
        \left( 1 + |{z_2}|^2 \right)  \right) $

        $ +
    \frac{3\,r_1\,{\left( \frac{\zeta}
           {r_1\,|z_3|^2} \right) }^{\frac{1}{6}}\,
       \left( 4\,{{z_1}}^{17} - 4\,\phi\,{{z_1}}^5\,{{z_2}}^6 +
         3\,{{z_1}}^{18}\,{\bar z_1} -
         \phi\,{{z_1}}^6\,{{z_2}}^6\,{\bar z_1} -
         {{z_2}}^{18}\,{\bar z_1} +
         4\,{{z_1}}^{17}\,{z_2}\,{\bar z_2} -
         4\,\phi\,{{z_1}}^5\,{{z_2}}^7\,{\bar z_2} -
         \frac{3\,r_2\,{{z_1}}^{17}}
          {{\left( \frac{\zeta}
               {r_1\,|z_3|^2} \right) }^{\frac{1}{6}}} +
         \frac{3\,\phi\,r_2\,{{z_1}}^5\,{{z_2}}^6}
          {{\left( \frac{\zeta}
               {r_1\,|z_3|^2} \right) }^{\frac{1}{6}}} \right) }{-
         r_2 + {\left( \frac{\zeta}
           {r_1\,|z_3|^2} \right) }^{\frac{1}{6}} +
       {z_1}\,{\bar z_1}\,
        {\left( \frac{\zeta}
            {r_1\,|z_3|^2} \right) }^{\frac{1}{6}} +
       {z_2}\,{\bar z_2}\,
        {\left( \frac{\zeta}
            {r_1|z_3|^2} \right) }^{\frac{1}{6}}}$

            $ +
    \frac{2\,\zeta\,{\left( \frac{\zeta}
           {r_1\,|z_3|^2} \right) }^{\frac{1}{6}}\,
       \left( \left( 2\,\phi\,{{z_1}}^6\,{{z_2}}^6 - {{z_2}}^{18} \right) \,
          {\bar z_1} + {{z_1}}^5\,
          \left( {{z_1}}^{12} - \phi\,{{z_2}}^6 \right) \,
          \left( 1 + |{z_2}|^2 \right)  \right) \,
       \left( r_2 - \left( 1 + |{z_1}|^2 +
            |{z_2}|^2 \right) \,
          {\left( \frac{\zeta}
              {r_1\,|z_3|^2} \right) }^{\frac{1}{6}} \right) }{
       r_1}$

$\sim {\cal V}^{\frac{11}{36}}
+ \left(ln {\cal V}\right)^{-\frac{1}{12}}{\cal V}^{\frac{17}{36}}
+ \sqrt{ln {\cal V}}{\cal V}^{\frac{17}{36}}
+ \left(ln {\cal V}\right)^{-\frac{7}{12}}{\cal V}^{\frac{29}{36}}
 \sim \left(ln {\cal V}\right)^{-\frac{7}{12}}{\cal V}^{\frac{29}{36}}$

\item
$\partial_{z_2}K_{\rm geom}=\partial_{z_1}K_{\rm geom}(z_1\leftrightarrow z_2)\sim \left(ln {\cal V}\right)^{-\frac{7}{12}}{\cal V}^{\frac{29}{36}}$

\item
$\partial_{z_1}{\bar\partial_{\bar z_1}}K_{\rm geom}=$

$
\frac{-1}{3\,
    \left( z_3^3 \right) }{\left( \frac{\zeta}{r_1\,|z_3|^2} \right) }^{\frac{1}{6}}\,
    \Biggl( 4\,\left( -2\,\phi\,{{z_1}}^6\,{{z_2}}^6 + {{z_2}}^{18} \right)  +
      \frac{4\,\left( {{\bar z_1}}^{17} -
           \phi\,{{\bar z_1}}^5\,{{\bar z_2}}^6 \right) \,
         \left( \left( 2\,\phi\,{{z_1}}^6\,{{z_2}}^6 - {{z_2}}^{18} \right) \,
            {\bar z_1} + {{z_1}}^5\,
            \left( {{z_1}}^{12} - \phi\,{{z_2}}^6 \right) \,
            \left( 1 + |{z_2}|^2 \right)  \right) }{-3z_3^3}$

            $ +
      \frac{2\,\zeta\,{\left( \frac{\zeta}
             {r_1\,|z_3|^2} \right) }^{\frac{1}{6}}\,
         \left( \left( 2\,\phi\,{{z_1}}^6\,{{z_2}}^6 - {{z_2}}^{18} \right) \,
            {\bar z_1} + {{z_1}}^5\,
            \left( {{z_1}}^{12} - \phi\,{{z_2}}^6 \right) \,
            \left( 1 + |{z_2}|^2 \right)  \right) \,
         \left( 2\,\phi\,{z_1}\,{{\bar z_1}}^6\,{{\bar z_2}}^6 -
           {z_1}\,{{\bar z_2}}^{18} +
           {{\bar z_1}}^{17}\,
            \left( 1 + |{z_2}|^2 \right)  -
           \phi\,{{\bar z_1}}^5\,{{\bar z_2}}^6\,
            \left( 1 + |{z_2}|^2 \right)  \right) }{r_1\,
         \left( {\bar z_3}^3 \right) }$

         $ -
      \frac{3\,r_1\,{{\bar z_1}}^5\,
         \left( {{\bar z_1}}^{12} - \phi\,{{\bar z_2}}^6 \right) \,
         \left( 4\,{{z_1}}^{17} - 4\,\phi\,{{z_1}}^5\,{{z_2}}^6 +
           3\,{{z_1}}^{18}\,{\bar z_1} -
           \phi\,{{z_1}}^6\,{{z_2}}^6\,{\bar z_1} -
           {{z_2}}^{18}\,{\bar z_1} +
           4\,{{z_1}}^{17}\,{z_2}\,{\bar z_2} -
           4\,\phi\,{{z_1}}^5\,{{z_2}}^7\,{\bar z_2} -
           \frac{3\,r_2\,{{z_1}}^{17}}
            {{\left( \frac{\zeta}
                 {r_1\,|z_3|^2} \right) }^{\frac{1}{6}}} +
           \frac{3\,\phi\,r_2\,{{z_1}}^5\,{{z_2}}^6}
            {{\left( \frac{\zeta}
                 {r_1\,|z_3|^2} \right) }^{\frac{1}{6}}} \right) }
         {\left( -{\bar z_3}^3 \right) \,
         \left( -r_2 + {\left( \frac{\zeta}
               {r_1\,|z_3|^2} \right) }^{\frac{1}{6}} +
           |{z_1}|^2\,
            {\left( \frac{\zeta}
                {r_1\,|z_3|^2} \right) }^{\frac{1}{6}} +
           |{z_2}|^2\,
            {\left( \frac{\zeta}
                {r_1\,|z_3|^2} \right) }^{\frac{1}{6}} \right) } $

                $-
      \frac{3\,r_1\,\left( 2\,\phi\,{z_1}\,{{\bar z_1}}^6\,
            {{\bar z_2}}^6 - {z_1}\,{{\bar z_2}}^{18} +
           {{\bar z_1}}^{17}\,
            \left( 1 + |{z_2}|^2 \right)  -
           \phi\,{{\bar z_1}}^5\,{{\bar z_2}}^6\,
            \left( 1 + |{z_2}|^2 \right)  \right) \,
         \left( \Sigma_1 \right) }{
         \left(-{\bar z_3}^3 \right) \,
         {\left( -r_2 + {\left( \frac{\zeta}
                 {r_1\,|z_3|^2} \right) }^{\frac{1}{6}} +
             |{z_1}|^2\,
              {\left( \frac{\zeta}
                  {r_1\,|z_3|^2} \right) }^{\frac{1}{6}} +
             |{z_2}|^2\,
              {\left( \frac{\zeta}
                  {r_1\,|z_3|^2} \right) }^{\frac{1}{6}} \right) }^2} +
       \frac{2\,\left( 2\,\phi\,{{z_1}}^6\,{{z_2}}^6 - {{z_2}}^{18} \right) \,\zeta\,
         \left( r_2 - \left( 1 + |{z_1}|^2 +
              |{z_2}|^2 \right) \,
            {\left( \frac{\zeta}
                {r_1\,|z_3|^2} \right) }^{\frac{1}{6}} \right) }{
         r_1}$

         $ - \frac{2\,\zeta\,{{\bar z_1}}^5\,
         \left( {{\bar z_1}}^{12} - \phi\,{{\bar z_2}}^6 \right) \,
         \left( \left( 2\,\phi\,{{z_1}}^6\,{{z_2}}^6 - {{z_2}}^{18} \right) \,
            {\bar z_1} + {{z_1}}^5\,
            \left( {{z_1}}^{12} - \phi\,{{z_2}}^6 \right) \,
            \left( 1 + |{z_2}|^2 \right)  \right) \,
         \left( r_2 - \left( 1 + |{z_1}|^2 +
              |{z_2}|^2 \right) \,
            {\left( \frac{\zeta}
                {r_1\,|z_3|^2} \right) }^{\frac{1}{6}} \right) }{
         r_1\,\left( -{\bar z_3}^3\right) }$

         $ +
      \frac{3\,r_1\,\left( 3\,{{z_1}}^{18} - \phi\,{{z_1}}^6\,{{z_2}}^6 -
           {{z_2}}^{18} - \frac{3\,r_2\,{{z_1}}^{17}\,
              \left( {{\bar z_1}}^{17} -
                \phi\,{{\bar z_1}}^5\,{{\bar z_2}}^6 \right) }{
              {\left( \frac{\zeta}
                  {r_1\,|z_3|^2} \right) }^{\frac{1}{6}}\,
              \left( -{\bar z_3}^3\right) } -
           \frac{3\,\phi\,r_2\,{{z_1}}^5\,{{z_2}}^6\,
              \left( -{{\bar z_1}}^{17} +
                \phi\,{{\bar z_1}}^5\,{{\bar z_2}}^6 \right) }{
              {\left( \frac{\zeta}
                  {r_1\,|z_3|^2} \right) }^{\frac{1}{6}}\,
              \left( {{\bar z_1}}^{18} -
                3\,\phi\,{{\bar z_1}}^6\,{{\bar z_2}}^6 +
                {{\bar z_2}}^{18} \right) } \right) }{-r_2 +
         {\left( \frac{\zeta}
             {r_1\,|z_3|^2} \right) }^{\frac{1}{6}} +
         |{z_1}|^2\,
          {\left( \frac{\zeta}
              {r_1\,|z_3|^2} \right) }^{\frac{1}{6}} +
         |{z_2}|^2\,
          {\left( \frac{\zeta}
              {r_1\,|z_3|^2} \right) }^{\frac{1}{6}}} \Biggr) $

$\sim \left(ln {\cal V}\right)^{-\frac{1}{12}}{\cal V}^{-\frac{5}{9}}\Biggl(
\sqrt{\cal V}
+ \sqrt{\cal V}
+ \left(ln {\cal V}\right)^{-\frac{7}{12}}{\cal V}^{-\frac{1}{36}}
+ \left(ln {\cal V}\right)^{\frac{7}{12}}\sqrt{\cal V}
+ \sqrt{{\cal V} ln {\cal V}}
+ \frac{1}{\sqrt{ln {\cal V}}}{\cal V}^{\frac{5}{6}}
+ \frac{1}{\sqrt{ln {\cal V}}}{\cal V}^{\frac{5}{6}}
+ \left(ln {\cal V}\right)^{\frac{7}{12}}\sqrt{\cal V}
\Biggr)\sim \left(ln {\cal V}\right)^{-\frac{7}{12}}{\cal V}^{\frac{5}{18}}$

\item
$
\partial_{z_2}{\bar\partial_{\bar z_2}}K_{\rm geom}=\partial_{z_1}{\bar\partial_{\bar z_1}}K_{\rm geom}(z_1\leftrightarrow z_2)\sim \left(ln {\cal V}\right)^{-\frac{7}{12}}{\cal V}^{\frac{5}{18}}$

\item
$\partial_{z_1}{\bar\partial_{\bar z_2}}K_{\rm geom}=$

$
-\frac{1}{3\,
    \left( -z_3^3 \right) }\Biggl( 4\,{{z1}}^5\,{z_2}\,\left( {{z_1}}^{12} - \phi\,{{z_2}}^6 \right) \,
       {\left( \frac{\zeta}
           {r_1\,|z_3|^2} \right) }^{\frac{1}{6}} +
      \frac{4\,\left( \phi\,{{\bar z_1}}^6\,{{\bar z_2}}^5 -
           {{\bar z_2}}^{17} \right) \,
         {\left( \frac{\zeta}
             {r_1\,|z_3|^2} \right) }^{\frac{1}{6}}\,
         \left( \left( 2\,\phi\,{{z1}}^6\,{{z_2}}^6 - {{z_2}}^{18} \right) \,
            {\bar z_1} + {{z_1}}^5\,
            \left( {{z_1}}^{12} - \phi\,{{z_2}}^6 \right) \,
            \left( 1 + |{z_2}|^2 \right)  \right) }{-{\bar z_3}^3}$

            $ +
      \frac{2\,\zeta\,{\left( \frac{\zeta}
             {r_1\,|z_3|^2} \right) }^{\frac{1}{3}}\,
         \left( {z_2}\,{{\bar z_1}}^{18} +
           \phi\,{z_1}\,{{\bar z_1}}^7\,{{\bar z_2}}^5 -
           {{\bar z_2}}^{17} -
           {z1}\,{\bar z_1}\,{{\bar z_2}}^{17} +
           \phi\,{{\bar z_1}}^6\,{{\bar z_2}}^5\,
            \left( 1 - 2\,{z2}\,{\bar z_2} \right)  \right) \,
         \left( \left( 2\,\phi\,{{z_1}}^6\,{{z_2}}^6 - {{z_2}}^{18} \right) \,
            {\bar z_1} + {{z_1}}^5\,
            \left( {{z_1}}^{12} - \phi\,{{z_2}}^6 \right) \,
            \left( 1 + |{z_2}|^2 \right)  \right) }{r_1\,
         \left( -{\bar z_3}^3\right) } $

         $\hskip-0.8in+
      \frac{3\,r_1\,{{\bar z_2}}^5\,
         \left( -\left( \phi\,{{\bar z_1}}^6 \right)  + {{\bar z_2}}^{12}
           \right) \,{\left( \frac{\zeta}
             {r_1\,|z_3|^2} \right) }^{\frac{1}{6}}\,
         \left( 4\,{{z_1}}^{17} - 4\,\phi\,{{z_1}}^5\,{{z_2}}^6 +
           3\,{{z_1}}^{18}\,{\bar z_1} -
           \phi\,{{z_1}}^6\,{{z_2}}^6\,{\bar z_1} -
           {{z_2}}^{18}\,{\bar z_1} +
           4\,{{z_1}}^{17}\,{z2}\,{\bar z_2} -
           4\,\phi\,{{z_1}}^5\,{{z_2}}^7\,{\bar z_2} -
           \frac{3\,r_2\,{{z_1}}^{17}}
            {{\left( \frac{\zeta}
                 {r_1\,|z_3|^2} \right) }^{\frac{1}{6}}} +
           \frac{3\,\phi\,r_2\,{{z1}}^5\,{{z_2}}^6}
            {{\left( \frac{\zeta}
                 {r_1\,|z_3|^2} \right) }^{\frac{1}{6}}} \right) }
         {\left( -{\bar z_3}^3 \right) \,
         \left( -r_2 + {\left( \frac{\zeta}
               {r_1\,|z_3|^2} \right) }^{\frac{1}{6}} +
           |{z_1}|^2\,
            {\left( \frac{\zeta}
                {r_1\,|z_3|^2} \right) }^{\frac{1}{6}} +
           |{z_2}|^2\,
            {\left( \frac{\zeta}
                {r_1\,|z_3|^2} \right) }^{\frac{1}{6}} \right) } +$

                $
      \frac{3\,r_1\,{\left( \frac{\zeta}
             {r_1\,|z_3|^2} \right) }^{\frac{1}{6}}\,
         \left( -\left( {z_2}\,{{\bar z_1}}^{18} \right)  -
           \phi\,{z_1}\,{{\bar z_1}}^7\,{{\bar z_2}}^5 +
           {{\bar z_2}}^{17} +
           {z_1}\,{\bar z_1}\,{{\bar z_2}}^{17} +
           \phi\,{{\bar z_1}}^6\,{{\bar z_2}}^5\,
            \left( -1 + 2\,|{z_2}|^2 \right)  \right) \,
         \left( \Sigma_1\right) }{
         \left( -{\bar z_3}^3\right) \,
         {\left( -r_2 + {\left( \frac{\zeta}
                 {r_1\,|z_3|^2} \right) }^{\frac{1}{6}} +
             {z_1}\,{\bar z_1}\,
              {\left( \frac{\zeta}
                  {r_1\,|z_3|^2} \right) }^{\frac{1}{6}} +
             {z_2}\,{\bar z_2}\,
              {\left( \frac{\zeta}
                  {r_1\,|z_3|^2} \right) }^{\frac{1}{6}} \right) }^2} $

                  $+
       \frac{3\,r_1\,{{z_1}}^5\,\left( {{z_1}}^{12} - \phi\,{{z_2}}^6 \right) \,
         \left( -3\,\phi\,r_2\,{{\bar z_1}}^6\,{{\bar z_2}}^5 +
           3\,r_2\,{{\bar z_2}}^{17} -
           4\,{z_2}\,{{\bar z_1}}^{18}\,
            {\left( \frac{\zeta}
                {r_1\,|z_3|^2} \right) }^{\frac{1}{6}} +
           12\,\phi\,{z_2}\,{{\bar z_1}}^6\,{{\bar z_2}}^6\,
            {\left( \frac{\zeta}
                {r_1\,|z_3|^2} \right) }^{\frac{1}{6}} -
           4\,{z_2}\,{{\bar z_2}}^{18}\,
            {\left( \frac{\zeta}
                {r_1\,|z_3|^2} \right) }^{\frac{1}{6}} \right) }{
         \left( -{\bar z_3}^3\right) \,
         \left( -r_2 + {\left( \frac{\zeta}
               {r_1\,|z_3|^2} \right) }^{\frac{1}{6}} +
           |{z_1}|^2\,
            {\left( \frac{\zeta}
                {r_1\,|z_3|^2} \right) }^{\frac{1}{6}} +
           |{z_2}|^2\,
            {\left( \frac{\zeta}
                {r_1\,|z_3|^2} \right) }^{\frac{1}{6}} \right) }$

                $ -
      \frac{2\,{{z_1}}^5\,z_2\,\left( {{z_1}}^{12} - \phi\,{z_2}^6 \right) \,\zeta\,
         {\left( \frac{\zeta}
             {r_1\,|z_3|^2} \right) }^{\frac{1}{6}}\,
         \left( r_2 - \left( 1 + |{z_1}|^2 +
              |{z_2}|^2 \right) \,
            {\left( \frac{\zeta}
                {r_1\,|z_3|^2} \right) }^{\frac{1}{6}} \right) }{
         r_1} $

         $+ \frac{2\,\zeta\,{{\bar z_2}}^5\,
         \left( -\left( \phi\,{{\bar z_1}}^6 \right)  + {{\bar z_2}}^{12}
           \right) \,{\left( \frac{\zeta}
             {r_1\,|z_3|^2} \right) }^{\frac{1}{6}}\,
         \left( \left( 2\,\phi\,{{z_1}}^6\,{z_2}^6 - {z_2}^{18} \right) \,
            {\bar z_1} + {{z_1}}^5\,
            \left( {{z_1}}^{12} - \phi\,{z_2}^6 \right) \,
            \left( 1 + |{z_2}|^2 \right)  \right) \,
         \left( r_2 - \left( 1 + |{z_1}|^2 +
              |{z_2}|^2 \right) \,
            {\left( \frac{\zeta}
                {r_1\,|z_3|^2} \right) }^{\frac{1}{6}} \right) }{
         r_1\,\left( -{\bar z_3}^3 \right) } \Biggr), $

$\sim\frac{1}{\sqrt{{\cal V}}}\Biggl(
\left(ln {\cal V}\right)^{-\frac{1}{12}}{\cal V}^{\frac{11}{18}}
+ \left(ln {\cal V}\right)^{-\frac{1}{12}}{\cal V}^{\frac{4}{9}}
+ \left(ln {\cal V}\right)^{-\frac{2}{3}}{\cal V}^{\frac{4}{9}}
+ \left(ln {\cal V}\right)^{\frac{1}{12}}{\cal V}^{\frac{4}{9}}
+ \left(ln {\cal V}\right)^{\frac{5}{12}}{\cal V}^{\frac{4}{9}}
+ \sqrt{ln {\cal V}}{\cal V}^{\frac{4}{9}}
+ \left(ln {\cal V}\right)^{-\frac{7}{12}}{\cal V}^{\frac{7}{9}}
+ \left(ln {\cal V}\right)^{-\frac{7}{12}}{\cal V}^{\frac{7}{9}}
\Biggr) \sim \left(ln {\cal V}\right)^{-\frac{7}{12}}{\cal V}^{\frac{5}{18}}$

         where
         $\Sigma_1\equiv 3\,r_2\,{{z_1}}^{17} - 3\,\phi\,r_2\,{{z_1}}^5\,{{z_2}}^6 -
           4\,{{z_1}}^{17}\,{\left( \frac{\zeta}
                {r_1\,|z_3|^2} \right) }^{\frac{1}{6}} +
           4\,\phi\,{{z_1}}^5\,{{z_2}}^6\,
            {\left( \frac{\zeta}
                {r_1\,|z_3|^2} \right) }^{\frac{1}{6}} -
           3\,{{z_1}}^{18}\,{\bar z_1}\,
            {\left( \frac{\zeta}
                {r_1\,|z_3|^2} \right) }^{\frac{1}{6}} +
           \phi\,{{z_1}}^6\,{{z_2}}^6\,{\bar z_1}\,
            {\left( \frac{\zeta}
                {r_1\,|z_3|^2} \right) }^{\frac{1}{6}} +
           {{z_2}}^{18}\,{\bar z_1}\,
            {\left( \frac{\zeta}
                {r_1\,|z_3|^2} \right) }^{\frac{1}{6}} -
           4\,{{z_1}}^{17}\,|{z_2}|^2\,
            {\left( \frac{\zeta}
                {r_1\,|z_3|^2} \right) }^{\frac{1}{6}} +
           4\,\phi\,{{z_1}}^5\,{{z_2}}^7\,{\bar z_2}\,
            {\left( \frac{\zeta}
                {r_1\,|z_3|^2} \right) }^{\frac{1}{6}}\sim{\cal V}^{\frac{29}{36}},$
and

$\Sigma_2=\Sigma_1(z_1\leftrightarrow z_2)\sim{\cal V}^{\frac{29}{36}}$

Hence, in the LVS limit, the $D_5$-metric components will scale with ${\cal V}$ as
follows:
\begin{equation}
\label{eq:D_5_metric}
G_{i{\bar j}}|_{D_5}(z_1,z_2)=\left(\begin{array}{cc}
\partial_{z_1}{\bar\partial_{\bar z_1}}K_{\rm geom} &
\partial_{z_1}{\bar\partial_{\bar z_2}}K_{\rm geom}\\
\partial_{z_2}{\bar\partial_{\bar z_1}}K_{\rm geom} &
\partial_{z_2}{\bar\partial_{\bar z_2}}K_{\rm geom}
\end{array}\right)\sim
 \left(ln {\cal V}\right)^{-\frac{7}{12}}{\cal V}^{\frac{5}{18}}\left(\begin{array}{cc}
{\cal O}(1) & {\cal O}(1) \\
{\cal O}(1) & {\cal O}(1)
\end{array}\right).
\end{equation}
\end{itemize}

The mixed double derivatives of the K\"{a}hler potential restricted to $D_4$
are given as under:

$\partial_2{\bar\partial_2}K|_{D_4}=
\frac{1}{3}\Biggl\{\frac{-12\,3^{\frac{1}{9}}\,{{r_2}}^2\,
       \left( \phi\,{{z_1}}^6\,{{z_2}}^5 - {{z_2}}^{17} \right) \,
       {{\bar z_2}}^5\,\left( 1 + |{z_1}|^2 +
         |{z_2}|^2 \right) \,
       \left( -\left( \phi\,{{\bar z_1}}^6 \right)  + {{\bar z_2}}^{12}
         \right) }{{\left( {{r_2}}^2\,|z_1^{18}+z_2^{18}-3\phi z_1^6z_2^6|^2 \right) }^{\frac{10}{9}}}$

            $ -
    \frac{2\,3^{\frac{1}{9}}\,{{r_2}}^2\,{{\bar z_2}}^5\,
       \left( -2\,\phi\,{{z_1}}^6\,{{z_2}}^5 + 2\,{{z_2}}^{17} +
         \left( -2\,\phi\,{{z_1}}^7\,{{z_2}}^5 + 2\,{z_1}\,{{z_2}}^{17} \right) \,
          {\bar z_1} + \left( -{{z_1}}^{18} + \phi\,{{z_1}}^6\,{{z_2}}^6 +
            {{z_2}}^{18} \right) \,{\bar z_2} \right) \,
       \left( -\left( \phi\,{{\bar z_1}}^6 \right)  + {{\bar z_2}}^{12}
         \right) }{{\left( {{r_2}}^2\,|z_1^{18}+z_2^{18}-3\phi z_1^6z_2^6|^2\right) }^{\frac{10}{9}}} - $

            $
    \frac{6\,3^{\frac{1}{9}}\,{{r_2}}^2\,\left( {{z_1}}^{18} - 3\,\phi\,{{z_1}}^6\,{{z_2}}^6 +
         {{z_2}}^{18} \right) \,{\bar z_2}\,
       \left( -\left( \phi\,{{\bar z_1}}^6\,{{\bar z_2}}^5 \right)  +
         {{\bar z_2}}^{17} \right) }{{\left( {{r_2}}^2\,
          |z_1^{18}+z_2^{18}-3\phi z_1^6z_2^6|^2 \right) }^{\frac{10}{9}}} +
    \frac{3\,3^{\frac{1}{9}}}
     {{\left( {{r_2}}^2\,|z_1^{18}+z_2^{18}-3\phi z_1^6z_2^6|^2  \right) }^{\frac{1}{9}}} -$

            $
    \frac{6\,3^{\frac{1}{9}}\,{z_2}\,\left( -\left( \phi\,{{z_1}}^6\,{{z_2}}^5 \right)  +
         {{z_2}}^{17} \right) }{\left( {{z_1}}^{18} - 3\,\phi\,{{z_1}}^6\,{{z_2}}^6 +
         {{z_2}}^{18} \right) \,{\left( {{r_2}}^2\,
           |z_1^{18}+z_2^{18}-3\phi z_1^6z_2^6|^2\right) }^{\frac{1}{9}}} + \frac{3^{\frac{1}{9}}\,\left( -{{z_1}}^{18} + \phi\,{{z_1}}^6\,{{z_2}}^6 + {{z_2}}^{18} \right) }{\left( {{z_1}}^{18} - 3\,\phi\,{{z_1}}^6\,{{z_2}}^6 + {{z_2}}^{18} \right) \,
       {\left( {{r_2}}^2\,|z_1^{18}+z_2^{18}-3\phi z_1^6z_2^6|^2\right) }^{\frac{1}{9}}} -$

             $
    \frac{3\,{r_1}\,\left( 3^{\frac{1}{9}}\,
          \left( {{z_1}}^{18} + 5\,\phi\,{{z_1}}^6\,{{z_2}}^6 - 7\,{{z_2}}^{18} \right)  -
         \frac{12\,{{r_2}}^3\,{{z_2}}^5\,
            \left( -\left( \phi\,{{z_1}}^6 \right)  + {{z_2}}^{12} \right) \,
            \left( {{z_1}}^{18} - 3\,\phi\,{{z_1}}^6\,{{z_2}}^6 + {{z_2}}^{18} \right) \,
            \left( \phi\,{{\bar z_1}}^6\,{{\bar z_2}}^5 -
              {{\bar z_2}}^{17} \right) }{{\left( {{r_2}}^2\,
               |z_1^{18}+z_2^{18}-3\phi z_1^6z_2^6|^2\right) }^{\frac{8}{9}}} \right) }{
       \left( {{z_1}}^{18} - 3\,\phi\,{{z_1}}^6\,{{z_2}}^6 + {{z_2}}^{18} \right) \,
       \left( 3^{\frac{1}{9}} + 3^{\frac{1}{9}}\,|{z_1}|^2 +
         3^{\frac{1}{9}}\,|{z_2}|^2 -
         {r_2}\,{\left( {{r_2}}^2\,
              |z_1^{18}+z_2^{18}-3\phi z_1^6z_2^6|^2 \right) }^{\frac{1}{9}} \right) } -$

                $
    \frac{2\,3^{\frac{1}{9}}\,\left( 2\,\phi\,{{z_1}}^6\,{{z_2}}^5 - 2\,{{z_2}}^{17} +
         2\,\left( \phi\,{{z_1}}^7\,{{z_2}}^5 - {z_1}\,{{z_2}}^{17} \right) \,
          {\bar z_1} + \left( {{z_1}}^{18} - \phi\,{{z_1}}^6\,{{z_2}}^6 -
            {{z_2}}^{18} \right) \,{\bar z_2} \right) \,
       \left( {{\bar z_1}}^{18} -
         3\,\phi\,{{\bar z_1}}^6\,{{\bar z_2}}^6 +
         {{\bar z_2}}^{18} \right) \,
       \left(\Sigma_2\right) \,
       \left( \Sigma_1 \right) }{{\left( {{r_2}}^2\,
          |z_1^{18}+z_2^{18}-3\phi z_1^6z_2^6|^2\right) }^{\frac{4}{3}}} +$

            $
    \frac{6\,3^{\frac{1}{9}}\,{{\bar z_2}}^5\,
       \left( 2\,\phi\,{{z_1}}^6\,{{z_2}}^5 - 2\,{{z_2}}^{17} +
         2\,\left( \phi\,{{z_1}}^7\,{{z_2}}^5 - {z_1}\,{{z_2}}^{17} \right) \,
          {\bar z_1} + \left( {{z_1}}^{18} - \phi\,{{z_1}}^6\,{{z_2}}^6 -
            {{z_2}}^{18} \right) \,{\bar z_2} \right) \,
       \left( -\left( \phi\,{{\bar z_1}}^6 \right)  + {{\bar z_2}}^{12}
         \right) \,{\left( \Sigma_1 \right) }^2}{{\left(
          {{r_2}}^2\,|z_1^{18}+z_2^{18}-3\phi z_1^6z_2^6|^2\right) }^{\frac{4}{3}}} -$

            $
    \frac{3^{\frac{1}{9}}\,\left( {{z_1}}^{18} - \phi\,{{z_1}}^6\,{{z_2}}^6 - {{z_2}}^{18}
         \right) \,\left( {{\bar z_1}}^{18} -
         3\,\phi\,{{\bar z_1}}^6\,{{\bar z_2}}^6 +
         {{\bar z_2}}^{18} \right) \,
       {\left( \Sigma_1\right) }^2}{{\left(
          {{r_2}}^2\,|z_1^{18}+z_2^{18}-3\phi z_1^6z_2^6|^2\right) }^{\frac{4}{3}}} +
    \frac{1}{
       \left( {{z_1}}^{18} - 3\,\phi\,{{z_1}}^6\,{{z_2}}^6 + {{z_2}}^{18} \right) \,
       {\left( \Sigma \right) }^2}$

       $\times\Biggl[3\,{r_1}\,\Biggl( \Sigma_2\Biggr) \,
       \Biggl( 8\,3^{\frac{1}{9}}\,{z_1}\,{{z_2}}^5\,
          \left( \phi\,{{z_1}}^6 - {{z_2}}^{12} \right) \,{\bar z_1} +
         3^{\frac{1}{9}}\,\left( {{z_1}}^{18} + 5\,\phi\,{{z_1}}^6\,{{z_2}}^6 -
            7\,{{z_2}}^{18} \right) \,{\bar z_2} -$

            $
         2\,{{z_2}}^5\,\left( -\left( \phi\,{{z_1}}^6 \right)  + {{z_2}}^{12} \right) \,
          \left( 4\,3^{\frac{1}{9}} - 3\,{r_2}\,
             {\left( {{r_2}}^2\,|z_1^{18}+z_2^{18}-3\phi z_1^6z_2^6|^2\right) }^{\frac{1}{9}} \right)  \Biggr) \Biggr]\Biggr\}
$

$\stackrel{LVS}{\sim}\frac{3\,{r_1}\,\left( 3^{\frac{1}{9}}\,
          \left( {{z_1}}^{18} + 5\,\phi\,{{z_1}}^6\,{{z_2}}^6 - 7\,{{z_2}}^{18} \right)  -
         \frac{12\,{{r_2}}^3\,{{z_2}}^5\,
            \left( -\left( \phi\,{{z_1}}^6 \right)  + {{z_2}}^{12} \right) \,
            \left( {{z_1}}^{18} - 3\,\phi\,{{z_1}}^6\,{{z_2}}^6 + {{z_2}}^{18} \right) \,
            \left( \phi\,{{\bar z_1}}^6\,{{\bar z_2}}^5 -
              {{\bar z_2}}^{17} \right) }{{\left( {{r_2}}^2\,
               |z_1^{18}+z_2^{18}-3\phi z_1^6z_2^6|^2\right) }^{\frac{8}{9}}} \right) }{
       \left( {{z_1}}^{18} - 3\,\phi\,{{z_1}}^6\,{{z_2}}^6 + {{z_2}}^{18} \right) \,
       \left( 3^{\frac{1}{9}} + 3^{\frac{1}{9}}\,|{z_1}|^2 +
         3^{\frac{1}{9}}\,|{z_2}|^2 -
         {r_2}\,{\left( {{r_2}}^2\,
              |z_1^{18}+z_2^{18}-3\phi z_1^6z_2^6|^2\right) }^{\frac{1}{9}} \right) }$

                $+\frac{1}{
       \left( {{z_1}}^{18} - 3\,\phi\,{{z_1}}^6\,{{z_2}}^6 + {{z_2}}^{18} \right) \,
       {\left( \Sigma_1 \right) }^2}$

       $\times\Biggl[3\,{r_1}\,\Biggl( \Sigma_2\Biggr) \,
       \Biggl( 8\,3^{\frac{1}{9}}\,{z_1}\,{{z_2}}^5\,
          \left( \phi\,{{z_1}}^6 - {{z_2}}^{12} \right) \,{\bar z_1} +
         3^{\frac{1}{9}}\,\left( {{z_1}}^{18} + 5\,\phi\,{{z_1}}^6\,{{z_2}}^6 -
            7\,{{z_2}}^{18} \right) \,{\bar z_2} -$

            $
         2\,{{z_2}}^5\,\left( -\left( \phi\,{{z_1}}^6 \right)  + {{z_2}}^{12} \right) \,
          \left( 4\,3^{\frac{1}{9}} - 3\,{r_2}\,
             {\left( {{r_2}}^2\,|z_1^{18}+z_2^{18}-3\phi z_1^6z_2^6|^2\right) }^{\frac{1}{9}} \right)  \Biggr) \Biggr]$

         $\sim\sqrt{ln{\cal V}} {\cal V}^{-\frac{1}{18}}.$
where

$\Sigma_1\equiv 3^{\frac{1}{9}} + 3^{\frac{1}{9}}\,|{z_1}|^2 +
         3^{\frac{1}{9}}\,|{z_2}|^2 -
         {r_2}\,{\left( {{r_2}}^2\,
              |z_1^{18}+z_2^{18}-3\phi z_1^6z_2^6|^2\right) }^{\frac{1}{9}},$

               $\Sigma_2\equiv  3^{\frac{1}{9}}\,{z_2} + \frac{2\,{{r_2}}^3\,
            \left( {{z_1}}^{18} - 3\,\phi\,{{z_1}}^6\,{{z_2}}^6 + {{z_2}}^{18} \right) \,
            \left( \phi\,{{\bar z_1}}^6\,{{\bar z_2}}^5 -
              {{\bar z_2}}^{17} \right) }{{\left( {{r_2}}^2\,
               |z_1^{18}+z_2^{18}-3\phi z_1^6z_2^6|^2\right) }^{\frac{8}{9}}} $

$\partial_1{\bar\partial_2}K_s=
\frac{1}{3}\Biggl\{\frac{12\,3^{\frac{1}{9}}\,{{r_2}}^2\,
       \left( {{z_1}}^{17} - \phi\,{{z_1}}^5\,{{z_2}}^6 \right) \,
       {{\bar z_2}}^5\,\left( 1 + |{z_1}|^2 +
         |{z_2}|^2 \right) \,
       \left( -\left( \phi\,{{\bar z_1}}^6 \right)  + {{\bar z_2}}^{12}
         \right) }{{\left( {{r_2}}^2\,|z_1^{18}+z_2^{18}-3\phi z_1^6z_2^6|^2 \right) }^{\frac{10}{9}}} +
    \frac{6\,3^{\frac{1}{9}}\,{{r_2}}^2\,\left( {{z_1}}^{18} - 3\,\phi\,{{z_1}}^6\,{{z_2}}^6 +
         {{z_2}}^{18} \right) \,{\bar z_1}\,
       \left( \phi\,{{\bar z_1}}^6\,{{\bar z_2}}^5 -
         {{\bar z_2}}^{17} \right) }{{\left( {{r_2}}^2\,
          |z_1^{18}+z_2^{18}-3\phi z_1^6z_2^6|^2 \right) }^{\frac{10}{9}}} +$

            $
    \frac{2\,3^{\frac{1}{9}}\,{{z_1}}^5\,{z_2}\,\left( {{z_1}}^{12} - \phi\,{{z_2}}^6 \right) }
     {\left( {{z_1}}^{18} - 3\,\phi\,{{z_1}}^6\,{{z_2}}^6 + {{z_2}}^{18} \right) \,
       {\left( {{r_2}}^2\,|z_1^{18}+z_2^{18}-3\phi z_1^6z_2^6|^2\right) }^{\frac{1}{9}}} +$

             $
    \frac{6\,3^{\frac{1}{9}}\,{z_2}\,\left( -{{z_1}}^{17} + \phi\,{{z_1}}^5\,{{z_2}}^6 \right) }
     {\left( {{z_1}}^{18} - 3\,\phi\,{{z_1}}^6\,{{z_2}}^6 + {{z_2}}^{18} \right) \,
       {\left( {{r_2}}^2\,|z_1^{18}+z_2^{18}-3\phi z_1^6z_2^6|^2\right) }^{\frac{1}{9}}} -\frac{2\,3^{\frac{1}{9}}\,{{r_2}}^2\,{{\bar z_2}}^5\,
       \left( -\left( \phi\,{{\bar z_1}}^6 \right)  + {{\bar z_2}}^{12}
         \right) \,\left( \left( {{z_1}}^{18} + \phi\,{{z_1}}^6\,{{z_2}}^6 - {{z_2}}^{18}
            \right) \,{\bar z_1} +
         2\,{{z_1}}^5\,\left( {{z_1}}^{12} - \phi\,{{z_2}}^6 \right) \,
          \left( 1 + |{z_2}|^2 \right)  \right) }{{\left( {{r_2}}^2\,
          |z_1^{18}+z_2^{18}-3\phi z_1^6z_2^6|^2 \right) }^{\frac{10}{9}}} +$

            $
    \frac{6\,{r_1}\,{{z_1}}^5\,\left( {{z_1}}^{12} - \phi\,{{z_2}}^6 \right) \,
       \left( 4\,3^{\frac{1}{9}}\,{z_2} + \frac{6\,{{r_2}}^3\,
            \left( {{z_1}}^{18} - 3\,\phi\,{{z_1}}^6\,{{z_2}}^6 + {{z_2}}^{18} \right) \,
            \left( \phi\,{{\bar z_1}}^6\,{{\bar z_2}}^5 -
              {{\bar z_2}}^{17} \right) }{{\left( {{r_2}}^2\,
               |z_1^{18}+z_2^{18}-3\phi z_1^6z_2^6|^2\right) }^{\frac{8}{9}}} \right) }{
       \left( {{z_1}}^{18} - 3\,\phi\,{{z_1}}^6\,{{z_2}}^6 + {{z_2}}^{18} \right) \,
       \left( \Sigma_1\right) } +\frac{2\,3^{\frac{1}{9}}\,{{z_1}}^5\,{z_2}\,\left( {{z_1}}^{12} - \phi\,{{z_2}}^6 \right) \,
       \left( {{\bar z_1}}^{18} -
         3\,\phi\,{{\bar z_1}}^6\,{{\bar z_2}}^6 +
         {{\bar z_2}}^{18} \right) \,
       {\left( \Sigma_1 \right) }^2}{{\left(
          {{r_2}}^2\,|z_1^{18}+z_2^{18}-3\phi z_1^6z_2^6|^2\right) }^{\frac{4}{3}}}+$

                $
    \frac{2\,3^{\frac{1}{9}}\,\left( {{\bar z_1}}^{18} -
         3\,\phi\,{{\bar z_1}}^6\,{{\bar z_2}}^6 +
         {{\bar z_2}}^{18} \right) \,
       \left( \left( {{z_1}}^{18} + \phi\,{{z_1}}^6\,{{z_2}}^6 - {{z_2}}^{18} \right) \,
          {\bar z_1} + 2\,{{z_1}}^5\,
          \left( {{z_1}}^{12} - \phi\,{{z_2}}^6 \right) \,
          \left( 1 + |{z_2}|^2 \right)  \right) \,
       \left( \Sigma_2\right) \,
       \left( \Sigma_1 \right) }{{\left( {{r_2}}^2\,
          |z_1^{18}+z_2^{18}-3\phi z_1^6z_2^6|^2\right) }^{\frac{4}{3}}} -$

            $
    \frac{6\,3^{\frac{1}{9}}\,{{\bar z_2}}^5\,
       \left( -\left( \phi\,{{\bar z_1}}^6 \right)  + {{\bar z_2}}^{12}
         \right) \,\left( \left( {{z_1}}^{18} + \phi\,{{z_1}}^6\,{{z_2}}^6 - {{z_2}}^{18}
            \right) \,{\bar z_1} +
         2\,{{z_1}}^5\,\left( {{z_1}}^{12} - \phi\,{{z_2}}^6 \right) \,
          \left( 1 + |{z_2}|^2 \right)  \right) \,
       {\left( \Sigma_1 \right) }^2}{{\left(
          {{r_2}}^2\,|z_1^{18}+z_2^{18}-3\phi z_1^6z_2^6|^2\right) }^{\frac{4}{3}}} - \frac{1}{
       \left( {{z_1}}^{18} - 3\,\phi\,{{z_1}}^6\,{{z_2}}^6 + {{z_2}}^{18} \right) \,
       {\left(\Sigma_1\right) }^2}$

       $\hskip-0.6in\times\Biggl[3\,{r_1}\,\left( 3^{\frac{1}{9}}\,{z_2} +
         \frac{2\,{{r_2}}^3\,\left( {{z_1}}^{18} - 3\,\phi\,{{z_1}}^6\,{{z_2}}^6 +
              {{z_2}}^{18} \right) \,\left( \phi\,{{\bar z_1}}^6\,
               {{\bar z_2}}^5 - {{\bar z_2}}^{17} \right) }{{\left(
               {{r_2}}^2\,|z_1^{18}+z_2^{18}-3\phi z_1^6z_2^6|^2 \right) }^{\frac{8}{9}}} \right) \,
       \Biggl( 3^{\frac{1}{9}}\,\left( 7\,{{z_1}}^{18} - 5\,\phi\,{{z_1}}^6\,{{z_2}}^6 -
            {{z_2}}^{18} \right) \,{\bar z_1} +
         2\,{{z_1}}^5\,\left( {{z_1}}^{12} - \phi\,{{z_2}}^6 \right) \,$

         $\times
          \left( 4\,3^{\frac{1}{9}} + 4\,3^{\frac{1}{9}}\,|{z_2}|^2 -
            3\,{r_2}\,{\left( {{r_2}}^2\,
                 |z_1^{18}+z_2^{18}-3\phi z_1^6z_2^6|^2 \right) }^{\frac{1}{9}} \right)  \Biggr)\Biggr] \Biggr\}$

$\stackrel{LVS}{\sim}\frac{6\,{r_1}\,{{z_1}}^5\,\left( {{z_1}}^{12} - \phi\,{{z_2}}^6 \right) \,
       \left( 4\,3^{\frac{1}{9}}\,{z_2} + \frac{6\,{{r_2}}^3\,
            \left( {{z_1}}^{18} - 3\,\phi\,{{z_1}}^6\,{{z_2}}^6 + {{z_2}}^{18} \right) \,
            \left( \phi\,{{\bar z_1}}^6\,{{\bar z_2}}^5 -
              {{\bar z_2}}^{17} \right) }{{\left( {{r_2}}^2\,
               |z_1^{18}+z_2^{18}-3\phi z_1^6z_2^6|^2 \right) }^{\frac{8}{9}}} \right) }{
       \left( {{z_1}}^{18} - 3\,\phi\,{{z_1}}^6\,{{z_2}}^6 + {{z_2}}^{18} \right) \,
       \left( \Sigma_1\right) } -\frac{1}{
       \left( {{z_1}}^{18} - 3\,\phi\,{{z_1}}^6\,{{z_2}}^6 + {{z_2}}^{18} \right) \,
       {\left(\Sigma_1\right) }^2}$

       $\hskip-0.6in\times\Biggl[3\,{r_1}\,\left( 3^{\frac{1}{9}}\,{z_2} +
         \frac{2\,{{r_2}}^3\,\left( {{z_1}}^{18} - 3\,\phi\,{{z_1}}^6\,{{z_2}}^6 +
              {{z_2}}^{18} \right) \,\left( \phi\,{{\bar z_1}}^6\,
               {{\bar z_2}}^5 - {{\bar z_2}}^{17} \right) }{{\left(
               {{r_2}}^2\,|z_1^{18}+z_2^{18}-3\phi z_1^6z_2^6|^2\right) }^{\frac{8}{9}}} \right) \,
       \Biggl( 3^{\frac{1}{9}}\,\left( 7\,{{z_1}}^{18} - 5\,\phi\,{{z_1}}^6\,{{z_2}}^6 -
            {{z_2}}^{18} \right) \,{\bar z_1} +
         2\,{{z_1}}^5\,\left( {{z_1}}^{12} - \phi\,{{z_2}}^6 \right) \,$

         $\times
          \left( 4\,3^{\frac{1}{9}} + 4\,3^{\frac{1}{9}}\,|{z_2}|^2 -
            3\,{r_2}\,{\left( {{r_2}}^2\,
                 |z_1^{18}+z_2^{18}-3\phi z_1^6z_2^6|^2\right) }^{\frac{1}{9}} \right)  \Biggr)\Biggr]$

$\sim\sqrt{ln {\cal V}}{\cal V}^{-\frac{1}{18}}.$

\section{Intermediate Expansions Relevant to Evaluation of the Complete K\"{a}hler Potential as a Power Series in the Matter Fields}
\setcounter{equation}{0} \seceqcc

The following are relevant to the expansion of the geometric K\"{a}hler potential in $\delta z_i$:
\begin{itemize}
\item
\begin{eqnarray}
\label{eq:K_geom_fluc}
& & 3\phi_0z_1^6z_2^6 - z_1^{18} - z_2^{18}\sim\sqrt{\cal V}\left[1 - {\cal V}^{-\frac{1}{36}}(\delta z_1 + \delta z_2) - {\cal V}^{-\frac{1}{18}}((\delta z_1)^2 + (\delta z_2)^2) + ...\right],\nonumber\\
& & r_2 - \left(\frac{\zeta}{r_1}\right)^{\frac{1}{6}}\left(1 + |z_1|^2 + |z_2|^2\right)\frac{1}{\left(\left|3\phi_0z_1^6z_2^6 - z_1^{18} - z_2^{18}\right|^2\right)^{\frac{1}{18}}}\sim\nonumber\\
& & \hskip -1in r_2 - \left(\frac{\zeta}{r_1}\right)^{\frac{1}{6}}{\cal V}^{-\frac{1}{18}}\left[{\cal V}^{\frac{1}{18}} + {\cal V}^{\frac{1}{36}}(\delta z_1 + \delta z_2 + c.c.) + |\delta z_1|^2 + |\delta z_2|^2 + (\delta z_1)^2 + (\delta z_2)^2 + \delta z_1 \delta z_2 + c.c. + |\delta z_1 + \delta z_2|^2 + ...\right],\nonumber\\
& & -r_2ln\left(\frac{\zeta}{r_1\left|3\phi_0z_1^6z_2^6 - z_1^{18} - z_2^{18}\right|^2}\right)^{\frac{1}{6}}\sim\nonumber\\
& & \hskip-1.3in -r_2ln\left\{\left(\frac{\zeta}{r_1}\right)^{\frac{1}{6}}{\cal V}^{-\frac{1}{18}}\right\} + r_2\frac{\delta z-1 + \delta z_2 + c.c.}{{\cal V}^{\frac{1}{36}}} + r_2\frac{(\delta z_1)^2 + (\delta z_2)^2 + \delta z_1\delta z_2 + c.c. + |\delta z_1 + \delta z_2|^2}{{\cal V}^{\frac{1}{18}}}
- r_2\frac{(\delta z_1 + \delta z_2 + c.c.)^2}{{\cal V}^{\frac{1}{18}}} + ...,\nonumber\\
& & \frac{\zeta}{9r_1}\left[r_2 - \left(\frac{\zeta}{r_1}\right)^{\frac{1}{6}}\left(1 + |z_1|^2 + |z_2|^2\right)\frac{1}{\left(\left|3\phi_0z_1^6z_2^6 - z_1^{18} - z_2^{18}\right|^2\right)^{\frac{1}{18}}}\right]^2\sim\nonumber\\
& & \hskip-1in\frac{\zeta r_2^2}{9r_1}\left[1 - \frac{1}{r_2}\left(\frac{\zeta}{r_1}\right)^{\frac{1}{6}}\frac{\delta z_1 + \delta z_2}{{\cal V}^{\frac{1}{36}}} + \frac{1}{r_2}\frac{(\delta z_1)^2 + (\delta z_2)^2 + \delta z_1\delta z_2 + c.c. + |\delta z_1 + \delta z_2|^2}{{\cal V}^{\frac{1}{18}}} + \frac{1}{r_2^2}\left(\frac{\zeta}{r_1}\right)^{\frac{1}{3}}\frac{(\delta z_1 + \delta z_2 + c.c.)^2}{{\cal V}^{\frac{1}{18}}} + ...\right],\nonumber\\
& & -r_1ln\left[\frac{1}{3}\left(\frac{\zeta}{r_1}\right)^{\frac{1}{2}}\frac{r_2 - \left(\frac{\zeta}{r_1}\right)^{\frac{1}{6}}\left(1 + |z_1|^2 + |z_2|^2\right)\frac{1}{\left(\left|3\phi_0z_1^6z_2^6 - z_1^{18} - z_2^{18}\right|^2\right)^{\frac{1}{18}}}}{\sqrt{\left|3\phi_0z_1^6z_2^6 - z_1^{18} - z_2^{18}\right|^2}}\right]\nonumber\\
& & \sim - r_2ln\left(r_2\sqrt{\frac{\zeta}{r_1}}\right)\nonumber\\
 & & - r_2\Biggl[\sqrt{\frac{\zeta}{r_1}}\frac{(|\delta z_1|^2 + |\delta z_2|^2 + (\delta z_1)^2 + (\delta z_2)^2 + \delta z_1 \delta z_2 + c.c. + |\delta z_1 + \delta z_2|^2)}{r_2{\cal V}^{\frac{1}{18}}} - \frac{1}{r_2}\sqrt{\frac{\zeta}{r_1}}\frac{(\delta z_1 + \delta z_2 + c.c.}{{\cal V}^{\frac{1}{36}}}\nonumber\\
  & & - \left(\frac{\zeta}{r_1}\right)^{\frac{1}{3}}\frac{((\delta z_1 + \delta z_2 + c.c.)^2}{{\cal V}^{\frac{1}{18}}}\Biggr]\nonumber\\
& & \hskip-1in+ r_1\left[-\frac{(\delta z_1 + \delta z_2 + c.c.)}{{\cal V}^{\frac{1}{36}}} - \frac{(|\delta z_1|^2 + |\delta z_2|^2 + (\delta z_1)^2 + (\delta z_2)^2 + \delta z_1 \delta z_2 + c.c. + |\delta z_1 + \delta z_2|^2)}{r_2{\cal V}^{\frac{1}{18}}} - \frac{((\delta z_1 + \delta z_2 + c.c.)^2}{{\cal V}^{\frac{1}{18}}}\right] + ...\nonumber\\
& & \sim  - r_2ln\left(r_2\sqrt{\frac{\zeta}{r_1}}\right) + \nonumber\\
& & \hskip -1inr_1\left[-\frac{(\delta z_1 + \delta z_2 + c.c.)}{{\cal V}^{\frac{1}{36}}} - \frac{(|\delta z_1|^2 + |\delta z_2|^2 + (\delta z_1)^2 + (\delta z_2)^2 + \delta z_1 \delta z_2 + c.c. + |\delta z_1 + \delta z_2|^2)}{r_2{\cal V}^{\frac{1}{18}}} - \frac{((\delta z_1 + \delta z_2 + c.c.)^2}{{\cal V}^{\frac{1}{18}}}\right] + ...\nonumber\\
& &
\end{eqnarray}
Using (\ref{eq:K_geom_fluc}), one otains:
\begin{equation}
\label{eq:Kexp}
\hskip -1inK_{\rm geom}|_{D_5}\sim\frac{r_2^2\zeta}{r_1} + \frac{r_2(\delta z_1 + \delta z_2 + c.c.)}{{\cal V}^{\frac{1}{36}}} + r_2\frac{((\delta z_1)^2 + (\delta z_2)^2 + \delta z_1 \delta z_2 + c.c.)}{{\cal V}^{\frac{1}{18}}} + r_2\frac{(|\delta z_1|^2 + |\delta z_2|^2 + \delta z_1 {\bar\delta z_2} + \delta z_2{\bar\delta z_1}}{{\cal V}^{\frac{1}{18}}} + ...
\end{equation}

\item

\begin{eqnarray}
\label{eq:K_fluc_1}
& & -2ln\Biggl[\Biggl({\cal T}_B(\sigma^B,{\bar\sigma ^B};{\cal G}^a,{\bar{\cal G}^a};\tau,{\bar\tau}) \nonumber\\
& & \hskip-1in+ \mu_3\Biggl\{{\cal V}^{\frac{1}{18}} + {\cal V}^{\frac{1}{36}}(\delta z_1 + \delta z_2) + (\delta z_1)^2 + (\delta z_2)^2 + \delta z_1\delta z_2 + {\cal V}^{\frac{1}{36}}(\delta z_1 + \delta z_2 + c.c.) + |\delta z_1|^2 + |\delta z_2|^2 + \delta z_1{\bar\delta z_2} + \delta z_2{\bar\delta z_1})\Biggr\}\nonumber\\
& & + i\kappa_4^2\mu_7C_{1{\bar 1}}\left[{\cal V}^{-\frac{1}{2}} + {\cal V}^{-\frac{1}{4}}(\delta a_1 + {\bar\delta a_1}) + |\delta a_1|^2\right]\nonumber\\
& & -\gamma\Biggl[\frac{\zeta r_2^2}{r_1} + r_2\frac{(\delta z_1 + \delta z_2 + c.c.)}{{\cal V}^{\frac{1}{36}}} + r_2\frac{(|\delta z_1|^2 + |\delta z_2|^2 + (\delta z_1)^2 + (\delta z_2)^2 + \delta z_1 \delta z_2 + c.c. + \delta z_1{\bar\delta z_2} + \delta z_2{\bar\delta z_1} }{{\cal V}^{\frac{1}{18}}}\Biggr]\Biggr)^{\frac{3}{2}}\nonumber\\
& & - \Biggl({\cal T}_S(\sigma^S,{\bar\sigma^S};{\cal G}^a,{\bar{\cal G}^a};\tau,{\bar\tau})+ \mu_3\Biggl\{{\cal V}^{\frac{1}{18}} + {\cal V}^{\frac{1}{36}}(\delta z_1 + \delta z_2) + (\delta z_1)^2 + (\delta z_2)^2 + \delta z_1\delta z_2 + {\cal V}^{\frac{1}{36}}(\delta z_1 + \delta z_2 + c.c.)\nonumber\\
& &  + |\delta z_1|^2 + |\delta z_2|^2 + \delta z_1{\bar\delta z_2} + \delta z_2{\bar\delta z_1})\Biggr\} \nonumber\\
& & -\gamma\Biggl[\frac{\zeta r_2^2}{r_1} + r_2\frac{(\delta z_1 + \delta z_2 + c.c.)}{{\cal V}^{\frac{1}{36}}} + r_2\frac{(|\delta z_1|^2 + |\delta z_2|^2 + (\delta z_1)^2 + (\delta z_2)^2 + \delta z_1 \delta z_2 + c.c. + \delta z_1{\bar\delta z_2} + \delta z_2{\bar\delta z_1} }{{\cal V}^{\frac{1}{18}}}\Biggr]\Biggr)^{\frac{3}{2}}\nonumber\\
& & +\sum_\beta n^0_\beta f({\cal G}^a,{\bar{\cal G}^a};\tau,{\bar\tau})\Biggr]\nonumber\\
& & \sim-2ln\Biggl[\sum_\beta n^0_\beta f({\cal G}^a,{\bar{\cal G}^a};\tau,{\bar\tau}) + \Biggl\{{\cal T}_B(\sigma^B,{\bar\sigma ^B};{\cal G}^a,{\bar{\cal G}^a};\tau,{\bar\tau}) + \mu_3{\cal V}^{\frac{1}{18}} + i\kappa_4^2\mu_7C_{1{\bar 1}}{\cal V}^{-\frac{1}{2}} - \gamma\left(r_2 + \frac{r_2^2\zeta}{r_1}\right)\Biggr\}^{\frac{3}{2}}\nonumber\\
 & & - \Biggl\{{\cal T}_S(\sigma^S,{\bar\sigma ^S};{\cal G}^a,{\bar{\cal G}^a};\tau,{\bar\tau}) + \mu_3{\cal V}^{\frac{1}{18}}  - \gamma\left(r_2 + \frac{r_2^2\zeta}{r_1}\right)\Biggr\}^{\frac{3}{2}}\Biggr]\nonumber\\
& & + (\delta z_1 + \delta z_2 + {\bar\delta z_1} + {\bar\delta z_2})\Biggl(\mu_3{\cal V}^{\frac{1}{36}}\Biggl[\sqrt{{\cal T}_B(\sigma^B,{\bar\sigma ^B};{\cal G}^a,{\bar{\cal G}^a};\tau,{\bar\tau}) + \mu_3{\cal V}^{\frac{1}{18}} + i\kappa_4^2\mu_7C_{1{\bar 1}}{\cal V}^{-\frac{1}{2}} - \gamma\left(r_2 + \frac{r_2^2\zeta}{r_1}\right)}\nonumber\\
 & & - \sqrt{{\cal T}_S(\sigma^S,{\bar\sigma ^S};{\cal G}^a,{\bar{\cal G}^a};\tau,{\bar\tau}) + \mu_3{\cal V}^{\frac{1}{18}}  - \gamma\left(r_2 + \frac{r_2^2\zeta}{r_1}\right)}\Biggr]\nonumber\\
  & & + \gamma r_2{\cal V}^{-\frac{1}{36}}\Biggl[\sqrt{{\cal T}_B(\sigma^B,{\bar\sigma ^B};{\cal G}^a,{\bar{\cal G}^a};\tau,{\bar\tau}) + \mu_3{\cal V}^{\frac{1}{18}} + i\kappa_4^2\mu_7C_{1{\bar 1}}{\cal V}^{-\frac{1}{2}} - \gamma\left(r_2 + \frac{r_2^2\zeta}{r_1}\right)}\nonumber\\
 & & - \sqrt{{\cal T}_S(\sigma^S,{\bar\sigma ^S};{\cal G}^a,{\bar{\cal G}^a};\tau,{\bar\tau}) + \mu_3{\cal V}^{\frac{1}{18}}  - \gamma\left(r_2 + \frac{r_2^2\zeta}{r_1}\right)}\Biggr]\Biggr)\nonumber\\
 & & + \left(\delta {\cal A}_1 + \delta{\bar{\cal A}_1}\right)\Biggl(i\kappa^2_4\mu_7C_{1{\bar 1}}{\cal V}^{-\frac{1}{2}}\sqrt{{\cal T}_B(\sigma^B,{\bar\sigma ^B};{\cal G}^a,{\bar{\cal G}^a};\tau,{\bar\tau}) + \mu_3{\cal V}^{\frac{1}{18}} + i\kappa_4^2\mu_7C_{1{\bar 1}}{\cal V}^{-\frac{1}{2}} - \gamma\left(r_2 + \frac{r_2^2\zeta}{r_1}\right)}\Biggr)\nonumber\\
 & & \hskip -1in\left(|\delta z_1|^2 + |\delta z_2|^2 + \delta z_1{\bar\delta z_2} + \delta z_2{\bar\delta z_1}\right)\Biggl((\mu_3{\cal V}^{\frac{1}{36}})^2\Biggl[\frac{1}{\sqrt{{\cal T}_B(\sigma^B,{\bar\sigma ^B};{\cal G}^a,{\bar{\cal G}^a};\tau,{\bar\tau}) + \mu_3{\cal V}^{\frac{1}{18}} + i\kappa_4^2\mu_7C_{1{\bar 1}}{\cal V}^{-\frac{1}{2}} - \gamma\left(r_2 + \frac{r_2^2\zeta}{r_1}\right)}}\nonumber\\
 & & -\frac{1}{\sqrt{{\cal T}_S(\sigma^S,{\bar\sigma ^S};{\cal G}^a,{\bar{\cal G}^a};\tau,{\bar\tau}) + \mu_3{\cal V}^{\frac{1}{18}}  - \gamma\left(r_2 + \frac{r_2^2\zeta}{r_1}\right)}}\Biggr] \nonumber\\
 & & + (\gamma r_2{\cal V}^{-\frac{1}{36}})^2\Biggl[\sqrt{{\cal T}_B(\sigma^B,{\bar\sigma ^B};{\cal G}^a,{\bar{\cal G}^a};\tau,{\bar\tau}) + \mu_3{\cal V}^{\frac{1}{18}} + i\kappa_4^2\mu_7C_{1{\bar 1}}{\cal V}^{-\frac{1}{2}} - \gamma\left(r_2 + \frac{r_2^2\zeta}{r_1}\right)}\nonumber\\
 & & - \sqrt{{\cal T}_S(\sigma^S,{\bar\sigma ^S};{\cal G}^a,{\bar{\cal G}^a};\tau,{\bar\tau}) + \mu_3{\cal V}^{\frac{1}{18}}  - \gamma\left(r_2 + \frac{r_2^2\zeta}{r_1}\right)}\Biggr]\nonumber\\
 & & +\mu_3\gamma r_2\Biggl[\frac{1}{\sqrt{{\cal T}_B(\sigma^B,{\bar\sigma ^B};{\cal G}^a,{\bar{\cal G}^a};\tau,{\bar\tau}) + \mu_3{\cal V}^{\frac{1}{18}} + i\kappa_4^2\mu_7C_{1{\bar 1}}{\cal V}^{-\frac{1}{2}} - \gamma\left(r_2 + \frac{r_2^2\zeta}{r_1}\right)}}\nonumber\\
 & & -\frac{1}{\sqrt{{\cal T}_S(\sigma^S,{\bar\sigma ^S};{\cal G}^a,{\bar{\cal G}^a};\tau,{\bar\tau}) + \mu_3{\cal V}^{\frac{1}{18}}  - \gamma\left(r_2 + \frac{r_2^2\zeta}{r_1}\right)}}\Biggr]\nonumber\\
 & & +\mu_3\Biggl[\sqrt{{\cal T}_B(\sigma^B,{\bar\sigma ^B};{\cal G}^a,{\bar{\cal G}^a};\tau,{\bar\tau}) + \mu_3{\cal V}^{\frac{1}{18}} + i\kappa_4^2\mu_7C_{1{\bar 1}}{\cal V}^{-\frac{1}{2}} - \gamma\left(r_2 + \frac{r_2^2\zeta}{r_1}\right)}\nonumber\\
 & & - \sqrt{{\cal T}_S(\sigma^S,{\bar\sigma ^S};{\cal G}^a,{\bar{\cal G}^a};\tau,{\bar\tau}) + \mu_3{\cal V}^{\frac{1}{18}}  - \gamma\left(r_2 + \frac{r_2^2\zeta}{r_1}\right)}\Biggr]\Biggr)\nonumber\\
 & & +|{\cal A}_1|^2\Biggl\{\frac{i\kappa_4^2\mu_7C_{1{\bar 1}}{\cal V}^{-\frac{1}{2}}}{\sqrt{{\cal T}_B(\sigma^B,{\bar\sigma ^B};{\cal G}^a,{\bar{\cal G}^a};\tau,{\bar\tau}) + \mu_3{\cal V}^{\frac{1}{18}} + i\kappa_4^2\mu_7C_{1{\bar 1}}{\cal V}^{-\frac{1}{2}} - \gamma\left(r_2 + \frac{r_2^2\zeta}{r_1}\right)}} \nonumber\\
 & & + i\kappa_4^2\mu_7C_{1{\bar 1}}\sqrt{{\cal T}_B(\sigma^B,{\bar\sigma ^B};{\cal G}^a,{\bar{\cal G}^a};\tau,{\bar\tau}) + \mu_3{\cal V}^{\frac{1}{18}} + i\kappa_4^2\mu_7C_{1{\bar 1}}{\cal V}^{-\frac{1}{2}} - \gamma\left(r_2 + \frac{r_2^2\zeta}{r_1}\right)}\Biggr\}
 \nonumber\\
 & & \hskip-1in+ \left(\delta z_1{\delta\bar{\cal A}_1} + \delta z_2{\delta\bar{\cal A}_1} + \delta{\bar z_1}\delta{\cal A}_1 + \delta{\bar z_2}\delta{\cal A}_1\right)
 \Biggl\{\frac{i\kappa_4^2\mu_7{\cal V}^{-\frac{1}{4}}\mu_3{\cal V}^{\frac{1}{36}}}{\sqrt{{\cal T}_B(\sigma^B,{\bar\sigma ^B};{\cal G}^a,{\bar{\cal G}^a};\tau,{\bar\tau}) + \mu_3{\cal V}^{\frac{1}{18}} + i\kappa_4^2\mu_7C_{1{\bar 1}}{\cal V}^{-\frac{1}{2}} - \gamma\left(r_2 + \frac{r_2^2\zeta}{r_1}\right)}}\Biggr\}\Biggr]\nonumber\\
 & &
\end{eqnarray}

\item
Using (\ref{eq:K_geom_fluc}) and (\ref{eq:K_fluc_1}), one arrives at (\ref{eq:K2}), wherein:
\begin{eqnarray}
\label{eq:K2defs}
& & K_{z_i{\bar z}_j}\sim\Biggl[\frac{(\mu_3{\cal V}^{\frac{1}{36}}+\gamma r_2{\cal V}^{-\frac{1}{36}})^2}{\Xi^2}\times\nonumber\\
& & \hskip-1.3in\left(\sqrt{{\cal T}_B(\sigma^B,{\bar\sigma ^B};{\cal G}^a,{\bar{\cal G}^a};\tau,{\bar\tau}) + \mu_3{\cal V}^{\frac{1}{18}} + i\kappa_4^2\mu_7C_{1{\bar 1}}{\cal V}^{-\frac{1}{2}} - \gamma\left(r_2 + \frac{r_2^2\zeta}{r_1}\right)}
   -\sqrt{{\cal T}_S(\sigma^S,{\bar\sigma ^S};{\cal G}^a,{\bar{\cal G}^a};\tau,{\bar\tau}) + \mu_3{\cal V}^{\frac{1}{18}}  - \gamma\left(r_2 + \frac{r_2^2\zeta}{r_1}\right)}\right)^2\nonumber\\
   & & + \left(\frac{(\mu_3{\cal V}^{\frac{1}{36}})^2 + \mu_3\gamma r_2}{\Xi}  \right)\Biggl\{\frac{1}{\sqrt{{\cal T}_B(\sigma^B,{\bar\sigma ^B};{\cal G}^a,{\bar{\cal G}^a};\tau,{\bar\tau}) + \mu_3{\cal V}^{\frac{1}{18}} + i\kappa_4^2\mu_7C_{1{\bar 1}}{\cal V}^{-\frac{1}{2}} - \gamma\left(r_2 + \frac{r_2^2\zeta}{r_1}\right)}}\nonumber\\
 & & -\frac{1}{\sqrt{{\cal T}_S(\sigma^S,{\bar\sigma ^S};{\cal G}^a,{\bar{\cal G}^a};\tau,{\bar\tau}) + \mu_3{\cal V}^{\frac{1}{18}}  - \gamma\left(r_2 + \frac{r_2^2\zeta}{r_1}\right)}}\Biggr\}\nonumber\\
 & & +\frac{(\gamma r_2{\cal V}^{-\frac{1}{36}})^2 + \mu_3}{\Xi}\Biggl\{\sqrt{{\cal T}_B(\sigma^B,{\bar\sigma ^B};{\cal G}^a,{\bar{\cal G}^a};\tau,{\bar\tau}) + \mu_3{\cal V}^{\frac{1}{18}} + i\kappa_4^2\mu_7C_{1{\bar 1}}{\cal V}^{-\frac{1}{2}} - \gamma\left(r_2 + \frac{r_2^2\zeta}{r_1}\right)}\nonumber\\
 & & - \sqrt{{\cal T}_S(\sigma^S,{\bar\sigma ^S};{\cal G}^a,{\bar{\cal G}^a};\tau,{\bar\tau}) + \mu_3{\cal V}^{\frac{1}{18}}  - \gamma\left(r_2 + \frac{r_2^2\zeta}{r_1}\right)}\Biggr\}\Biggr];\nonumber\\
   & & K_{{\cal A}_1\bar{\cal A}_1}\sim\Biggl\{\frac{(i\kappa_4^2\mu_7C_{1{\bar 1}}{\cal V}^{-\frac{1}{4}})^2}{\Xi\sqrt{{\cal T}_B(\sigma^B,{\bar\sigma ^B};{\cal G}^a,{\bar{\cal G}^a};\tau,{\bar\tau}) + \mu_3{\cal V}^{\frac{1}{18}} + i\kappa_4^2\mu_7C_{1{\bar 1}}{\cal V}^{-\frac{1}{2}} - \gamma\left(r_2 + \frac{r_2^2\zeta}{r_1}\right)}} \nonumber\\
   & & + \frac{i\kappa_4^2\mu_7C_{1{\bar 1}}}{\Xi}\sqrt{{\cal T}_B(\sigma^B,{\bar\sigma ^B};{\cal G}^a,{\bar{\cal G}^a};\tau,{\bar\tau}) + \mu_3{\cal V}^{\frac{1}{18}} + i\kappa_4^2\mu_7C_{1{\bar 1}}{\cal V}^{-\frac{1}{2}} - \gamma\left(r_2 + \frac{r_2^2\zeta}{r_1}\right)}\nonumber\\
   & & + \frac{\left(i\kappa_4^2\mu_7C_{1{\bar 1}}{\cal V}^{-\frac{1}{4}}\sqrt{{\cal T}_B(\sigma^B,{\bar\sigma ^B};{\cal G}^a,{\bar{\cal G}^a};\tau,{\bar\tau}) + \mu_3{\cal V}^{\frac{1}{18}} + i\kappa_4^2\mu_7C_{1{\bar 1}}{\cal V}^{-\frac{1}{2}} - \gamma\left(r_2 + \frac{r_2^2\zeta}{r_1}\right)}\right)^2}{\Xi^2}\Biggr\};\nonumber\\
   & & K_{z_i\bar{\cal A}_1}\sim
\Biggl\{\frac{(i\kappa_4^2\mu_7C_{1{\bar 1}}{\cal V}^{-\frac{1}{4}})(\mu_3{\cal V}^{\frac{1}{36}})}{\Xi\sqrt{{\cal T}_B(\sigma^B,{\bar\sigma ^B};{\cal G}^a,{\bar{\cal G}^a};\tau,{\bar\tau}) + \mu_3{\cal V}^{\frac{1}{18}} + i\kappa_4^2\mu_7C_{1{\bar 1}}{\cal V}^{-\frac{1}{2}} - \gamma\left(r_2 + \frac{r_2^2\zeta}{r_1}\right)}} \nonumber\\
   & & + \frac{i\kappa_4^2\mu_7C_{1{\bar 1}}{\cal V}^{-\frac{1}{4}}(\mu_3{\cal V}^{\frac{1}{36}}+\gamma r_2{\cal V}^{-\frac{1}{36}})\sqrt{{\cal T}_B(\sigma^B,{\bar\sigma ^B};{\cal G}^a,{\bar{\cal G}^a};\tau,{\bar\tau}) + \mu_3{\cal V}^{\frac{1}{18}} + i\kappa_4^2\mu_7C_{1{\bar 1}}{\cal V}^{-\frac{1}{2}} - \gamma\left(r_2 + \frac{r_2^2\zeta}{r_1}\right)}}{\Xi^2}\times\nonumber\\
   & & \hskip-1.3in\left(\sqrt{{\cal T}_B(\sigma^B,{\bar\sigma ^B};{\cal G}^a,{\bar{\cal G}^a};\tau,{\bar\tau}) + \mu_3{\cal V}^{\frac{1}{18}} + i\kappa_4^2\mu_7C_{1{\bar 1}}{\cal V}^{-\frac{1}{2}} - \gamma\left(r_2 + \frac{r_2^2\zeta}{r_1}\right)}
   -\sqrt{{\cal T}_S(\sigma^S,{\bar\sigma ^S};{\cal G}^a,{\bar{\cal G}^a};\tau,{\bar\tau}) + \mu_3{\cal V}^{\frac{1}{18}}  - \gamma\left(r_2 + \frac{r_2^2\zeta}{r_1}\right)}\right)\Biggr\};\nonumber\\
   & &
\end{eqnarray}
and
\begin{eqnarray}
\label{eq:Xi_def}
& &
\Xi\equiv \left({\cal T}_B(\sigma^B,{\bar\sigma ^B};{\cal G}^a,{\bar{\cal G}^a};\tau,{\bar\tau}) + \mu_3{\cal V}^{\frac{1}{18}} + i\kappa_4^2\mu_7C_{1{\bar 1}}{\cal V}^{-\frac{1}{2}} - \gamma\left(r_2 + \frac{r_2^2\zeta}{r_1}\right)\right)^\frac{3}{2}\nonumber\\
& &
-\left({\cal T}_S(\sigma^S,{\bar\sigma ^S};{\cal G}^a,{\bar{\cal G}^a};\tau,{\bar\tau}) + \mu_3{\cal V}^{\frac{1}{18}}  - \gamma\left(r_2 + \frac{r_2^2\zeta}{r_1}\right)\right)^{\frac{3}{2}}
+\sum_\beta n^0_\beta f({\cal G}^a,{\bar{\cal G}^a};\tau,{\bar\tau}).
\end{eqnarray}

\end{itemize}

\section{First and Second Derivatives of $\hat{K}_{{\cal Z}_i{\bar{\cal Z}}_i}$ and $\hat{K}_{\tilde{{\cal A}_1}{\bar{\tilde{\cal A}_1}}}$ and First Derivatives of det$\hat{K}_{i{\bar j}}$ with respect to Closed String Moduli $\sigma^\alpha,{\cal G}^a$}
\setcounter{equation}{0} \seceqdd

The first and second derivatives of $\hat{K}_{{\cal Z}_i{\bar{\cal Z}}_i}$ and $\hat{K}_{\tilde{{\cal A}_1}{\bar{\tilde{\cal A}_1}}}$ are relevant to the calculation of the soft SUSY breaking parameters in section {\bf 5}.

We can show that:
\begin{eqnarray}
\label{eq:dKhat_z}
& & \partial_{\sigma^\alpha}\hat{K}_{{\cal Z}_i{\bar{\cal Z}}_i}\sim\frac{(\mu_3{\cal V}^{\frac{1}{36}}+\gamma r_2{\cal V}^{-\frac{1}{36}})\sqrt{{\cal T}_\alpha(\sigma^\alpha,{\bar\sigma ^\alpha};{\cal G}^a,{\bar{\cal G}^a};\tau,{\bar\tau}) + \mu_3{\cal V}^{\frac{1}{18}} + i\kappa_4^2\mu_7C_{1{\bar 1}}{\cal V}^{-\frac{1}{2}} - \gamma\left(r_2 + \frac{r_2^2\zeta}{r_1}\right)}}{\Xi^2}\nonumber\\
& & \times\Biggl\{\frac{1}{\sqrt{{\cal T}_B(\sigma^B,{\bar\sigma ^B};{\cal G}^a,{\bar{\cal G}^a};\tau,{\bar\tau}) + \mu_3{\cal V}^{\frac{1}{18}} + i\kappa_4^2\mu_7C_{1{\bar 1}}{\cal V}^{-\frac{1}{2}} - \gamma\left(r_2 + \frac{r_2^2\zeta}{r_1}\right)}}\nonumber\\
 & & -\frac{1}{\sqrt{{\cal T}_S(\sigma^S,{\bar\sigma ^S};{\cal G}^a,{\bar{\cal G}^a};\tau,{\bar\tau}) + \mu_3{\cal V}^{\frac{1}{18}}  - \gamma\left(r_2 + \frac{r_2^2\zeta}{r_1}\right)}}\Biggr\}\nonumber\\
 & & + \frac{(\mu_3{\cal V}^{\frac{1}{36}}+\gamma r_2{\cal V}^{-\frac{1}{36}})}{\left({\cal T}_\alpha(\sigma^\alpha,{\bar\sigma ^\alpha};{\cal G}^a,{\bar{\cal G}^a};\tau,{\bar\tau}) + \mu_3{\cal V}^{\frac{1}{18}} + i\kappa_4^2\mu_7C_{1{\bar 1}}{\cal V}^{-\frac{1}{2}} - \gamma\left(r_2 + \frac{r_2^2\zeta}{r_1}\right)\right)^{\frac{3}{2}}\Xi}\nonumber\\
 & & + \frac{\sqrt{{\cal T}_\alpha(\sigma^\alpha,{\bar\sigma ^\alpha};{\cal G}^a,{\bar{\cal G}^a};\tau,{\bar\tau}) + \mu_3{\cal V}^{\frac{1}{18}} + i\kappa_4^2\mu_7C_{1{\bar 1}}{\cal V}^{-\frac{1}{2}} - \gamma\left(r_2 + \frac{r_2^2\zeta}{r_1}\right)}\left((\gamma r_2{\cal V}^{-\frac{1}{36}})^2 + \mu_3\right)}{\Xi^2}\nonumber\\
 & & \frac{\left((\gamma r_2{\cal V}^{-\frac{1}{36}})^2 + \mu_3\right)}{\Xi\sqrt{{\cal T}_\alpha(\sigma^\alpha,{\bar\sigma ^\alpha};{\cal G}^a,{\bar{\cal G}^a};\tau,{\bar\tau}) + \mu_3{\cal V}^{\frac{1}{18}} + i\kappa_4^2\mu_7C_{1{\bar 1}}{\cal V}^{-\frac{1}{2}} - \gamma\left(r_2 + \frac{r_2^2\zeta}{r_1}\right)}}
 \sim{\cal V}^{-\frac{35}{18}};\nonumber\\
 & & \partial_{{\cal G}^a}\hat{K}_{{\cal Z}_i{\bar{\cal Z}}_i}\sim\frac{(\mu_3{\cal V}^{\frac{1}{36}}+\gamma r_2{\cal V}^{-\frac{1}{36}})}{\Xi^2}
 \nonumber\\
 & & \times
 \Biggl[\sum_\beta k^a n^0_\beta sin(...) + ({\cal G}^a,{\bar{\cal G}^a})\Biggl\{\sqrt{{\cal T}_B(\sigma^B,{\bar\sigma ^B};{\cal G}^a,{\bar{\cal G}^a};\tau,{\bar\tau}) + \mu_3{\cal V}^{\frac{1}{18}} + i\kappa_4^2\mu_7C_{1{\bar 1}}{\cal V}^{-\frac{1}{2}} - \gamma\left(r_2 + \frac{r_2^2\zeta}{r_1}\right)}\nonumber\\
  & & - \sqrt{{\cal T}_S(\sigma^S,{\bar\sigma ^S};{\cal G}^a,{\bar{\cal G}^a};\tau,{\bar\tau}) + \mu_3{\cal V}^{\frac{1}{18}}  - \gamma\left(r_2 + \frac{r_2^2\zeta}{r_1}\right)}\Biggr\}\Biggr]\nonumber\\
 & & \times\Biggl\{\frac{1}{\sqrt{{\cal T}_B(\sigma^B,{\bar\sigma ^B};{\cal G}^a,{\bar{\cal G}^a};\tau,{\bar\tau}) + \mu_3{\cal V}^{\frac{1}{18}} + i\kappa_4^2\mu_7C_{1{\bar 1}}{\cal V}^{-\frac{1}{2}} - \gamma\left(r_2 + \frac{r_2^2\zeta}{r_1}\right)}}\nonumber\\
 & & -\frac{1}{\sqrt{{\cal T}_S(\sigma^S,{\bar\sigma ^S};{\cal G}^a,{\bar{\cal G}^a};\tau,{\bar\tau}) + \mu_3{\cal V}^{\frac{1}{18}}  - \gamma\left(r_2 + \frac{r_2^2\zeta}{r_1}\right)}}\Biggr\}\nonumber\\
 & & \frac{\left((\gamma r_2{\cal V}^{-\frac{1}{36}})^2 + \mu_3\right)}{\Xi^2}\nonumber\\
 & & \times\Biggl[\sum_\beta k^a n^0_\beta sin(...) + ({\cal G}^a,{\bar{\cal G}^a})\Biggl\{\sqrt{{\cal T}_B(\sigma^B,{\bar\sigma ^B};{\cal G}^a,{\bar{\cal G}^a};\tau,{\bar\tau}) + \mu_3{\cal V}^{\frac{1}{18}} + i\kappa_4^2\mu_7C_{1{\bar 1}}{\cal V}^{-\frac{1}{2}} - \gamma\left(r_2 + \frac{r_2^2\zeta}{r_1}\right)} - \nonumber\\
 & & \sqrt{{\cal T}_S(\sigma^S,{\bar\sigma ^S};{\cal G}^a,{\bar{\cal G}^a};\tau,{\bar\tau}) + \mu_3{\cal V}^{\frac{1}{18}}  - \gamma\left(r_2 + \frac{r_2^2\zeta}{r_1}\right)}\Biggr\}\Biggr]\nonumber\\
 & & \hskip-0.9in\times \Biggl\{\sqrt{{\cal T}_B(\sigma^B,{\bar\sigma ^B};{\cal G}^a,{\bar{\cal G}^a};\tau,{\bar\tau}) + \mu_3{\cal V}^{\frac{1}{18}} + i\kappa_4^2\mu_7C_{1{\bar 1}}{\cal V}^{-\frac{1}{2}} - \gamma\left(r_2 + \frac{r_2^2\zeta}{r_1}\right)}
   -\sqrt{{\cal T}_S(\sigma^S,{\bar\sigma ^S};{\cal G}^a,{\bar{\cal G}^a};\tau,{\bar\tau}) + \mu_3{\cal V}^{\frac{1}{18}}  - \gamma\left(r_2 + \frac{r_2^2\zeta}{r_1}\right)}\Biggr\}\nonumber\\
   & & \sim{\cal V}^{\frac{41}{36}}.
\end{eqnarray}
Hence,
\begin{eqnarray}
\label{eq:ddKhat_z}
& & \partial_{\sigma^B{\bar\sigma}^B}\hat{K}_{{\cal Z}_i{\bar{\cal Z}}_i}\sim\frac{(\mu_3{\cal V}^{\frac{1}{36}}+\gamma r_2{\cal V}^{-\frac{1}{36}})\sqrt{{\cal T}_B(\sigma^B,{\bar\sigma ^B};{\cal G}^a,{\bar{\cal G}^a};\tau,{\bar\tau}) + \mu_3{\cal V}^{\frac{1}{18}} + i\kappa_4^2\mu_7C_{1{\bar 1}}{\cal V}^{-\frac{1}{2}} - \gamma\left(r_2 + \frac{r_2^2\zeta}{r_1}\right)}}{\Xi^3}\nonumber\\
& & \times\Biggl\{\frac{1}{\sqrt{{\cal T}_B(\sigma^B,{\bar\sigma ^B};{\cal G}^a,{\bar{\cal G}^a};\tau,{\bar\tau}) + \mu_3{\cal V}^{\frac{1}{18}} + i\kappa_4^2\mu_7C_{1{\bar 1}}{\cal V}^{-\frac{1}{2}} - \gamma\left(r_2 + \frac{r_2^2\zeta}{r_1}\right)}}\nonumber\\
 & & -\frac{1}{\sqrt{{\cal T}_S(\sigma^S,{\bar\sigma ^S};{\cal G}^a,{\bar{\cal G}^a};\tau,{\bar\tau}) + \mu_3{\cal V}^{\frac{1}{18}}  - \gamma\left(r_2 + \frac{r_2^2\zeta}{r_1}\right)}}\Biggr\}\nonumber\\
 & & + \frac{(\mu_3{\cal V}^{\frac{1}{36}}+\gamma r_2{\cal V}^{-\frac{1}{36}})}{\Xi\left({\cal T}_B(\sigma^B,{\bar\sigma ^B};{\cal G}^a,{\bar{\cal G}^a};\tau,{\bar\tau}) + \mu_3{\cal V}^{\frac{1}{18}} + i\kappa_4^2\mu_7C_{1{\bar 1}}{\cal V}^{-\frac{1}{2}} - \gamma\left(r_2 + \frac{r_2^2\zeta}{r_1}\right)\right)}\nonumber\\
 & & \times\Biggl\{\frac{1}{\sqrt{{\cal T}_B(\sigma^B,{\bar\sigma ^B};{\cal G}^a,{\bar{\cal G}^a};\tau,{\bar\tau}) + \mu_3{\cal V}^{\frac{1}{18}} + i\kappa_4^2\mu_7C_{1{\bar 1}}{\cal V}^{-\frac{1}{2}} - \gamma\left(r_2 + \frac{r_2^2\zeta}{r_1}\right)}}\nonumber\\
 & & -\frac{1}{\sqrt{{\cal T}_S(\sigma^S,{\bar\sigma ^S};{\cal G}^a,{\bar{\cal G}^a};\tau,{\bar\tau}) + \mu_3{\cal V}^{\frac{1}{18}}  - \gamma\left(r_2 + \frac{r_2^2\zeta}{r_1}\right)}}\Biggr\}\nonumber\\
 & & \frac{(\mu_3{\cal V}^{\frac{1}{36}}+\gamma r_2{\cal V}^{-\frac{1}{36}})}{\Xi^2\left({\cal T}_B(\sigma^B,{\bar\sigma ^B};{\cal G}^a,{\bar{\cal G}^a};\tau,{\bar\tau}) + \mu_3{\cal V}^{\frac{1}{18}} + i\kappa_4^2\mu_7C_{1{\bar 1}}{\cal V}^{-\frac{1}{2}} - \gamma\left(r_2 + \frac{r_2^2\zeta}{r_1}\right)\right)}\nonumber\\
 & & \frac{(\mu_3{\cal V}^{\frac{1}{36}}+\gamma r_2{\cal V}^{-\frac{1}{36}})}{\Xi\left({\cal T}_B(\sigma^B,{\bar\sigma ^B};{\cal G}^a,{\bar{\cal G}^a};\tau,{\bar\tau}) + \mu_3{\cal V}^{\frac{1}{18}} + i\kappa_4^2\mu_7C_{1{\bar 1}}{\cal V}^{-\frac{1}{2}} - \gamma\left(r_2 + \frac{r_2^2\zeta}{r_1}\right)\right)^{\frac{5}{2}}}\nonumber\\
 & & + \frac{\left(\mu_3+\left\{\gamma r_2{\cal V}^{-\frac{1}{36}}\right\}\right)}{\Xi^2}+\frac{\left(\mu_3+\left\{\gamma r_2{\cal V}^{-\frac{1}{36}}\right\}\right)\left[{\cal T}_B(\sigma^B,{\bar\sigma ^B};{\cal G}^a,{\bar{\cal G}^a};\tau,{\bar\tau}) + \mu_3{\cal V}^{\frac{1}{18}} + i\kappa_4^2\mu_7C_{1{\bar 1}}{\cal V}^{-\frac{1}{2}} - \gamma\left(r_2 + \frac{r_2^2\zeta}{r_1}\right)\right]^{\frac{3}{2}}}{\Xi^3}\nonumber\\
 & & \frac{\left(\mu_3+\left\{\gamma r_2{\cal V}^{-\frac{1}{36}}\right\}\right)}{\Xi^2} + \frac{\left(\mu_3+\left\{\gamma r_2{\cal V}^{-\frac{1}{36}}\right\}\right)}{\Xi\left[{\cal T}_B(\sigma^B,{\bar\sigma ^B};{\cal G}^a,{\bar{\cal G}^a};\tau,{\bar\tau}) + \mu_3{\cal V}^{\frac{1}{18}} + i\kappa_4^2\mu_7C_{1{\bar 1}}{\cal V}^{-\frac{1}{2}} - \gamma\left(r_2 + \frac{r_2^2\zeta}{r_1}\right)\right]^{\frac{3}{2}}}
 \nonumber\\
 & & \sim {\cal V}^{-\frac{19}{18}}\sim\partial_{\sigma^S{\bar\sigma}^S}\hat{K}_{{\cal Z}_i{\bar{\cal Z}}_i};\nonumber\\
& & \partial_{\sigma^B}{\bar\partial}_{{\bar\sigma}^S}\hat{K}_{{\cal Z}_i{\bar{\cal Z}}_i}\sim \frac{(\mu_3{\cal V}^{\frac{1}{36}}+\gamma r_2{\cal V}^{-\frac{1}{36}})}{\Xi^3}\nonumber\\
& & \times \hskip -0.8in \times\sqrt{{\cal T}_B(\sigma^B,{\bar\sigma ^B};{\cal G}^a,{\bar{\cal G}^a};\tau,{\bar\tau}) + \mu_3{\cal V}^{\frac{1}{18}} + i\kappa_4^2\mu_7C_{1{\bar 1}}{\cal V}^{-\frac{1}{2}} - \gamma\left(r_2 + \frac{r_2^2\zeta}{r_1}\right)}\sqrt{{\cal T}_S(\sigma^S,{\bar\sigma ^S};{\cal G}^a,{\bar{\cal G}^a};\tau,{\bar\tau}) + \mu_3{\cal V}^{\frac{1}{18}}  - \gamma\left(r_2 + \frac{r_2^2\zeta}{r_1}\right)}\nonumber\\
& & \times\Biggl\{\frac{1}{\sqrt{{\cal T}_B(\sigma^B,{\bar\sigma ^B};{\cal G}^a,{\bar{\cal G}^a};\tau,{\bar\tau}) + \mu_3{\cal V}^{\frac{1}{18}} + i\kappa_4^2\mu_7C_{1{\bar 1}}{\cal V}^{-\frac{1}{2}} - \gamma\left(r_2 + \frac{r_2^2\zeta}{r_1}\right)}}\nonumber\\
 & & -\frac{1}{\sqrt{{\cal T}_S(\sigma^S,{\bar\sigma ^S};{\cal G}^a,{\bar{\cal G}^a};\tau,{\bar\tau}) + \mu_3{\cal V}^{\frac{1}{18}}  - \gamma\left(r_2 + \frac{r_2^2\zeta}{r_1}\right)}}\Biggr\}\nonumber\\
 & & \hskip-0.8in+ \frac{(\mu_3{\cal V}^{\frac{1}{36}}+\gamma r_2{\cal V}^{-\frac{1}{36}})}{\Xi^2\left[{\cal T}_S(\sigma^S,{\bar\sigma ^S};{\cal G}^a,{\bar{\cal G}^a};\tau,{\bar\tau}) + \mu_3{\cal V}^{\frac{1}{18}}  - \gamma\left(r_2 + \frac{r_2^2\zeta}{r_1}\right)\right]}\sqrt{\frac{{\cal T}_B(\sigma^B,{\bar\sigma ^B};{\cal G}^a,{\bar{\cal G}^a};\tau,{\bar\tau}) + \mu_3{\cal V}^{\frac{1}{18}} + i\kappa_4^2\mu_7C_{1{\bar 1}}{\cal V}^{-\frac{1}{2}} - \gamma\left(r_2 + \frac{r_2^2\zeta}{r_1}\right)}{{\cal T}_S(\sigma^S,{\bar\sigma ^S};{\cal G}^a,{\bar{\cal G}^a};\tau,{\bar\tau}) + \mu_3{\cal V}^{\frac{1}{18}}  - \gamma\left(r_2 + \frac{r_2^2\zeta}{r_1}\right)}}\nonumber\\
 & & \hskip-0.8in+ \frac{(\mu_3{\cal V}^{\frac{1}{36}}+\gamma r_2{\cal V}^{-\frac{1}{36}})}{\Xi^2\left[{\cal T}_B(\sigma^B,{\bar\sigma^B};{\cal G}^a,{\bar{\cal G}^a};\tau,{\bar\tau}) + \mu_3{\cal V}^{\frac{1}{18}}  - \gamma\left(r_2 + \frac{r_2^2\zeta}{r_1}\right)\right]}\sqrt{\frac{{\cal T}_S(\sigma^S,{\bar\sigma ^S};{\cal G}^a,{\bar{\cal G}^a};\tau,{\bar\tau}) + \mu_3{\cal V}^{\frac{1}{18}} + i\kappa_4^2\mu_7C_{1{\bar 1}}{\cal V}^{-\frac{1}{2}} - \gamma\left(r_2 + \frac{r_2^2\zeta}{r_1}\right)}{{\cal T}_B(\sigma^B,{\bar\sigma ^B};{\cal G}^a,{\bar{\cal G}^a};\tau,{\bar\tau}) + \mu_3{\cal V}^{\frac{1}{18}}  - \gamma\left(r_2 + \frac{r_2^2\zeta}{r_1}\right)}}\nonumber\\
 & & + \frac{(\mu_3{\cal V}^{\frac{1}{36}}+\gamma r_2{\cal V}^{-\frac{1}{36}})\sqrt{{\cal T}_S(\sigma^S,{\bar\sigma ^S};{\cal G}^a,{\bar{\cal G}^a};\tau,{\bar\tau}) + \mu_3{\cal V}^{\frac{1}{18}}  - \gamma\left(r_2 + \frac{r_2^2\zeta}{r_1}\right)}}{\Xi^2\left({\cal T}_B(\sigma^B,{\bar\sigma ^B};{\cal G}^a,{\bar{\cal G}^a};\tau,{\bar\tau}) + \mu_3{\cal V}^{\frac{1}{18}} + i\kappa_4^2\mu_7C_{1{\bar 1}}{\cal V}^{-\frac{1}{2}} - \gamma\left(r_2 + \frac{r_2^2\zeta}{r_1}\right)\right)^{\frac{3}{2}}}\nonumber\\
 & & + \frac{\left(\mu_3+\left\{\gamma r_2{\cal V}^{-\frac{1}{36}}\right\}\right)\sqrt{{\cal T}_S(\sigma^S,{\bar\sigma ^S};{\cal G}^a,{\bar{\cal G}^a};\tau,{\bar\tau}) + \mu_3{\cal V}^{\frac{1}{18}}  - \gamma\left(r_2 + \frac{r_2^2\zeta}{r_1}\right)}}{\Xi^3}\nonumber\\
 & & \times\left[{\cal T}_B(\sigma^B,{\bar\sigma ^B};{\cal G}^a,{\bar{\cal G}^a};\tau,{\bar\tau}) + \mu_3{\cal V}^{\frac{1}{18}} + i\kappa_4^2\mu_7C_{1{\bar 1}}{\cal V}^{-\frac{1}{2}} - \gamma\left(r_2 + \frac{r_2^2\zeta}{r_1}\right)\right]\nonumber\\
 & & + \frac{\left(\mu_3+\left\{\gamma r_2{\cal V}^{-\frac{1}{36}}\right\}\right)}{\Xi^2}\sqrt{\frac{{\cal T}_B(\sigma^B,{\bar\sigma ^B};{\cal G}^a,{\bar{\cal G}^a};\tau,{\bar\tau}) + \mu_3{\cal V}^{\frac{1}{18}} + i\kappa_4^2\mu_7C_{1{\bar 1}}{\cal V}^{-\frac{1}{2}} - \gamma\left(r_2 + \frac{r_2^2\zeta}{r_1}\right)}{{\cal T}_S(\sigma^S,{\bar\sigma ^S};{\cal G}^a,{\bar{\cal G}^a};\tau,{\bar\tau}) + \mu_3{\cal V}^{\frac{1}{18}}  - \gamma\left(r_2 + \frac{r_2^2\zeta}{r_1}\right)}}\nonumber\\
 & & \sim {\cal V}^{-2};\nonumber\\
 & & \partial_{\sigma^\alpha}{\bar\partial}_{{\bar{\cal G}}^a}\hat{K}_{{\cal Z}_i{\bar{\cal Z}}_i}
 \sim\frac{\left(\mu_3{\cal V}^{\frac{1}{36}}+\gamma r_2{\cal V}^{-\frac{1}{36}}\right)\sqrt{{\cal T}_\alpha(\sigma^\alpha,{\bar\sigma^\alpha};{\cal G}^a,{\bar{\cal G}^a};\tau,{\bar\tau}) + \mu_3{\cal V}^{\frac{1}{18}}  - \gamma\left(r_2 + \frac{r_2^2\zeta}{r_1}\right)}{\cal V}^{\frac{5}{6}}}{\Xi^3}\nonumber\\
 & & \times\Biggl\{\frac{1}{\sqrt{{\cal T}_B(\sigma^B,{\bar\sigma ^B};{\cal G}^a,{\bar{\cal G}^a};\tau,{\bar\tau}) + \mu_3{\cal V}^{\frac{1}{18}} + i\kappa_4^2\mu_7C_{1{\bar 1}}{\cal V}^{-\frac{1}{2}} - \gamma\left(r_2 + \frac{r_2^2\zeta}{r_1}\right)}}\nonumber\\
 & & -\frac{1}{\sqrt{{\cal T}_S(\sigma^S,{\bar\sigma ^S};{\cal G}^a,{\bar{\cal G}^a};\tau,{\bar\tau}) + \mu_3{\cal V}^{\frac{1}{18}}  - \gamma\left(r_2 + \frac{r_2^2\zeta}{r_1}\right)}}\Biggr\}+ \frac{\left(\mu_3{\cal V}^{\frac{1}{36}}+\gamma r_2{\cal V}^{-\frac{1}{36}}\right){\cal V}^{\frac{5}{6}}}{\Xi^2\left[{\cal T}_\alpha(\sigma^\alpha,{\bar\sigma^\alpha};{\cal G}^a,{\bar{\cal G}^a};\tau,{\bar\tau}) + \mu_3{\cal V}^{\frac{1}{18}}  - \gamma\left(r_2 + \frac{r_2^2\zeta}{r_1}\right)\right]^{\frac{3}{2}}}\nonumber\\
 & & + \frac{\left({\cal T}_\alpha(\sigma^\alpha,{\bar\sigma^\alpha};{\cal G}^a,{\bar{\cal G}^a};\tau,{\bar\tau}) + \mu_3{\cal V}^{\frac{1}{18}}  - \gamma\left(r_2 + \frac{r_2^2\zeta}{r_1}\right)\right)\left(\mu_3+\left\{\gamma r_2{\cal V}^{-\frac{1}{36}}\right\}\right){\cal V}^{\frac{5}{6}}}{\Xi^3}\nonumber\\
 & & + \frac{\left(\mu_3{\cal V}^{\frac{1}{36}}+\gamma r_2{\cal V}^{-\frac{1}{36}}\right){\cal V}^{\frac{5}{6}}}{\Xi^2\sqrt{{\cal T}_\alpha(\sigma^\alpha,{\bar\sigma^\alpha};{\cal G}^a,{\bar{\cal G}^a};\tau,{\bar\tau}) + \mu_3{\cal V}^{\frac{1}{18}}  - \gamma\left(r_2 + \frac{r_2^2\zeta}{r_1}\right)}}\sim {\cal V}^{-\frac{43}{36}};\nonumber\\
 & & \partial_{{\cal G}^a}{\bar\partial}_{{\bar{\cal G}}^a}\hat{K}_{{\cal Z}_i{\bar{\cal Z}}_i}\sim\frac{\left((\mu_3{\cal V}^{\frac{1}{36}}+\gamma r_2{\cal V}^{-\frac{1}{36}})\right)}{Xi^3}\nonumber\\
 & & \times\Biggl[\sum_\beta k^a n^0_\beta sin(...) + ({\cal G}^a,{\bar{\cal G}^a})\Biggl\{\sqrt{{\cal T}_B(\sigma^B,{\bar\sigma ^B};{\cal G}^a,{\bar{\cal G}^a};\tau,{\bar\tau}) + \mu_3{\cal V}^{\frac{1}{18}} + i\kappa_4^2\mu_7C_{1{\bar 1}}{\cal V}^{-\frac{1}{2}} - \gamma\left(r_2 + \frac{r_2^2\zeta}{r_1}\right)} \nonumber\\
 & & - \sqrt{{\cal T}_S(\sigma^S,{\bar\sigma ^S};{\cal G}^a,{\bar{\cal G}^a};\tau,{\bar\tau}) + \mu_3{\cal V}^{\frac{1}{18}}  - \gamma\left(r_2 + \frac{r_2^2\zeta}{r_1}\right)}\Biggr\}\Biggr]^2\nonumber\\
 & & \times\Biggl\{\frac{1}{\sqrt{{\cal T}_B(\sigma^B,{\bar\sigma ^B};{\cal G}^a,{\bar{\cal G}^a};\tau,{\bar\tau}) + \mu_3{\cal V}^{\frac{1}{18}} + i\kappa_4^2\mu_7C_{1{\bar 1}}{\cal V}^{-\frac{1}{2}} - \gamma\left(r_2 + \frac{r_2^2\zeta}{r_1}\right)}}\nonumber\\
 & & -\frac{1}{\sqrt{{\cal T}_S(\sigma^S,{\bar\sigma ^S};{\cal G}^a,{\bar{\cal G}^a};\tau,{\bar\tau}) + \mu_3{\cal V}^{\frac{1}{18}}  - \gamma\left(r_2 + \frac{r_2^2\zeta}{r_1}\right)}}\Biggr\}\nonumber\\
 & & + \frac{\left(\mu_3{\cal V}^{\frac{1}{36}}+\gamma r_2{\cal V}^{-\frac{1}{36}}\right)}{\Xi^2}\nonumber\\
 & & \times \Biggl[\sum_\beta n^0_\beta cos(...)k^ak^b + \delta_{ab}\Biggl\{\sqrt{{\cal T}_B(\sigma^B,{\bar\sigma ^B};{\cal G}^a,{\bar{\cal G}^a};\tau,{\bar\tau}) + \mu_3{\cal V}^{\frac{1}{18}} + i\kappa_4^2\mu_7C_{1{\bar 1}}{\cal V}^{-\frac{1}{2}} - \gamma\left(r_2 + \frac{r_2^2\zeta}{r_1}\right)}\nonumber\\
  & & - \sqrt{{\cal T}_S(\sigma^S,{\bar\sigma ^S};{\cal G}^a,{\bar{\cal G}^a};\tau,{\bar\tau}) + \mu_3{\cal V}^{\frac{1}{18}}  - \gamma\left(r_2 + \frac{r_2^2\zeta}{r_1}\right)}\Biggr\}\Biggr]\nonumber\\
 & & \times\Biggl\{\frac{1}{\sqrt{{\cal T}_B(\sigma^B,{\bar\sigma ^B};{\cal G}^a,{\bar{\cal G}^a};\tau,{\bar\tau}) + \mu_3{\cal V}^{\frac{1}{18}} + i\kappa_4^2\mu_7C_{1{\bar 1}}{\cal V}^{-\frac{1}{2}} - \gamma\left(r_2 + \frac{r_2^2\zeta}{r_1}\right)}}\nonumber\\
 & & -\frac{1}{\sqrt{{\cal T}_S(\sigma^S,{\bar\sigma ^S};{\cal G}^a,{\bar{\cal G}^a};\tau,{\bar\tau}) + \mu_3{\cal V}^{\frac{1}{18}}  - \gamma\left(r_2 + \frac{r_2^2\zeta}{r_1}\right)}}\Biggr\}\nonumber\\
 & & + \frac{\left(\mu_3+\left\{\gamma r_2{\cal V}^{-\frac{1}{36}}\right\}\right)}{\Xi^3}\nonumber\\
 & & \times\Biggl[\sum_\beta k^a n^0_\beta sin(...) + ({\cal G}^a,{\bar{\cal G}^a})\Biggl\{\sqrt{{\cal T}_B(\sigma^B,{\bar\sigma ^B};{\cal G}^a,{\bar{\cal G}^a};\tau,{\bar\tau}) + \mu_3{\cal V}^{\frac{1}{18}} + i\kappa_4^2\mu_7C_{1{\bar 1}}{\cal V}^{-\frac{1}{2}} - \gamma\left(r_2 + \frac{r_2^2\zeta}{r_1}\right)} \nonumber\\
  & & - \sqrt{{\cal T}_S(\sigma^S,{\bar\sigma ^S};{\cal G}^a,{\bar{\cal G}^a};\tau,{\bar\tau}) + \mu_3{\cal V}^{\frac{1}{18}}  - \gamma\left(r_2 + \frac{r_2^2\zeta}{r_1}\right)}\Biggr\}\Biggr]^2\nonumber\\
 & & \hskip-0.9in\times\Biggl\{\sqrt{{\cal T}_B(\sigma^B,{\bar\sigma ^B};{\cal G}^a,{\bar{\cal G}^a};\tau,{\bar\tau}) + \mu_3{\cal V}^{\frac{1}{18}} + i\kappa_4^2\mu_7C_{1{\bar 1}}{\cal V}^{-\frac{1}{2}} - \gamma\left(r_2 + \frac{r_2^2\zeta}{r_1}\right)}
   -\sqrt{{\cal T}_S(\sigma^S,{\bar\sigma ^S};{\cal G}^a,{\bar{\cal G}^a};\tau,{\bar\tau}) + \mu_3{\cal V}^{\frac{1}{18}}  - \gamma\left(r_2 + \frac{r_2^2\zeta}{r_1}\right)}\Biggr\}\nonumber\\
   & & + \frac{\left(\mu_3+\left\{\gamma r_2{\cal V}^{-\frac{1}{36}}\right\}\right)}{\Xi^2}\nonumber\\
   & & \Biggl[\sum_\beta n^0_\beta cos(...)k^ak^b + \delta_{ab}\Biggl\{\sqrt{{\cal T}_B(\sigma^B,{\bar\sigma ^B};{\cal G}^a,{\bar{\cal G}^a};\tau,{\bar\tau}) + \mu_3{\cal V}^{\frac{1}{18}} + i\kappa_4^2\mu_7C_{1{\bar 1}}{\cal V}^{-\frac{1}{2}} - \gamma\left(r_2 + \frac{r_2^2\zeta}{r_1}\right)}\nonumber\\
    & & - \sqrt{{\cal T}_S(\sigma^S,{\bar\sigma ^S};{\cal G}^a,{\bar{\cal G}^a};\tau,{\bar\tau}) + \mu_3{\cal V}^{\frac{1}{18}}  - \gamma\left(r_2 + \frac{r_2^2\zeta}{r_1}\right)}\Biggr\}\Biggr]\nonumber\\
 & & \hskip-0.9in\times\Biggl\{\sqrt{{\cal T}_B(\sigma^B,{\bar\sigma ^B};{\cal G}^a,{\bar{\cal G}^a};\tau,{\bar\tau}) + \mu_3{\cal V}^{\frac{1}{18}} + i\kappa_4^2\mu_7C_{1{\bar 1}}{\cal V}^{-\frac{1}{2}} - \gamma\left(r_2 + \frac{r_2^2\zeta}{r_1}\right)}
   -\sqrt{{\cal T}_S(\sigma^S,{\bar\sigma ^S};{\cal G}^a,{\bar{\cal G}^a};\tau,{\bar\tau}) + \mu_3{\cal V}^{\frac{1}{18}}  - \gamma\left(r_2 + \frac{r_2^2\zeta}{r_1}\right)}\Biggr\}\nonumber\\
   & & \sim {\cal V}^{-\frac{35}{36}}.
\end{eqnarray}
Similarly,
\begin{eqnarray}
\label{eq:dKhat_a}
& & \partial_{\sigma^B}\hat{K}_{{\tilde{{\cal A}_1}{\bar{\tilde{\cal A}_1}}}}\sim\frac{{\cal V}^{\frac{11}{6}}}{\Xi\left({\cal T}_B(\sigma^B,{\bar\sigma ^B};{\cal G}^a,{\bar{\cal G}^a};\tau,{\bar\tau}) + \mu_3{\cal V}^{\frac{1}{18}} + i\kappa_4^2\mu_7C_{1{\bar 1}}{\cal V}^{-\frac{1}{2}} - \gamma\left(r_2 + \frac{r_2^2\zeta}{r_1}\right)\right)^{\frac{3}{2}}} +\frac{{\cal V}^{\frac{11}{6}}}{\Xi^2}\sim{\cal V}^{\frac{3}{4}},\nonumber\\
& & \partial_{\sigma^S}\hat{K}_{{\tilde{{\cal A}_1}{\bar{\tilde{\cal A}_1}}}}\sim\frac{{\cal V}^{\frac{11}{6}}}{\Xi^2}\sqrt{\frac{{\cal T}_B(\sigma^B,{\bar\sigma ^B};{\cal G}^a,{\bar{\cal G}^a};\tau,{\bar\tau}) + \mu_3{\cal V}^{\frac{1}{18}} + i\kappa_4^2\mu_7C_{1{\bar 1}}{\cal V}^{-\frac{1}{2}} - \gamma\left(r_2 + \frac{r_2^2\zeta}{r_1}\right)}{{\cal T}_S(\sigma^S,{\bar\sigma ^S};{\cal G}^a,{\bar{\cal G}^a};\tau,{\bar\tau}) + \mu_3{\cal V}^{\frac{1}{18}}  - \gamma\left(r_2 + \frac{r_2^2\zeta}{r_1}\right)}}\sim{\cal V}^{-\frac{1}{6}};\nonumber\\
& & \partial_{{\cal G}^a}\hat{K}_{{\tilde{{\cal A}_1}{\bar{\tilde{\cal A}_1}}}}\sim \frac{{\cal V}^{\frac{11}{6}}}{\Xi^2\sqrt{{\cal T}_B(\sigma^B,{\bar\sigma ^B};{\cal G}^a,{\bar{\cal G}^a};\tau,{\bar\tau}) + \mu_3{\cal V}^{\frac{1}{18}} + i\kappa_4^2\mu_7C_{1{\bar 1}}{\cal V}^{-\frac{1}{2}} - \gamma\left(r_2 + \frac{r_2^2\zeta}{r_1}\right)}}\nonumber\\
& & \times\Biggl[\sum_\beta k^a n^0_\beta sin(...) + ({\cal G}^a,{\bar{\cal G}^a})\Biggl\{\sqrt{{\cal T}_B(\sigma^B,{\bar\sigma ^B};{\cal G}^a,{\bar{\cal G}^a};\tau,{\bar\tau}) + \mu_3{\cal V}^{\frac{1}{18}} + i\kappa_4^2\mu_7C_{1{\bar 1}}{\cal V}^{-\frac{1}{2}} - \gamma\left(r_2 + \frac{r_2^2\zeta}{r_1}\right)} - \nonumber\\
 & & \sqrt{{\cal T}_S(\sigma^S,{\bar\sigma ^S};{\cal G}^a,{\bar{\cal G}^a};\tau,{\bar\tau}) + \mu_3{\cal V}^{\frac{1}{18}}  - \gamma\left(r_2 + \frac{r_2^2\zeta}{r_1}\right)}\Biggr\}\Biggr]\nonumber\\
 & & \sim{\cal V}^{\frac{23}{36}},
\end{eqnarray}
from where one concludes:
\begin{eqnarray}
\label{eq:ddKhat_a}
& & \partial_{\sigma^B}{\bar\partial}_{{\bar\sigma}^B}\hat{K}_{{\tilde{{\cal A}_1}{\bar{\tilde{\cal A}_1}}}}\sim
\frac{{\cal V}^{\frac{11}{6}}}{\Xi^2\left({\cal T}_B(\sigma^B,{\bar\sigma ^B};{\cal G}^a,{\bar{\cal G}^a};\tau,{\bar\tau}) + \mu_3{\cal V}^{\frac{1}{18}} + i\kappa_4^2\mu_7C_{1{\bar 1}}{\cal V}^{-\frac{1}{2}} - \gamma\left(r_2 + \frac{r_2^2\zeta}{r_1}\right)\right)}\nonumber\\
& & + \frac{{\cal V}^{\frac{11}{6}}}{\Xi\left({\cal T}_B(\sigma^B,{\bar\sigma ^B};{\cal G}^a,{\bar{\cal G}^a};\tau,{\bar\tau}) + \mu_3{\cal V}^{\frac{1}{18}} + i\kappa_4^2\mu_7C_{1{\bar 1}}{\cal V}^{-\frac{1}{2}} - \gamma\left(r_2 + \frac{r_2^2\zeta}{r_1}\right)\right)^{\frac{5}{2}}}
\nonumber\\
& & + \frac{{\cal V}^{\frac{11}{6}}\sqrt{{\cal T}_B(\sigma^B,{\bar\sigma ^B};{\cal G}^a,{\bar{\cal G}^a};\tau,{\bar\tau}) + \mu_3{\cal V}^{\frac{1}{18}} + i\kappa_4^2\mu_7C_{1{\bar 1}}{\cal V}^{-\frac{1}{2}} - \gamma\left(r_2 + \frac{r_2^2\zeta}{r_1}\right)}}{\Xi^3}
\sim {\cal V}^{\frac{25}{36}};\nonumber\\
& & \partial_{\sigma^S}{\bar\partial}_{{\bar\sigma}^S}\hat{K}_{{\tilde{{\cal A}_1}{\bar{\tilde{\cal A}_1}}}}\sim
\frac{{\cal V}^{\frac{11}{6}}\left({\cal T}_S(\sigma^S,{\bar\sigma ^S};{\cal G}^a,{\bar{\cal G}^a};\tau,{\bar\tau}) + \mu_3{\cal V}^{\frac{1}{18}}  - \gamma\left(r_2 + \frac{r_2^2\zeta}{r_1}\right)\right)}{\sqrt{{\cal T}_B(\sigma^B,{\bar\sigma ^B};{\cal G}^a,{\bar{\cal G}^a};\tau,{\bar\tau}) + \mu_3{\cal V}^{\frac{1}{18}} + i\kappa_4^2\mu_7C_{1{\bar 1}}{\cal V}^{-\frac{1}{2}} - \gamma\left(r_2 + \frac{r_2^2\zeta}{r_1}\right)}}
\nonumber\\
& & \hskip-0.8in+ \frac{{\cal V}^{\frac{11}{6}}}{\Xi\sqrt{\left({\cal T}_S(\sigma^S,{\bar\sigma ^S};{\cal G}^a,{\bar{\cal G}^a};\tau,{\bar\tau}) + \mu_3{\cal V}^{\frac{1}{18}}  - \gamma\left(r_2 + \frac{r_2^2\zeta}{r_1}\right)\right)\left({\cal T}_B(\sigma^B,{\bar\sigma ^B};{\cal G}^a,{\bar{\cal G}^a};\tau,{\bar\tau}) + \mu_3{\cal V}^{\frac{1}{18}} + i\kappa_4^2\mu_7C_{1{\bar 1}}{\cal V}^{-\frac{1}{2}} - \gamma\left(r_2 + \frac{r_2^2\zeta}{r_1}\right)\right)}}\nonumber\\
& & \sim {\cal V}^{\frac{7}{9}};\nonumber\\
& & \partial_{\sigma^B}{\bar\partial}_{{\bar\sigma}^S}\hat{K}_{{\tilde{{\cal A}_1}{\bar{\tilde{\cal A}_1}}}}\sim
\frac{{\cal V}^{\frac{11}{6}}\sqrt{{\cal T}_S(\sigma^S,{\bar\sigma ^S};{\cal G}^a,{\bar{\cal G}^a};\tau,{\bar\tau}) + \mu_3{\cal V}^{\frac{1}{18}}  - \gamma\left(r_2 + \frac{r_2^2\zeta}{r_1}\right)}}{\Xi^2\left({\cal T}_B(\sigma^B,{\bar\sigma ^B};{\cal G}^a,{\bar{\cal G}^a};\tau,{\bar\tau}) + \mu_3{\cal V}^{\frac{1}{18}} + i\kappa_4^2\mu_7C_{1{\bar 1}}{\cal V}^{-\frac{1}{2}} - \gamma\left(r_2 + \frac{r_2^2\zeta}{r_1}\right)\right)^{\frac{3}{2}}}\nonumber\\
& & + \frac{{\cal V}^{\frac{11}{6}}\sqrt{{\cal T}_S(\sigma^S,{\bar\sigma ^S};{\cal G}^a,{\bar{\cal G}^a};\tau,{\bar\tau}) + \mu_3{\cal V}^{\frac{1}{18}}  - \gamma\left(r_2 + \frac{r_2^2\zeta}{r_1}\right)}}{\Xi^3}\sim {\cal V}^{-\frac{2}{9}};\nonumber\\
& & \partial_{\sigma^B}{\bar\partial}_{{\bar{\cal G}}^a}\hat{K}_{{\tilde{{\cal A}_1}{\bar{\tilde{\cal A}_1}}}}\sim
\frac{{\cal V}^{\frac{11}{6}+\frac{5}{6}}}{\Xi^2\left({\cal T}_B(\sigma^B,{\bar\sigma ^B};{\cal G}^a,{\bar{\cal G}^a};\tau,{\bar\tau}) + \mu_3{\cal V}^{\frac{1}{18}} + i\kappa_4^2\mu_7C_{1{\bar 1}}{\cal V}^{-\frac{1}{2}} - \gamma\left(r_2 + \frac{r_2^2\zeta}{r_1}\right)\right)^{\frac{3}{2}}}
 + \frac{{\cal V}^{\frac{11}{6}+\frac{5}{6}}}{\Xi^3}\sim{\cal V}^{\frac{7}{12}};\nonumber\\
& & \partial_{\sigma^S}{\bar\partial}_{{\bar{\cal G}}^a}\hat{K}_{{\tilde{{\cal A}_1}{\bar{\tilde{\cal A}_1}}}}\sim
\frac{{\cal V}^{\frac{11}{6}+\frac{5}{6}}}{\Xi^3}\nonumber\\
& & \times\sqrt{\frac{{\cal T}_B(\sigma^B,{\bar\sigma ^B};{\cal G}^a,{\bar{\cal G}^a};\tau,{\bar\tau}) + \mu_3{\cal V}^{\frac{1}{18}} + i\kappa_4^2\mu_7C_{1{\bar 1}}{\cal V}^{-\frac{1}{2}} - \gamma\left(r_2 + \frac{r_2^2\zeta}{r_1}\right)}{{\cal T}_S(\sigma^S,{\bar\sigma ^S};{\cal G}^a,{\bar{\cal G}^a};\tau,{\bar\tau}) + \mu_3{\cal V}^{\frac{1}{18}}  - \gamma\left(r_2 + \frac{r_2^2\zeta}{r_1}\right)}}\sim{\cal V}^{-\frac{1}{3}};
\nonumber\\
& & \partial_{{\cal G}^a}{\bar\partial}_{{\bar{\cal G}}^a}\hat{K}_{{\tilde{{\cal A}_1}{\bar{\tilde{\cal A}_1}}}}\sim{\cal V}^{\frac{29}{36}}.
\end{eqnarray}
In the above $\left(i\kappa_4^2\mu_7C_{1{\bar 1}}{\cal V}^{-\frac{1}{4}}\right)^2\sim{\cal V}^{\frac{11}{6}}$.

Now,
\begin{eqnarray}
\label{eq:detKhat}
& & {\rm det}\ \left(\hat{K}_{i{\bar j}}\right)=\left(\hat{K}_{{\cal Z}_1{\bar{\cal Z}}_1}\right)^2\left(\hat{K}_{{\tilde{{\cal A}_1}{\bar{\tilde{\cal A}_1}}}}\right)\nonumber\\
& & \sim\Biggl[\frac{\left(\left(\mu_3{\cal V}^{\frac{1}{36}}\right)^2+\mu_3\gamma r_2\right)}{\Xi}\nonumber\\
& & \times\Biggl\{\frac{1}{\sqrt{{\cal T}_B(\sigma^B,{\bar\sigma ^B};{\cal G}^a,{\bar{\cal G}^a};\tau,{\bar\tau}) + \mu_3{\cal V}^{\frac{1}{18}} + i\kappa_4^2\mu_7C_{1{\bar 1}}{\cal V}^{-\frac{1}{2}} - \gamma\left(r_2 + \frac{r_2^2\zeta}{r_1}\right)}}\nonumber\\
 & & -\frac{1}{\sqrt{{\cal T}_S(\sigma^S,{\bar\sigma ^S};{\cal G}^a,{\bar{\cal G}^a};\tau,{\bar\tau}) + \mu_3{\cal V}^{\frac{1}{18}}  - \gamma\left(r_2 + \frac{r_2^2\zeta}{r_1}\right)}}\Biggr\}\nonumber\\
 & & +\frac{\left(\left(\gamma r_2{\cal V}^{-\frac{1}{36}}\right)^2\right)}{\Xi}\nonumber\\
 & & \times\Biggl\{\sqrt{{\cal T}_B(\sigma^B,{\bar\sigma ^B};{\cal G}^a,{\bar{\cal G}^a};\tau,{\bar\tau}) + \mu_3{\cal V}^{\frac{1}{18}} + i\kappa_4^2\mu_7C_{1{\bar 1}}{\cal V}^{-\frac{1}{2}} - \gamma\left(r_2 + \frac{r_2^2\zeta}{r_1}\right)} - \nonumber\\
 & & \sqrt{{\cal T}_S(\sigma^S,{\bar\sigma ^S};{\cal G}^a,{\bar{\cal G}^a};\tau,{\bar\tau}) + \mu_3{\cal V}^{\frac{1}{18}}  - \gamma\left(r_2 + \frac{r_2^2\zeta}{r_1}\right)}\Biggr\}\Biggr]^2\nonumber\\
 & & \times\frac{\left(i\kappa_4^2\mu_7C_{1{\bar 1}}{\cal V}^{-\frac{1}{4}}\right)^2}{\Xi\sqrt{{\cal T}_B(\sigma^B,{\bar\sigma ^B};{\cal G}^a,{\bar{\cal G}^a};\tau,{\bar\tau}) + \mu_3{\cal V}^{\frac{1}{18}} + i\kappa_4^2\mu_7C_{1{\bar 1}}{\cal V}^{-\frac{1}{2}} - \gamma\left(r_2 + \frac{r_2^2\zeta}{r_1}\right)}}\sim{\cal V}^{-\frac{41}{36}}.
\end{eqnarray}
Hence, we see that:
\begin{eqnarray}
\label{eq:ddetKhat}
& & \partial_{\sigma^B}{\rm det}\ \left(\hat{K}_{i{\bar j}}\right)\sim\frac{{\cal V}^{\frac{11}{6}}}{\Xi^4}\nonumber\\
& & \hskip-0.8in\times\Biggl[{\cal V}^{\frac{1}{18}}\Biggl\{\frac{1}{\sqrt{{\cal T}_B(\sigma^B,{\bar\sigma ^B};{\cal G}^a,{\bar{\cal G}^a};\tau,{\bar\tau}) + \mu_3{\cal V}^{\frac{1}{18}} + i\kappa_4^2\mu_7C_{1{\bar 1}}{\cal V}^{-\frac{1}{2}} - \gamma\left(r_2 + \frac{r_2^2\zeta}{r_1}\right)}}\nonumber\\
 & & -\frac{1}{\sqrt{{\cal T}_S(\sigma^S,{\bar\sigma ^S};{\cal G}^a,{\bar{\cal G}^a};\tau,{\bar\tau}) + \mu_3{\cal V}^{\frac{1}{18}}  - \gamma\left(r_2 + \frac{r_2^2\zeta}{r_1}\right)}}\Biggr\}\nonumber\\
 & &\hskip-0.8in +\mu_3\Biggl\{\sqrt{{\cal T}_B(\sigma^B,{\bar\sigma ^B};{\cal G}^a,{\bar{\cal G}^a};\tau,{\bar\tau}) + \mu_3{\cal V}^{\frac{1}{18}} + i\kappa_4^2\mu_7C_{1{\bar 1}}{\cal V}^{-\frac{1}{2}} - \gamma\left(r_2 + \frac{r_2^2\zeta}{r_1}\right)} - \nonumber\\
 & & \sqrt{{\cal T}_S(\sigma^S,{\bar\sigma ^S};{\cal G}^a,{\bar{\cal G}^a};\tau,{\bar\tau}) + \mu_3{\cal V}^{\frac{1}{18}}  - \gamma\left(r_2 + \frac{r_2^2\zeta}{r_1}\right)}\Biggr\}\Biggr]^2\nonumber\\
 & & + \frac{{\cal V}^{\frac{11}{6}}}{\Xi^3\Biggl[{\cal T}_B(\sigma^B,{\bar\sigma ^B};{\cal G}^a,{\bar{\cal G}^a};\tau,{\bar\tau}) + \mu_3{\cal V}^{\frac{1}{18}} + i\kappa_4^2\mu_7C_{1{\bar 1}}{\cal V}^{-\frac{1}{2}} - \gamma\left(r_2 + \frac{r_2^2\zeta}{r_1}\right)\Biggr]^{\frac{3}{2}}}\nonumber\\
 & & \hskip-0.8in\times\Biggl[{\cal V}^{\frac{1}{18}}\Biggl\{\frac{1}{\sqrt{{\cal T}_B(\sigma^B,{\bar\sigma ^B};{\cal G}^a,{\bar{\cal G}^a};\tau,{\bar\tau}) + \mu_3{\cal V}^{\frac{1}{18}} + i\kappa_4^2\mu_7C_{1{\bar 1}}{\cal V}^{-\frac{1}{2}} - \gamma\left(r_2 + \frac{r_2^2\zeta}{r_1}\right)}}\nonumber\\
 & & -\frac{1}{\sqrt{{\cal T}_S(\sigma^S,{\bar\sigma ^S};{\cal G}^a,{\bar{\cal G}^a};\tau,{\bar\tau}) + \mu_3{\cal V}^{\frac{1}{18}}  - \gamma\left(r_2 + \frac{r_2^2\zeta}{r_1}\right)}}\Biggr\}\nonumber\\
 & &\hskip-0.8in +\mu_3\Biggl\{\sqrt{{\cal T}_B(\sigma^B,{\bar\sigma ^B};{\cal G}^a,{\bar{\cal G}^a};\tau,{\bar\tau}) + \mu_3{\cal V}^{\frac{1}{18}} + i\kappa_4^2\mu_7C_{1{\bar 1}}{\cal V}^{-\frac{1}{2}} - \gamma\left(r_2 + \frac{r_2^2\zeta}{r_1}\right)} - \nonumber\\
 & & \sqrt{{\cal T}_S(\sigma^S,{\bar\sigma ^S};{\cal G}^a,{\bar{\cal G}^a};\tau,{\bar\tau}) + \mu_3{\cal V}^{\frac{1}{18}}  - \gamma\left(r_2 + \frac{r_2^2\zeta}{r_1}\right)}\Biggr\}\Biggr]^2\nonumber\\
 & & + \frac{{\cal V}^{\frac{11}{6}}}{\Xi^3\sqrt{{\cal T}_B(\sigma^B,{\bar\sigma ^B};{\cal G}^a,{\bar{\cal G}^a};\tau,{\bar\tau}) + \mu_3{\cal V}^{\frac{1}{18}} + i\kappa_4^2\mu_7C_{1{\bar 1}}{\cal V}^{-\frac{1}{2}} - \gamma\left(r_2 + \frac{r_2^2\zeta}{r_1}\right)}}\nonumber\\
 & & \hskip-0.8in\times\Biggl[{\cal V}^{\frac{1}{18}}\Biggl\{\frac{1}{\sqrt{{\cal T}_B(\sigma^B,{\bar\sigma ^B};{\cal G}^a,{\bar{\cal G}^a};\tau,{\bar\tau}) + \mu_3{\cal V}^{\frac{1}{18}} + i\kappa_4^2\mu_7C_{1{\bar 1}}{\cal V}^{-\frac{1}{2}} - \gamma\left(r_2 + \frac{r_2^2\zeta}{r_1}\right)}}\nonumber\\
 & & -\frac{1}{\sqrt{{\cal T}_S(\sigma^S,{\bar\sigma ^S};{\cal G}^a,{\bar{\cal G}^a};\tau,{\bar\tau}) + \mu_3{\cal V}^{\frac{1}{18}}  - \gamma\left(r_2 + \frac{r_2^2\zeta}{r_1}\right)}}\Biggr\}\nonumber\\
 & &\hskip-0.8in +\mu_3\Biggl\{\sqrt{{\cal T}_B(\sigma^B,{\bar\sigma ^B};{\cal G}^a,{\bar{\cal G}^a};\tau,{\bar\tau}) + \mu_3{\cal V}^{\frac{1}{18}} + i\kappa_4^2\mu_7C_{1{\bar 1}}{\cal V}^{-\frac{1}{2}} - \gamma\left(r_2 + \frac{r_2^2\zeta}{r_1}\right)} - \nonumber\\
 & & \sqrt{{\cal T}_S(\sigma^S,{\bar\sigma ^S};{\cal G}^a,{\bar{\cal G}^a};\tau,{\bar\tau}) + \mu_3{\cal V}^{\frac{1}{18}}  - \gamma\left(r_2 + \frac{r_2^2\zeta}{r_1}\right)}\Biggr\}\Biggr]\nonumber\\
 & & \hskip-0.8in\times\Biggl[\frac{{\cal V}^{\frac{1}{18}}}{\left({\cal T}_B(\sigma^B,{\bar\sigma ^B};{\cal G}^a,{\bar{\cal G}^a};\tau,{\bar\tau}) + \mu_3{\cal V}^{\frac{1}{18}} + i\kappa_4^2\mu_7C_{1{\bar 1}}{\cal V}^{-\frac{1}{2}} - \gamma\left(r_2 + \frac{r_2^2\zeta}{r_1}\right)\right)^{\frac{3}{2}}}\nonumber\\
 & & + \frac{\mu_3}{\sqrt{{\cal T}_B(\sigma^B,{\bar\sigma ^B};{\cal G}^a,{\bar{\cal G}^a};\tau,{\bar\tau}) + \mu_3{\cal V}^{\frac{1}{18}} + i\kappa_4^2\mu_7C_{1{\bar 1}}{\cal V}^{-\frac{1}{2}} - \gamma\left(r_2 + \frac{r_2^2\zeta}{r_1}\right)}}\Biggr]\sim {\cal V}^{-\frac{43}{36}};
 \nonumber\\
 & & \partial_{\sigma^S}{\rm det}\ \left(\hat{K}_{i{\bar j}}\right)\sim\frac{{\cal V}^{\frac{11}{6}}\sqrt{{\cal T}_B(\sigma^B,{\bar\sigma ^B};{\cal G}^a,{\bar{\cal G}^a};\tau,{\bar\tau}) + \mu_3{\cal V}^{\frac{1}{18}} + i\kappa_4^2\mu_7C_{1{\bar 1}}{\cal V}^{-\frac{1}{2}} - \gamma\left(r_2 + \frac{r_2^2\zeta}{r_1}\right)}}{\Xi^4\sqrt{{\cal T}_S(\sigma^S,{\bar\sigma ^S};{\cal G}^a,{\bar{\cal G}^a};\tau,{\bar\tau}) + \mu_3{\cal V}^{\frac{1}{18}}  - \gamma\left(r_2 + \frac{r_2^2\zeta}{r_1}\right)}}\nonumber\\
& & \hskip-0.8in\times\Biggl[{\cal V}^{\frac{1}{18}}\Biggl\{\frac{1}{\sqrt{{\cal T}_B(\sigma^B,{\bar\sigma ^B};{\cal G}^a,{\bar{\cal G}^a};\tau,{\bar\tau}) + \mu_3{\cal V}^{\frac{1}{18}} + i\kappa_4^2\mu_7C_{1{\bar 1}}{\cal V}^{-\frac{1}{2}} - \gamma\left(r_2 + \frac{r_2^2\zeta}{r_1}\right)}}\nonumber\\
 & & -\frac{1}{\sqrt{{\cal T}_S(\sigma^S,{\bar\sigma ^S};{\cal G}^a,{\bar{\cal G}^a};\tau,{\bar\tau}) + \mu_3{\cal V}^{\frac{1}{18}}  - \gamma\left(r_2 + \frac{r_2^2\zeta}{r_1}\right)}}\Biggr\}\nonumber\\
 & &\hskip-0.8in +\mu_3\Biggl\{\sqrt{{\cal T}_B(\sigma^B,{\bar\sigma ^B};{\cal G}^a,{\bar{\cal G}^a};\tau,{\bar\tau}) + \mu_3{\cal V}^{\frac{1}{18}} + i\kappa_4^2\mu_7C_{1{\bar 1}}{\cal V}^{-\frac{1}{2}} - \gamma\left(r_2 + \frac{r_2^2\zeta}{r_1}\right)} - \nonumber\\
 & & \sqrt{{\cal T}_S(\sigma^S,{\bar\sigma ^S};{\cal G}^a,{\bar{\cal G}^a};\tau,{\bar\tau}) + \mu_3{\cal V}^{\frac{1}{18}}  - \gamma\left(r_2 + \frac{r_2^2\zeta}{r_1}\right)}\Biggr\}\Biggr]^2\nonumber\\
 & & + \frac{{\cal V}^{\frac{11}{6}}}{\Xi^3\sqrt{{\cal T}_B(\sigma^B,{\bar\sigma ^B};{\cal G}^a,{\bar{\cal G}^a};\tau,{\bar\tau}) + \mu_3{\cal V}^{\frac{1}{18}} + i\kappa_4^2\mu_7C_{1{\bar 1}}{\cal V}^{-\frac{1}{2}} - \gamma\left(r_2 + \frac{r_2^2\zeta}{r_1}\right)}}\nonumber\\
 & & \hskip-0.8in\times\Biggl[{\cal V}^{\frac{1}{18}}\Biggl\{\frac{1}{\sqrt{{\cal T}_B(\sigma^B,{\bar\sigma ^B};{\cal G}^a,{\bar{\cal G}^a};\tau,{\bar\tau}) + \mu_3{\cal V}^{\frac{1}{18}} + i\kappa_4^2\mu_7C_{1{\bar 1}}{\cal V}^{-\frac{1}{2}} - \gamma\left(r_2 + \frac{r_2^2\zeta}{r_1}\right)}}\nonumber\\
 & & -\frac{1}{\sqrt{{\cal T}_S(\sigma^S,{\bar\sigma ^S};{\cal G}^a,{\bar{\cal G}^a};\tau,{\bar\tau}) + \mu_3{\cal V}^{\frac{1}{18}}  - \gamma\left(r_2 + \frac{r_2^2\zeta}{r_1}\right)}}\Biggr\}\nonumber\\
 & &\hskip-0.8in +\mu_3\Biggl\{\sqrt{{\cal T}_B(\sigma^B,{\bar\sigma ^B};{\cal G}^a,{\bar{\cal G}^a};\tau,{\bar\tau}) + \mu_3{\cal V}^{\frac{1}{18}} + i\kappa_4^2\mu_7C_{1{\bar 1}}{\cal V}^{-\frac{1}{2}} - \gamma\left(r_2 + \frac{r_2^2\zeta}{r_1}\right)} - \nonumber\\
 & & \sqrt{{\cal T}_S(\sigma^S,{\bar\sigma ^S};{\cal G}^a,{\bar{\cal G}^a};\tau,{\bar\tau}) + \mu_3{\cal V}^{\frac{1}{18}}  - \gamma\left(r_2 + \frac{r_2^2\zeta}{r_1}\right)}\Biggr\}\Biggr]\nonumber\\
 & & \hskip-0.8in\times\Biggl[\frac{{\cal V}^{\frac{1}{18}}}{\left({\cal T}_S(\sigma^S,{\bar\sigma ^S};{\cal G}^a,{\bar{\cal G}^a};\tau,{\bar\tau}) + \mu_3{\cal V}^{\frac{1}{18}} + i\kappa_4^2\mu_7C_{1{\bar 1}}{\cal V}^{-\frac{1}{2}} - \gamma\left(r_2 + \frac{r_2^2\zeta}{r_1}\right)\right)^{\frac{3}{2}}}\nonumber\\
 & & + \frac{\mu_3}{\sqrt{{\cal T}_S(\sigma^S,{\bar\sigma ^S};{\cal G}^a,{\bar{\cal G}^a};\tau,{\bar\tau}) + \mu_3{\cal V}^{\frac{1}{18}} + i\kappa_4^2\mu_7C_{1{\bar 1}}{\cal V}^{-\frac{1}{2}} - \gamma\left(r_2 + \frac{r_2^2\zeta}{r_1}\right)}}\Biggr]\sim {\cal V}^{-\frac{43}{36}};
 \nonumber\\
 & & \partial_{{\cal G}^a}{\rm det}\ \left(\hat{K}_{i{\bar j}}\right)\sim\frac{{\cal V}^{\frac{11}{6}}}{\Xi^4}\nonumber\\
 & &  \times\Biggl[\sum_\beta n^0_\beta k^a sin(...)+({\cal G}^a,{\bar{\cal G}}^a)\times\nonumber\\
   & & \hskip-0.9in\left(\sqrt{{\cal T}_B(\sigma^B,{\bar\sigma ^B};{\cal G}^a,{\bar{\cal G}^a};\tau,{\bar\tau}) + \mu_3{\cal V}^{\frac{1}{18}} + i\kappa_4^2\mu_7C_{1{\bar 1}}{\cal V}^{-\frac{1}{2}} - \gamma\left(r_2 + \frac{r_2^2\zeta}{r_1}\right)}
   -\sqrt{{\cal T}_S(\sigma^S,{\bar\sigma ^S};{\cal G}^a,{\bar{\cal G}^a};\tau,{\bar\tau}) + \mu_3{\cal V}^{\frac{1}{18}}  - \gamma\left(r_2 + \frac{r_2^2\zeta}{r_1}\right)}\right)\Biggr]\nonumber\\
 & &  \hskip-0.8in\times\Biggl[{\cal V}^{\frac{1}{18}}\Biggl\{\frac{1}{\sqrt{{\cal T}_B(\sigma^B,{\bar\sigma ^B};{\cal G}^a,{\bar{\cal G}^a};\tau,{\bar\tau}) + \mu_3{\cal V}^{\frac{1}{18}} + i\kappa_4^2\mu_7C_{1{\bar 1}}{\cal V}^{-\frac{1}{2}} - \gamma\left(r_2 + \frac{r_2^2\zeta}{r_1}\right)}}\nonumber\\
 & & -\frac{1}{\sqrt{{\cal T}_S(\sigma^S,{\bar\sigma ^S};{\cal G}^a,{\bar{\cal G}^a};\tau,{\bar\tau}) + \mu_3{\cal V}^{\frac{1}{18}}  - \gamma\left(r_2 + \frac{r_2^2\zeta}{r_1}\right)}}\Biggr\}\nonumber\\
 & &\hskip-0.8in +\mu_3\Biggl\{\sqrt{{\cal T}_B(\sigma^B,{\bar\sigma ^B};{\cal G}^a,{\bar{\cal G}^a};\tau,{\bar\tau}) + \mu_3{\cal V}^{\frac{1}{18}} + i\kappa_4^2\mu_7C_{1{\bar 1}}{\cal V}^{-\frac{1}{2}} - \gamma\left(r_2 + \frac{r_2^2\zeta}{r_1}\right)} - \nonumber\\
 & & \sqrt{{\cal T}_S(\sigma^S,{\bar\sigma ^S};{\cal G}^a,{\bar{\cal G}^a};\tau,{\bar\tau}) + \mu_3{\cal V}^{\frac{1}{18}}  - \gamma\left(r_2 + \frac{r_2^2\zeta}{r_1}\right)}\Biggr\}\Biggr]^2\nonumber\\
 & & + \frac{{\cal V}^{\frac{11}{6}}\left({\cal G}^a,{\bar{\cal G}}^a\right)}{\Xi^3\left({\cal T}_\alpha(\sigma^\alpha,{\bar\sigma ^\alpha};{\cal G}^a,{\bar{\cal G}^a};\tau,{\bar\tau}) + \mu_3{\cal V}^{\frac{1}{18}}  - \gamma\left(r_2 + \frac{r_2^2\zeta}{r_1}\right)\right)^{\frac{3}{2}}}
 \nonumber\\
 & &  \hskip-0.8in\times\Biggl[{\cal V}^{\frac{1}{18}}\Biggl\{\frac{1}{\sqrt{{\cal T}_B(\sigma^B,{\bar\sigma ^B};{\cal G}^a,{\bar{\cal G}^a};\tau,{\bar\tau}) + \mu_3{\cal V}^{\frac{1}{18}} + i\kappa_4^2\mu_7C_{1{\bar 1}}{\cal V}^{-\frac{1}{2}} - \gamma\left(r_2 + \frac{r_2^2\zeta}{r_1}\right)}}\nonumber\\
 & & -\frac{1}{\sqrt{{\cal T}_S(\sigma^S,{\bar\sigma ^S};{\cal G}^a,{\bar{\cal G}^a};\tau,{\bar\tau}) + \mu_3{\cal V}^{\frac{1}{18}}  - \gamma\left(r_2 + \frac{r_2^2\zeta}{r_1}\right)}}\Biggr\}\nonumber\\
 & &\hskip-0.8in +\mu_3\Biggl\{\sqrt{{\cal T}_B(\sigma^B,{\bar\sigma ^B};{\cal G}^a,{\bar{\cal G}^a};\tau,{\bar\tau}) + \mu_3{\cal V}^{\frac{1}{18}} + i\kappa_4^2\mu_7C_{1{\bar 1}}{\cal V}^{-\frac{1}{2}} - \gamma\left(r_2 + \frac{r_2^2\zeta}{r_1}\right)} - \nonumber\\
 & & \sqrt{{\cal T}_S(\sigma^S,{\bar\sigma ^S};{\cal G}^a,{\bar{\cal G}^a};\tau,{\bar\tau}) + \mu_3{\cal V}^{\frac{1}{18}}  - \gamma\left(r_2 + \frac{r_2^2\zeta}{r_1}\right)}\Biggr\}\Biggr]^2\nonumber\\
 & & + \frac{{\cal V}^{\frac{11}{6}}}{\Xi^3\sqrt{{\cal T}_B(\sigma^B,{\bar\sigma ^B};{\cal G}^a,{\bar{\cal G}^a};\tau,{\bar\tau}) + \mu_3{\cal V}^{\frac{1}{18}} + i\kappa_4^2\mu_7C_{1{\bar 1}}{\cal V}^{-\frac{1}{2}} - \gamma\left(r_2 + \frac{r_2^2\zeta}{r_1}\right)}}\nonumber\\
 & &
 \hskip-0.8in\times\Biggl[{\cal V}^{\frac{1}{18}}\Biggl\{\frac{1}{\sqrt{{\cal T}_B(\sigma^B,{\bar\sigma ^B};{\cal G}^a,{\bar{\cal G}^a};\tau,{\bar\tau}) + \mu_3{\cal V}^{\frac{1}{18}} + i\kappa_4^2\mu_7C_{1{\bar 1}}{\cal V}^{-\frac{1}{2}} - \gamma\left(r_2 + \frac{r_2^2\zeta}{r_1}\right)}}\nonumber\\
 & & -\frac{1}{\sqrt{{\cal T}_S(\sigma^S,{\bar\sigma ^S};{\cal G}^a,{\bar{\cal G}^a};\tau,{\bar\tau}) + \mu_3{\cal V}^{\frac{1}{18}}  - \gamma\left(r_2 + \frac{r_2^2\zeta}{r_1}\right)}}\Biggr\}\nonumber\\
 & &\hskip-0.8in +\mu_3\Biggl\{\sqrt{{\cal T}_B(\sigma^B,{\bar\sigma ^B};{\cal G}^a,{\bar{\cal G}^a};\tau,{\bar\tau}) + \mu_3{\cal V}^{\frac{1}{18}} + i\kappa_4^2\mu_7C_{1{\bar 1}}{\cal V}^{-\frac{1}{2}} - \gamma\left(r_2 + \frac{r_2^2\zeta}{r_1}\right)} - \nonumber\\
 & & \sqrt{{\cal T}_S(\sigma^S,{\bar\sigma ^S};{\cal G}^a,{\bar{\cal G}^a};\tau,{\bar\tau}) + \mu_3{\cal V}^{\frac{1}{18}}  - \gamma\left(r_2 + \frac{r_2^2\zeta}{r_1}\right)}\Biggr\}\Biggr]\left({\cal G}^a,{\bar{\cal G}}^a\right)\nonumber\\
& & \hskip-0.8in\times\Biggl[\frac{{\cal V}^{\frac{1}{18}}}{\left({\cal T}_B(\sigma^B,{\bar\sigma ^B};{\cal G}^a,{\bar{\cal G}^a};\tau,{\bar\tau}) + \mu_3{\cal V}^{\frac{1}{18}} + i\kappa_4^2\mu_7C_{1{\bar 1}}{\cal V}^{-\frac{1}{2}} - \gamma\left(r_2 + \frac{r_2^2\zeta}{r_1}\right)\right)^{\frac{3}{2}}}\nonumber\\
 & & + \frac{\mu_3}{\sqrt{{\cal T}_B(\sigma^B,{\bar\sigma ^B};{\cal G}^a,{\bar{\cal G}^a};\tau,{\bar\tau}) + \mu_3{\cal V}^{\frac{1}{18}} + i\kappa_4^2\mu_7C_{1{\bar 1}}{\cal V}^{-\frac{1}{2}} - \gamma\left(r_2 + \frac{r_2^2\zeta}{r_1}\right)}}\Biggr]\sim{\cal V}^{-\frac{109}{36} - g_a}.
 \end{eqnarray}
Using (\ref{eq:ddetKhat}) and (\ref{eq:Fs}), one obtains:
\begin{equation}
\label{eq:F.dlndetKhat}
F^m\partial_m\ ln\ {\rm det}\left(\hat{K}_{i{\bar j}}\right)\sim m_{\frac{3}{2}}.
\end{equation}

\section{First and Second Derivatives of $Z$ with respect to Closed String Moduli $\sigma^\alpha,{\cal G}^a$}
\setcounter{equation}{0}\seceqee

From (\ref{eq:K2}), one sees:
\begin{eqnarray}
\label{eq:Zcoeffs}
& & Z_{z_1z_2}\sim\frac{(\mu_3{\cal V}^{\frac{1}{36}}+\gamma r_2{\cal V}^{-\frac{1}{36}})^2}{\Xi^2}\times\nonumber\\
& & \hskip-0.9in\left(\sqrt{{\cal T}_B(\sigma^B,{\bar\sigma ^B};{\cal G}^a,{\bar{\cal G}^a};\tau,{\bar\tau}) + \mu_3{\cal V}^{\frac{1}{18}} + i\kappa_4^2\mu_7C_{1{\bar 1}}{\cal V}^{-\frac{1}{2}} - \gamma\left(r_2 + \frac{r_2^2\zeta}{r_1}\right)}
   -\sqrt{{\cal T}_S(\sigma^S,{\bar\sigma ^S};{\cal G}^a,{\bar{\cal G}^a};\tau,{\bar\tau}) + \mu_3{\cal V}^{\frac{1}{18}}  - \gamma\left(r_2 + \frac{r_2^2\zeta}{r_1}\right)}\right)^2;\nonumber\\
   & & Z_{{\cal A}_1{\cal A}_1}\sim\frac{\left(i\kappa_4^2\mu_7C_{1{\bar 1}}{\cal V}^{-\frac{1}{4}}\sqrt{{\cal T}_B(\sigma^B,{\bar\sigma ^B};{\cal G}^a,{\bar{\cal G}^a};\tau,{\bar\tau}) + \mu_3{\cal V}^{\frac{1}{18}} + i\kappa_4^2\mu_7C_{1{\bar 1}}{\cal V}^{-\frac{1}{2}} - \gamma\left(r_2 + \frac{r_2^2\zeta}{r_1}\right)}\right)^2}{\Xi^2};\nonumber\\
   & & Z_{z_i{\cal A}_1}\sim\Biggl\{\frac{i\kappa_4^2\mu_7C_{1{\bar 1}}{\cal V}^{-\frac{1}{4}}(\mu_3{\cal V}^{\frac{1}{36}}+\gamma r_2{\cal V}^{-\frac{1}{36}})\sqrt{{\cal T}_B(\sigma^B,{\bar\sigma ^B};{\cal G}^a,{\bar{\cal G}^a};\tau,{\bar\tau}) + \mu_3{\cal V}^{\frac{1}{18}} + i\kappa_4^2\mu_7C_{1{\bar 1}}{\cal V}^{-\frac{1}{2}} - \gamma\left(r_2 + \frac{r_2^2\zeta}{r_1}\right)}}{\Xi^2}\nonumber\\
   & & \hskip-0.9in\left(\sqrt{{\cal T}_B(\sigma^B,{\bar\sigma ^B};{\cal G}^a,{\bar{\cal G}^a};\tau,{\bar\tau}) + \mu_3{\cal V}^{\frac{1}{18}} + i\kappa_4^2\mu_7C_{1{\bar 1}}{\cal V}^{-\frac{1}{2}} - \gamma\left(r_2 + \frac{r_2^2\zeta}{r_1}\right)}
   -\sqrt{{\cal T}_S(\sigma^S,{\bar\sigma ^S};{\cal G}^a,{\bar{\cal G}^a};\tau,{\bar\tau}) + \mu_3{\cal V}^{\frac{1}{18}}  - \gamma\left(r_2 + \frac{r_2^2\zeta}{r_1}\right)}\right)\Biggr\}.\nonumber\\
   & &
\end{eqnarray}
The first and second derivatives of $Z$ are also relevant to the evaluation of the soft SUSY breaking parameters in section {\bf 5}. The same are given as under:
\begin{eqnarray}
\label{eq:dZ_z}
& & \partial_{\sigma^\alpha}Z_{{\cal Z}_i{\cal Z}_i}\sim\nonumber\\
& & \hskip-0.9in\frac{{\cal V}^{\frac{1}{18}}\left(\sqrt{{\cal T}_B(\sigma^B,{\bar\sigma ^B};{\cal G}^a,{\bar{\cal G}^a};\tau,{\bar\tau}) + \mu_3{\cal V}^{\frac{1}{18}} + i\kappa_4^2\mu_7C_{1{\bar 1}}{\cal V}^{-\frac{1}{2}} - \gamma\left(r_2 + \frac{r_2^2\zeta}{r_1}\right)}
   -\sqrt{{\cal T}_S(\sigma^S,{\bar\sigma ^S};{\cal G}^a,{\bar{\cal G}^a};\tau,{\bar\tau}) + \mu_3{\cal V}^{\frac{1}{18}}  - \gamma\left(r_2 + \frac{r_2^2\zeta}{r_1}\right)}\right)}{\Xi^2} \nonumber\\
   & & + \nonumber\\
   & & \hskip-0.9in \frac{{\cal V}^{\frac{1}{18}}\left(\sqrt{{\cal T}_B(\sigma^B,{\bar\sigma ^B};{\cal G}^a,{\bar{\cal G}^a};\tau,{\bar\tau}) + \mu_3{\cal V}^{\frac{1}{18}} + i\kappa_4^2\mu_7C_{1{\bar 1}}{\cal V}^{-\frac{1}{2}} - \gamma\left(r_2 + \frac{r_2^2\zeta}{r_1}\right)}
   -\sqrt{{\cal T}_S(\sigma^S,{\bar\sigma ^S};{\cal G}^a,{\bar{\cal G}^a};\tau,{\bar\tau}) + \mu_3{\cal V}^{\frac{1}{18}}  - \gamma\left(r_2 + \frac{r_2^2\zeta}{r_1}\right)}\right)^2}{\Xi^3}\nonumber\\
   & & \times\sqrt{{\cal T}_\alpha(\sigma^S,{\bar\sigma ^S};{\cal G}^a,{\bar{\cal G}^a};\tau,{\bar\tau}) + \mu_3{\cal V}^{\frac{1}{18}}  - \gamma\left(r_2 + \frac{r_2^2\zeta}{r_1}\right)}\sim{\cal V}^{-\frac{23}{12}};\nonumber\\
   & & \partial_{{\cal G}^a}Z_{{\cal Z}_i{\cal Z}_i}\sim\nonumber\\
   & & \hskip-0.9in\frac{{\cal V}^{\frac{1}{18}}\left(\sqrt{{\cal T}_B(\sigma^B,{\bar\sigma ^B};{\cal G}^a,{\bar{\cal G}^a};\tau,{\bar\tau}) + \mu_3{\cal V}^{\frac{1}{18}} + i\kappa_4^2\mu_7C_{1{\bar 1}}{\cal V}^{-\frac{1}{2}} - \gamma\left(r_2 + \frac{r_2^2\zeta}{r_1}\right)}  -\sqrt{{\cal T}_S(\sigma^S,{\bar\sigma ^S};{\cal G}^a,{\bar{\cal G}^a};\tau,{\bar\tau}) + \mu_3{\cal V}^{\frac{1}{18}}  - \gamma\left(r_2 + \frac{r_2^2\zeta}{r_1}\right)}\right)^2}{\Xi^3}\nonumber\\
   & & \times\Biggl[\sum_\beta n^0_\beta k^a sin(...)+({\cal G}^a,{\bar{\cal G}}^a)\times\nonumber\\
   & & \hskip-0.9in\left(\sqrt{{\cal T}_B(\sigma^B,{\bar\sigma ^B};{\cal G}^a,{\bar{\cal G}^a};\tau,{\bar\tau}) + \mu_3{\cal V}^{\frac{1}{18}} + i\kappa_4^2\mu_7C_{1{\bar 1}}{\cal V}^{-\frac{1}{2}} - \gamma\left(r_2 + \frac{r_2^2\zeta}{r_1}\right)}
   -\sqrt{{\cal T}_S(\sigma^S,{\bar\sigma ^S};{\cal G}^a,{\bar{\cal G}^a};\tau,{\bar\tau}) + \mu_3{\cal V}^{\frac{1}{18}}  - \gamma\left(r_2 + \frac{r_2^2\zeta}{r_1}\right)}\right)\Biggr]\nonumber\\
   & & \sim{\cal V}^{-\frac{7}{54}}.
\end{eqnarray}

\begin{eqnarray}
\label{eq:ddZ_z}
& & \partial_{\sigma^\alpha}{\bar\partial}_{{\bar\sigma}^{\bar\beta}}Z_{{\cal Z}_i{\cal Z}_i}\sim\frac{{\cal V}^{\frac{1}{18}}}{\Xi^2\left({\cal T}_\alpha(\sigma^\alpha,{\bar\sigma ^\alpha};{\cal G}^a,{\bar{\cal G}^a};\tau,{\bar\tau}) + \mu_3{\cal V}^{\frac{1}{18}} + i\kappa_4^2\mu_7C_{1{\bar 1}}{\cal V}^{-\frac{1}{2}} - \gamma\left(r_2 + \frac{r_2^2\zeta}{r_1}\right)\right)}\nonumber\\
& & \hskip-0.9in + \frac{{\cal V}^{\frac{1}{18}}\left(\sqrt{{\cal T}_B(\sigma^B,{\bar\sigma ^B};{\cal G}^a,{\bar{\cal G}^a};\tau,{\bar\tau}) + \mu_3{\cal V}^{\frac{1}{18}} + i\kappa_4^2\mu_7C_{1{\bar 1}}{\cal V}^{-\frac{1}{2}} - \gamma\left(r_2 + \frac{r_2^2\zeta}{r_1}\right)}
   -\sqrt{{\cal T}_S(\sigma^S,{\bar\sigma ^S};{\cal G}^a,{\bar{\cal G}^a};\tau,{\bar\tau}) + \mu_3{\cal V}^{\frac{1}{18}}  - \gamma\left(r_2 + \frac{r_2^2\zeta}{r_1}\right)}\right)}{\Xi^3}\nonumber\\
   & & \times \sqrt{{\cal T}_\alpha(\sigma^\alpha,{\bar\sigma ^\alpha};{\cal G}^a,{\bar{\cal G}^a};\tau,{\bar\tau}) + \mu_3{\cal V}^{\frac{1}{18}} + i\kappa_4^2\mu_7C_{1{\bar 1}}{\cal V}^{-\frac{1}{2}} - \gamma\left(r_2 + \frac{r_2^2\zeta}{r_1}\right)}\sim {\cal V}^{-\frac{26}{27}};\nonumber\\
   & & \partial_{{\cal G}^a}{\bar\partial}_{{\bar\sigma}^\alpha}Z_{{\cal Z}_i{\cal Z}_i}\sim\frac{{\cal V}^{\frac{1}{18}}}{\Xi^2}\nonumber\\
   & & \hskip-0.9in\left(
   \frac{\left({\cal G}^a,{\bar{\cal G}^a}\right)}{\sqrt{{\cal T}_B(\sigma^B,{\bar\sigma ^B};{\cal G}^a,{\bar{\cal G}^a};\tau,{\bar\tau}) + \mu_3{\cal V}^{\frac{1}{18}} + i\kappa_4^2\mu_7C_{1{\bar 1}}{\cal V}^{-\frac{1}{2}} - \gamma\left(r_2 + \frac{r_2^2\zeta}{r_1}\right)}}
   - \frac{\left({\cal G}^a,{\bar{\cal G}^a}\right)}{\sqrt{{\cal T}_S(\sigma^S,{\bar\sigma ^S};{\cal G}^a,{\bar{\cal G}^a};\tau,{\bar\tau}) + \mu_3{\cal V}^{\frac{1}{18}}  - \gamma\left(r_2 + \frac{r_2^2\zeta}{r_1}\right)}}\right)+\nonumber\\
   & &\hskip-0.93in  \frac{{\cal V}^{\frac{1}{18}}\left(\sqrt{{\cal T}_B(\sigma^B,{\bar\sigma ^B};{\cal G}^a,{\bar{\cal G}^a};\tau,{\bar\tau}) + \mu_3{\cal V}^{\frac{1}{18}} + i\kappa_4^2\mu_7C_{1{\bar 1}}{\cal V}^{-\frac{1}{2}} - \gamma\left(r_2 + \frac{r_2^2\zeta}{r_1}\right)}
   -\sqrt{{\cal T}_S(\sigma^S,{\bar\sigma ^S};{\cal G}^a,{\bar{\cal G}^a};\tau,{\bar\tau}) + \mu_3{\cal V}^{\frac{1}{18}}  - \gamma\left(r_2 + \frac{r_2^2\zeta}{r_1}\right)}\right)}{\Xi^3}\nonumber\\
   & & \times {\cal V}^{\frac{5}{6}}k_a\sim{\cal V}^{-\frac{25}{36}}k_a;\nonumber\\
   & & \partial_{{\cal G}^a}{\bar\partial}_{{\bar{\cal G}^a}}Z_{{\cal Z}_i{\cal Z}_i}\sim\nonumber\\
   & & \hskip-0.9in\frac{{\cal V}^{\frac{1}{18}}\left(\sqrt{{\cal T}_B(\sigma^B,{\bar\sigma ^B};{\cal G}^a,{\bar{\cal G}^a};\tau,{\bar\tau}) + \mu_3{\cal V}^{\frac{1}{18}} + i\kappa_4^2\mu_7C_{1{\bar 1}}{\cal V}^{-\frac{1}{2}} - \gamma\left(r_2 + \frac{r_2^2\zeta}{r_1}\right)}
   -\sqrt{{\cal T}_S(\sigma^S,{\bar\sigma ^S};{\cal G}^a,{\bar{\cal G}^a};\tau,{\bar\tau}) + \mu_3{\cal V}^{\frac{1}{18}}  - \gamma\left(r_2 + \frac{r_2^2\zeta}{r_1}\right)}\right)}{\Xi^3}\nonumber\\
   & & \times \left({\cal G}^a,{\cal G}^b\right)k_a{\cal V}^{\frac{5}{6}}+\nonumber\\
   & & \hskip-0.93in  \frac{{\cal V}^{\frac{1}{18}}\left(\sqrt{{\cal T}_B(\sigma^B,{\bar\sigma ^B};{\cal G}^a,{\bar{\cal G}^a};\tau,{\bar\tau}) + \mu_3{\cal V}^{\frac{1}{18}} + i\kappa_4^2\mu_7C_{1{\bar 1}}{\cal V}^{-\frac{1}{2}} - \gamma\left(r_2 + \frac{r_2^2\zeta}{r_1}\right)}
   -\sqrt{{\cal T}_S(\sigma^S,{\bar\sigma ^S};{\cal G}^a,{\bar{\cal G}^a};\tau,{\bar\tau}) + \mu_3{\cal V}^{\frac{1}{18}}  - \gamma\left(r_2 + \frac{r_2^2\zeta}{r_1}\right)}\right)^2}{\Xi^3}\nonumber\\
   & & \times {\cal V}k_ak_b+\nonumber\\
    & & \hskip-0.93in \frac{{\cal V}^{\frac{1}{18}}\left(\sqrt{{\cal T}_B(\sigma^B,{\bar\sigma ^B};{\cal G}^a,{\bar{\cal G}^a};\tau,{\bar\tau}) + \mu_3{\cal V}^{\frac{1}{18}} + i\kappa_4^2\mu_7C_{1{\bar 1}}{\cal V}^{-\frac{1}{2}} - \gamma\left(r_2 + \frac{r_2^2\zeta}{r_1}\right)}
   -\sqrt{{\cal T}_S(\sigma^S,{\bar\sigma ^S};{\cal G}^a,{\bar{\cal G}^a};\tau,{\bar\tau}) + \mu_3{\cal V}^{\frac{1}{18}}  - \gamma\left(r_2 + \frac{r_2^2\zeta}{r_1}\right)}\right)^2}{\Xi^4}\nonumber\\
   & & \times k_ak_b{\cal V}^{\frac{5}{6}}\sim{\cal V}^{-\frac{17}{9}}.\nonumber\\
   & & \hskip -1.3in
   \end{eqnarray}
Similarly,
\begin{eqnarray}
\label{eq:dZ_a}
& & \partial_{\sigma^B}Z_{{\cal A}_1{\cal A}_1}\sim\frac{{\cal V}^{\frac{11}{6}}}{\Xi^2} + \frac{{\cal V}^{\frac{11}{6}}\left({\cal T}_B(\sigma^B,{\bar\sigma ^B};{\cal G}^a,{\bar{\cal G}^a};\tau,{\bar\tau}) + \mu_3{\cal V}^{\frac{1}{18}} + i\kappa_4^2\mu_7C_{1{\bar 1}}{\cal V}^{-\frac{1}{2}} - \gamma\left(r_2 + \frac{r_2^2\zeta}{r_1}\right)\right)^{\frac{3}{2}}}{\Xi^3} \sim{\cal V}^{-\frac{1}{6}};\nonumber\\
& & \partial_{\sigma^S}Z_{{\cal A}_1{\cal A}_1}\sim\frac{{\cal V}^{\frac{11}{6}}\left({\cal T}_B(\sigma^B,{\bar\sigma ^B};{\cal G}^a,{\bar{\cal G}^a};\tau,{\bar\tau}) + \mu_3{\cal V}^{\frac{1}{18}} + i\kappa_4^2\mu_7C_{1{\bar 1}}{\cal V}^{-\frac{1}{2}} - \gamma\left(r_2 + \frac{r_2^2\zeta}{r_1}\right)\right)}{\Xi^3}\nonumber\\
& & \times\sqrt{\left({\cal T}_S(\sigma^B,{\bar\sigma ^B};{\cal G}^a,{\bar{\cal G}^a};\tau,{\bar\tau}) + \mu_3{\cal V}^{\frac{1}{18}} + i\kappa_4^2\mu_7C_{1{\bar 1}}{\cal V}^{-\frac{1}{2}} - \gamma\left(r_2 + \frac{r_2^2\zeta}{r_1}\right)\right)}\sim {\cal V}^{-\frac{13}{12}};
\nonumber\\
& & \partial_{{\cal G}^a}Z_{{\cal A}_1{\cal A}_1}\sim\frac{{\cal V}^{\frac{11}{6}}({\cal G}^a,{\bar{\cal G}^a})}{\Xi^2}+\frac{{\cal V}^{\frac{11}{6}}\left({\cal T}_B(\sigma^B,{\bar\sigma ^B};{\cal G}^a,{\bar{\cal G}^a};\tau,{\bar\tau}) + \mu_3{\cal V}^{\frac{1}{18}} + i\kappa_4^2\mu_7C_{1{\bar 1}}{\cal V}^{-\frac{1}{2}} - \gamma\left(r_2 + \frac{r_2^2\zeta}{r_1}\right)\right)}{\Xi^3}\nonumber\\
& & \times\Biggl[\sum_\beta n^0_\beta k^a sin(...)+({\cal G}^a,{\bar{\cal G}}^a)\times\nonumber\\
   & & \hskip-0.9in\left(\sqrt{{\cal T}_B(\sigma^B,{\bar\sigma ^B};{\cal G}^a,{\bar{\cal G}^a};\tau,{\bar\tau}) + \mu_3{\cal V}^{\frac{1}{18}} + i\kappa_4^2\mu_7C_{1{\bar 1}}{\cal V}^{-\frac{1}{2}} - \gamma\left(r_2 + \frac{r_2^2\zeta}{r_1}\right)}
   -\sqrt{{\cal T}_S(\sigma^S,{\bar\sigma ^S};{\cal G}^a,{\bar{\cal G}^a};\tau,{\bar\tau}) + \mu_3{\cal V}^{\frac{1}{18}}  - \gamma\left(r_2 + \frac{r_2^2\zeta}{r_1}\right)}\right)\Biggr]\nonumber\\
   & & \sim{\cal V}^{-\frac{1}{6}-g_a},
\end{eqnarray}
where ${\cal G}^a\sim{\cal V}^{-g_a},\ 0<g_a<\frac{1}{9}$, and
\begin{eqnarray}
\label{eq:ddZ_a}
& & \partial_{\sigma^B}{\bar\partial}_{{\bar\sigma}^B}Z_{\tilde{\cal A}_1\tilde{\cal A}_1}\sim\frac{{\cal V}^{\frac{11}{6}}\sqrt{{\cal T}_B(\sigma^B,{\bar\sigma ^B};{\cal G}^a,{\bar{\cal G}^a};\tau,{\bar\tau}) + \mu_3{\cal V}^{\frac{1}{18}} + i\kappa_4^2\mu_7C_{1{\bar 1}}{\cal V}^{-\frac{1}{2}} - \gamma\left(r_2 + \frac{r_2^2\zeta}{r_1}\right)}}{\Xi^3}\nonumber\\
& & +\frac{{\cal V}^{\frac{11}{6}}\left({\cal T}_B(\sigma^B,{\bar\sigma ^B};{\cal G}^a,{\bar{\cal G}^a};\tau,{\bar\tau}) + \mu_3{\cal V}^{\frac{1}{18}} + i\kappa_4^2\mu_7C_{1{\bar 1}}{\cal V}^{-\frac{1}{2}} - \gamma\left(r_2 + \frac{r_2^2\zeta}{r_1}\right)\right)^2}{\Xi^4}\nonumber\\
& & \sim {\cal V}^{-\frac{-41}{36}};\nonumber\\
& & \partial_{\sigma^S}{\bar\partial}_{{\bar\sigma}^S}Z_{\tilde{\cal A}_1\tilde{\cal A}_1}\sim\frac{{\cal V}^{\frac{11}{6}}\left({\cal T}_B(\sigma^B,{\bar\sigma ^B};{\cal G}^a,{\bar{\cal G}^a};\tau,{\bar\tau}) + \mu_3{\cal V}^{\frac{1}{18}} + i\kappa_4^2\mu_7C_{1{\bar 1}}{\cal V}^{-\frac{1}{2}} - \gamma\left(r_2 + \frac{r_2^2\zeta}{r_1}\right)\right)}{\Xi^3\sqrt{{\cal T}_S(\sigma^S,{\bar\sigma ^S};{\cal G}^a,{\bar{\cal G}^a};\tau,{\bar\tau}) + \mu_3{\cal V}^{\frac{1}{18}}  - \gamma\left(r_2 + \frac{r_2^2\zeta}{r_1}\right)}}\nonumber\\
& & + \frac{{\cal V}^{\frac{11}{6}}\left({\cal T}_B(\sigma^B,{\bar\sigma ^B};{\cal G}^a,{\bar{\cal G}^a};\tau,{\bar\tau}) + \mu_3{\cal V}^{\frac{1}{18}} + i\kappa_4^2\mu_7C_{1{\bar 1}}{\cal V}^{-\frac{1}{2}} - \gamma\left(r_2 + \frac{r_2^2\zeta}{r_1}\right)\right)}{\Xi^3\sqrt{{\cal T}_S(\sigma^S,{\bar\sigma ^S};{\cal G}^a,{\bar{\cal G}^a};\tau,{\bar\tau}) + \mu_3{\cal V}^{\frac{1}{18}}  - \gamma\left(r_2 + \frac{r_2^2\zeta}{r_1}\right)}}\nonumber\\
& & \hskip-0.93in + \frac{{\cal V}^{\frac{11}{6}}\left({\cal T}_B(\sigma^B,{\bar\sigma ^B};{\cal G}^a,{\bar{\cal G}^a};\tau,{\bar\tau}) + \mu_3{\cal V}^{\frac{1}{18}} + i\kappa_4^2\mu_7C_{1{\bar 1}}{\cal V}^{-\frac{1}{2}} - \gamma\left(r_2 + \frac{r_2^2\zeta}{r_1}\right)\right)\left({\cal T}_S(\sigma^S,{\bar\sigma ^S};{\cal G}^a,{\bar{\cal G}^a};\tau,{\bar\tau}) + \mu_3{\cal V}^{\frac{1}{18}}  - \gamma\left(r_2 + \frac{r_2^2\zeta}{r_1}\right)\right)}{\Xi^4}\nonumber\\
& & \sim {\cal V}^{-\frac{41}{36}};\nonumber\\
& & \partial_{\sigma^B}{\bar\partial}_{{\bar\sigma}^S}Z_{\tilde{\cal A}_1\tilde{\cal A}_1}\sim\frac{{\cal V}^{\frac{11}{6}}\sqrt{{\cal T}_S(\sigma^S,{\bar\sigma ^S};{\cal G}^a,{\bar{\cal G}^a};\tau,{\bar\tau}) + \mu_3{\cal V}^{\frac{1}{18}}  - \gamma\left(r_2 + \frac{r_2^2\zeta}{r_1}\right)}}{\Xi^3}\nonumber\\
& & \hskip-0.93in+\frac{{\cal V}^{\frac{11}{6}}\left({\cal T}_B(\sigma^B,{\bar\sigma ^B};{\cal G}^a,{\bar{\cal G}^a};\tau,{\bar\tau}) + \mu_3{\cal V}^{\frac{1}{18}} + i\kappa_4^2\mu_7C_{1{\bar 1}}{\cal V}^{-\frac{1}{2}} - \gamma\left(r_2 + \frac{r_2^2\zeta}{r_1}\right)\right)^{\frac{3}{2}}\sqrt{{\cal T}_S(\sigma^S,{\bar\sigma ^S};{\cal G}^a,{\bar{\cal G}^a};\tau,{\bar\tau}) + \mu_3{\cal V}^{\frac{1}{18}}  - \gamma\left(r_2 + \frac{r_2^2\zeta}{r_1}\right)}}{\Xi^4}\nonumber\\
& & \sim {\cal V}^{-\frac{41}{36}};\nonumber\\
& & \partial_{\sigma^B}{\bar\partial}_{{\bar{\cal G}^a}}Z_{\tilde{\cal A}_1\tilde{\cal A}_1}\sim\frac{{\cal V}^{\frac{11}{6}}\left({\cal G}^a,{\bar{\cal G}^a}\right)\sqrt{{\cal T}_B(\sigma^B,{\bar\sigma ^B};{\cal G}^a,{\bar{\cal G}^a};\tau,{\bar\tau}) + \mu_3{\cal V}^{\frac{1}{18}} + i\kappa_4^2\mu_7C_{1{\bar 1}}{\cal V}^{-\frac{1}{2}} - \gamma\left(r_2 + \frac{r_2^2\zeta}{r_1}\right)}}{\Xi^3}\nonumber\\
& & + \frac{{\cal V}^{\frac{11}{6}+\frac{5}{6}}k_a}{\Xi^3}+\frac{{\cal V}^{\frac{11}{6}+\frac{5}{6}}k_a\left({\cal T}_B(\sigma^B,{\bar\sigma ^B};{\cal G}^a,{\bar{\cal G}^a};\tau,{\bar\tau}) + \mu_3{\cal V}^{\frac{1}{18}} + i\kappa_4^2\mu_7C_{1{\bar 1}}{\cal V}^{-\frac{1}{2}} - \gamma\left(r_2 + \frac{r_2^2\zeta}{r_1}\right)\right)^{\frac{3}{2}}}{\Xi^4}\sim{\cal V}^{-\frac{13}{6}};\nonumber\\
& & {\rm Similarly,}\ \partial_{\sigma^S}{\bar\partial}_{{\bar{\cal G}^a}}Z_{\tilde{\cal A}_1\tilde{\cal A}_1}\sim{\cal V}^{-\frac{5}{4}};
\nonumber\\
& & \partial_{{\cal G}^a}{\bar\partial}_{{\bar{\cal G}^b}}Z_{\tilde{\cal A}_1\tilde{\cal A}_1}\sim\frac{{\cal V}^{\frac{11}{6}}\delta_{ab}}{\Xi^2}
+ \frac{{\cal V}^{\frac{11}{6}+\frac{5}{6}}\left({\cal G}^a,{\bar{\cal G}^a}\right)k_b}{\Xi^3}\nonumber\\
& & + \frac{{\cal V}^{\frac{11}{6}+1}\left({\cal T}_B(\sigma^B,{\bar\sigma ^B};{\cal G}^a,{\bar{\cal G}^a};\tau,{\bar\tau}) + \mu_3{\cal V}^{\frac{1}{18}} + i\kappa_4^2\mu_7C_{1{\bar 1}}{\cal V}^{-\frac{1}{2}} - \gamma\left(r_2 + \frac{r_2^2\zeta}{r_1}\right)\right)}{\Xi^3}
\nonumber\\
& & + \frac{{\cal V}^{\frac{11}{6}+\frac{10}{6}}k_ak_b\left({\cal T}_B(\sigma^B,{\bar\sigma ^B};{\cal G}^a,{\bar{\cal G}^a};\tau,{\bar\tau}) + \mu_3{\cal V}^{\frac{1}{18}} + i\kappa_4^2\mu_7C_{1{\bar 1}}{\cal V}^{-\frac{1}{2}} - \gamma\left(r_2 + \frac{r_2^2\zeta}{r_1}\right)\right)}{\Xi^4}
\sim{\cal V}^{-\frac{4}{9}}.
\end{eqnarray}


\begin{thebibliography}{99}

\bibitem{fluxesGiddindsetal}
S.~B.~Giddings, S.~Kachru and J.~Polchinski, {\it Hierarchies from fluxes in string compactifications}, Phys. Rev. D {\bf 66}, 106006 (2002),
  [arXiv:hep-th/0105097].

\bibitem{Granafluxreview}
 M.~Grana, {\it Flux compactifications in string theory: A comprehensive review}, Phys. Rept.  {\bf 423}, 91 (2006), [arXiv:hep-th/0509003].

\bibitem{KKLT} S.~Kachru, R.~Kallosh, A.~Linde and S.~P.~Trivedi,{\it De Sitter vacua in string theory}, Phys. Rev. D {\bf 68} (2003) 046005
  [arXiv:hep-th/0301240].

\bibitem{otherupliftings} C.~P.~Burgess, R.~Kallosh and F.~Quevedo, {\it de Sitter string vacua from supersymmetric D-terms}, JHEP 10 (2003) 056, [arXiv:hep-th/0309187].
%
A.~Saltman and E.~Silverstein, {\it The scaling of the no-scale potential and de Sitter model building}, JHEP 11 (2004) 066, [arXiv:hep-th/0402135].


\bibitem{Balaetal2} V.~Balasubramanian, P.~Berglund, J.~P.~Conlon and F.~Quevedo,
{\it Systematics of moduli stabilisation in Calabi-Yau flux
compactifications}, JHEP {\bf 0503}, (2005) 007 [arXiv:hep-th/0502058].

\bibitem{dSetal} A.~Misra and P.~Shukla, {\it Moduli Stabilization, Large-Volume dS Minimum Without anti-D3-Branes, (Non-)Supersymmetric Black Hole Attractors and Two-Parameter Swiss Cheese Calabi-Yau's}, Nuclear Physics B {\bf 799} (2008) 165-198, [arXiv:0707.0105].






\bibitem{mirrormediation} J.~P.~Conlon, {\it Mirror Mediation},JHEP 0803:025, 2008, [arXiv:0710.0873].

\bibitem{conloncal} J.~P.~Conlon, S.~S.~Abdussalam, F.~Quevedo and  K.~Suruliz, {\it Soft SUSY Breaking Terms for Chiral Matter in IIB String Compactifications}, JHEP 0701: 032, 2007 , [arXiv:hep-th/0610129].

\bibitem{FCNC} D.~Choudhury, F.~Eberlein, A.~K\"{o}nig, j.~Louis and S.~Pokorski, {\it Constraints on non-universal soft terms
from flavor changing neutral currents}, Physics Letters B {\bf 342} (1995) 180-188.

\bibitem{susyinitials}
V.~S.~Kaplunovsky and J.~Louis, {\it Model independent analysis of soft terms in effective supergravity and in string theory}, Phys. Lett.B {\bf 306} (1993) 269, [arXiv:hep-th/9303040]; A.~Brignole, L.~E.~Ib\'a\~nez and C.~Munoz, {\it Towards a theory of soft terms for the supersymmetric Standard Model}, Nucl. Phys. B {\bf 422} (1994) 125, [arXiv:hep-ph/9308271]; C. Bachas, {\it A Way to break supersymmetry}, [arXiv:hep-th/9503030].




\bibitem{Quevedosusy2} S.~A.~Abel, B.~C.~Allanach, F.~Quevedo, L.~Ib\'a\~nez and M.~Klein, {\it Soft SUSY breaking, dilaton domination and intermediate scale string models}, JHEP {\bf 0012}, 026 (2000), [arXiv:hep-ph/0005260].

\bibitem{beckerhack} K.~Becker, M.~Becker, M.~Haack and J.~Louis, {\it Supersymmetry breaking and alpha'-corrections to flux induced potentials}, JHEP {\bf 0206} (2002) 060, [arXiv:hep-th/0204254].

\bibitem{towardsrealvacua} Joseph P.~Conlon, Anshuman Maharana and Fernando Quevedo, {\it Towards Realistic String Vacua From Branes At Singularities}, [arXiv:0810.5660].



\bibitem{susybreakingKKLT}
K.~Choi, A.~Falkowski, H.~P.~Nilles and M.~Olechowski, {\it Soft supersymmetry breaking in KKLT flux compactification}, Nucl.\ Phys.\ B {\bf 718}, 113 (2005), [arXiv:hep-th/0503216];

\bibitem{susykklt2} A.~Falkowski, O.~Lebedev and Y.~Mambrini, {\it SUSY phenomenology of KKLT flux compactifications}, JHEP {\bf 0511}, 034 (2005), [arXiv:hep-ph/0507110].


\bibitem{SMreview}
For a review see, E. Kiritsis, {\it D-branes in standard model building, gravity and cosmology}, Fortsch. Phys. 52 (2004) 200 [arXiv:hep-th/0310001].

\bibitem{SM2}
  D.~Berenstein, V.~Jejjala and R.~G.~Leigh, {\it The standard model on a D-brane}, Phys.\ Rev.\ Lett.\  {\bf 88}, 071602 (2002), [arXiv:hep-ph/0105042];  D. L¨ust, {\it Intersecting brane worlds: A path to the standard model}, Class. Quant.Grav. 21 (2004) S1399, [arXiv:hep-th/0401156].





\bibitem{QuevedoMSSM}
G.~Aldazabal, L.~E.~Ibanez and F.~Quevedo, {\it A D-brane alternative to the MSSM}, JHEP {\bf 0002} (2000) 015, [arXiv:hep-ph/0001083]; B.~C.~Allanach, F.~Quevedo and K.~Suruliz, {\it Low-energy supersymmetry breaking from string flux compactifications: Benchmark scenarios}, JHEP {\bf 0604}, 040 (2006), [arXiv:hep-ph/0512081].

\bibitem{SM3}
  H.~Verlinde and M.~Wijnholt, {\it Building the standard model on a D3-brane}, [arXiv:hep-th/0508089]; J.~F.~G.~Cascales, M.~P.~Garcia del Moral, F.~Quevedo and A.~M.~Uranga, {\it Realistic D-brane models on warped throats: Fluxes, hierarchies and  moduli stabilization}, JHEP {\bf 0402}, 031 (2004), [arXiv:hep-th/0312051].



\bibitem{granagrimm} M.~Grana, T.~W.~Grimm, H.~Jockers and J.~Louis, {\it Soft supersymmetry breaking in Calabi-Yau orientifolds with D-branes and fluxes}, Nucl. Phys. B 690, 21 (2004), [arXiv:hep-th/0312232].

\bibitem{ibanezuranga} P.~G.~ Camara, L.E. Ibanez and A.~Uranga, {\it Flux-induced SUSY-breaking soft terms}, Nucl.Phys.B689 (2004) 195. [arXiv:hep-th/0311241].



\bibitem{Ibanez} L.~E.~Ib\'a\~nez, {\it Strings, unification and dilaton/moduli induced SUSY-breaking}, [arXiv:hep-th/9505098]; B.~C.~Allanach, A.~Brignole and L.~E.~Ib\'a\~nez, {\it Phenomenology of a fluxed MSSM}, JHEP {\bf 0505}, 030 (2005), [arXiv:hep-ph/0502151].

\bibitem{Lustetal}
  D.~Lust, S.~Reffert and S.~Stieberger, {\it Flux-induced soft supersymmetry breaking in chiral type IIb  orientifolds with D3/D7-branes}, Nucl.\ Phys.\ B {\bf 706}, 3 (2005), [arXiv:hep-th/0406092].

\bibitem{balaetal2}
 V.~Balasubramanian and P.~Berglund, {\it Stringy corrections to Kaehler potentials, SUSY breaking, and the cosmological constant problem},JHEP {\bf 0411}, 085 (2004), [arXiv:hep-th/0408054].

\bibitem{conlonLVSsusy}
  J.~P.~Conlon, F.~Quevedo and K.~Suruliz, {\it Large-volume flux compactifications: Moduli spectrum and D3/D7 soft
  supersymmetry breaking}, JHEP {\bf 0508} (2005) 007, [arXiv:hep-th/0505076];


\bibitem{tension1} J.~P.~Conlon, R.~Kallosh, A.~Linde and F. Quevedo, {\it Volume Modulus Inflation and the Gravitino Mass
Problem}, [arXiv:0806.0809[hep-th]].

\bibitem{quevedojan09} S.~Krippendorf and F.~Quevedo, {\it Metastable SUSY Breaking, de Sitter Moduli Stabilisation and K\"{a}hler Moduli Inflation}, [arXiv:0901.0683].


\bibitem{ibanezfont} A.~Font and L.~E.~Ib\'a\~nez, {\it SUSY-breaking soft terms in a MSSM magnetized D7-brane model}, JHEP {\bf 0503} (2005) 040,
   [arXiv:hep-th/0412150].

\bibitem{jockersetal} H.~Jockers and J.~Louis, {\it The effective action of D7-branes in N = 1 Calabi-Yau orientifolds}, Nucl.\ Phys.\ B {\bf 705} (2005) 167, [arXiv:hep-th/0409098].

\bibitem{berghack} M.~Berg, M.~Haack, E.~Pajer, {\it Jumping Through Loops: On Soft Terms from Large Volume Compactifications}, JHEP 0709 : 031, 2007, [arXiv:0704.0737].

\bibitem{kaehlerinflation} J.~P.~Conlon and F.~Quevedo,
  {\it Kaehler moduli inflation},
  JHEP {\bf 0601}, 146 (2006)
  [arXiv:hep-th/0509012].



\bibitem{abdussalam2} S.~S.~AbdusSalam, B.~C.~Allanach, F.~Quevedo, {\it Fitting the Phenomenological MSSM}, [arXiv:0904.2548].


\bibitem{LHCpheno} G.~L.~Kane, P.~Kumar and J.~Shao, {\it LHC String Phenomenology}, [arXiv:hep-ph/0610038]; L.~Aparicio, D.~G.~Cerdeno and L.~E.~Ibanez, {\it Modulus-dominated SUSY-breaking soft terms in F-theory and their test at LHC}, JHEP 07 (2008) 099, [arXiv:0805.2943].

\bibitem{KKLMMT} S.~ Kachru, R.~ Kallosh, A.~ Linde, J.~ Maldacena, L.~ McAllister and S. P.~ Trivedi, {\it Towards inflation in string theory},
 JCAP 0310, 013 (2003), [arXiv:hep-th/0308055]; R.~Kallosh, A.~Linde, {\it Testing String Theory with CMB},
 JCAP 0704, 017 (2007), [arXiv:hep-th/0704.0647].

\bibitem{LargeVcons}J.~P.~Conlon and F.~Quevedo, {\it Astrophysical and Cosmological Implications of Large Volume String
Compactifications}, JCAP {\bf 0708}, 019 (2007)[arXiv:0705.3460 [hep-ph]].





\bibitem{largefNL_r}A.~Misra and P.~Shukla,
{\it `Finite' Non-Gaussianities and Tensor-Scalar Ratio in Large Volume
  Swiss-Cheese Compactifications},
  Nucl.\ Phys.\  B {\bf 810}, 174 (2009)
  [arXiv:0807.0996 [hep-th]].

\bibitem{quevedoftheorysusy} R.~Blumenhagen, J.~P.~ Conlon, S.~Krippendorf and F.~Quevedo, {\it Supersymmetry Breaking in Local String/F-Theory Models}, [arXiv:0906.3297].


\bibitem{axionicswisscheese}A.~Misra and P.~Shukla,
{\it Large Volume Axionic Swiss-Cheese Inflation},
  Nucl.\ Phys.\  B {\bf 800}, 384 (2008)
  [arXiv:0712.1260 [hep-th]].

\bibitem{Kachruetal}S.~Kachru, S.~H.~Katz, A.~E.~Lawrence and J.~McGreevy,
{\it Open string instantons and superpotentials},
  Phys.\ Rev.\  D {\bf 62}, 026001 (2000)
  [arXiv:hep-th/9912151].

\bibitem{Candelasetal} P.~Candelas, A.~Font, S.~H.~Katz and D.~R.~Morrison,
{\it Mirror symmetry for two parameter models. 2},
Nucl.\ Phys.\  B {\bf 429}, (1994) 626  [arXiv:hep-th/9403187].

\bibitem{BBHL} K.~Becker, M.~Becker, M.~Haack and J.~Louis,
{\it Supersymmetry breaking and alpha'-corrections to flux induced
 potentials}, JHEP {\bf 0206}, 060 (2002) [arXiv:hep-th/0204254].


\bibitem{Grimm}T.~W.~Grimm,
{\it Non-Perturbative Corrections and Modularity in N=1 Type IIB
  Compactifications},
  JHEP {\bf 0710}, 004 (2007)
  [arXiv:0705.3253 [hep-th]].

\bibitem{loops} M.~Cicoli, J.~P.~Conlon and F.~Quevedo,
{\it Systematics of String Loop Corrections in Type IIB Calabi-Yau Flux
Compactifications}, [arXiv:hep-th/0708.1873] .

\bibitem{Curio+Spillner}G.~Curio and V.~Spillner,
{\it On the modified KKLT procedure: A case study for the P(11169)(18) model},
  Int.\ J.\ Mod.\ Phys.\  A {\bf 22}, 3463 (2007)
  [arXiv:hep-th/0606047].



\bibitem{Klemm_GV}M.~x.~Huang, A.~Klemm and S.~Quackenbush,
  {\it Topological String Theory on Compact Calabi-Yau: Modularity and Boundary
  Conditions},  Lect.\ Notes Phys.\  {\bf 757}, 45 (2009)
  [arXiv:hep-th/0612125].


\bibitem{Yokoyamafnl}S.~Yokoyama, T.~Suyama and T.~Tanaka, {\it Primordial Non-Gaussianity in Multi-Scalar Slow-Roll Inflation} [arXiv:astro-ph/0705.3178].
\bibitem{Yokoyama}  S.~Yokoyama, T.~Suyama and T.~Tanaka,{\it Primordial Non-Gaussianity in Multi-Scalar Inflation}, Physical Review D 77 (2008) 083511 [arXiv:astro-ph/0711.2920].







\bibitem{Jockers_thesis}H.~Jockers,
 {\it The effective action of D-branes in Calabi-Yau orientifold
  compactifications},
  Fortsch.\ Phys.\  {\bf 53}, 1087 (2005)
  [arXiv:hep-th/0507042].

\bibitem{V_D7_fl}M.~Haack, D.~Krefl, D.~Lust, A.~Van Proeyen and M.~Zagermann,
{\it Gaugino condensates and D-terms from D7-branes},
  JHEP {\bf 0701}, 078 (2007)
  [arXiv:hep-th/0609211].


\bibitem{Kimura}T.~Kimura, {\it Kaehler potentials on toric varieties}, arXiv:hep-th/0305072.


\bibitem{Umemura}H.~Umemura, {\it Resolution of algebraic equations by theta constants}, in D. Mumford, Tata Lectures
on Theta II, Progress in Math. Vol. 43, Birkh\"{a}user, 1984, 261-272.

\bibitem{Zhivkov}A.~Zhivkov,{\it Resolution of degree$\leq$6 algebraic
equations by genus two theta constants}, Journal of Geometry and Symmetry in Physics, {\bf 11}, 77 (2008).



\bibitem{Quevedo_etal_explicit_Y}J.~P.~Conlon, A.~Maharana and F.~Quevedo,
{\it Wave Functions and Yukawa Couplings in Local String Compactifications},
  JHEP {\bf 0809}, 104 (2008)
  [arXiv:0807.0789 [hep-th]].

\bibitem{Bergshoeff_etal} E.~Bergshoeff, J.~Hartong, T.~Ortin and D.~Roest,
{\it IIb Seven-Branes Revisited},
  J.\ Phys.\ Conf.\ Ser.\  {\bf 66}, 012054 (2007).


\bibitem{Ganor1_2}O.~J.~Ganor,
 {\it A note on zeroes of superpotentials in F-theory},
  Nucl.\ Phys.\  B {\bf 499}, 55 (1997)
  [arXiv:hep-th/9612077]; {\it On Zeroes Of Superpotentials In F-Theory}, Nucl.\ Phys.\ B (Proc. Suppl ) {\bf 67}, 25 (1998).

\bibitem{Husemoeller} {\it Elliptic Curves}, D.Husem\"{o}ller, Springer-Verlag New York (2004).

\bibitem{mirage} O.~Loaiza-Brito, J.~Martin, H.~P.~Nilles and M.~Ratz, ``log(M(Pl/m(3/2))),'' AIP Conf.\ Proc.\  {\bf 805}, 198 (2006) [arXiv:hep-th/0509158].

\bibitem{Maldaetal_Wnp_pref}D.~Baumann, A.~Dymarsky, I.~R.~Klebanov, J.~M.~Maldacena, L.~P.~McAllister and A.~Murugan,
{\it On D3-brane potentials in compactifications with fluxes and wrapped D-branes},  JHEP {\bf 0611}, 031 (2006) [arXiv:hep-th/0607050].




\bibitem{Bagger_et_al}J.~A.~Bagger, T.~Moroi and E.~Poppitz,
 {\it Anomaly mediation in supergravity theories},
  JHEP {\bf 0004}, 009 (2000)
  [arXiv:hep-th/9911029].

\bibitem{Alwis} S.~P.~de Alwis,
{\it On Anomaly Mediated SUSY Breaking},
  Phys.\ Rev.\  D {\bf 77}, 105020 (2008)
  [arXiv:0801.0578 [hep-th]].


\bibitem{Quevedo+Conlon_supp_gaugino_mass}J.~P.~Conlon and F.~Quevedo,
{\it Gaugino and scalar masses in the landscape},
  JHEP {\bf 0606}, 029 (2006)
  [arXiv:hep-th/0605141].

\bibitem{Green_Weigand}D.~Green and T.~Weigand,
{\it Retrofitting and the mu Problem},
  arXiv:0906.0595 [hep-th].


\bibitem{mu1} A.~Brignole,L.~E.~Ib\'a\~nez and C.~M\~{u}noz, {\it Soft Supersymmetry-Breaking Terms from Supergravity and Superstring Models},
[arXiv:hep-th/9707209].



\end{thebibliography}
   \end{document}